\documentclass[onecolumn,pra,aps]{revtex4-1}
\usepackage{bm}
\usepackage{mathrsfs}
\usepackage{natbib}
\usepackage{amsmath}
\usepackage{amssymb}
\usepackage{graphicx}
\usepackage{amsfonts}
\usepackage{amsthm}
\usepackage{color}
\usepackage{dcolumn}
\usepackage{txfonts}
\newtheorem{lemma}{Lemma}
\newtheorem{corollary}{Corollary}
\newtheorem{theorem}{Theorem}

\begin{document}

\title{Quantum computing by optical control of electron spins}

\author{Ren-Bao Liu}
\thanks{Email: rbliu@cuhk.edu.hk}
\affiliation{Department of Physics, The Chinese University of Hong Kong, Hong Kong, China}
\author{Wang Yao}
\thanks{Email: wangyao@hkucc.hku.hk}
\affiliation{Department of Physics, The University of Hong Kong, Hong Kong, China}
\author{L. J. Sham}
\thanks{Corresponding author. Email: lsham@ucsd.edu}
\affiliation{Department of Physics, University of California San Diego, La Jolla, California 92093-0319}

\begin{abstract}
We review the progress and main challenges in implementing
large-scale quantum computing by optical control of electron
spins in quantum dots (QDs). Relevant systems include
self-assembled QDs of III-V or II-VI compound semiconductors
(such as InGaAs and CdSe), monolayer fluctuation QDs in compound
semiconductor quantum wells, and impurity centers in solids
such as P-donors in silicon and nitrogen-vacancy centers in
diamond. The decoherence of the electron spin qubits is
discussed and various schemes for countering the decoherence
problem are reviewed. We put forward designs of local nodes
consisting of a few qubits which can be individually addressed
and controlled. Remotely separated local nodes are connected by
photonic structures (microcavities and waveguides) to form
a large-scale distributed quantum system or a quantum network.
The operation of the quantum network consists of optical control
of a single electron spin, coupling of two spins in a local
nodes, optically controlled quantum interfacing between
stationary spin qubits in QDs and flying photon qubits in
waveguides, rapid initialization of spin qubits, and
qubit-specific single-shot non-demolition quantum measurement.
The rapid qubit initialization may be realized by selectively
enhancing certain entropy dumping channels via phonon or photon
baths. The single-shot quantum measurement may be in-situ
implemented through the integrated photonic network. The
relevance of quantum non-demolition measurement to large-scale
quantum computation is discussed. To illustrate the feasibility
and demand, the resources are estimated for the benchmark
problem of factorizing 15 with Shor's algorithm.
\end{abstract}

\maketitle

\tableofcontents

\section{Introduction \label{sec-intro}}

Based on the quantum parallelism rooted in
the superposition principle of quantum mechanics, quantum computers
are expected to dramatically outperform their classical counterpart,
particularly with exponential speedup in solving some hard problems such as factoring~\cite{Shor94}.
Toward the ambitious realization of
practical quantum computation such as factoring a million-bit number,
enormous efforts are still to be put on both the design and invention of
software (quantum algorithms) and the hardware development (physical implementation).
Here we are mostly interested in the latter part.
In principle, all quantum systems (which arguably amount to all physical systems)
could be considered for the physical realization. But certain qualifications
(such as the DiVincenzo criteria~\cite{DiVincenzo_Criteria_7}) are to be fulfilled for them to be brought
into consideration. Still there is a vast range of systems in the candidate pool,
including nuclear spins in liquids~\cite{Chuang_factoring15,Gershenfeld:1997Science,Chuang:1998Nature,Jones:1998JCP,Cory:1998PhysicaD},
trapped ions or atoms~\cite{Cirac:1995IonTrap,Monroe:1995PRL,Sorensen:1999Nature,Cirac:2000Nature,Folman:2000PRL,Kielpinski:2002Nature,
Gulde:2003Nature,Leibfried:2003Nature,Schmidt-Kler:2003Nature,Leibfried:2005Nature,Benhelm:2008NaturePhysics},
atoms in optical lattices~\cite{Brennen:1999PRL,Jaksch:1999PRL,Jaksch:2000PRL,Mandel:2003Nature,Anderlini:2007Nature,Trotzky:2008Science},
photons~\cite{Knill:2001LOQC,Obrien:2003Nature,Raussendorf:2003PRA,Nielsen:2004Cluster,Walther:2005Nature,Browne:2005PRL,
Lim:2005PRL_repeat_until_success,Prevedel:2007Nature,Lu:2007PRLShor,Varnava:2008PRL},
superconducting circuits~\cite{Shnirman:1997PRL,Averin:1999SSC,Makhlin:1999Nature,Nakamura:1999Nature,
Makhlin:2001RMP,Yu:2002Science,Martinis:2002PRL,Chiorescu:2003Science,
Yamamoto:2003Nature,Strauch:2003PRL,Blais:2003PRL,Grajcar:2006PRL,
Wallraff04Nature,niskanen_quantum_2007,Majer07Nature,Neeley09Science,Hofheinz09Nature,dicarlo_demonstration_2009,
You:2005PhysTod,Clarke:2008NatureReview,Martinis09Rev},
electrons suspended over liquid helium surfaces~\cite{Platzman:1999Science,Lea:2000FdP,Dahm:2002JLTP,Lyon:2006PRA},
molecular magnets~\cite{Leuenberger:2001Nature}, nuclear spins in solids~\cite{Kane_QC_nuclei,Ladd:2002PRL},
electron spins in semiconductor quantum dots (QDs)~\cite{Loss_QDspinQC,Imamoglu_CQED_Spin,Burkard:1999PRB,Vrijen:2000PRA,Loss_SpinQC_recipes,
Calarco_OpticalQC,Pazy_QDQC,Nazir_PRL2004_SpinCouplebyXton,Lovett_PRB05_SpinCouplebyXton,Chen_Raman,
ORKKY,Reinecke_indirect,ORKKY_vertical,Yamamoto_OpticalQC,Economou07PRLGeoControl,
Ramsay08PRLSpinControl,Yamamoto_08NatureSpinControl},
hole spins in QDs~\cite{Chamarro_holespin,Loss_holespin},
electron spins in impurity centers in semiconductors such as
phosphorus donors in silicon~\cite{Silicon_P_QC_review,Schofield:2003,Bradbury:2006,Huebl:2006,Stegner:2006}
and nitrogen-vacancy (NV) centers in
diamonds~\cite{Wrachtrup:2001,Jelezko:2002,Kennedy:2003,Jelezko2004_PRLNV,Jelezko2004_PRL93,Santori:2006,
Epstein05NPhysNVN,Gaebel2006_NaturephysicsNV,Hanson2006_PRLNV,Hanson:2006RT,Childress2006_ScienceNV,
Dutt2007_ScienceNV,Hanson2008_ScienceNV,Wrachtrup:2006},
and non-Abelian anyon excitations in quantum matters with
topological orders~\cite{Kitaev:2003AP,Freedman:2003BAMS,DasSarma:2005PRL,Dolev:2008Nature,Radu:2008Science}.
Here we concentrate on solid-state systems, and in particular
on electron spins in semiconductors, while nuclear spins therein are also considered either as an
adverse noise source or as a beneficial information storage~\cite{Lukin_NucleiMemory,Witzel2007_PRB}.
Electrical, magnetic and/or optical means may be employed to access and/or manipulate
the spins. In this review, we discuss the optical operations
which may be applied to electron spins in III-V compound semiconductors
where the direct band-gaps facilitate controllable optical transitions.

We shall not give an overall review of different
schemes under current investigation for quantum computation. A
comprehensive review of progresses and challenges in research of
different systems may be found, e.g., in the Quantum Computation
Roadmap published by Quantum Institute, LANL. It, however, would
be useful to have a perspective of the position and connections of the
systems under review in the global picture of quantum computation research. As compared with their ``soft''
counterparts such as trapped atoms or ions, cavity-trapped or
flying photons, electrons floating on liquid helium surfaces, and
nuclear spins in liquids, the ``hard'' solid-state systems as candidates
for quantum computers have the advantages of stability and integratability, but
have the disadvantages of relatively short coherence time due to interactions with complex
environments in solids. The solid-state systems under current
investigation include superconducting circuits, nuclear spins in
solids, and electron spins in semiconductors. While the qubits in
superconducting circuits are made of excitations with macroscopic coherence
in superconductors under designed confinement, nuclear or electron
spins are natural qubit carriers since the information can be encoded in an intrinsic degree of
freedom of elementary particles and thus are very stable. For
example, while a superconducting qubit may be lost during the
measurement and control processes, a spin qubit would always
exist unless the hosting particle like a nucleus or an electron
disappears (by decay, ionization, thermal activation, etc.). Also,
a spin does not feel an electrical field directly, which makes
spins less vulnerable than superconducting qubits to charge or
current noises from environments or operating devices. The
decoherence of spin qubits may be caused by coupling to other
environmental spins, local magnetic field fluctuations, or phonon
scattering via spin-orbital interaction, all of which are usually
rather weak in semiconductors. So the coherence time of spins are
usually very long at low temperature and under a moderately strong magnetic field,
varying from microseconds to milliseconds for electron
spins (excluding the inhomogeneous broadening effect)~\cite{Marcus_T2,Greilich_lock,Lyon_echo,ESR_Si29_Abe},
and longer than seconds for nuclear spins~\cite{Gammon_t2star,Takahashi:2008PRL,Ramon_ZamboniEffect,Reilly:2008Science}.
The weak coupling of spins to environments, of course, has also
detrimental effects -- the control, initialization and measurement
of spins are all challenging tasks. In this regard, electron spins
are more tractable than nuclear spins. Nuclear spins have longer
coherence memory time but slow operating rate and low detection
efficiency. Electron spins are relatively more controllable but
less resilient to decoherence. Schemes have been pursued to
combine the advantages of the two kinds of systems by using
nuclei as storage~\cite{Lukin_NucleiMemory,Espin_HF_dipole_Hu,Witzel2007_PRB} and
electrons as operating units and interfaces~\cite{Jelezko:2004}.
The coupling between nuclear and electron spins is the hyperfine
interaction. The hyperfine interaction is a main mechanism
causing electron spin decoherence, but
it can also be utilized to realize coupled qubit systems.

Electron spin qubits
can be formed in various structures, such as doped electrons in
QDs and impurities in solids. The fabrication of
such systems with designed patterns and structures is
possible because of the advances in modern semiconductor technologies
and nano-technologies -- compatibility with the modern semiconductor industry
is an extra advantage of using electron spins
as qubits in quantum computation. The direct pump, control and
probe of electron spins may be done by electron spin resonance
techniques with microwave pulses~\cite{Jelezko:2002,Jelezko:2004,Lyon_echo,ESR_Si29_Abe}.
For faster operation clock as desired, the electron spin states
may be converted to other degrees of freedom through quantum
interfacing (the same as one does from nuclear spin states to electron
spin states). Then electron spins may be accessed indirectly by
control of agents such as excitons in semiconductors and photons
generated by recombination of
excitons~\cite{Imamoglu_CQED_Spin,Calarco_OpticalQC,Pazy_QDQC,Nazir_PRL2004_SpinCouplebyXton,
Lovett_PRB05_SpinCouplebyXton,Chen_Raman,ORKKY,Reinecke_indirect,
ORKKY_vertical,Yamamoto_OpticalQC,Yao_phasegate,yao_network,Liu_readwrite,Economou07PRLGeoControl,
Ramsay08PRLSpinControl,Yamamoto_08NatureSpinControl}.
Quantum interfacing between spins and photons~\cite{yao_network}
makes also possible quantum communication between distributed
quantum nodes which is required for scalable quantum computation.
In this sense, the direct-gap semiconductors such as InAs and
GaAs, where ultrafast optical control and interfacing are
possible, have some advantages over the indirect gap materials
such as silicon, where electrical gating and microwave pulses may
be the only possible means of control.
As compared with silicon, a main concern with the III-V
materials is the much shorter spin coherence time due to the
abundance of nuclear spins as noisy environments (in GaAs, e.g.,
the electron spin coherence time is in the order of microseconds,
while in silicon, it is in milliseconds, excluding the effects of
phonon scattering and inhomogeneous broadening)~\cite{Espin_HF_2_Loss,Schliemann:2002,Espin_HF_dipole_1_DaSSarma,
Espin_HF_1_Loss,Espin_HF_2_deSousa,Espin_HF_3_DasSarma,Yao_Decoherence,Espin_HF_Hu,
Saikin_Cluster,Liu_spinlong,Coish08PRB,Cywinski09PRL}. Fortunately, there already exist
various schemes to elongate the spin coherence time by orders of
magnitude, via dynamical
control~\cite{concatenation_Lidar,Yao_DecoherenceControl,Liu_spinlong,
Witzel2007PRL_control,Witzel2007_PRBCDD,Zhang:2007PRB,Zhang:PRB2008,Uhrig2007UDD,Lee2008_PRL,Yang2008PRL,Uhrig2009PRL},
or nuclear state preparation~\cite{Takahashi:2008PRL,Ramon_ZamboniEffect,Reilly:2008Science,Levitov_selfdnp,
Burkard_NuclearPreparation,Yao_doubledotsqueeze,Greilich_nuclearfocusing,Steel_locking,
Vandersypen_NuclearLocking,Giedke_spinmeasure,Burkard_spinprep,Klauser_spinmeasure}.

All these said, we would like to remark that at this point, it
would be premature to discourage effort
in exploring different physical systems, existing or emerging. It is
conceivable that the future quantum computers will be realized by
combination of innovative technologies, ideas, concepts, and
synthesis of materials and systems. For instance, the idea of
topological quantum computation may be implemented with trapped
atoms in optical lattice~\cite{Duan:2003PRLAtomLattice};
the systems under the focus of this review involve both stationary electron spins and flying
photons~\cite{yao_network}, which may also be applied to coupled systems of photons and superconducting
qubits~\cite{Wallraff04Nature,Majer07Nature,Neeley09Science,Hofheinz09Nature};
semiconductor chips may provide micro-trap for ions;
photon-based quantum computation may use quantum lights from
trapped ions~\cite{Keller_Ion_cavity}, atoms~\cite{Kimple_atom_cavity,Darquie_photon_atom},
or QDs~\cite{Michler_science_SinglePhotonSource,Yamamoto_QDcavityPhoton};
and so on. In the present initial stage of quantum computation technology, it would be
highly risky to exclude certain candidates just because of
difficulties encountered in the beginning of the adventure, since
different systems may have their bottleneck problems at different
stages. In particular, solid-state systems, while promising with
their large-scale stability and integratability in the future,
are still facing severe obstacles of environmental noise and
control errors for one or a few qubits. In this review, based on
many experimentally demonstrated elements and theoretically proposed
schemes, we would like to put forward blue prints of relatively
large-scale quantum computing via optical control of electron
spins in QDs. We should point out that such targets are by no means easy
and still require significant advances of technologies and concepts.
Also, although our discussions, to be specific, will be based on electron spins in InAs
or GaAs QDs, the schemes, with certain modifications, can be applied to a few
emerging novel systems such as hole spins in QDs and NV centers in diamond,
where the physics is similar to electron spins in QDs.

\begin{figure}[b]
\begin{center}
\includegraphics[width=10cm]{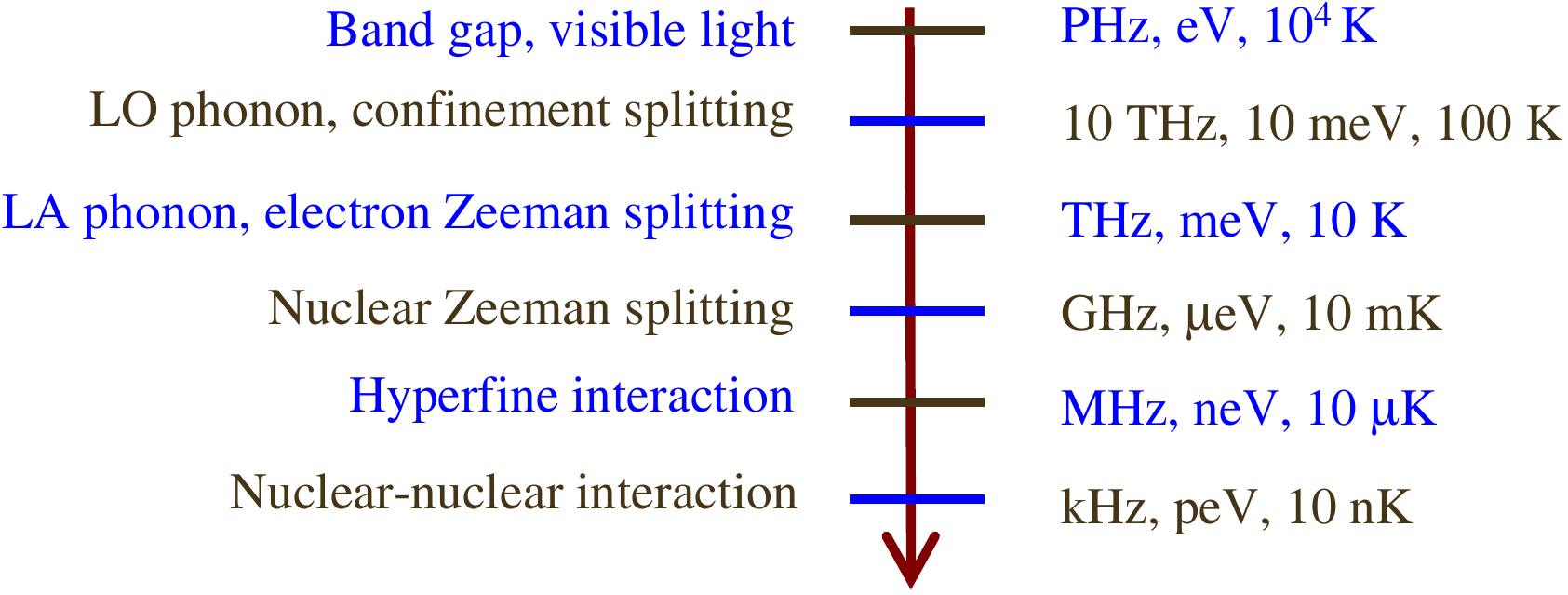}
\end{center}
\caption{Relevant energy scales for a typical GaAs semiconductor QD, in terms of frequency, energy,
and temperature. The hyperfine constant refers to the interaction between a single nucleus and an electron.}
\label{energyscale}
\end{figure}

There are two main concerns with solid-state systems for quantum computation, when compared with their
atomic peers like trapped ions or atoms. One is how to fabricate and construct a large
array of reproducible or identical qubit units (such as QDs). And the other one is the
many-body problem.

The fabrication issue is even worse for the systems to be discussed in this review, namely,
QDs under optical control. Most likely, the QDs are formed by molecular beam epitaxy (MBE) growth
process such as the self-assembled QDs and the fluctuation QDs.
Effort is being made on growth process so well controlled that all the QDs are almost identical
and regularly located. There are also such promising systems as
impurities in semiconductors such as phosphorus donors in silicon~\cite{Schofield:2003} and
the  NV centers in diamond~\cite{Meijer:2005,Meijer:2006}
where the nanometer-precision ion implantation technology may allow a patterned array of qubits represented
by the electron spins in the impurity centers. In the
foreseeable future, however, we may have to live with the problem of the irregularity.
On the positive side, we could take advantage of the irregularity produced by
the system fabrication. In small systems for demonstration purposes, the varying
size and position of QDs may be used as fingerprints by which different qubits may be addressed
and individually controlled by light beams which usually have resolution no better than half the wavelength.

The many-body problem raises two related questions: Can an isolated qubit be properly defined
at all among so many particles, and will a coherent superposition state of a qubit last
long enough for quantum gates before it collapses due to decoherence in the noisy environment?

It turns out~\cite{ShamRice66} that large gaps between valence bands and conduction bands in semiconductors
protect elementary excitations such as excitons and electrons so well that an extra electron
doped in a QD is well-defined as a single particle moving with a renormalized effective mass
and coupling constants such as the g-factor, in analogy to a low-energy electron in the
Dirac sea. The protective gap sets a fundamental limit on the operation speed of quantum computation in solid-systems.
Such a limit is far from being  approached in current experiments,
whether in typical semiconductors where the gap is about 1~eV (for fs-order operation time)
or a few meV in superconductors.

In optical control of spins in QDs, decoherence is caused mainly by three mechanisms, namely,
coupling to nuclear spins of the host lattice~\cite{Marcus_T2,Koppens_Rabi,Espin_HF_2_Loss,
Schliemann:2002,Espin_HF_dipole_1_DaSSarma, Espin_HF_1_Loss,Espin_HF_2_deSousa,
Espin_HF_3_DasSarma,Yao_Decoherence,Espin_HF_Hu,Saikin_Cluster,Liu_spinlong,Coish08PRB,Cywinski09PRL},
phonon scattering~\cite{Nazarov_Spinflip1,Nazarov_Spinflip2,Lyanda-Geller_Spinflip,Wu_spinT1,
Loss_SpinT2_phonon,Semenov_SpinT2_phonon,San-Jose06PRL},
and spontaneous photon emission during the optical control~\cite{Calarco_OpticalQC,Piermarocchi05Control}.
For some pseudo-spin qubits with orbital-state dependence (such as the singlet-triplet qubits~\cite{Marcus_T2}),
the charge fluctuation may also contribute to the environment noise~\cite{Hu_ChargeFluctuation,Coish_ST}.
Besides the standard quantum error correction protocols~\cite{NielsenChuang}, there are
specific strategies to deal with various decoherence mechanisms, which are needed anyway to achieve
fidelity of quantum gates above the threshold for quantum error correction.
The spontaneous photon emission could be suppressed by completing the optical control rapidly
or via off-resonance (virtual) excitation~\cite{Chen_Raman,Calarco_OpticalQC,Saikin_adiabatic,Yamamoto_OpticalQC}.
The phonon scattering may be quenched simply by lowering
the temperature to a few kelvins~\cite{Nazarov_Spinflip1,Nazarov_Spinflip2,Lyanda-Geller_Spinflip,
Wu_spinT1,1shot_r_Kouwenhoven,Fujisawa_T1,Finley_spin_memory,Cheng_spinT1_phasemodulation}
or by using light-element materials such as diamond~\cite{Wrachtrup:2001,Jelezko:2002,Kennedy:2003,Jelezko2004_PRLNV,Jelezko2004_PRL93,Santori:2006,
Epstein05NPhysNVN,Gaebel2006_NaturephysicsNV,Hanson2006_PRLNV,Hanson:2006RT,Childress2006_ScienceNV,
Dutt2007_ScienceNV,Hanson2008_ScienceNV,Wrachtrup:2006} or organic materials~\cite{CoryMalonic,Du:2009NatureUDD}
where the spin-orbital coupling is weak.
The nuclear spins, being a slow bath, may have their decoherence controlled by certain
dynamical decoupling or disentanglement control~\cite{concatenation_Lidar,Yao_DecoherenceControl,Liu_spinlong,
Witzel2007PRL_control,Witzel2007_PRBCDD,Zhang:2007PRB,Zhang:PRB2008,Uhrig2007UDD,Lee2008_PRL,Yang2008PRL,Uhrig2009PRL}.
Again, the normally harmful noise sources could be made useful by design.
The photon and phonon baths are rapid entropy dumping pools when
certain quantum channels are selectively enhanced~\cite{Liu_readwrite}.
The photon emission, when enhanced by cavities and guided by quantum channels, is an important basis for quantum
communication between remotely separated qubits~\cite{yao_network,Yamamoto_OpticalQC}.
The nuclear spins, having very slow dynamics, are considered as good local quantum
memories~\cite{Lukin_NucleiMemory,Espin_HF_dipole_Hu,Witzel2007_PRB}
with an electron spin in contact acting as a mediator for quantum information operation and transfer.

To achieve large-scale quantum computation, consensus has been reached on several criteria to be
fulfilled, known as the ``DiVincenzo criteria''~\cite{DiVincenzo_Criteria_7}.
We quote these criteria below as the guidelines for reviewing the progresses and main challenges toward the realization
of quantum computation by optical control of electron spins in QDs:
{\it
\begin{enumerate}
\item A scalable physical system with well characterized qubits;
\item The ability to initialize the state of the qubits to a simple fiducial state;
\item Long relevant decoherence times, much longer than the gate operation time;
\item A universal set of quantum gates;
\item A qubit-specific measurement capability;
\item The ability to interconvert stationary and flying qubits;
\item The ability faithfully to transmit flying qubits between specified locations.
\end{enumerate}
}
The stationary qubits under our focus are well defined by electron spins in QDs,
and the flying qubits carrying quantum information between distributed nodes
are photons flying in waveguides. We will discuss in more details
the decoherence of the spin qubits and show that the decoherence time
($\sim 10^{-6}$~sec in a typical GaAs QD
with the inhomogeneous broadening excluded~\cite{Marcus_T2,Greilich_lock,Reilly:2008Science,
Steel_locking,Koppens_echo,Yamamoto_echo})
is indeed much longer than the quantum gate operation time
($\sim 10^{-11}$~sec~\cite{Yamamoto_08NatureSpinControl,Yamamoto_echo,Greilich2009NPhys}).
The one- and two-qubit gates, which form a universal
set~\cite{DiVincenzo_UniversalQC,universal_quantum_gate,NielsenChuang},
are realized by optical excitation of charged excitons.
Some fundamental physics issues with the initialization and quantum measurement of qubits
will be reviewed.
Measurement and initialization are put together because they are related to the same physical process.
Initialization disposes of entropy to the environment while
in measurement the environment acts as part of a readout device.
For a large-scale quantum computation blueprint, we use
designs of local nodes of a few qubits and structures of distributed nodes connected by quantum
channels which may be realized by photonic elements such as waveguides and microcavities.
Control schemes of quantum interfacing are also an important topic to be covered.
To illustrate the feasibility and demands of the quantum computation in the discussed systems, the resources, in
terms of the number of optical pulses and operation time (compared with the spin decoherence time),
will be estimated for the benchmark problem of factoring 15 with Shor's algorithm.

\section{Spin Qubits in Quantum Dots}
\label{sec-single qubit}

In this section, we begin (in Sec.~\ref{subsec_spinQubit_confinement}) with a brief review of the
confinement of single electrons in optically controllable
semiconductor QDs, followed (in Sec.~\ref{subsec_spinQubit_levels}) by discussions of QD energy
level structures and optical properties. In Sec.~\ref{subsec_spinQubit_generaldecoherence}, we briefly outline recent
theoretical and experimental results on the spin coherence
properties of single electrons confined in QDs. Both
theories and experiments show that, as phonon mechanisms are
suppressed at low temperature ($\sim~4$~K and below), lattice nuclear
spins become the dominant cause
for the electron spin decoherence. In Sec.~\ref{subsec_spinQubit_nucdecoherence}, we review the theory of
electron spin decoherence by interacting nuclear spins in a QD. Coherence protection of electron spin in the interacting
nuclear spin bath is possible by applying a sequence of $\pi $ pulses
to the electron, as discussed in Sec.~\ref{subsec_spinQubit_restoration}.
In Sec.~\ref{subsec_spinQubit_summary}, an overview of QD
electron spins as qubits is given from the perspective of
fault-tolerance requirement for scalable quantum computation, and two
other promising spin qubit systems, namely, hole spins and NV center
spins, are also discussed.

\subsection{Confinement of  a single electron in a quantum dot \label{subsec_spinQubit_confinement}}

\begin{figure}[b]
\begin{center}
\includegraphics[width=8cm]{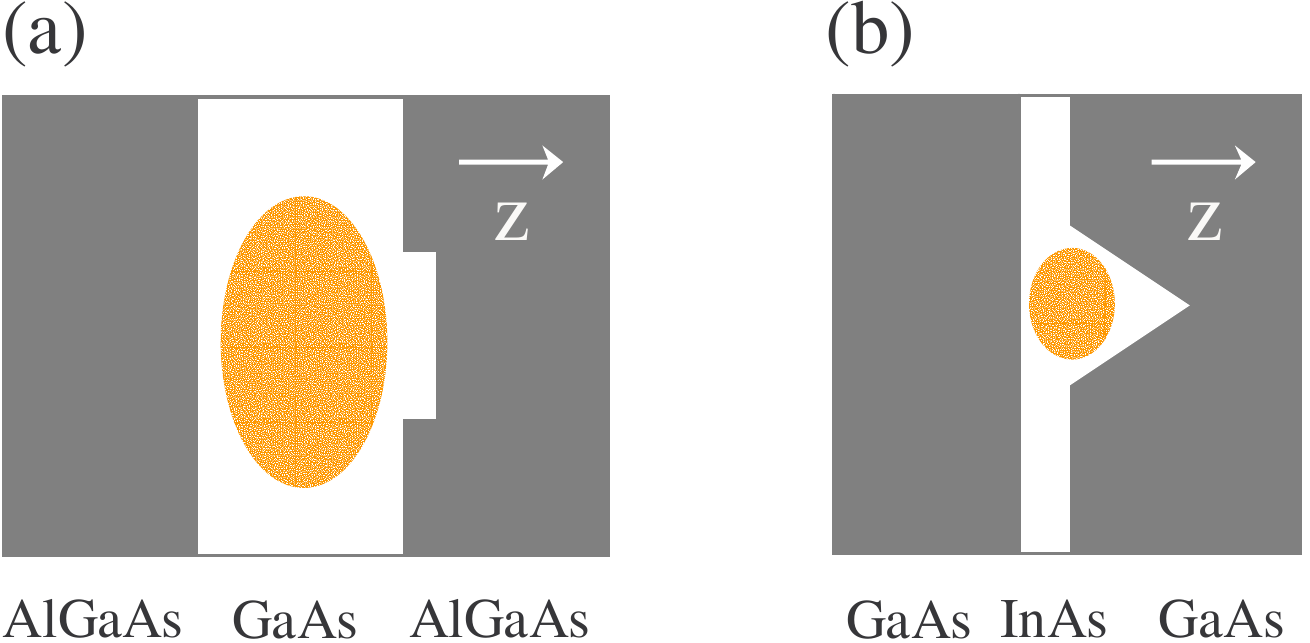}
\caption{Illustration of two types of optically controllable QDs. The arrows indicate the growth
direction. The ellipse regions show schematically the confinement of electrons in the dots.
(a) A GaAs fluctuation QD. (b) An InAs self-assembled QD.} \label{fluc_dot and SAD}
\end{center}
\end{figure}

Two types of MBE grown QDs formed in direct bandgap III-V
compounds offer a great deal of controllability by ultra-fast optics and
are being investigated as building blocks for optically manipulated
quantum computers.

\begin{figure}[b]
\begin{center}
\includegraphics[width=8cm]{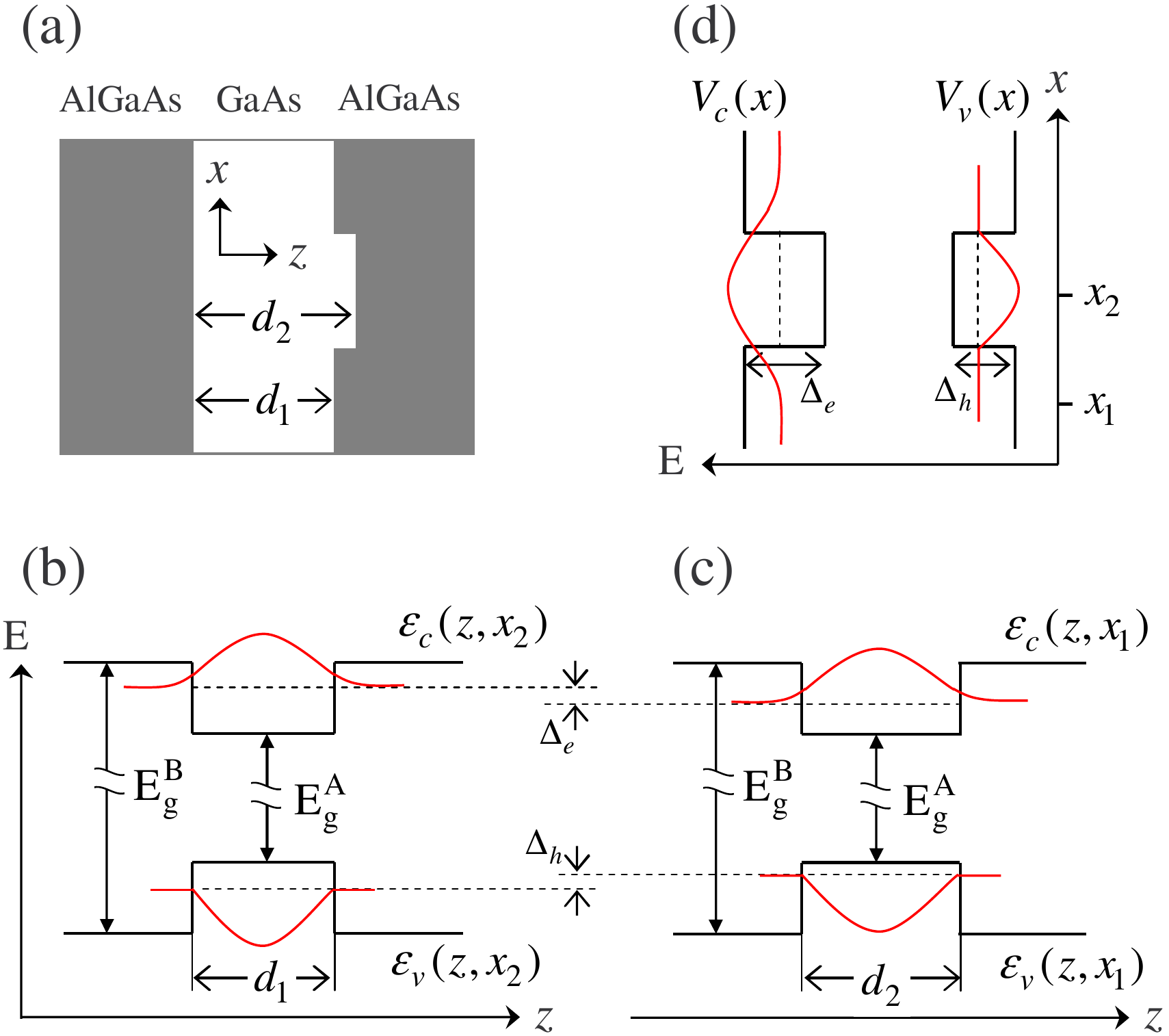}
\end{center}
\caption{Schematics of the three-dimensional confinement of
electrons and holes in a GaAs fluctuation QD. (a) Interface
fluctuation, typically of one monolayer, forms a fluctuation dot. The
bright region is the GaAs material and the dark regions are AlGaAs.
(b) Vertical confinement for electrons and holes in the growth
direction (the $z$ direction) at $x=x_1$ where the quantum well is thinner
$d=d_1$. (c) Vertical confinement in the growth direction at $x=x_2$
where the quantum well is one monolayer thicker $d=d_2$. (d) The
difference of the energy lift by vertical confinement in regions of
different thickness forms a lateral confinement for the electrons and
holes.} \label{QD_confinement}
\end{figure}

The first is referred to as interface fluctuation formed QDs in GaAs/AlGaAs quantum well
structures~\cite{Abstreiter_FlucD0,Gammon_FlucD1,Stievater:2001,Abstreiter_FlucD1,
Abstreiter_FlucD2,Abstreiter_FlucD3,Gammon_FlucD2,Gammon_FlucD3,Gammon_FlucD4}.
We will refer them in short as GaAs fluctuation QDs. As
illustrated in Fig.~\ref{fluc_dot and SAD}(a) and \ref{QD_confinement}, an
electron in such a structure is confined in the growth direction along the $z$ axis in
a low bandgap GaAs layer, with a thickness of tens of \AA, between two higher
bandgap Al$_{x}$Ga$_{1-x}$As layers. In the III-V materials,
the conduction band minimum occurs at the $\Gamma$ point in the
momentum space and the heterostructure wavefunction of
the electron is constructed from the conduction band Bloch
functions in the vicinity of the $\Gamma$ point. Thus the difference
in energy of the conduction band minimum in the high and low
bandgap materials form a square well potential for electron in
the growth direction. The band discontinuity in the interface of
the low and high bandgap material is typically hundreds of meV so
that the vertical confinement is strong. Growth interruption
leads to roughness at the material interfaces, usually a thickness
fluctuation of one monolayer. Electron confinement within the plane of the quantum
well (in the $x$ and $y$ directions) is caused by the quantum well thickness fluctuation.
The energy lift by the quantum confinement in the $z$ direction is roughly $\frac{\hbar^2}{2 m^{\ast}}
(\frac{\pi}{d})^2$ for the lowest energy state,
where $m^{\ast}$ is the effective mass of conduction electron and $d$ is the quantum well thickness.
This energy lift is larger where the quantum well is thinner. Therefore lateral confinement
is formed where the quantum well has an island.
Fig.~\ref{QD_confinement} shows schematically how monolayer-size
fluctuation in a quantum well gives rise to the localized
envelope function in the plane. The energy scale of this lateral
confinement is typically of several to tens of meV in GaAs
fluctuation dot. A GaAs fluctuation dot with lateral size $\sim
40~$nm can hold several localized energy levels with level
spacing of several meV.

\begin{figure}[b]
\begin{center}
\includegraphics[width=12cm]{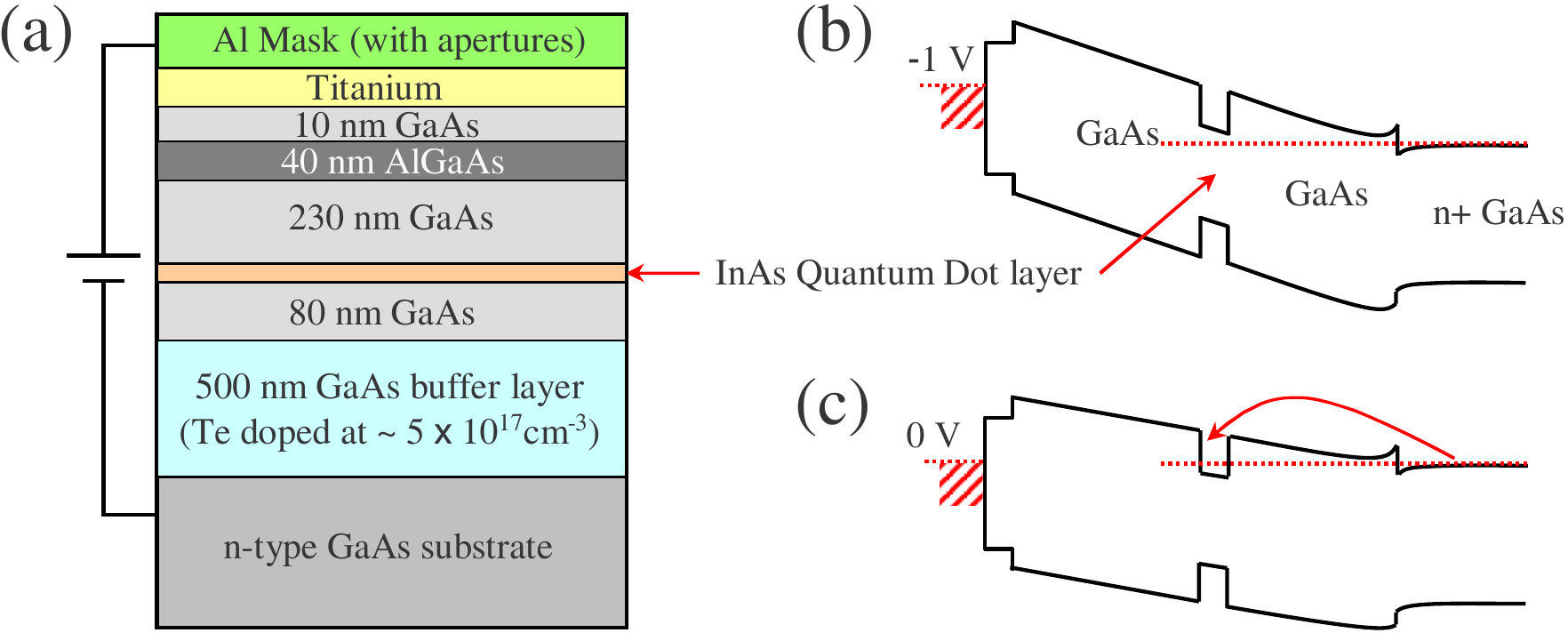}
\end{center}
\caption{Gate controlled charging of an InAs self-assembled QD
in an n$^+$-intrinsic-Schottky (NIS) diode structure. (a) Schematic
illustration of the NIS diode structure given in~\cite{DanPRL2005,Gammon06SciQDs}.
(b) Band diagram of the NIS diode structure where the InAs QD is uncharged. (c)
An InAs QD charged with a single electron.} \label{NISdiode}
\end{figure}

The second type of QDs, referred as InAs self-assembled
dots, are formed using the Stranski-Krastanow growth mode which
utilizes the strain caused by the lattice mismatch between InAs
layers and GaAs substrates. InAs self-assembles into
islands which are primarily in the shape of a pyramid [see
Fig.~\ref{fluc_dot and SAD} (b)], with a height of tens of \AA
and a base size of tens of nanometers~\cite{Petroff_SAD1,Richard_SAD1,ChenGang:2002,Petroff_SAD2}.
In an InAs self-assembled QD, the lateral confinement is much stronger
due to the pyramid structure, and the smaller dot size leads to
level spacing of $\sim10$~meV or larger.

For a GaAs fluctuation QD, a single conduction band electron
can be incorporated in the dot by modulation Si doping in the
barriers~\cite{Gammon_PRB2002}. For an InAs self-assembled dot, gate
voltage tuning in an n$^+$-intrinsic-Schottky diode structure is a
more controllable way to charge and discharge the QD with
a single electron~\cite{DanPRL2005,Gammon06SciQDs} (see the schematic
illustration in Fig.~\ref{NISdiode}). The qubit is typically encoded
in the spin subspace of the lowest energy level of the single
electron in the QD.

There are other notable systems where single electrons
are localized in nanoscale regions in semiconductors.
These include the confinement by electric gates on top of
two-dimensional electron gas in
GaAs~\cite{Loss_QDspinQC,tarucha03,T2star_Marcus}, and the
localization by impurities such as phosphorus donors in
silicon~\cite{Silicon_P_QC_review,Schofield:2003,Bradbury:2006,Huebl:2006,Stegner:2006}
or NV centers in diamond~\cite{Jelezko:2002,Jelezko:2004,Jelezko:2004b,
Santori:2006,Tamarat:2006,Epstein05NPhysNVN,Gaebel2006_NaturephysicsNV,
Hanson2006_PRLNV,Hanson:2006RT,Childress2006_ScienceNV,Dutt2007_ScienceNV,Wrachtrup:2006}.
The spins of single electrons localized in these systems are also
under extensive investigation as qubit carriers.

\subsection{Energy levels in a charged quantum dot \label{subsec_spinQubit_levels}}

Control of a spin qubit makes use of a larger Hilbert space in a
QD, involving optical transitions from valance bands to conduction
bands. In this subsection, we briefly describe the
relevant energy level structures and the corresponding optical
transition selection rules.

\begin{figure}[b]
\begin{center}
\includegraphics[width=8cm]{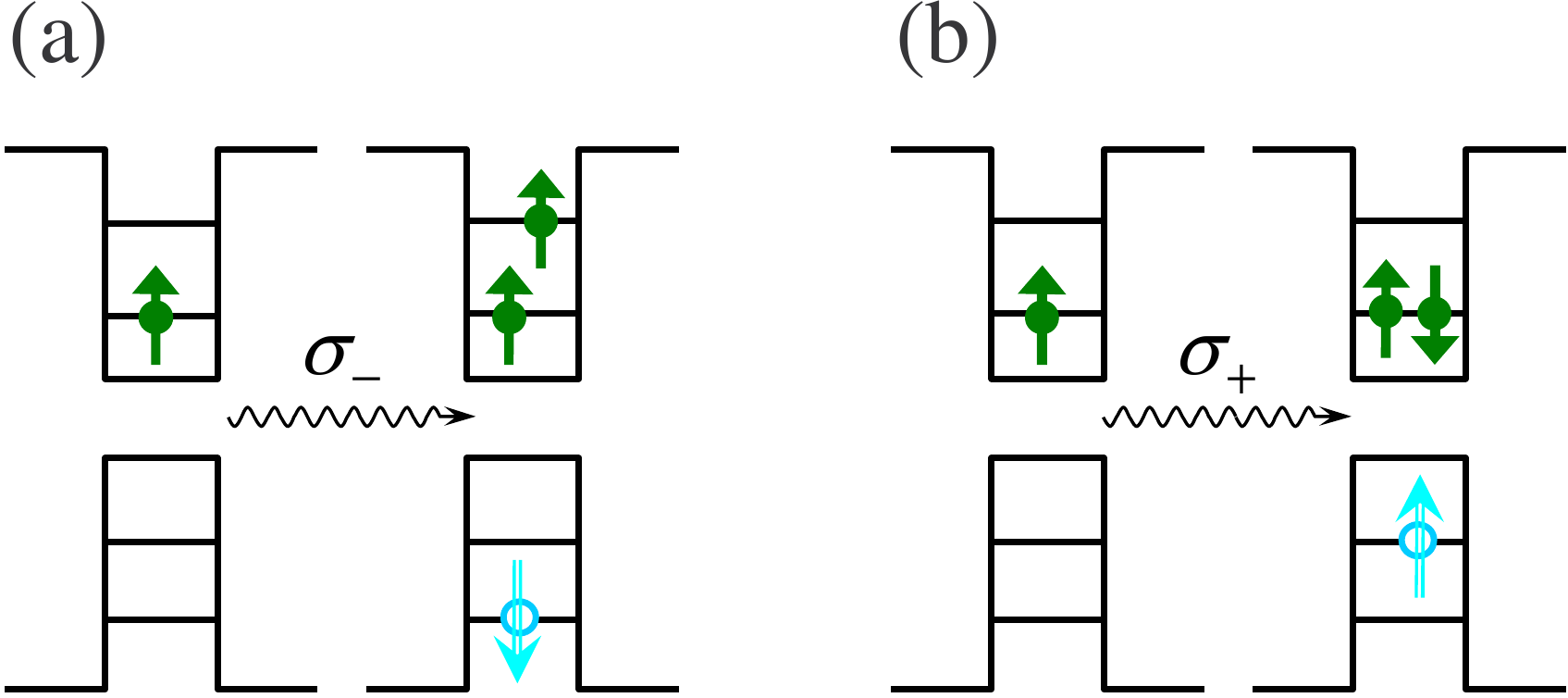}
\caption{Schematic illustration of the creation of
trion states in a QD. The QD initially holds a
single conduction band electron with spin up in the lowest state of
the confinement potential. (a) By the Pauli exclusion principle,
a $\sigma-$ circularly polarized light can create an additional exciton to a higher
excited state of the QD. (b) $\sigma +$
polarized light can create the lowest energy trion state with two
electrons forming a singlet and a hole in the spin up state. }
\label{trion}
\end{center}
\end{figure}

A direct interband transition creates an additional electron in the conduction band by
leaving a hole in the valance band. At the $\rm{\Gamma}$ point, the fourfold
degeneracy of the bulk $\Gamma_8$ valance band is lifted by the quantum well confinement effect. The
top valance subband is a doublet derived from the $J_z=\pm3/2$ bulk
band, which is also denoted as the ``heavy hole'' band (see
Appendix.~\ref{mixing}). Other valance bands are irrelevant in our
control schemes as they are always far off-resonance. Electrons can
be excited from the valance band states of angular momentum $J_z$ to conduction
band states of spin $S_z$ by absorbing a photon, with the selection rule
$S_z=J_z+\sigma$ where $\sigma=\pm 1$ corresponds to
the circular polarization of light. Therefore, the single electron states are optically
coupled to the charged exciton states (also known as trion states)
composed of two conduction electrons and one heavy hole.
Fig.~\ref{trion} shows schematically two different types of trion
states. The two electrons in Fig.~\ref{trion}(a) occupy different
electronic levels of the QD and the two
electrons in Fig.~\ref{trion}(b) are in a spin singlet configuration
on the same electronic level. The energy spacing between different
trion configurations is $\sim 1-10$~meV depending on the QD
size. The bandwidth and Rabi frequency of the optical field is much
smaller than this energy spacing. So when the frequency of the
optical field is near the resonance of one type of trion
configuration, the remaining trion configurations can be neglected.

\begin{figure}[b]
\begin{center}
\includegraphics[width=10cm]{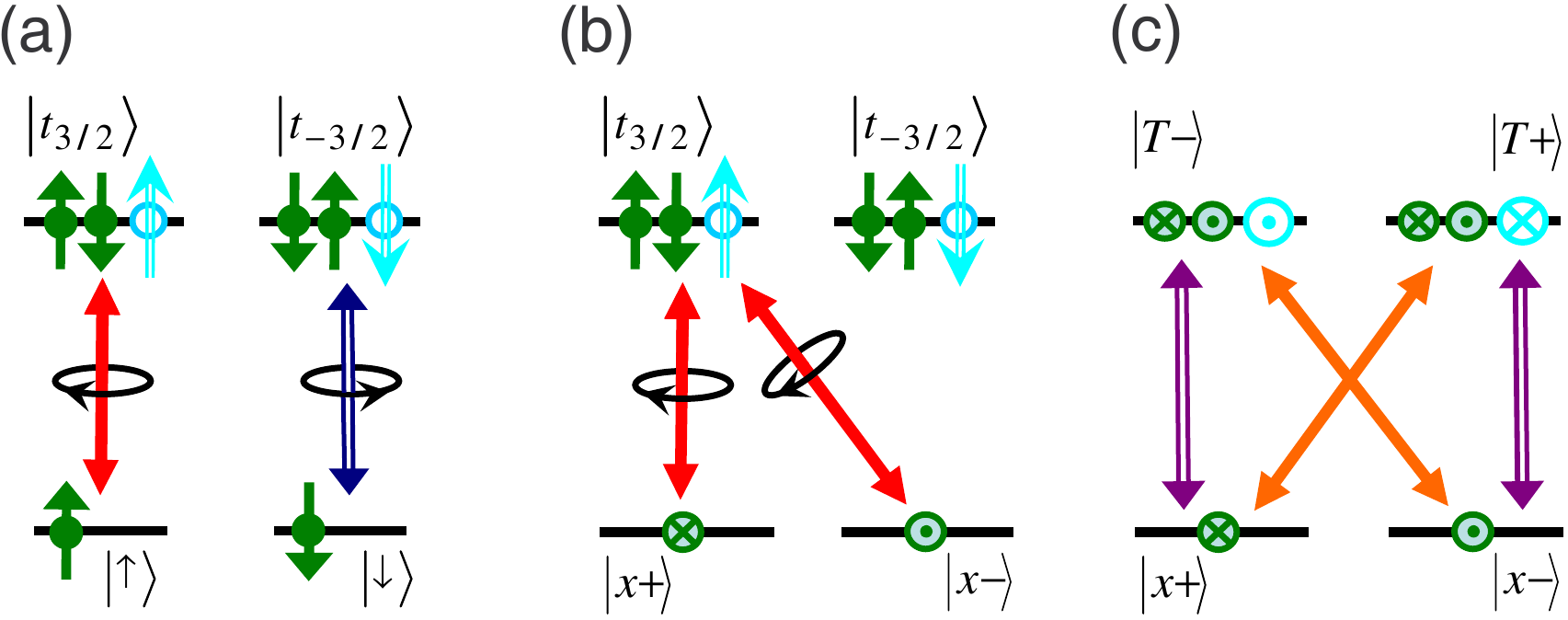}
\caption{Optical transition selection rules in a QD illustrated in various basis sets.
(a) The basis is the eigenstates of $\hat{J}_z$. The solid two-headed arrow
denotes $\sigma_+$ polarized light and the hollow two-headed one $\sigma_-$ polarized light.
(b) The two electron spin states are transformed to the basis in the $x$ direction: $|x\pm \rangle \equiv
(|\uparrow \rangle \pm | \downarrow \rangle)/\sqrt{2}$. The solid two-headed arrow
denotes $\sigma_+$ polarized light. (c) The two trion
states are also transformed to the basis in the $x$ direction: $|T \pm
\rangle \equiv (|t_{\frac{3}{2}} \rangle \pm | |t_{-\frac{3}{2}} \rangle)/\sqrt{2}$.
Here, the hollow two-headed arrow denotes $X$-polarized light and
the solid one $Y$-polarized light.}
\label{selectionrule_ideal}
\end{center}
\end{figure}

Most manipulation schemes use the ground state trions
[Fig.~\ref{trion}(b)] to mediate optical control of the spin. In
such cases, the relevant Hilbert space is composed of the two single
electron spin states and the two ground state trions (see
Fig.~\ref{selectionrule_ideal}). For simplicity, we use
below the $e^{\dagger}_{\pm}$ to denote the creation of
a conduction band electron of $S_z=\pm1/2$ in the lowest energy
level of the QD and similarly
$h^{\dagger}_{\pm}$ to denote the creation of a heavy hole with
$J_z=\pm 3/2$ (annihilation of an electron with $J_z= \mp 3/2$). The
relevant Hilbert space for optical control of spin is: $| \uparrow
\rangle \equiv e^{\dagger}_{+}| G \rangle $, $| \downarrow \rangle
\equiv e^{\dagger}_{-}| G \rangle $, $| t_{\frac{3}{2}} \rangle
\equiv e^{\dagger}_{+} e^{\dagger}_{-} h^{\dagger}_{+}| G \rangle $
and $| t_{-\frac{3}{2}} \rangle \equiv e^{\dagger}_{+}
e^{\dagger}_{-} h^{\dagger}_{-}| G \rangle $ where $|G\rangle$
denotes the configuration with empty conduction bands and full
valence bands. The transition selection rule between these four
states is shown in Fig.~\ref{selectionrule_ideal}(a). For different
control schemes, the selection rules represented in other basis sets
are also useful. In Fig.~\ref{selectionrule_ideal}(b) and (c), we
have changed the basis for the spin states and the trion states
so that the electron spin states are eigenstates along $x$-direction and the trion states are
$|T\pm\rangle\equiv \left(|t_{3/2}\rangle \pm|t_{-3/2}\rangle\right)/\sqrt{2}$.
The selection rule can also be represented with the linearly polarized basis for
the optical field as shown in
Fig.~\ref{selectionrule_ideal}(c)~\cite{Yao_phasegate}. The
selective coupling of the QD transitions with light of
different polarizations offers sufficient freedom
for optical control of the spin. Depending on the polarization and
frequency of the optical field, various optical pathways can be
established to realize the control in the spin subspace via a second
order process through the trion. For example, if the optical field
is $\sigma +$ circularly polarized [Fig.~\ref{selectionrule_ideal}(b)], the $|t_{-\frac{3}{2}}\rangle$
trion state is decoupled from optical field and the relevant
dynamics is in a $\Lambda$-type three-level system. Raman processes
in such three-level systems are central to the optical control of
spin dynamics as will be discussed in details later.

\subsection{Spin relaxation and decoherence in quantum dots \label{subsec_spinQubit_generaldecoherence}}

Coherence properties of qubits are crucial to
quantum information processing. Spin decoherence of a single electron
in a solid results from the coupling to various environmental modes.
Typical optical manipulations of spin qubits in QDs
are performed in the Voigt or Faraday geometry with a strong
magnetic field ($\sim 1-10$~T), and at low temperature ($\sim$ or
$<$~K) to suppress thermal excitations in the
environment~\cite{Gurudev,Imamoglu_coolingNuclear}. In this subsection, we
single out the environmental effects that dominate the
spin decoherence under these experimental conditions.

We first show the quantum mechanics of decoherence due to the
coupling to a general environment
\cite{Zurek_decoherence_RMP,Decoherence_classicalWorld,Schlosshauer_decoherence}.
The initial state of the electron spin,
$|\varphi^s(0)\rangle=C_+|+\rangle+C_-|-\rangle$, is prepared as a
coherent superposition of the spin up and down states $|\pm \rangle$
in an external magnetic field. The state of the total system of the
spin plus bath at that instant forms an {\it unentangled} state,
$|\Psi(0)\rangle=|\varphi^s(0)\rangle\otimes|{ \mathcal J}\rangle$.
It evolves over time $t$ to $|\Psi(t)\rangle=C_+
(t)|+\rangle\otimes|\mathcal{J}^+(t)\rangle+C_-(t)|-\rangle\otimes|\mathcal{J}^-(t)\rangle$
where the bath states $|{\mathcal J}^+(t)\rangle$ and $|{\mathcal J}^-(t)\rangle$
are generally different. The mixed state
of the electron spin is determined by its reduced density matrix obtained
by tracing over the environment states $\rho^s_{\sigma, \sigma'}(t)
= C^{\ast}_{\sigma'}(t) C_{\sigma}(t) \langle \mathcal{J}^{\sigma'}
(t) | \mathcal{J}^{\sigma}(t) \rangle$. The diagonal element of the
reduced density matrix $\rho^s_{\sigma, \sigma'}$ gives the
probability of finding the spin in state $|\sigma\rangle$. Either
off-diagonal element is a measure of the spin phase coherence. The
environment-driven transfer of the probability between the spin
states is known as the longitudinal relaxation, while the
loss of the off-diagonal element is known as the transverse decoherence.
The longitudinal relaxation also contributes to the transverse decoherence.
The spin decoherence without longitudinal relaxation
is called pure dephasing, characterized by the quantity $\langle
\mathcal{J}^+ (t) | \mathcal{J}^-(t) \rangle$. Pure dephasing is
thus a consequence of system-bath entanglement when the bath
evolution $| \mathcal{J}^{\pm}(t) \rangle$ is conditioned on the
system states $|\pm\rangle$. The above discussion can be generalized
to the situation where the environment is initially
in a mixed state. The single electron plus its environment is then described at
the initial time by the density matrix $\rho(0)=|\varphi^s(0)\rangle
\langle \varphi^s(0)| \otimes \sum_{\mathcal J} P_{\mathcal J} |{
\mathcal J}\rangle \langle { \mathcal J}| $, where the reduced
density matrix for the electron at any moment is $\rho^s_{\sigma,
\sigma'}(t) = \sum_{\mathcal J} P_{\mathcal J} C^{\ast}_{\sigma'}(t)
C_{\sigma}(t) \langle \mathcal{J}^{\sigma'} (t) |
\mathcal{J}^{\sigma}(t) \rangle$.

Novel experimental techniques have been developed in the past
several years for measurement of the spin relaxation and decoherence
in QDs. In GaAs fluctuation dots and InAs
self-assembled QDs, optical techniques are commonly used for
measuring the spin $T_1$ time. These include the time domain
measurement by optical generation and detection of non-equilibrium
spin population~\cite{Finley_spin_memory}, and the frequency domain
approach of the coherent phase-modulation spectroscopy~\cite{Cheng_spinT1_phasemodulation}.
In gate-defined QDs, spin-to-charge conversion can be
implemented to trace the time evolution of the initially created
spin population~\cite{1shot_r_Kouwenhoven,Fujisawa_T1}. In strong
magnetic field of $\sim 1- 10~$T and at low temperature ($\lesssim$~1 K),
the experimentally measured value
$T_1 \sim 10^{-4}-10^{-2}~$s~\cite{1shot_r_Kouwenhoven,Fujisawa_T1,Finley_spin_memory,Cheng_spinT1_phasemodulation}
are in good agreement with the theoretical predictions of the spin
longitudinal relaxation induced by phonon~\cite{Nazarov_Spinflip1,Nazarov_Spinflip2,Lyanda-Geller_Spinflip,Wu_spinT1}.
In a recent experiment by Amasha {\it et al.}~\cite{amasha_PRL08}, spin relaxation rate of a single electron in a lateral
QD is studied when the orbital wavefunction is manipulated using gate voltages.
The measured dependence of of $T_1$ on orbital confinement and magnetic field is in excellent agreement
with theory~\cite{Nazarov_Spinflip2,Loss_SpinT2_phonon}, which confirms that phonon scattering in the
presence of spin-orbit coupling is the dominant cause for spin relaxation in magnetic field down to 1~T.
In low magnetic field and when spin-orbit coupling strength is weak, $T_1$ as long as 1~sec is observed~\cite{amasha_PRL08}.

Transverse decoherence may be measured in optically accessible QDs by
the frequency domain approach utilizing
the Hanle effect~\cite{Gammon_t2star}, or by the time domain pump-probe
measurement. In the later approach, a circularly polarized pump
pulse initiates spin polarization in the growth direction, whose
precession about an in-plane magnetic field is tracked by the
differential transmission of a circularly polarized probe
beam~\cite{Gurudev} or the Faraday rotation angle of a linearly
polarized probe beam~\cite{Greilich_FR}. In a gate-defined double-dot
structure, the spin-to-charge conversion process can also be
implemented to probe the relative coherence between states of a coupled spin pair,
which provides information about the decoherence of a single
spin~\cite{T2star_Marcus,Kouwenhoven_singlet_triplet}. In these
experiments, transverse decoherence times of $T_2^{\ast}\sim 1 -
10~$ns were obtained either from measurements on spatial ensembles
of QDs~\cite{Gurudev,Greilich_FR,Steel_spinholeburning} or from time-ensemble
measurements of single
dots~\cite{T2star_Marcus,Gammon_t2star,Kouwenhoven_singlet_triplet}.
Spin echo type of measurement was also performed on the gate-defined
single GaAs dot~\cite{Marcus_T2,Koppens_echo}, which showed an echo decay time of
$T_H \sim ~\mu$s. The sharp difference between $T_H$ and
$T_2^{\ast}$ suggests that the ensemble dephasing is mainly affected
by the inhomogeneous broadening of the local environment of the
QDs, which is removed in the spin echo measurement. Greilich
{\it et al.} have shown that by using a periodic train of circularly
polarized light pulses to excite an ensemble of InAs self-assembled
QDs, spin polarization is amplified only in a subset of the
ensemble where the electron spin Zeeman frequency has to be an integer
multiple of the pulse repetition frequency~\cite{Greilich_lock}.
These quasi-discrete spectra leads to constructive interference of the spin precession at each pulse
arrival time. As dephasing by inhomogeneous broadening is thus
removed, single spin $T_2\sim ~\mu$s is obtained~\cite{Greilich_lock,Greilich_nuclearfocusing,
Espin_HF_1_Loss,Klauser_spinmeasure,Burkard_spinprep}.
Very recently, Hahn echo measurement was also made possible
for impurity spins in GaAs where rotations of spins were achieved by
ultrafast optical pulses~\cite{Yamamoto_echo}. The measured Hahn echo decay
time is $6~\mu$s, consistent with theories and other experiments.

These experiments all show that transverse decoherence times
($T_2^{\ast}$, $T_H$ and $T_2$) are orders of magnitude
faster than the longitudinal relaxation time $T_1$. On the other hand, theoretical
analysis of the phonon mechanisms concludes
that pure dephasing due to phonon is well suppressed at the
temperature where these experiments are performed
($\lesssim1~$ K)~\cite{Loss_SpinT2_phonon,Semenov_SpinT2_phonon}.
As phonon is unlikely to be responsible for the observed fast
transverse decoherence, the remaining possibility is then the
nuclear spins sitting on the lattice sites which are coupled to the
electrons through the hyperfine interaction~\cite{Espin_HF_2_Loss,Schliemann:2002,Espin_HF_dipole_1_DaSSarma,
Espin_HF_1_Loss,Espin_HF_2_deSousa,Espin_HF_3_DasSarma,Yao_Decoherence,
Yao_DecoherenceControl,Saikin_Cluster,Liu_spinlong,Cywinski09PRL,Cywinski09PRB,Marcus_T2,Koppens_Rabi,Coish08PRB}.

\subsection{Decoherence by an interacting nuclear spin bath \label{subsec_spinQubit_nucdecoherence}}

In this subsection, we discuss the effects of the lattice nuclear spins
on the electron spin coherence. In the relevant III-V materials, all
stable isotopes have non-zero nuclear spins. The nuclear magneton is
about 3 orders of magnitude smaller than the electron Bohr magneton. A strong
magnetic field of $10$~T only results in a nuclear zeeman energy of
$\sim~$mK, much smaller than the experimentally achievable
temperatures in cryostats ($\sim 4$~K) or even in dilution
refrigerators ($\gtrsim 50 $~mK) (see Fig.~\ref{energyscale}). Therefore, coupling to a thermalized
nuclear spin bath will be an inevitable source of decoherence for
quantum computation in III-V materials. As the longitudinal spin
relaxation of the electron is found to be much slower than the transverse decoherence, we will focus on the pure dephasing of electron
spins. In this Subsection, we set the $z$-direction along
the direction of the external magnetic field.

\subsubsection{Single electron in a mesoscopic bath of interacting nuclear spins}
\label{subsubsec-2}

\begin{figure}[b]
\begin{center}
\includegraphics[width=8cm]{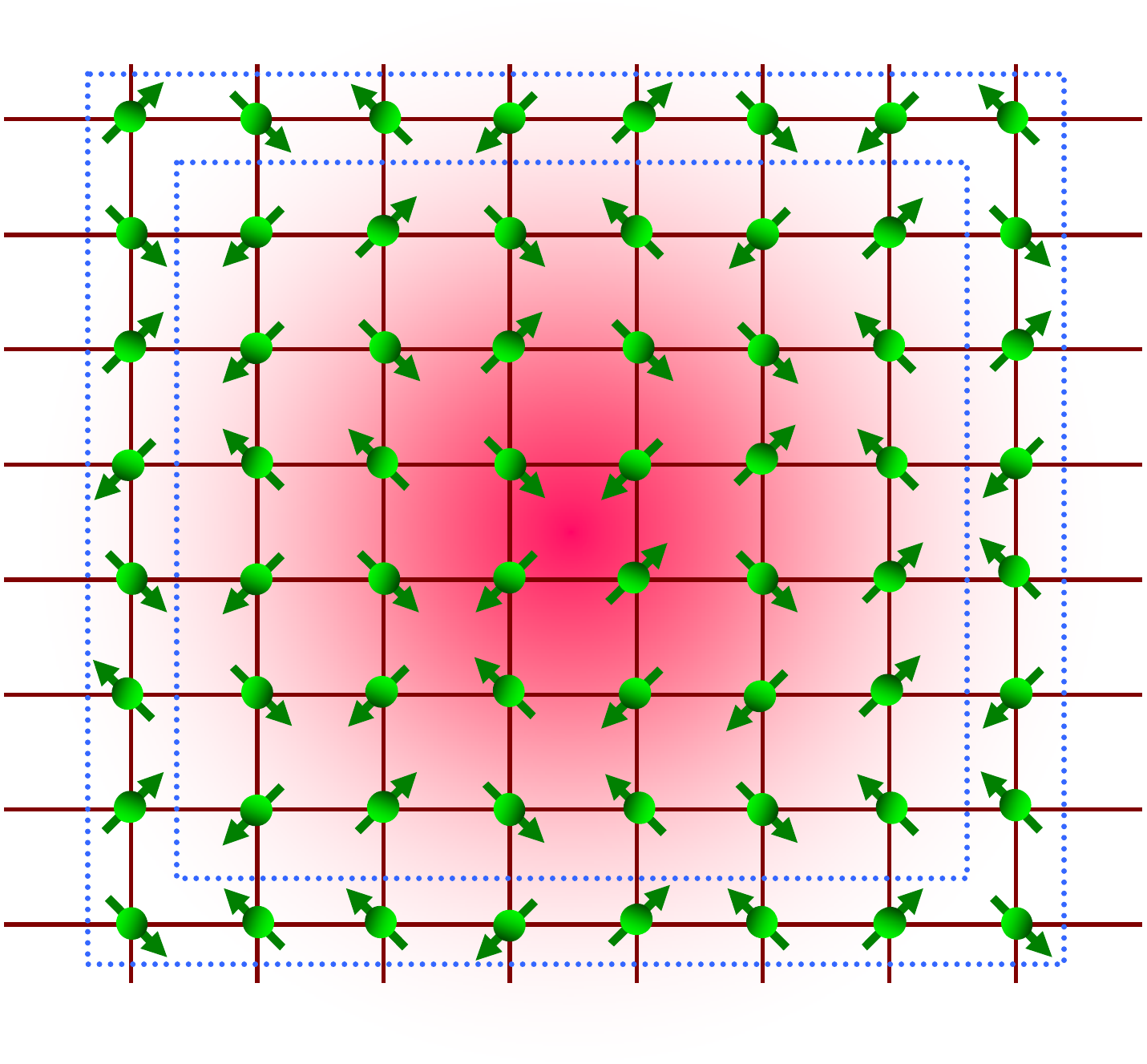}
\caption{{{}} Schematics of an
electron (the shadow) and one layer of lattice nuclear spins in a
QD. The two boxes in dotted lines indicate two possible
choices of boundary of the mesoscopic nuclear spin bath, which are
relatively arbitrary due to the interaction between nuclei within
and without the boundary. When the hyperfine interaction dominates
over the nuclear spin interaction, such arbitrariness has negligible
effects on calculation of the electron spin decoherence as long as
all the nuclei in direct contact with the electron spin have been
enclosed.} \label{mesobath}
\end{center}
\end{figure}

In III-V semiconductors, the electron spin is coupled to the lattice
nuclear spin through the contact hyperfine interaction. The averaged
magnitude of the coupling to a nucleus is inversely proportional to
the total number $N$ of nuclei in the QD. For QDs of all practical sizes ($N\sim10^5-10^7$),
this hyperfine coupling is much stronger
($\sim$~MHz) than the mutual interactions between the nuclear spins
($\lesssim$~kHz). Therefore, a mesoscopic bath consisting of all nuclear
spins within the QD (i.e. in direct contact with the single
electron) can be identified (see
Fig.~\ref{mesobath})~\cite{Liu_spinlong}. The coupling between the
mesoscopic bath and nuclei outside the boundary can be neglected since such dynamics occurs in a
much slower timescale as compared with the electron spin decoherence
caused by the mesoscopic bath. The decoherence problem can then
be solved by considering the quantum dynamics of the coupled mesoscopic
system of electron and nuclear spins. The assumption that this
mesoscopic system is well isolated from the back ground lattice has
been confirmed in numerical studies where the boundary of the
mesoscopic bath is systematically extended and the electron spin
coherence shows fast convergence~\cite{Liu_spinlong}. The
nuclear spin bath is typically of a randomized configuration as
schematically illustrated in Fig.~\ref{mesobath} since the
experimentally achievable temperature is always much higher than the
nuclear Zeeman energy.

\begin{figure}[b]
\begin{center}
\includegraphics[width=10cm]{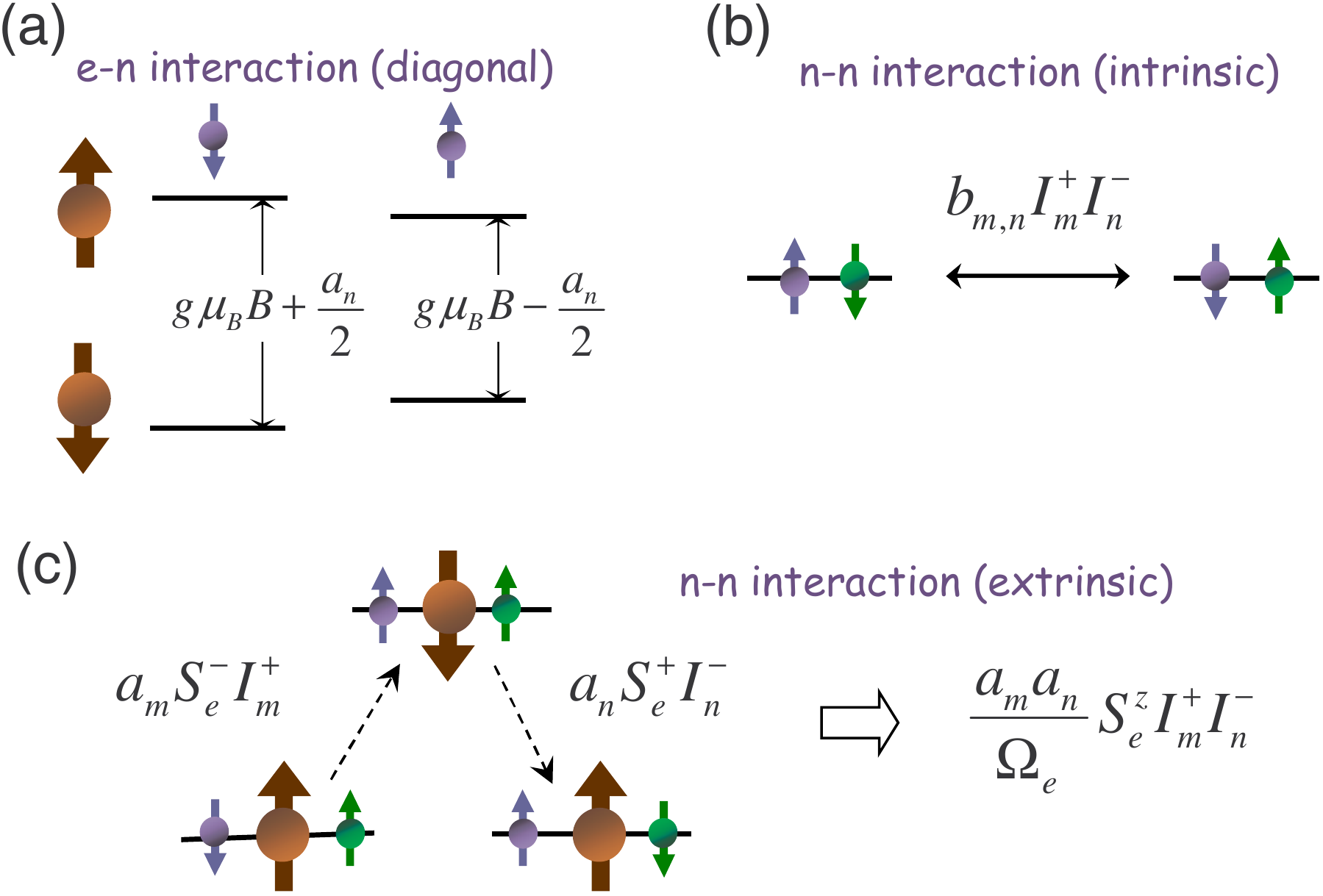}
\caption{{{}} Nuclear spin
processes relevant for electron spin decoherence. (a) By the
diagonal part of electron nuclear hyperfine interaction (which
involves only the spin vector components along the field $z$
direction), the electron Zeeman energy depends on the nuclear
configuration. (b) Nuclear pair-wise flip-flop by the intrinsic
nuclear interactions. (c) Nuclear pair-wise flip-flop mediated by
two virtual flips of the electron spin, which results in an
effective {\it extrinsic} nuclear interaction.
$\Omega_e$ is the electron Zeeman energy in the external magnetic field, and
$a_n$ is the hyperfine interaction strength between the electron and the $n$th
nuclear spin.} \label{e-n_process}
\end{center}
\end{figure}

We briefly describe below the key ingredients for electron spin
decoherence in the high field limit with details given in Appendix~\ref{decoherence_appendix}.
By the diagonal part of the electron-nuclear hyperfine
interaction, the electron Zeeman energy is conditioned on the
nuclear spin states (see Fig.\ref{e-n_process}(a)). In a QD
ensemble, the different nuclear spin configurations for each ensemble
member lead to inhomogeneous broadening of the Overhauser field for the electron spin.
This is the dominant cause of ensemble dephasing
in the timescale of $T_2^{\ast}$ which is inversely proportional to
the inhomogeneous broadening. Nuclear-nuclear interactions become
relevant due to the non-uniform hyperfine coupling strength between
the electron and different nuclear spins. Pair-wise nuclear
flip-flops can then lead to dynamical fluctuation of the nuclear
Overhauser field (see Fig.\ref{e-n_process}(b)). This is the cause
of single electron spin decoherence in the nuclear environment
which begins with a pure state (referred as {\it single-system dynamics} hereafter).
Electron-nuclear coupling also has an
off-diagonal part which tends to cause flip-flop between the
electron and a nuclear spin. Because the electron Zeeman energy is
much larger than the strength of the hyperfine interaction,
the real process of electron nuclear flip-flop is
suppressed~\cite{Yao_Decoherence,Espin_HF_1_Loss}. However, a second order process
which consists of two virtual flips of the single electron ends up
as a flip-flop between two nuclear spins (see
Fig.\ref{e-n_process}(c)). This effective nuclear interaction
due to the single electron is designated
as the {\it extrinsic} interaction~\cite{Yao_Decoherence,Yao_DecoherenceControl,Liu_spinlong},
as opposed to the {\it intrinsic} nuclear interactions that exist in
the semiconductor matrix, e.g. the dipole-dipole coupling and the
indirect coupling mediated by virtual interband transitions via the
hyperfine interaction~\cite{indirect_exchange_Bloembergen,indirect_exchange_Anderson,
indirect_exchange_Shulman1,indirect_exchange_Shulman,indirect_exchange_Sundfors}.

The {\it extrinsic} nuclear interaction couples any two nuclear
spins within the mesoscopic bath and is therefore infinitely-ranged,
i.e., throughout the entire mesoscopic region.
By contrast, the {\it intrinsic} nuclear interaction is finite-ranged.
For near neighbors, the intrinsic one is much
stronger than the extrinsic one for the field strength under
consideration. In addition, for the {\it extrinsic} nuclear
interaction, the magnitude is inversely proportional to the external
magnetic field and the sign is conditioned on the electron spin
states.

\subsubsection{Nuclear spin pair-flip excitations and pair correlation approximation}
\label{subsubsec-pca}

From the quantum mechanical picture of decoherence (see section
\ref{subsec_spinQubit_generaldecoherence}), pure dephasing of a
single quantum system is caused by the bifurcation of environmental
evolution under the drive by different system states, or
system-bath entanglement. Thus, the nuclear bath evolutions
conditioned on different spin states of the electron are key to the
solution of electron spin decoherence. The elementary
excitations in the nuclear spin bath are pair-flip excitations as
shown in Fig. \ref{e-n_process}(b) and Fig. \ref{e-n_process}(c).
The flip-pairs are independent of each other if they are well
separated. For a typical QD with $N\sim 10^6$ nuclear
spins, the number of pairs that can flip-flop in a random
configuration is large. We have $O(N)$ local flip-pairs whose
dynamics are dominated by the finite range {\it intrinsic} nuclear
interactions, and $O(N^2)$ non-local flip-pairs whose dynamics are
dominated by the infinite range {\it extrinsic} nuclear
interactions. On the other hand, the number of pair-flips that can
occur on the timescale of electron spin decoherence is negligibly
small as compared with $N$. This is due to the slowness in the nuclear
spin interacting dynamics. Thus, the probability of two pair-flips
occurring in the neighborhood of each other is negligibly small and
pair-flips as elementary excitations can be treated as independent
of each other (see Appendix~\ref{decoherence_appendix} for details).
This approximation is further confirmed by the linked cluster expansion approach~\cite{Saikin_Cluster}.
Independent pair-excitation approximation
corresponds to keeping the lowest (2nd) order linked diagrams in the
exponential factor. Higher order linked diagrams contain more
nuclear interaction lines and are negligible in the relevant
timescale because of the weakness of the nuclear interactions as
compared with the electron-nuclear coupling.
For long time evolution (such as for decoherence under pulse control) and
for relatively small spin baths, higher order correlations would be important.
The linked cluster expansion would be increasingly inefficient for calculating higher
order correlations. The density matrix cluster expansion is
an alternative method which requires no evaluation of higher order Feynman diagrams
and is thus quite convenient~\cite{Witzel:2006}. For small spin baths, however, it has been shown that
the cluster expansion may not converge to the exact results~\cite{Yang2008PRB}.
For a systematic and accurate account of the higher order correlations in spin baths
in the qubit decoherence problem, a cluster-correlation expansion method has been
developed~\cite{Yang2008PRB,Yang2009PRB}, which covers the valid ranges of all the methods mentioned above and
in particularly produces the exact results even for a relatively small bath where the
standard cluster expansion fails. The cluster correlation expansion is based on
the factorization of the bath dynamics into non-factorizable correlations of
certain groups of bath spins. The lowest order of the cluster correlation expansion
coincides with the independent pair excitation approximation.
We also note that recently the higher order effects of the hyperfine interaction
(beyond the pair-excitation considered here) has also been considered for
a relatively weak external magnetic field while the pure dephasing condition is
still satisfied~\cite{Cywinski09PRL,Cywinski09PRB}

\begin{figure}[b]
\begin{center}
\includegraphics[width=12cm]{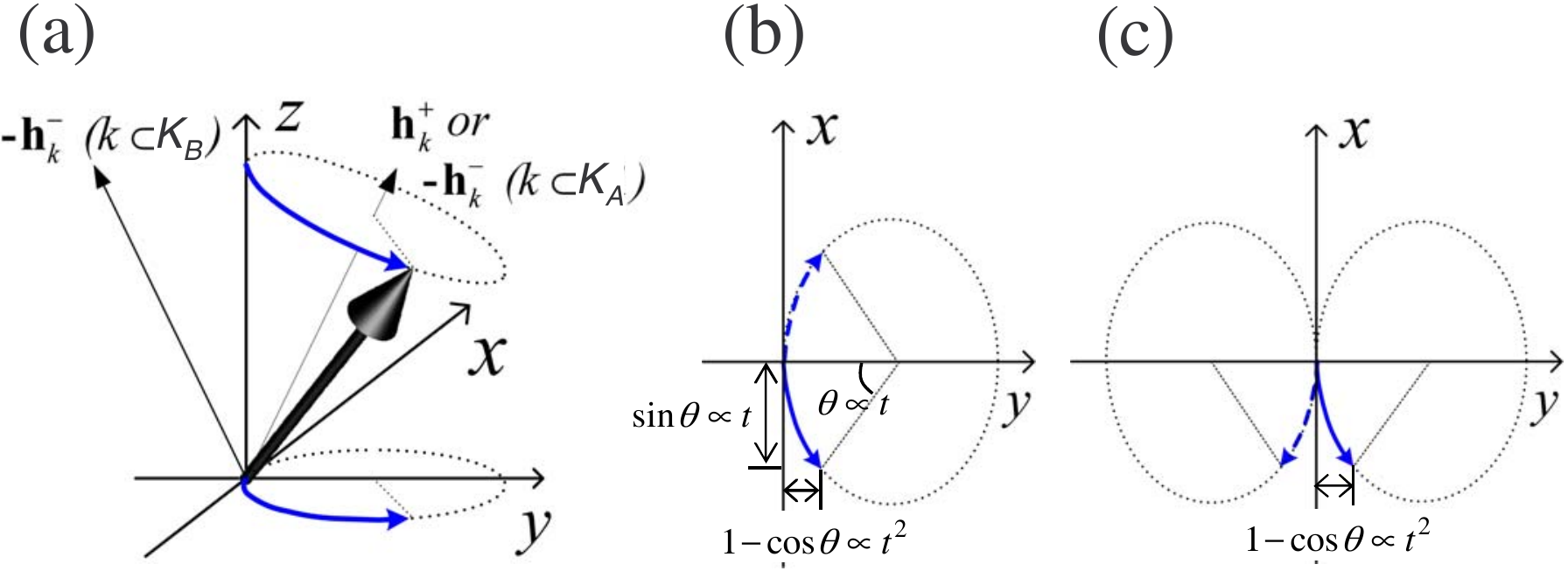}
\caption{A geometric picture for understanding the free-induction
decay. (a) The evolution of pair-excitation illustrated by the
rotation of a Bloch vector and the projected trajectory on the
pseudo $x$-$y$ plane. Direction of the effective pseudo-field
${\mathbf h}^{\pm}_k$ are indicated respectively for the set of
non-local pairs $K_A$ and the set of local pairs $K_B$. (b) The
projection of the Bloch vector trajectories to the pseudo $x$-$y$
plane for pair-excitation driven by extrinsic nuclear interaction.
The solid (dashed) line denotes the pair evolution conditioned on
the electron spin state $|+\rangle$ ($|-\rangle$). As the rotation
angle $\theta \propto t$, the distance between the conjugated
vectors $\delta_k \propto t$ at short time. (c) The projection of
the Bloch vector trajectories to the pseudo $x$-$y$ plane for
pair-excitations driven by intrinsic nuclear interactions. The
distance between the conjugated vectors $\delta_k \propto t^2$
at short time.} \label{Bloch_FID}
\end{center}
\end{figure}

The evolution of independent pair-correlations can be described
using a geometric representation. A pair-flip $k$ can be mapped to a
Bloch vector which precesses about a pseudo-field (see Fig.~\ref{Bloch_FID})
\begin{equation}
 {\mathbf h}^{\pm}_k \equiv (\pm 2A_k+2B_k,0, \pm E_k),
\end{equation}
where, for the electron spin state $|\pm\rangle$, $\pm A_k$ and
$B_k$ are the pair-flip transition amplitudes contributed by the
extrinsic nuclear interaction and the intrinsic nuclear interaction,
respectively, and $\pm E_k$ is the energy cost of the pair-flip
contributed by the hyperfine interaction (see Appendix~\ref{decoherence_appendix}).
At the initial time when the
electron spin coherence is prepared (see the general formulation of
decoherence process in Section~\ref{subsec_spinQubit_generaldecoherence}), the Bloch vector for
each pair-flip points in the pseudo $+z$ direction. As the
pseudo-field direction depends on the electron spin state
$|\pm \rangle$, the pair evolution takes different trajectories conditioned
on the electron spin state, and the distance $\delta_k$ (i.e., distinguishability)
between the two separated trajectories turns out to be a measure of the electron
spin decoherence,
\begin{eqnarray}
{\mathcal L}^s_{+,-}(t) \approx \prod_k e^{-\delta_k^2/2}.
\end{eqnarray}
where ${\mathcal L}^s_{+,-}(t)$ is the ratio between the electron
spin coherence at time $t$ and at time $0$ (see the general
formulation of decoherence in section
\ref{subsec_spinQubit_generaldecoherence}). $\delta_k$ also
quantifies the amount of entanglement between the electron spin and
the $k$th flip-pair. The electron spin decoherence is thus the
consequence of the entanglement with the pair-flip excitations in
the interacting nuclear spin bath.

With single system dynamics solved for an arbitrary random
configuration, ensemble dynamics is simply the statistical average
of the single system dynamics with the nuclear bath initially
in all possible configurations. As the number of flip-pairs is large
($O(N)$ for local pairs and $O(N^2)$ non-local pairs), the
central limit theorem of statistics leads to a factorized form for ensemble spin
coherence,
\begin{equation}
{\mathcal L}_{+,-}(t) = {\mathcal L}^{\rm s}_{+,-}(t)\times
{\mathcal L}^{(0)}_{+,-}(t), \label{e_dm_4a}
\end{equation}
where ${\mathcal L}^s_{+,-}(t)$ is the single-system decoherence in
a typical configuration of the nuclear bath, and
\begin{eqnarray}
{\mathcal L}^{(0)}_{+,-}(t)  &= & \sum_{\mathcal J} P_{\mathcal J}
e^{-i\phi_{\mathcal J}(t)},
\end{eqnarray}
where $\phi_{\mathcal J}(t) = (\Omega_e + \mathcal{E}_{\mathcal J})
t$ in free-induction decay,  and the summation runs over all possible
nuclear configurations $\mathcal{J}$. $\Omega_e$ and ${\mathcal E}_{\mathcal J}$
are the electron Zeeman energy resulting from the external magnetic field
and from the Overhauser field, respectively, with the latter dependent on
the nuclear configuration ${\mathcal J}$.
The ensemble effect resides entirely in the factor ${\mathcal L}^{(0)}_{+,-}(t)$, which may be
read as the inhomogeneous broadening of the Overhauser field
${\mathcal E}_\mathcal{J}$ with a distribution function $P_{\mathcal J}$.
The inhomogeneous broadening effect dominates the free-induction decay (FID)
in the ensemble dynamics in the form of ${\mathcal L}^{(0)}_{+,-}(t)=
e^{-i\Omega_{\rm e} t-(t/T_2^*)^2}$, with the dephasing time $T_2^*
\sim \sqrt{N} \mathcal{A}^{-1}\sim 10$~ns as
measured~\cite{Gammon_t2star,T2star_Marcus,Kouwenhoven_singlet_triplet,Gurudev,Espin_HF_2_Loss,Khaetskii_nuclear,Espin_HF_1_Loss},
where ${\mathcal A}\sim$THz is the hyperfine constant of the material.

A sequence of $\pi$ pulses can be applied to the electron spin to eliminate
the effects of the inhomogeneous broadening~\cite{Slichter_NMR,Lyon_echo,Koppens_echo,Cywinski09PRL}.
In a general scenario where the electron spin is flipped at time $\tau_1$, $\tau_2$, $\dots$, and
$\tau_n$ respectively, we have
$\phi_\mathcal{J}(t)
 = (\Omega_e + \mathcal{E}_{\mathcal J})
\left[\tau_1-(\tau_2-\tau_1)+\cdots+(-1)^n(t-\tau_n)\right]$. When
$\tau_1-(\tau_2-\tau_1)+\cdots+(-1)^n(t-\tau_n)=0$ is satisfied,
${\mathcal L}^{(0)}_{+,-}(t)=1$ and a spin echo is expected. The
echo magnitude will be determined by the dynamical part ${\mathcal
L}^{s}_{+,-}(t)$. Under the simplest scenario, a single $\pi$-pulse
is applied at $\tau$ and a spin echo is expected at $t=2\tau$, known
as Hahn echo~\cite{Hahn}. The spin echo profile, i.e., the echo
magnitude ${\mathcal L}^{\rm s}_{+,-}(2\tau)$ as a function of the
echo delay time $2\tau$, reveals the dynamical processes that leads
to decoherence.

It is worth noting that the factorized form of the ensemble spin coherence, Eq.~(\ref{e_dm_4a}),
allows direct observation of single-system dynamics behavior ${\mathcal L}^{\rm s}_{+,-}(t)$
from a spatial ensemble measurement when dephasing by inhomogeneous broadening is removed at
a general time $t$.  For example, the mode locking experiment reported in~\cite{Greilich_FR,Greilich_lock}
opens up such possibilities as discussed in section~\ref{subsec_spinQubit_generaldecoherence}, where
the single spin $T_2$ (defined here as the FID timescale of ${\mathcal L}^{\rm s}_{+,-}(t)$) has
been extracted from the experimental data.

\subsubsection{Timescales of single spin decoherence and ensemble spin echo decay} \label{subsubsec-results}

In FID, the conjugate Bloch vectors precess along opposite
directions for non-local pairs ($k\in K_A$), and symmetrically with
respect to the pseudo $y$-$z$ plane for the near-neighbor pairs
($k\in K_B$) [Fig.~\ref{Bloch_FID}]. The decoherence can be readily
grouped by the two different mechanisms as
\begin{eqnarray}
{\mathcal L}^{\rm s}_{+,-}\cong \prod_{k \in
K_B}e^{-\frac{t^4}{2}E_k^2 B^2_k
 {\rm sinc}^4\frac{h_k t }{2}}
 \prod_{k\in K_A}e^{-2 t^2 A_k^2 {\rm sinc}^2 (h_k t)},
 \label{separation}
\end{eqnarray}
where $h_k = | {\mathbf h}^{\pm}_k|$. We can see that the extrinsic
hyperfine-mediated and the intrinsic couplings lead to the
$e^{-(t/T_{2,A})^2}$ and the $e^{-(t/T_{2,B})^4}$ behavior
respectively in time shorter than the inverse pair-flip energy cost
(which corresponds to the width of the excitation spectrum),
\begin{equation}
T_{2,B}\approx b^{-1/2} \mathcal{A}^{-1/2} N^{1/4}; ~ T_{2,A}\approx
\Omega_e \mathcal{A}^{-2} N, \label{size_field_depend}
\end{equation}
where $b$ is the typical value of near neighbor intrinsic nuclear
coupling strength $B_k$ (see Appendix~\ref{decoherence_appendix}).
The super-exponential decay behaviour of the spin coherence indicates the
strong non-Markovian characteristic of the bath dynamics in the short time limit.
In the long-time limit, the super-exponential decay
will change to an exponential decay time first which indicates the onset of the Markovian
dynamics~\cite{Liu_spinlong,Coish08PRB,Cywinski09PRB}. The dynamics in the even longer time limit (which could occur, e.g.,
in a highly polarized spin bath), determined by the complex structure of the collective modes
of the bath, is rather complicated, and power-law decays have been predicted~\cite{Espin_HF_Hu,Cywinski09PRB}.

\begin{figure}[b]
\begin{center}
\includegraphics[width=13cm]{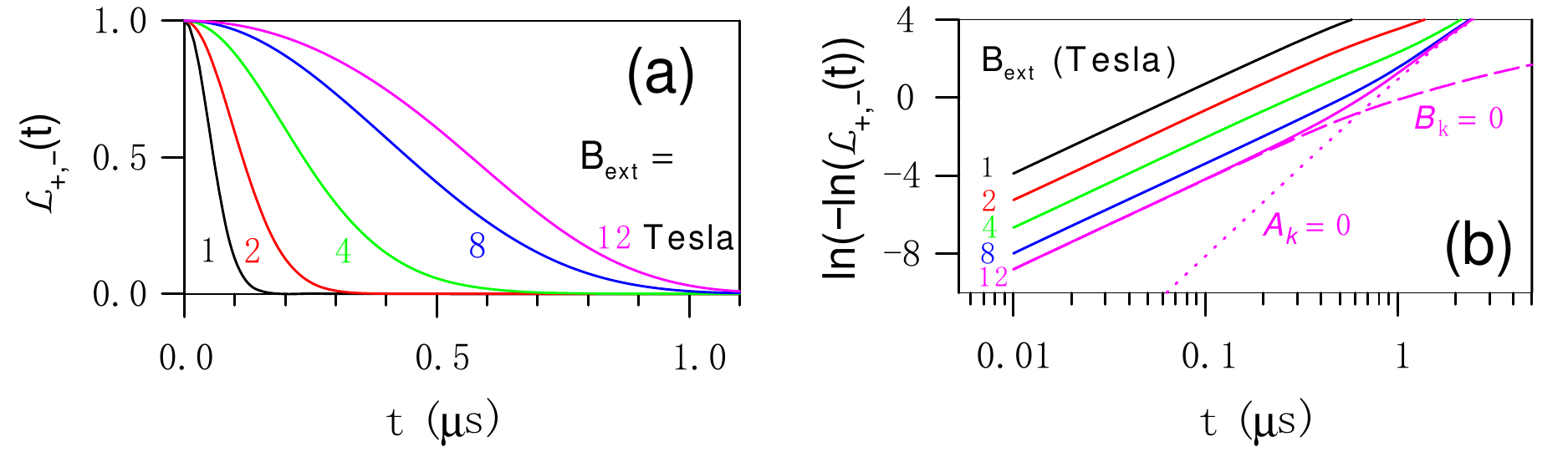}
\caption{(Extracted from Ref.~\cite{Liu_spinlong}) (a) Electron spin coherence
as functions of time for various field strengths. (b) The logarithm
plot of (a), in which the curve for $B_{\rm ext} = 12$ tesla is
compared with the contribution by the extrinsic nuclear interaction
(the dashed line) and that by the intrinsic nuclear interaction (the
dotted line), respectively. The size of the InAs dot is $33 \times
33 \times 3$ nm$^3$ and the nuclear-spin initial state
$|\mathcal{J}\rangle $ is randomly selected from an ensemble at
temperature 1 K. The field strength is indicated by the numbers for each curve.}
 \label{FID}
\end{center}
\end{figure}

\begin{figure}[h]
\begin{center}
\includegraphics[width=12cm]{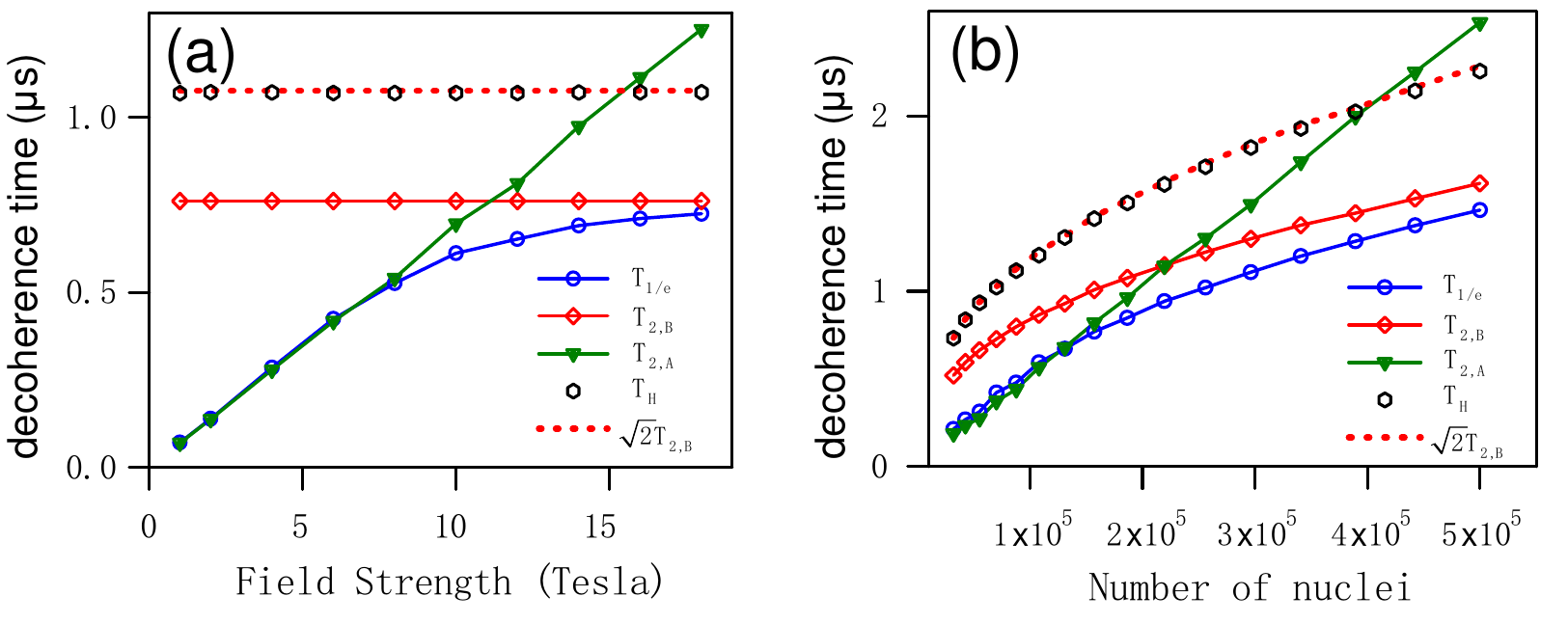}
\caption{(Extracted from Ref.~\cite{Liu_spinlong}) (a) Field dependence of decoherence times
with the inhomogeneous broadening effect excluded. The QD
is the same as in figure~\ref{FID} and the temperature is 1~K. (b) Dot-size
dependence of decoherence times with the inhomogeneous broadening
excluded. The QD size is varied with fixed width : depth :
height ratio 33 : 33 : 6. The field strength is 10 telsa and the
temperature is 1~K. $T_{1/e}$ (solid line with circle symbols) -- time for FID to $1/e$ of its initial value,
$T_{2,A}$ (solid line with triangle symbols)
-- FID decoherence time resulting solely from extrinsic hyperfine-mediated
pair-flips, $T_{2,B}$ (solid line with diamond symbols)
-- FID decoherence time resulting solely from the intrinsic nuclear
spin interaction, and $T_H$ (circle symbols) -- decay time of the
Hahn echo signal. The $\sqrt{2}T_{2,B}$ (dashed curve) is
plotted to compare with the Hahn echo decay time.}
\label{fieldNsize}
\end{center}
\end{figure}

Fig.~\ref{FID} shows the FID in single-system dynamics for a typical
dot under various field strengths $B_{\rm ext}$. The strong field
dependence of $T_2$ demonstrates the significance of the extrinsic
hyperfine mediated nuclear coupling up to a strong field
($\sim 10$~T). The short time $e^{-t^2}$ behavior of decoherence by
extrinsic nuclear interactions and the $e^{-t^4}$ behavior by
intrinsic nuclear interaction hold well within
the relevant timescale for single spin FID. The field and dot-size dependence shown in
Fig.~\ref{fieldNsize}, by the intrinsic and extrinsic mechanisms,
agrees well with the simple form given in Eqns.~(\ref{separation}) and (\ref{size_field_depend}). For
a small QD or in a small magnetic field,  the
extrinsic nuclear interaction dominates, while the
intrinsic nuclear interaction dominates otherwise. When
the two mechanisms are comparable, the single-system FID begins with
$e^{-t^2}$ behavior and may cross over towards $e^{-t^4}$ decay as
time increases (e.g., see the curve at $B_{\rm ext}=12$~T in
Fig.~\ref{FID}(b)). The timescale of single spin FID ranges from
$0.1~\mu$s to $10~ \mu$s depending on the dot size and external
magnetic field, which agrees well with the experimental observation
of $3 \mu$s by Greilich {\it et al.}~\cite{Greilich_lock}.

\begin{figure}[h]
\begin{center}
\includegraphics[width=10cm]{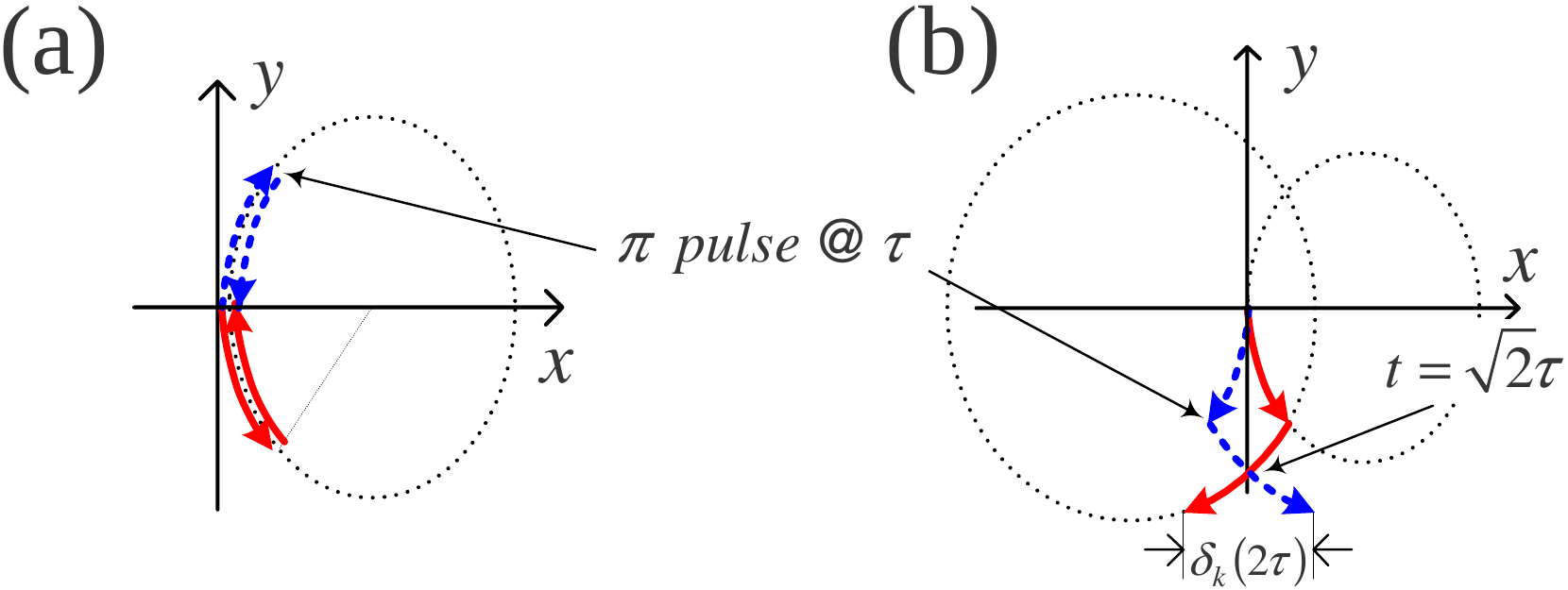}
\caption{(a)and (b) Evolution of the pair-excitations under the single-pulse control,
 driven by the extrinsic and intrinsic nuclear spin interactions, respectively.
 The red solid (blue dashed) trajectories denote the pair
evolutions conditioned on the electron spin state $|+\rangle$
($|-\rangle$).} \label{Bloch_1pulse}
\end{center}
\end{figure}

\begin{figure}[b]
\begin{center}
\includegraphics[width=12cm]{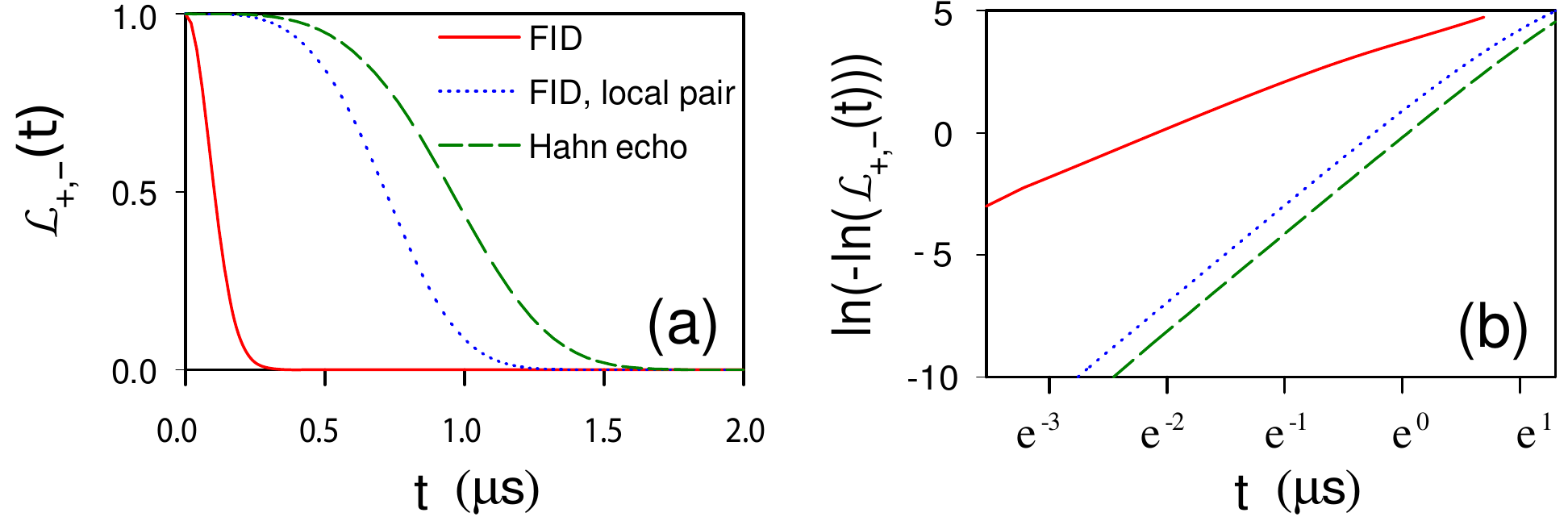}
\caption{(Extracted from Ref.~\cite{Liu_spinlong}) (a) Comparison of the
Hahn echo (dashed green line) and the FID (solid red line) signals.
The FID signal is also shown with the extrinsic hyperfine-mediated
pair-flips neglected (dotted blue line). (b) The logarithm plot of
(a). The QD is as in figure~\ref{FID}, with $B_{\rm ext} =
2$ tesla.} \label{HahnFID}
\end{center}
\end{figure}

In the spin echo scenario, as the electron spin is flipped by the
$\pi$-pulse, the transition amplitude by extrinsic nuclear
interaction $A_k$ and the hyperfine energy cost $E_k$ for each
pair-flip will change sign after the pulse. Thus, the
pair-excitations by the extrinsic hyperfine-mediated nuclear
coupling will reverse their precession after the pulse and return to
the origin at $t=2\tau$, disentangling the electron spin and the
pair excitations (see Fig.~\ref{Bloch_1pulse}(a)). So the
decoherence driven by the extrinsic hyperfine-mediated coupling is
largely eliminated in the spin-echo configuration as shown by the
calculation in Fig.~\ref{HahnFID}~\cite{Yao_Decoherence,Espin_HF_2_deSousa}. For
the pair-excitations driven by the intrinsic coupling, the conjugate
Bloch vectors will switch their precession axes
which also reverse the entanglement to some extent but no full recovery is
obtained at the echo time (see Fig.~\ref{Bloch_1pulse}(b)). Finally, the
electron spin coherence at the echo time is derived as,
\begin{eqnarray}
{\mathcal L}_{+,-}(2\tau) \cong \prod_{k\in K_B} e^{-{2 \tau^4
E_k^2B_k^2} {\rm sinc}^4\left(h_k^B\tau/2\right)} \label{d-echo}.
\end{eqnarray}
Similar to the analysis for single system FID, the spin echo signal
begins with the short-time behavior as $e^{-(2\tau /T_{\rm H}^{\rm sh})^4}$.

The ensemble spin echo profile is numerically calculated and compared
with the single-system FID for a typical QD in Fig.~\ref{HahnFID}.
While it has been a common practice to equal the spin echo decay time $T_H$
to the single-system FID time $T_2$~\cite{Espin_HF_dipole_1_DaSSarma,Witzel:2006}, the
two timescales can in fact be significantly different since the bath dynamics
is modified by the pulse control of the electron spin. And the spin echo decay and the single-system FID
follow different temporal behavior [Fig.~\ref{HahnFID}].
The spin echo decay time of $\mu$s from
calculation~\cite{Yao_Decoherence,Liu_spinlong} is in agreement with the
Hahn echo measurements by Clark et al~\cite{Yamamoto_echo} for impurity spins
in GaAs and by Petta et al~\cite{Marcus_T2} for gate-defined dot in GaAs.

\subsection{Coherence restoration and protection in the nuclear spin bath \label{subsec_spinQubit_restoration}}

Protection of the electron spin coherence by active physical
control is desired which can result in a better physical qubit
before the informatic approaches of quantum error correction may be
implemented. This is indeed possible for single electron spins in
interacting nuclear spin baths.

\begin{figure}[b]
\begin{center}
\includegraphics[width=12cm]{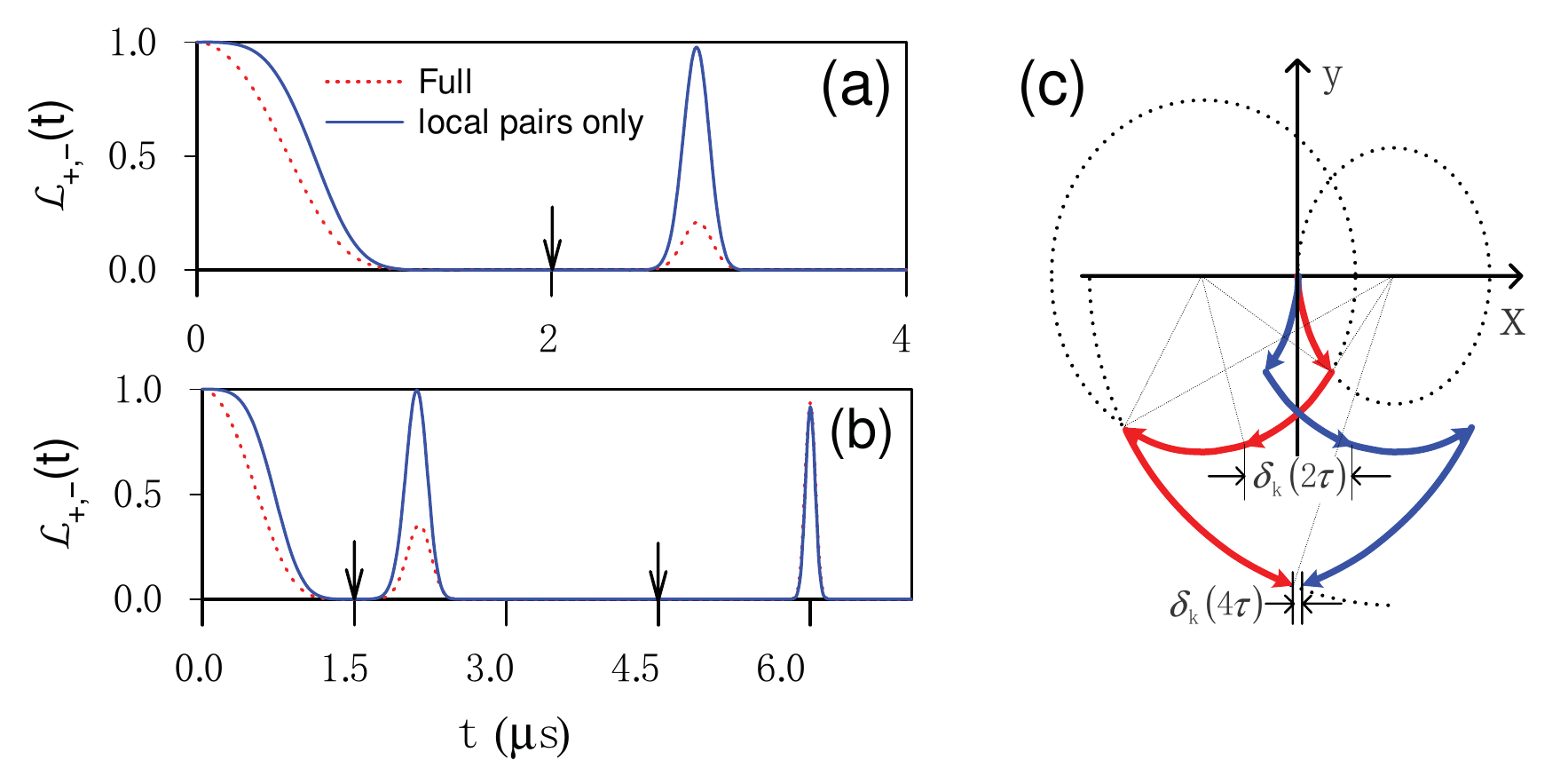}
\caption{(Extracted from Refs.~\cite{Yao_DecoherenceControl}
and~\cite{Liu_spinlong}). (a) The electron spin coherence under the
control of a short $\pi$-pulse applied at $\tau = 2~\mu$s (when the
FID signal has vanished), the recoherence at $\sqrt{2}\tau$ is
pronounced while no signal survives at the echo-time $2\tau$. (b)
The electron spin coherence under the control of a Carr-Purcell
pulse sequence. The arrows indicate positions of the $\pi$ pulses.
The solid blue (dotted red) lines are calculated with (without)
including the extrinsic hyperfine-mediated pair-flips. The QD is the same
as in figure~\ref{FID} with $B_{\rm ext} = 10$ tesla.
Inhomogeneous broadening is excluded. (c) The projection of the
Bloch vector trajectories to the pseudo $x$-$y$ plane for
intrinsically driven pair-excitations under the 2-pulse Carr-Purcell control.}
\label{1_2_pulse}
\end{center}
\end{figure}

By a sequence of $\pi$-rotations of the electron spin, the ensemble
dephasing by inhomogeneous broadening as well as decoherence by {\it
extrinsically} driven nuclear pair-dynamics is efficiently removed
at the classical spin echo time, as shown by the previous
discussions. Nuclear pair-dynamics driven by the {\it intrinsic}
interactions is also affected by such control
as shown in Fig.~\ref{Bloch_1pulse}(b): the two separated trajectories meet
again sometime after the $\pi$-pulse. This intersection signals
the disentanglement of the electron from the pair-excitation.
Surprisingly, even though different pair-excitations have very
different precession frequencies $h_k$, the trajectory separation
$\delta_k$ is eliminated for all local pairs in the leading order of
$B_k t$ at $t=\sqrt{2} \tau$. This leads to a recovery of the
electron spin coherence as illustrated by numerical evaluation shown
in Fig.~\ref{1_2_pulse}. Remarkably, even when the electron spin is
flipped after the coherence has completely vanished in single-system
dynamics, the coherence may be well recovered at time $\sqrt{2}\tau$
whereas no coherence is visible at the conventional spin echo time
$2 \tau$~\cite{Yao_DecoherenceControl}. Thus, in this context, the
decay of Hahn echo does not mean the irreversible lost of coherence
due to the nuclear interacting dynamics. It is simply because the
classical spin echo time for phase refocusing in ensemble does not
respect the quantum behavior of the interacting dynamics in a
mesoscopic bath.  When the extrinsic nuclear interaction is
significant, the $\sqrt{2}\tau$ echo can also be weakened by the
non-local pair-dynamics (see Fig.~\ref{1_2_pulse}).

\begin{figure}[b]
\begin{center}
\includegraphics[width=12cm]{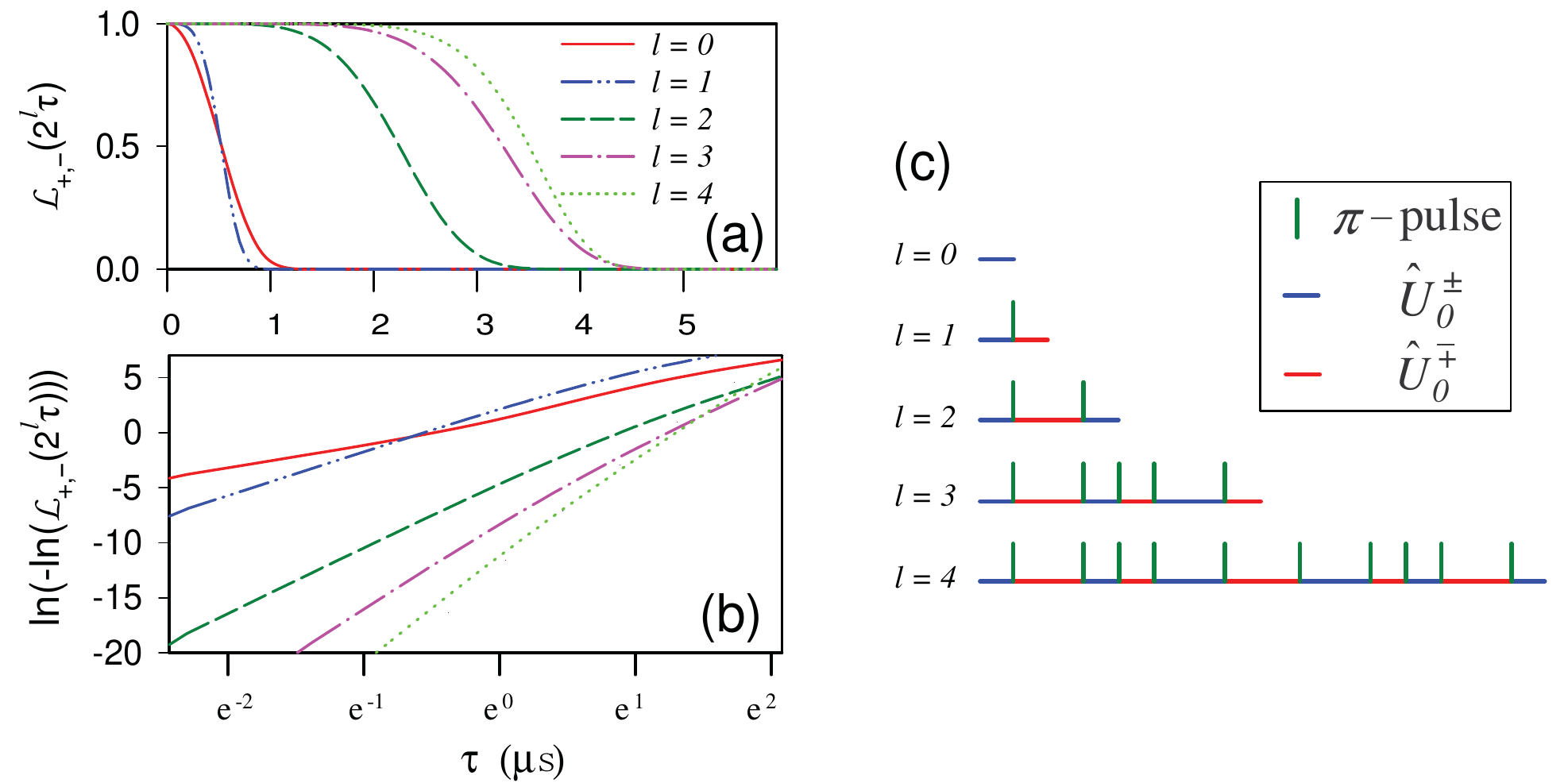}
\caption{(Extracted from Ref.~\cite{Liu_spinlong}) (a) The electron spin coherence under the
$l$th order concatenation control, $l = 0, . . . , 4$, as functions
of the pulse delay time $\tau$. (b) The logarithm plot of (a). The
QD is the same as in Fig.~\ref{FID} with $B_{\rm ext} = 10$ tesla.
Ensemble average is taken at $T=1$ K. (c) Concatenated sequences of
$\pi$-pulses flipping the electron spin, represented by vertical
bars.}
 \label{concan}
\end{center}
\end{figure}

The $\sqrt{2}\tau$ coherence echo is observable when the ensemble
factor $\mathcal{L}^{(0)}_{+,-}$ has a timescale longer or
comparable to the single spin $T_2$ time. This is possible with the
narrowing of inhomogeneous distribution by nuclear state
preparation~\cite{Takahashi:2008PRL,Ramon_ZamboniEffect,Reilly:2008Science,
Levitov_selfdnp,Burkard_NuclearPreparation,Yao_doubledotsqueeze,
Greilich_nuclearfocusing,Steel_locking,Vandersypen_NuclearLocking,
Giedke_spinmeasure,Burkard_spinprep,Klauser_spinmeasure} (see Sec.~\ref{subsec_spinQubit_summary}).
Furthermore, the unusual echo at
$\sqrt{2}\tau$ turns out to have a more general occurrence in other
echo processes. Months after its first
prediction~\cite{Yao_DecoherenceControl,Liu_spinlong}, such echo
behavior is observed in NMR experiments in a $^3$He gas undergoing
Brownian motion in a magnetic field gradient~\cite{Zanker_echo}.

{\em Disentanglement from bath} could be the guiding principle for
coherence protection with reduced overhead when the bath dynamics
is more or less understood. In the present case of protecting
electron spin from the nuclear spin bath, we need a pulse
sequence to produce a time where the decoherence from all three
sources can be removed. We give here a solution which is a
two-pulse control.
Fig.~\ref{1_2_pulse}(c) shows that, after a
second $\pi$ pulse at $3\tau$, the two conjugated paths
corresponding to the electron $|\pm\rangle$ states, driven by the
intrinsic nuclear interaction, cross again at $4\tau$, coinciding
with the secondary spin echo time for the other two causes. The
electron spin is thus disentangled from entire nuclear spin bath
to the leading order at $t=4 \tau$ (see numerical evaluation in
Fig.~\ref{1_2_pulse}(b)).

The power of concatenation design of pulse sequences has been
shown in the context of dynamical decoupling of quantum systems
from baths~\cite{concatenation_Lidar}. Similarly, the control
of quantum system for disentanglement from the bath may be
enhanced by concatenation design. We notice that the pair
evolution with the two-pulse control of the electron spin can be
constructed recursively from the free-induction evolution
$\hat{U}^{\pm}_0$, by the concatenation, $\hat{U}^{\pm}_l =
\hat{U}^{\mp}_{l-1}\hat{U}^{\pm}_{l-1}$,
$l=1,2$~\cite{Yao_DecoherenceControl,Liu_spinlong}. The process
can be extended by iteration to any level as shown in
Fig.~\ref{concan}(c). Disentanglement from local and non-local
pair-dynamics both occur at $\tau_l\equiv2^l \tau$ coinciding with
the classic spin echo. The decoherence is reduced by an order of
$b^2\tau_l^2$ at $\tau_l$ for each additional level of
concatenation till saturation at the level $l_0\approx -\log_2(b\tau)$~\cite{Liu_spinlong}. Hence, the
coherence echo magnitude scales with the echo delay time according
to $\exp\left(-(\tau_l /T_l)^{2l+2} \right)$ as shown in
Fig.~\ref{concan}(b), suggesting that short-time decoherence can
by arbitrarily suppressed with additional levels of concatenation.
Numerical calculation further shows that a proper level of
concatenation allows the protection of electron spin coherence by
pulse sequences with interpulse interval as large as $\sim \mu$s
(see Fig.~\ref{concan}(a)).

The concatenated control of the decoherence can also be optimized in terms of
the number of pulses so that the control errors due to imperfections in the
controlling pulses is minimized. The invention of the pulse sequences with
the minimum possible number of pulses to suppress the short-time decoherence
to a given order of the pulse delay time is due to Uhrig~\cite{Uhrig2007UDD}
in considering the qubit decoherence in a non-interacting boson bath. Later, the Uhrig decoherence control
was conjectured~\cite{Lee2008_PRL} and proved~\cite{Yang2008PRL} to be universal regardless of the bath Hamiltonian.
The Uhrig pulse sequences may also be interlaced with the concatenated pulse sequences
for control not only the pure dephasing but also the longitudinal spin relaxation~\cite{Uhrig2009PRL,LidarQDD}.
All these advances help to clear the obstacle of the qubit decoherence in solid environment.

\subsection{Summary: Quantum dot opportunity \label{subsec_spinQubit_summary}}

The three-dimensional confinement in QDs leads to the
quantized electronic and excitonic energy levels. As a consequence,
a QD resembles an atom in terms of the discrete energy levels
with long coherence times and well defined optical transition
selection rules, although it is essentially a mesoscopic system. The
atom-like electronic and optical properties have been well
established by experiments in the past
decades~\cite{ZrennerJChPh2000,Petroff_SAD2,Gammon_PRB2002,
Bayer_Linewidth_QD,Shih_mollow_g1,Gurudev,Imamoglu_coolingNuclear,Steel_spinholeburning},
including the initial demonstration of quantum coherent
control~\cite{Steel_Rabi,Shih_Rabi,Kamada_Rabi,Zrenner_rabi,Steel_Biexciton,Berezovsky08}.

Experiments also showed that, for single electrons in QDs,
spin polarization along an external magnetic field can be
preserved for a sufficiently long time ($T_1\sim 20$~ms reported
for InAs self-assembled dot~\cite{Finley_spin_memory}). A major
concern has been the inevitable cause of transverse decoherence
by the lattice nuclear spins in III-V materials. Due to the
extremely small energy scales even in a strong magnetic field,
the nuclear spin bath is of high entropy at experimentally
achievable temperature. Ensemble dephasing time $T_2^{\ast}\sim
1-10$~ns in different types of QDs~\cite{Gammon_t2star,T2star_Marcus}, spin echo decay time
$T_H$ in the order of $\mu$s in gate-defined QDs~\cite{Marcus_T2,Koppens_Rabi}
and for impurity spins in GaAs~\cite{Yamamoto_echo}, and single spin dephasing time
$T_2\sim 3~\mu$s in self-assembled QDs~\cite{Greilich_lock} have been extracted from
various experimental approaches, all in agreement with the theoretical analysis of nuclear spin
baths~\cite{Merkulov_2002,Yao_Decoherence,Liu_spinlong}. As
compared with the spin echo decay, FID in single-system dynamics is
subject to the additional cause of decoherence from the extrinsic
mechanism, which is dependent on the external magnetic field. In
a moderate field of $1$ tesla, $T_2\sim 100$~ns by
theory~\cite{Yao_Decoherence,Liu_spinlong},
is an order smaller than the spin echo decay time. Polarization of the nuclear spin bath
can partially suppress this decoherence channel~\cite{Imamoglu_coolingNuclear,Espin_HF_Hu,Bracker_pump}.
However, a substantial increase of electron spin coherence time would
require a nuclear polarization over $99\%$, which can be
extremely difficult. Current experimental capability on nuclear
spin polarization in III-V QDs is in the order of
10-70\%~\cite{Gammon_t2star,Imamoglu_nuclearpumping,tartakovskii:026806,eble:081306,
PhysRevLett.94.116601,PhysRevLett.92.256803}.

For spin qubits used as quantum memory, dynamical decoupling schemes
can be used to decouple system and environment for coherence
protection~\cite{DD_VL,DD_VKL1,DD_VKL2,DD_VKL3,concatenation_Lidar},
which typically requires frequent manipulations of the spin qubits.
As compared with these schemes designed to deal with a general
environment, the available solution to the dynamics of the nuclear
spin bath makes possible a different approach aiming at
disentanglement of the system from the bath only (their decoupling
is a sufficient but not necessary condition for
disentanglement)~\cite{Yao_DecoherenceControl,Liu_spinlong}.
Numerical calculations on realistic QD systems show that
designed sequence of $\pi$-pulses on the electron can efficiently
preserve its spin coherence up to $\sim 100~\mu$s in both
single-system and ensemble dynamics. As the interval between
adjacent pulses can be as long as $\mu$s, this disentanglement
approach not only substantially reduces the overhead but also avoids
the problem of unwanted heating of the system from the frequent
manipulations required by the dynamical decoupling schemes.

In the quantum logic control of the spin qubit, $T_2^{\ast}$ is the
shortest timescale one encounters. Even with the ultra-fast optical
manipulation timescale $T_{\rm op} \sim 10~$ps, $T_2^{\ast} \sim
1-10~$ns from FID in a thermal nuclear spin bath is not sufficient
to satisfy the current fault-tolerant threshold $T_{\rm coh} /
T_{\rm op} \sim 10^3-10^4$~\cite{Shor95_QEC,Steane_QEC0,bdsw,Laflamme_QEC,Gottesman_QEC,Steane_QEC}.
Considering the efficiency of spin echoes in eliminating the
dephasing by inhomogeneous broadening, the combination of the
desired control action with these coherence protect operations could
offer a promising route towards fault-tolerant quantum information
processing. Efforts are being devoted towards the search of
universal logic control strategies of coherence protected
qubits~\cite{DD_VKL2}.

An alternative approach is to pre-prepare the nuclear spin bath so that the
nuclear field inhomogeneous broadening can be squeezed below its thermal
value, as suggested by the various nuclear state preparation
schemes~\cite{Ramon_ZamboniEffect,Levitov_selfdnp,Burkard_NuclearPreparation,
Yao_doubledotsqueeze,Vandersypen_NuclearLocking,Giedke_spinmeasure,
Burkard_spinprep,Klauser_spinmeasure}.
The resultant enhancement on the $T_2^{\ast}$ time can last for seconds or
even longer as nuclear spin relaxation is extremely slow.
For optically controllable electron spin in self-assembled dot,
enhancement of $T_2^{\ast}$ up to microsecond by nuclear state
preparation has been achieved experimentally for a spin
ensemble~\cite{Greilich_nuclearfocusing}, and very recently
for a single spin~\cite{Steel_locking}. For electrically controllable
spin qubits, enhancement of $T_2^{\ast}$ of a coupled spin pair to
microsecond was also reported in a double-dot configuration~\cite{Reilly:2008Science}.

When the inhomogeneous broadening effect is suppressed (by spin echo
or bath state preparation), dephasing by nuclear interacting dynamics is the
limiting factor. Experiments~\cite{Greilich_lock} and theories~\cite{Liu_spinlong}
show that, in FID, $T_2 \sim 0.1 - 10~\mu$s for typical self-assembled QDs
under a moderate magnetic field, which is sufficiently long to satisfy the
fault-tolerant threshold. And the spin coherence time can be still elongated
further by dynamical decoupling.

As an alternative, the heavy hole spin in a positively charged
QD can also play the role of a qubit carrier. For the
$p$-type hole bands, the contact hyperfine interaction vanishes
and the hole spin is coupled to nuclear spins through the dipolar
hyperfine interaction. Theoretical studies shows that the
hole-nuclear hyperfine coupling strength is about one order
weaker as compared with the electron-nuclear hyperfine
coupling~\cite{Chamarro_holespin,Loss_holespin}. Furthermore,
unlike the isotropic electron-nuclear hyperfine interaction, the
hole-nuclear coupling is strongly anisotropic. In the absence of
heavy-light hole mixing, the hole-nuclear hyperfine interaction
is Ising-like with coupling only between spin components along the
growth direction~\cite{Chamarro_holespin,Loss_holespin}. Finite
hole-mixing effects can lead to the coupling between the hole and
nuclear spin components perpendicular to the growth direction. In
the absence of magnetic field, pump-probe and time-resolved
photoluminescence experiments on p-doped self-assembled dot have
revealed hole-spin ensemble dephasing time of 14~ns~\cite{Chamarro_holespin}.
Most significantly, in a magnetic field perpendicular to the growth direction,
coherent population trapping has been observed for a single $p$-doped dot which
suggests a transverse dephasing time longer than 100~ns~\cite{Warburton_holespin}.
These experimental findings are consistent with the theoretical studies of the hole-nuclear
hyperfine coupling~\cite{Chamarro_holespin,Loss_holespin}. Since
the $p$-doped QDs have similar energy level-schemes and
transition selection rules to the $n$-doped ones, hole spins in
QDs may be manipulated using optical schemes similar to
those for electron spins. For example, high fidelity hole spin
initialization has already been demonstrated~\cite{Warburton_holeInitialization},
using similar optical pumping schemes previously adopted for electron spin
initialization~\cite{Imamoglu_QDSpinPrep,Xu07PRLspinIni}.

Another kind of solid-state electron spin systems which are under exciting
development is NV centers in diamond. An NV center in diamond is a defect with a C-C bond
substituted with a negatively charged N atom, which has a spin-1 at the ground state.
NV center spins are a promising candidate for quantum computing for the following
virtues~\cite{Hason08NatureNVreview}: First, as deep-level defects, they
have chemical and thermal stability; Second, the spin-orbit coupling in the light
C and N atoms is very weak, so the spin decoherence by phonon scattering is negligible
even at room temperature~\cite{Kennedy:2003,Gaebel2006_NaturephysicsNV};
Third, the natural abundance of isotopes with non-zero spin
(C-13) is only about 1\% and also the hyperfine interaction between the center spins
and the bath nuclear spins is mostly dipolar which is highly anisotropic and decays
rapidly with the distance. Thus the electron spin decoherence by the hyperfine coupling
is very slow (coherence time $>50$~microseconds in natural samples~\cite{Kennedy:2003}
and $\gtrsim$~milliseconds in C-13 depleted diamond~\cite{Bala09NMatNVisotope});
Fourth, the material is optically transparent and the centers are optically active,
feasible for optical access~\cite{Jelezko:2002,Epstein05NPhysNVN,
Tamarat:2006,Santori:2006,Hanson2006_PRLNV} and coupling with cavities or
waveguides~\cite{Park:2006,Schietinger08NanoLettCavityNV,Fu08APLwaveguideNV,Barth09OLettCavityNV}.
The proposal of quantum computing with diamond defects~\cite{Wrachtrup:2001} exploded to a hot research
field after the experimental demonstration of electron spin Rabi oscillation~\cite{Jelezko2004_PRLNV} and
two-qubit gates for coupled electron and nuclear spins~\cite{Jelezko2004_PRL93} in single NV centers.
Awchalom group demonstrated coherent coupling between a ``bright'' NV center and
a ``dark'' nitrogen center~\cite{Epstein05NPhysNVN}. Gaebel {\it et al} realized strong coupling
between an NV spin and a nitrogen spin at room temperature~\cite{Gaebel2006_NaturephysicsNV}.
Lukin and colleagues showed spin echo of an NV spin and observed coherent coupling between the electron spin
and nuclear spins nearby~\cite{Childress2006_ScienceNV}. In 2007, Lukin group managed to isolate
and control an NV spin and a strongly coupled nuclear spin~\cite{Dutt2007_ScienceNV}.
In 2008, Neumann {\it et al} claimed multipartite entanglement among C-13 nuclear spin near an NV
center~\cite{Newmann08ScienceNVentangle}. Most recently, Lukin group and Wrachtrup group
independently demonstrated readout of an NV qubit improved by repetitive retrieval of
proximal nuclear spin ancillae~\cite{Jiang09ScienceNVreadout,Steiner10NVreadout}.

In the remaining part of this review, we will discuss the optical
manipulation schemes for QD electron spin qubits, and
we expect that most of them shall be applicable for hole spins
and NV centers as well, except the distinctive hyperfine effects on the
optical processes of the hole spins and of the NV center with its proximity nuclear spin.

\section{Physical Structure}

\subsection{Local nodes}

A local node is composed of a few QDs. The specific QD
systems of interest include In$_{1-x}$Ga$_{x}$As self-assembled
QDs and GaAs fluctuation QDs. These III-V compound semiconductors
have direct band-gaps and thus are suitable for optical control.
A fluctuation QD is formed by width fluctuation in a narrow
quantum well grown with certain procedure (such as interruption
for introducing interface roughness). This kind of dots has
lateral confinement size ($\gtrsim 10$~nm) much larger than the
growth direction size ($\lesssim 5$~nm) and the lateral
confinement potential is shallow (usually in the order of a few
meV). Thus a fluctuation QD would not host many bound electronic
states and often just one, which is still subject to ionization
due to thermal or optical excitation. Nonetheless, the loose
lateral confinement makes optical transitions in a fluctuation QD
well characterized by selection rules resulting from conservation
of angular momentum with respect to the growth direction. The
large size of fluctuation QDs also makes the dipole moment for
inter-band transitions to be large and therefore enhances the
optical coupling which is useful for strong coupling in cavity
quantum electrodynamics (QED) and has been utilized in
demonstrating optical control of excitonic
qubits~\cite{Steel_Rabi,Shih_Rabi}. A self-assembled QD is formed
by the spontaneous nucleation of one material (such as InAs) on
the surface of a substrate (such as GaAs) which has slightly
different lattice constant. Such a QD is relatively small (with
lateral size $\sim 10$~nm) and deep confinement potential (in the
order of hundreds of meV) which is defined by the offset between
the band-edge of the QD material and that of the substrate. As a
result, a self-assembled QD could host quite a few stable bound
states, providing extra flexibility for quantum control. There is
no obvious reason to exclude other types of QDs such as those
formed in II-VI materials and II-VI nanocrystals~\cite{IIVIQD_blink}, though they are
less comprehensively studied in experiments mostly because of
technical difficulties such as the requirement of UV lasers,
strong charge fluctuation around QD surfaces due to low material
mobility, and complex defects and impurity centers.
The optical
control schemes discussed in this review can be applied to the NV
centers in diamond with some modifications.

In classical electronic computers, the physical layout of logic circuits
is planarly extended. It is not hard to expect the planar layout to
be used in a quantum computer. In self-assembling growth processes,
vertically stacked QDs may be formed to several layers~\cite{Bayer01ScienceQDs}.
The vertical structure, however, is not extendable and furthermore,
limits the accessibility by optical pulses.
Planar distribution is naturally formed both in self-assembled QDs and in fluctuation
QDs. Formation of QD molecules or  arrays due to
lateral coupling is possible~\cite{Finley05PRL_controlQDs,Gammon06SciQDs}.
With proper growth art, clusters of QDs with certain patterns may be
fabricated~\cite{Songmung03APL,Barth05NanoSurface,SchmidbauerQDlattices}.
Remarkably, NV centers in diamond may be implanted by ion beams with position
precision in the order of 10~nm~\cite{Meijer:2005,Meijer:2006}. Having such technology within
the scope of our consideration, we assume a local node in a quantum computer
composed of a few (usually fewer than 10) QDs (or impurities), laterally distributed and coupled.
The control of electron spins in QDs
is to be designed after a local node consisting a cluster of QDs has
been fabricated and characterized. In this sense, the specific layout
of a local node, which could be a ring, a line, or any other graphs naturally formed,
does not make essential difference. But a linearly displaced array would be preferable
for its simplicity in coupling and practicality in manufacturing.

A scalable physical structure of a QD-based quantum computer
should have the following features:
(1)  the QDs are placed in an extendable layout;
(2)  the QDs are connected so that electron spins can be coupled to a common photonic or electronic state; and
(3)  the QDs are individually accessible so that the electron spins are individually controlled.
On the one hand, the spatial resolution of near-field optical devices is still not high enough to
identify each QD in a cluster. On the other hand, within the limit of current technologies,
it is impossible to control the growth of QDs so that they are almost identical.
Different sizes and shapes of the QDs would make the exciton transition energies in different QDs
different. In this
way, the near-field optics and the fingerprint transition frequencies of different QDs may be combined
together to individually address each dot or to selectively couple a pair of them.
The coupling between spins in general is mediated by
virtual tunneling between different QDs which may be activated
by virtual optical excitation of excitons in the presence of extra
electrons which bear the spins (More details will be discussed
in later sections). For NV centers in diamond, such mechanisms
are not yet considered, but other schemes may be applicable such as
coupling through virtual excitation of cavity modes~\cite{Imamoglu_CQED_Spin},
or hyperfine interaction with nuclear spin baths~\cite{Wrachtrup:2006}.

\begin{figure}[b]
\begin{center}
\includegraphics[width=8cm]{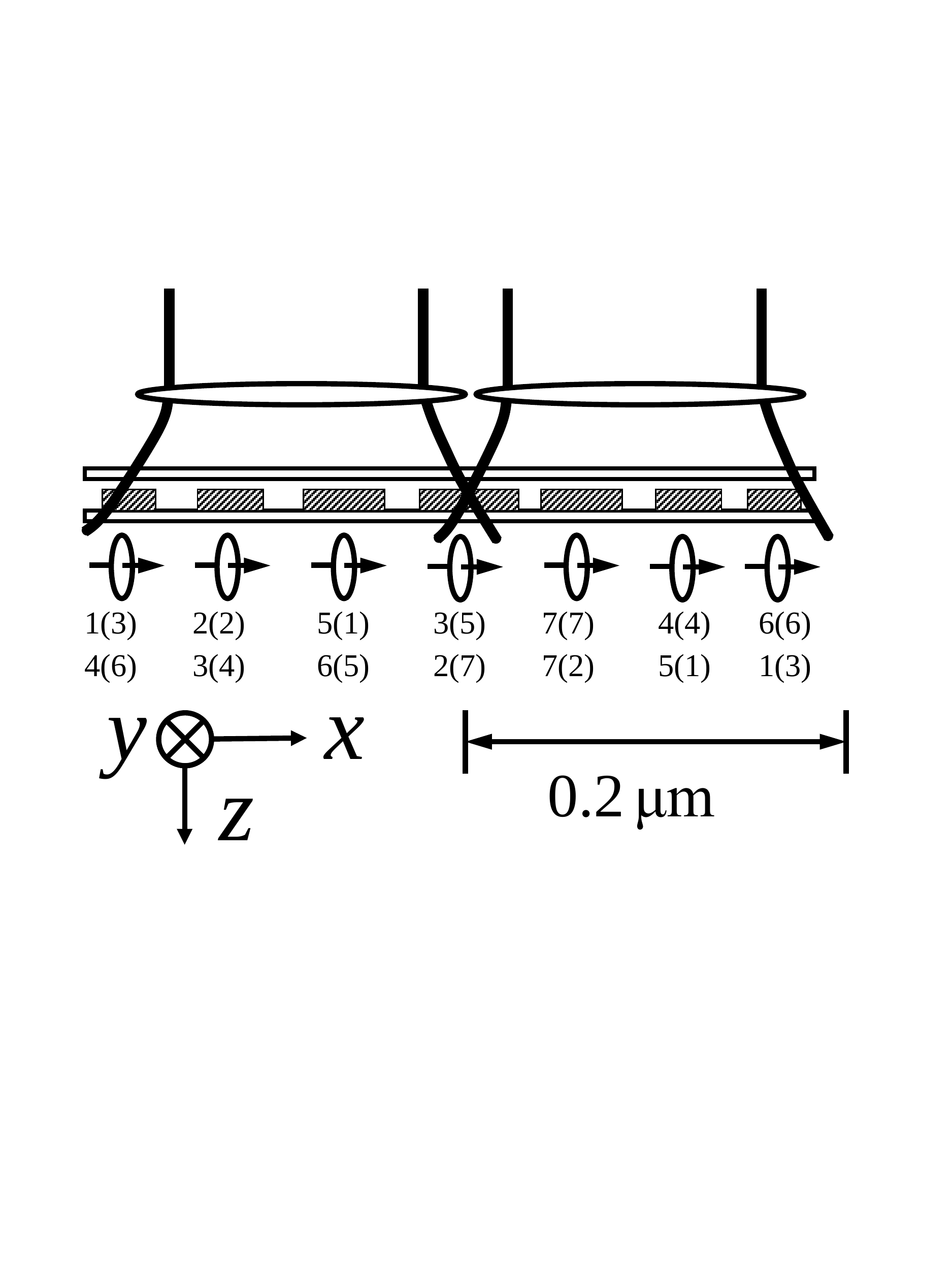}
\end{center}
\caption{
The physical structure of a local node consisting of seven electron
spins in a QD cluster. Each QD is identified by a near-field micro-lens as
well as by its characteristic transition energies. The  numbers
are used to label the qubits for factorizing 15 with the seed
number $a=4$ (top row) and $a=13$ (bottom row), respectively.
The numbers in the parentheses label the qubits at the end of
computation (see Sec.~\ref{factorize} for details).} \label{structure}
\end{figure}

\begin{figure}[b]
\begin{center}
\includegraphics[width=8cm]{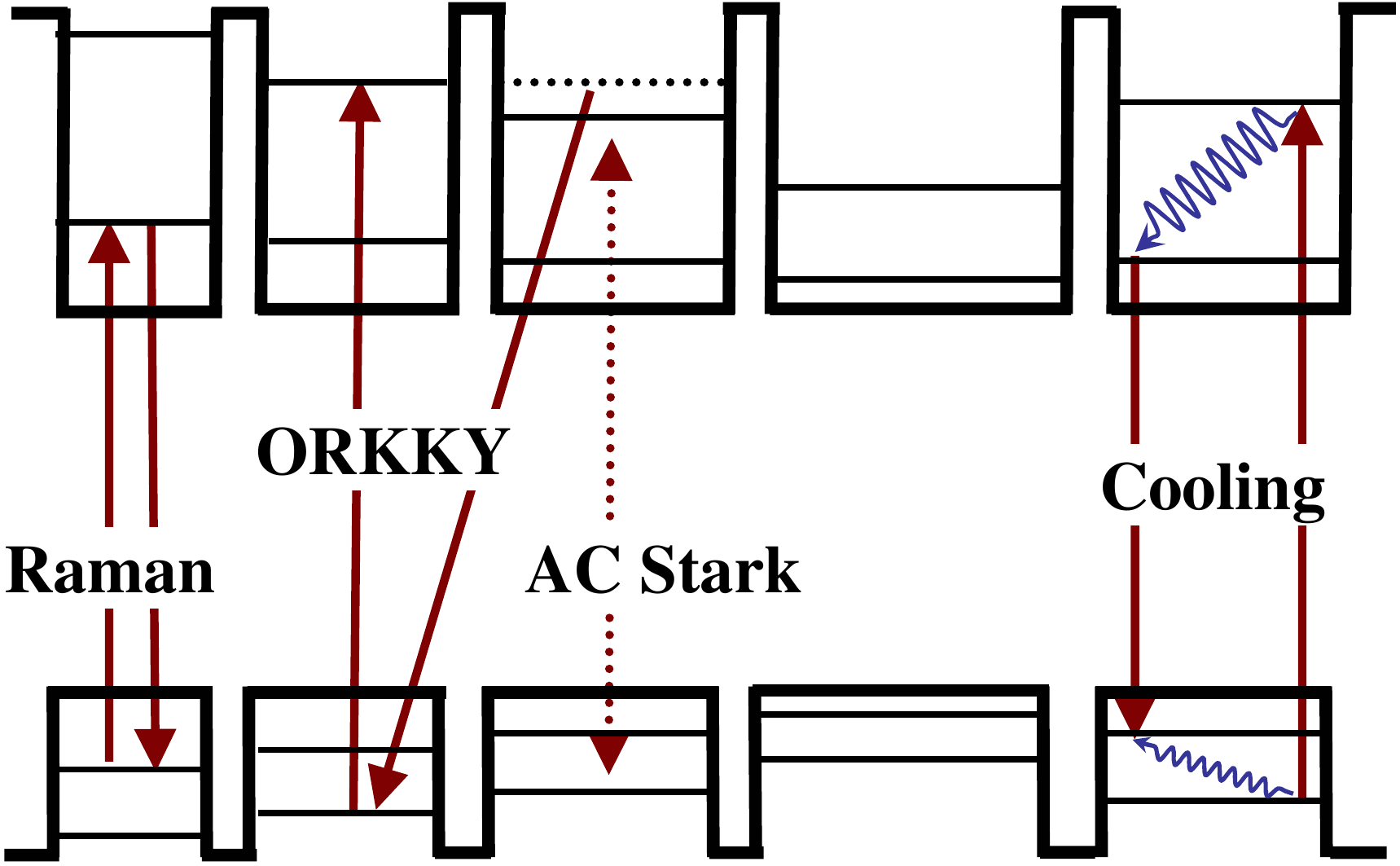}
\end{center}
\caption{The energy diagram of various optical processes for
quantum gates and initialization: ``Raman'' for single-spin control
or optical pumping and measurement of a single spin,
``ORKKY'' for two-spin control, ``AC Stark'' for transient shift of energy levels
to relize selective resonance, and ``Cooling'' for initialization of spins via phonon bath (phonon emission
as wavy lines).}
\label{optical_control}
\end{figure}

We show an example in Figure~\ref{structure} a working module of 7 qubits in a linearly displaced
array of QDs. With the QDs of the size around 50~nm and about 20~nm apart,
the 7 dots may be addressed with two micro-lenses attached to, e.g., optical fibers,
with resolution of about $0.2$~$\mu$m. The QDs addressed by the same optical fiber
are distinguished by their signature transition frequencies.
Using optical pulses with different frequencies and polarizations, each dot and each
adjacent dot pair may be (virtually) excited and various single- and
two-qubit gates and initialization could be realized, as
illustrated in Fig.~\ref{optical_control}. To realize arbitrary
single-spin rotation, a magnetic field is applied along the in-plane-direction (denoted as
the $z$-direction). We assume that only neighboring QDs are coupled, for simplicity. Coupling between spins in
farther separated dots is to be accomplished by recursively using the nearest neighbor coupling,
which increases the number of gates by an amount in the order of the number of dots between the two ends.
To minimize the number of operations between separated qubits, the quantum algorithms
are compiled by using optimum labeling of the qubits. Taking into account the size of the micro-photonic structures
(including cavities and waveguides connecting local nodes), a
working module of about 10 QDs could occupy an area of about $10$~$\mu$m$\times 10$~$\mu$m, so a quantum chip of size
$10$~cm$\times 10$~cm can in theory accommodate $10^8$ qubits.

For the purpose of addressing and optical control of a single QD, several near-field optics
technologies have been available, such as micro-optical masking and micro-fiber optics~\cite{dotSteel}.
For a cluster of QDs to function under optical control, however, there are still non-trivial
technical challenges, including at least fabrication of QD clusters with energy levels and
inter-dot couplings falling in the desired parameter ranges, assembling of micro-lenses
on the surface of QD clusters, and designing, shaping, and controlling complex laser pulses.
The micro-lens technology has been widely used in the digital imaging industry
(micro-lenses of similar sizes are routinely used to focus light into individual pixels
in commercial digital cameras).

\subsection{Distributed structure \label{subsec_distributed_sructure}}

The electron spin qubits are distinguished by the different optical
transition frequencies of the host QDs. Thus, the dot
density shall be rather dilute so that each laser spot contains only
a small number of QDs ($\sim\rm{O}(10)$, considering the
typical inhomogeneous broadening of the excitonic linewidth of
$\sim10-100~$meV, and the requirement of $\sim0.1-1~$meV frequency
separation for optically addressing individual QDs.
Therefore, the optical approach predetermines that a local node can
only have a limited number of qubits. In order to scale up, a
distributed architecture could be a
solution~\cite{DiVincenzo_Criteria_7,Cirac_DistributedQC}. In such
structures, clusters of QD electron spin qubits form
quantum nodes where logical operations can be performed locally, and connections
between clusters are through quantum channels in which the
flying qubits take information from one place to another.

The single photon wavepacket is an ideal candidate as the
carrier of flying qubits, being widely used in quantum cryptography~\cite{BBB92} and
linear optics quantum computation~\cite{LOQC}. The qubit can be
encoded in the photon-number subspace~\cite{StateTransfer_Cirac_Kimble} or
polarization subspace~\cite{LOQC}. While single photon propagation
in free space is un-channeled and inefficient,
optical waveguides in semiconductors and optical fibers provide
directional channels.

In the distributed architecture of optically controlled spin quantum
computation, flying photons in waveguides/fibers are responsible for
integrating the distributed stationary spin clusters into a
globally functioning quantum computer. This requires quantum
interfacing between single electron spins in QDs and single
photons in waveguide. As mentioned in Sec.~\ref{sec-single qubit},
QD electron spins interact with optical fields via the
intermediate states of trions. Such interface at the single photon
level requires strong light-matter interactions. As a QD
has a fixed optical transition dipole moment which is
limited by its size, one way to have such strong light-matter
interaction is to confine photons in optical cavity structures with
small volumes.

\begin{figure}[t]
\begin{center}
\includegraphics[width=12cm]{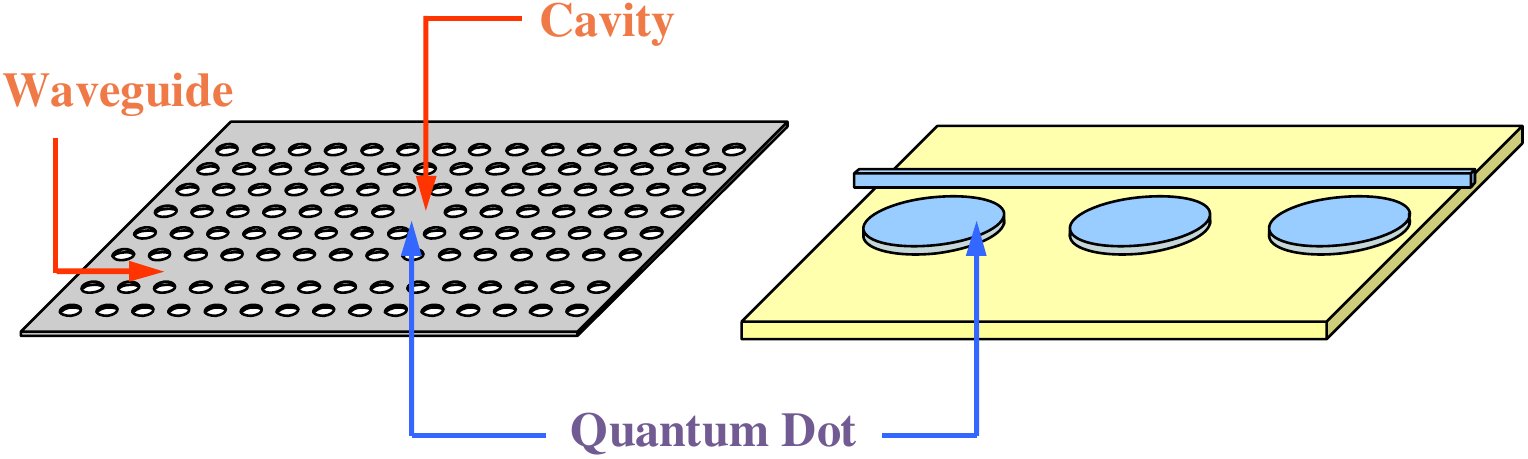}
\vskip 0.5cm
\caption{{{}} Left: coupled cavity and waveguide structure
formed by point and line defects in 2D photonic bandgap crystals.
Right: waveguide coupled micro-disk cavity etched on chip. Layers of
QDs can be embedded in the matrix slab where the cavities
are formed.} \label{DCW_chip}
\end{center}
\end{figure}

\begin{figure}[t]
\begin{center}
\includegraphics[width=12cm]{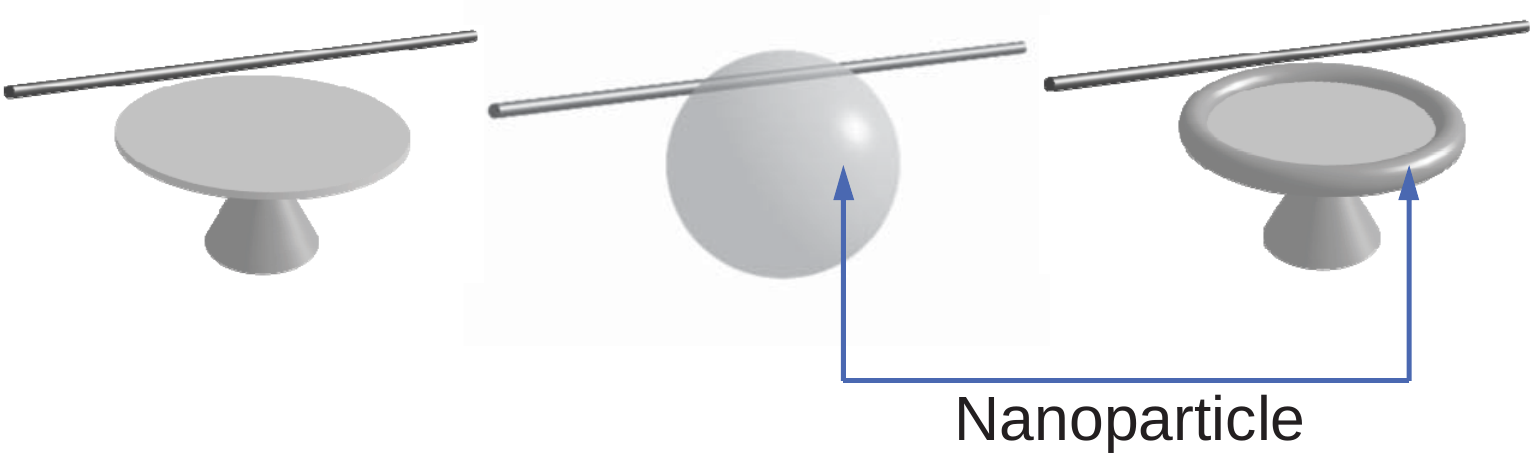}
\caption{{{}} Left: fiber coupled micro-disc cavity.
Middle: fiber coupled micro-sphere cavity. Right: fiber coupled
micro-toroid cavity. Semiconductor nanoparticles, e.g. diamond with
NV centers, can be adsorbed onto the latter two types of cavities.}
\label{DCW_2}
\end{center}
\end{figure}

Micro-cavities can be realized in a number of ways in semiconductor
structures. We list below the essential properties of a few
representative ones:
\begin{enumerate}
\item
{\it Microdisk} -- See Figs.~\ref{DCW_chip} and \ref{DCW_2}
for schematic illustrations of the geometry. Light is
confined by total internal reflection  in the inner wall and
the confined modes are known as the whispering gallery modes.
Quality factor $Q \sim {\rm O}(10^4)$ in III-V materials and $\sim
{\rm O}(10^5)$ in polymer; mode volume $V \sim 6 (\lambda_0 /n)^3$
where $n$ is the refractive index of the material and $\lambda_0$
the wavelength of cavity mode~\cite{vahala_review}. Strong coupling
regime for GaAs fluctuation QD embedded in micro-disk
structure similar to the illustration in Fig.~\ref{DCW_2} has been
achieved~\cite{Bloch_rabi_splitting}. Because of the large dipole moment of the fluctuation dot, the
measured cavity-dot coupling  constant $g_{\rm cav} \sim 0.2 ~$meV signifies strong coupling.
\item
{\it Defect cavity in 2D Photonic Crystal} -- Two dimensional photonic bandgap
crystals are an ideal structure to form a cavity resonator~\cite{2D_photoniclattice,Kress_PRB05}.
Propagation of light in the plane has a
forbidden bandgap for carefully designed periodical arrays of
air-holes drilled on the 2-D slab. As shown in Fig.~\ref{DCW_chip},
by forming a point defect in the 2D array of air holes, light can
be almost perfectly confined in the plane of the slab if its
frequency lies in the forbidden bandgap. The vertical confinement,
achieved by total internal reflection at the semiconductor-air interfaces, is imperfect,
in that light with small in-plane wavevectors can leak out of the
top and bottom. Vertical leakage can be greatly suppressed by proper
engineering of the
defect~\cite{Noda_PhotonicCrystal1,Noda_PhotonicCrystal2}. $Q \sim 6
\times 10^5$ and $V \sim 1.2 (\lambda_0 /n)^3 \sim 0.072~\mu$m$^3$ have been
achieved~\cite{Noda_PhotonicCrystal1,Noda_PhotonicCrystal2}.
Theoretically analysis shows that $Q$-factors greater than $2\times 10^7$ are realizable by optimizing the structure
\cite{Noda_PhotonicCrystal2}. The matrix of a 2D photonic crystal
can be either silicon or III-V compounds~\cite{khitrova_rabi_splitting}. Strong coupling with
single self-assembled InAs QDs has been demonstrated~\cite{khitrova_rabi_splitting,Imamoglu_deterministic_strongcouple},
where $g_{\rm cav} \sim 0.1 $ meV.
\item {\it Micro-pillar} -- Light is vertically confined by distributed Bragg refelector mirrors
and horizontally by total internal reflection. $Q \sim 10^4$ and $V \sim \mu$m$^3$ have been
achieved~\cite{forchel_rabi_splitting,stoltz:031105}. Strong coupling
with single self-assembled InAs QDs has been
demonstrated~\cite{forchel_rabi_splitting,Yamamoto_strongcoupling},
where $g_{\rm cav} \sim 0.1 ~$meV.
\item {\it Epitaxial cavity} -- Vertical confinement is by
distributed Bragg refelector mirrors and horizontal confinement by thickness variations,
similar to the confinement principle of the fluctuation QD~\cite{DotinCavity_Shih}.
$Q = 3 \times 10^4$, $V \sim ~\mu$m$^3$ have been achieved~\cite{DotinCavity_Shih}.
\item {\it Silicon microsphere} -- WGMs are confined by total internal reflection. $Q$ exceeding
$10^8$ and $V \sim 10^3\mu$m$^3$ have been achieved~\cite{WangHL_MCdot,vahala_MCfiberLaser}.
Nanocrystals (such as CdSe nanocrystals~\cite{Woggon_strongcoupling} and
diamond nano-crystals with NV centers~\cite{Park:2006})
deposited on the surface are usually used for coupling with the cavity photons.

\item {\it Microtoroid} -- See Fig.~\ref{DCW_2} for an illustration of the geometry.
WGMs  are confined by total internal reflection. $Q \sim 10^8$ is achieved with principal
diameter $D\sim 100~\mu$m and the minor diameter
$d\sim$~$\mu$m~\cite{Vahala_toroid}. Theoretical analysis
shows the possibility of realizing micro-toroid with $Q$ exceeding
$10^8$ and $V\sim {\rm O}(10) ~\mu$m$^3$~\cite{Vahala_Kimble_mctoroid}.
\end{enumerate}

\begin{figure}[b]
\begin{center}
\includegraphics[width=10cm]{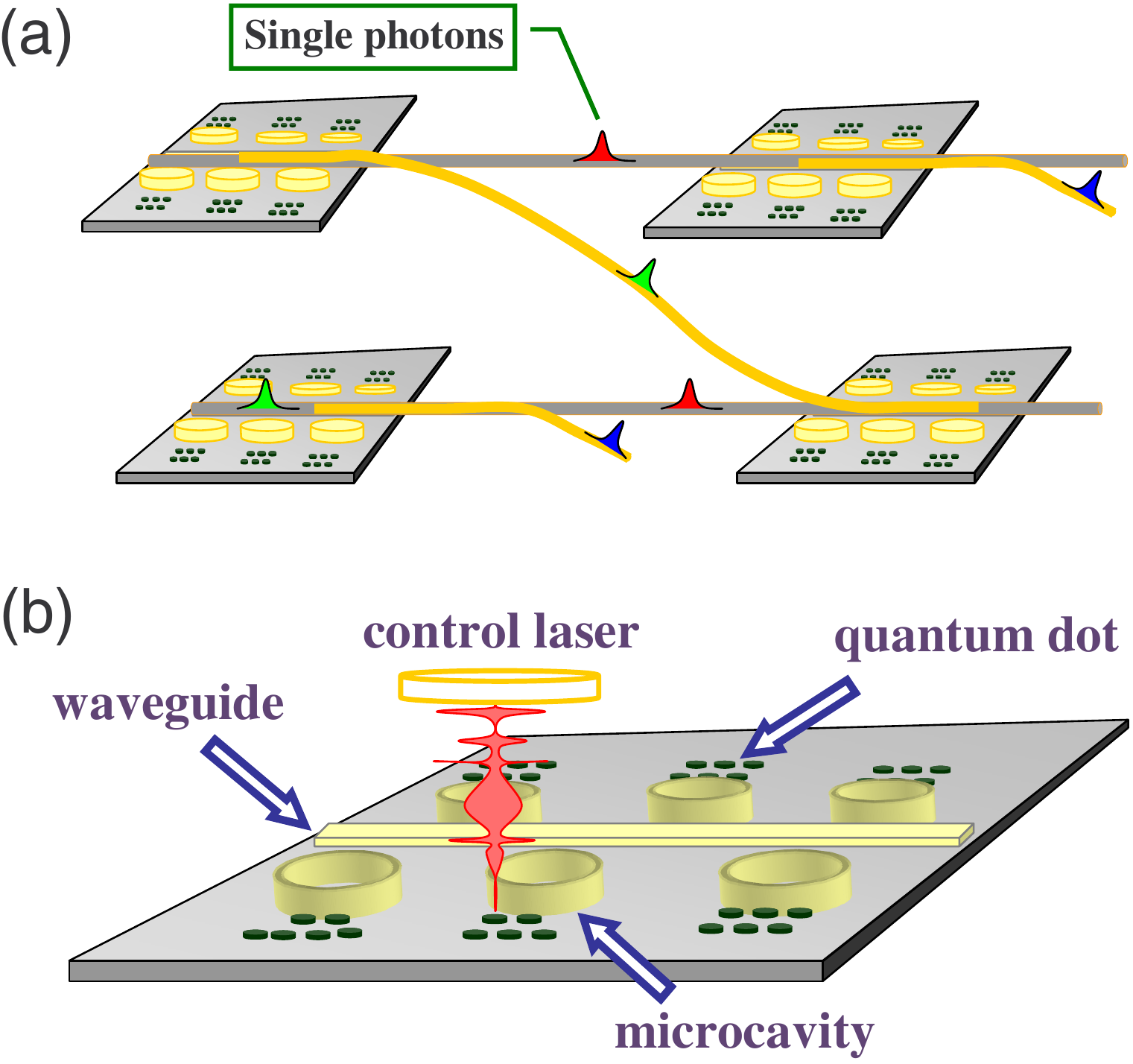}
\caption{Distributed quantum information processing in integrated
semiconductor structures. (a) Schematics of a distributed quantum
computer where communications between computation modules are
mediated by single photons in optical waveguides/fibers. (b) The
spin-based computation modules on a chip, controlled by ultrafast
optics.} \label{distributed_structure}
\end{center}
\end{figure}

The high quality factor allows photons to be confined for a sufficiently
long time inside a cavity and the small mode volume makes possible a large
intra-cavity electromagnetic field from a single photon. Both features are
critical for the strong dot-cavity coupling, defined by the
criteria that a single cavity photon can induce $\gtrsim 2 \pi$ Rabi
oscillation of an excitonic transition in a single dot within the lifetimes of
the cavity photon and the exciton. Strong coupling has already
been realized in several cavity-dot systems~\cite{khitrova_rabi_splitting,Imamoglu_deterministic_strongcouple,
Bloch_rabi_splitting,forchel_rabi_splitting,Yamamoto_strongcoupling,Park:2006,Woggon_strongcoupling}.
With micro-cavity being a playground for strong coupling of a single photon with a single dot,
the interface between a flying photon qubit in a waveguide and a stationary
spin qubit in a QD can be achieved by evanescent coupling
between an optical waveguide/fiber and a micro-cavity containing the QD
(see Figs.~\ref{DCW_chip} and \ref{DCW_2}). A single photon
wavepacket propagating in the waveguide can excite a cavity photon which
then significantly influences the QD spin dynamics, e.g.
through the optical Raman process via the trion states (see
Sec.~\ref{subsec_spinQubit_levels})~\cite{yao_network}. In addition to the exciting
advances in improving the $Q$-factor and reducing the mode volume of
semiconductor micro-cavities for strong coupling, many other
key ingredients towards the construction and control of such a
dot-cavity-waveguide coupled structure have been progressively
achieved in laboratories. These include the high-efficiency coupling
between micro-cavities and optical waveguides/fibers~\cite{microsphere1_vahala,microsphere2_vahala,
vahala_MCfiber,Painter_cavityFiber1,Painter_cavityFiber2,Painter_cavityFiber3,Vuckovic_cavityFiber},
precise control of intra-cavity location of QDs~\cite{Imamoglu_deterministic_strongcouple,Imamoglu_deterministic_couple,thon:111115},
fine-tuning of cavity modes into resonance with a given QD transition~\cite{Imamoglu_deterministic_strongcouple,Imamoglu_deterministic_couple,
Finley_NJP2009,Finley_NJP2009b,rakher:091118}, electrically controllable charging of intra-cavity QDs~\cite{rakher:097403},
and coherent resonant driving of QD excitonic transitions inside a cavity by an external laser~\cite{Shih_mollow_g1,Shih_mollow_g2}.

With the possibility of spin-photon interfacing in
dot-cavity-waveguide structures, we are able to outline the
construction of a distributed architecture for scalable quantum
information processing in an integrated semiconductor platform
composed of QDs, optical micro-cavities, and optical
waveguides/fibers (see Fig.~\ref{distributed_structure})~\cite{YLS_JAP}.
In this network structure, a local node contains a limited number of
charged QDs, distinguishable by their optical frequencies.
The stationary spin qubits form a basis for the quantum memory and
quantum logic modules. Optical waveguides/fibers connect distributed
nodes with single photon wavepackets as the flying qubits.
Micro-cavities offer the playground for the strong interaction between
the two types of qubits. Control of a local node will be the focus
of Sec.~\ref{quantumgates}. Control of the interfacing between single spins and
single photons will be the focus of section~\ref{sec-network}.
With the recent progresses in coupling NV centers in diamond with
photons in cavities and waveguides~\cite{Park:2006,Schietinger08NanoLettCavityNV,
Fu08APLwaveguideNV,Barth09OLettCavityNV}, similar distributed quantum computing
with NV centers is also foreseeable.

\section{One- and Two-Qubit Operations}
\label{quantumgates}

It has been established that universal quantum computation can be accomplished
by a set of single-qubit gates and one kind of entanglement gates
such as CNOT, $\sqrt{\text{SWAP}}$, or controlled phase-shift
gate~\cite{DiVincenzo_UniversalQC,universal_quantum_gate,NielsenChuang}.
While there are many alternatives, we follow a more traditional and
scalable approach of designing a system which has negligible interaction
between qubits when the system is idling, and creating transient interaction
between two qubits during the operation.

General speaking, there are two strategies for controlling electron spins
in QDs, namely, the direct and the indirect control.
As the terminology suggests, the direct control manipulates the electron spins
directly which may be realized by rotation of a single spin with an
AC magnetic field from, e.g., a microwave or a pulse DC magnetic
field~\cite{Vrijen:2000PRA,Jelezko:2002,Jelezko2004_PRLNV,
Childress2006_ScienceNV,Dutt2007_ScienceNV,Koppens_Rabi},
and by coupling two spins via exchange interaction mediated by
virtual tunneling between QDs switched by, e.g., gate
voltages~\cite{Loss_QDspinQC,Burkard:1999PRB,Silicon_P_QC_review}.
The direct control schemes are more applicable to electrically defined
QD systems made possible by the feasibility of in-situ electrical gates.
There is no fundamental obstacle to integrate electrical gates to
self-assembled or fluctuation QD systems and to apply the direct control
of electron spins. But the much smaller size of these QDs as compared with the
splitting-gate defined ones and the much stronger confinement
(or much less tunneling probability) cause non-trivial technical problems.
The indirect control schemes are based on indirect coupling between
electron spin states mediated by virtual excitation of auxiliary energy
levels which are modified by (or conditioned on) the states of the
spins~\cite{Imamoglu_CQED_Spin,Calarco_OpticalQC,Pazy_QDQC,Nazir_PRL2004_SpinCouplebyXton,
Lovett_PRB05_SpinCouplebyXton,Chen_Raman,ORKKY,
Reinecke_indirect,ORKKY_vertical,Yamamoto_OpticalQC,Economou07PRLGeoControl,
Ramsay08PRLSpinControl,Yamamoto_08NatureSpinControl,Yamamoto_echo,Greilich2009NPhys}.
Usually, such intermediate states are excitons which may be excited by
optical pulses. The optical control of electron spins is limited to
direct-gap semiconductors but otherwise have a great deal of merits
particularly due to the energy scale cascade. As bandgaps in semiconductors are
usually larger by orders of magnitude than the electron spin Zeeman energy,
the control of the electron spins could be made much faster by using excitation
cross the large bandgap. For comparison, electrical control of spins
on the nanosecond scale is already the state-of-the-art technique~\cite{Marcus_T2,Koppens_Rabi},
while optical manipulation of electron spins can be completed on the picosecond
timescale~\cite{Ramsay08PRLSpinControl,Yamamoto_08NatureSpinControl,Yamamoto_echo,Greilich2009NPhys}.

In essence, all schemes of optical control of the electron spin
states~\cite{Imamoglu_CQED_Spin,Calarco_OpticalQC,Pazy_QDQC,Nazir_PRL2004_SpinCouplebyXton,
Lovett_PRB05_SpinCouplebyXton,Chen_Raman,ORKKY,Reinecke_indirect,ORKKY_vertical,Yamamoto_OpticalQC,Saikin_adiabatic}
are realized by the Raman processes where the virtual
excitation of excitons plays the central role. In one-qubit
operations, the intermediate states are excitons with one excess
electron, i.e., the trion states~\cite{Calarco_OpticalQC,Chen_Raman,Yamamoto_OpticalQC}.
In two-qubit operations~\cite{Imamoglu_CQED_Spin,Calarco_OpticalQC,Pazy_QDQC,Nazir_PRL2004_SpinCouplebyXton,
Lovett_PRB05_SpinCouplebyXton,ORKKY,Reinecke_indirect,ORKKY_vertical,Yamamoto_OpticalQC,Saikin_adiabatic},
the intermediate states are excitons with two excess
electrons. The effective exchange interaction between two spins
may be induced by the virtual excitation of one exciton tunneling
back and forth between two dots, which is similar to the RKKY process
and is dubbed optical RKKY interaction to indicate the role of the
optical excitation~\cite{ORKKY,Reinecke_indirect,ORKKY_vertical}.
Alternatively, even if there is no tunneling between QDs,
the interplay between Coulomb interaction and the spin blocking effect
would induce effective interaction between two separated spins when
two excitons are (virtually) excited in two
dots~\cite{Calarco_OpticalQC,Pazy_QDQC,Nazir_PRL2004_SpinCouplebyXton,Lovett_PRB05_SpinCouplebyXton,Saikin_adiabatic}.
The spin interaction may also be mediated by virtual photon exchange
during the virtual excitation of excitons~\cite{Imamoglu_CQED_Spin,Yamamoto_OpticalQC}.

More details about the optical control of one-spin and two-spin gates follow.
Since the essential physics of various optical control schemes is similar,
we will focus on two specific examples, namely, Raman control of single spins, and
optical RKKY control of two spins, as an illustration of the physics and the
operation conditions.

\subsection{Single-spin rotation by Raman process}
\label{Singlebitgate}

Arbitrary single-spin rotation can be realized via adiabatic Raman
processes mediated by trion states~\cite{Chen_Raman,Piermarocchi05Control}.
We will illustrate the basic
idea of such a scheme and discuss the non-adiabatic generalization
and certain limits of the method. It is worth pointing out that
the stimulated Raman adiabatic passage (STIRAP)~\cite{Bergmann_1998} involving only
dark states could not realize an arbitrary rotation but only a
spin flip from a known initial state. The adiabatic Raman processes
in arbitrary spin rotation involves both dark states and
bright states~\cite{Chen_Raman}.

The control of an electron spin relies critically on the spin states of
the excited excitons which are determined by the optical
selection rules in the semiconductor QDs. In the III-V
semiconductors of zinc-blende crystal structure, the selection
rules of optical excitations at the band edge are well defined by
the angular momentum conservation in terms of the electron spin
in the conduction band and the hole spin in the valence band,
which have angular momentum $s=1/2$ and $J=3/2$, respectively. In
QDs of large lateral sizes and strong confinement in the growth
direction (defined as $z$-axis), the hole states will be split
into two sets of degenerate states, with spin states $J_z=\pm
3/2$ and $\pm 1/2$, respectively, designated as heavy and light
holes, respectively, according to their effective mass along the
$z$-axis. When the lateral sizes of the QD are much larger than
the confinement size in the growth direction, the mixing of
different angular momentum states by the lateral confinement is
small~\cite{Loss_holespin}. Thus the optical excitation is restricted by the angular
momentum conservation along the $z$-axis. Now if the controlling
optical pulses are applied normal to the sample surface, the
conservation of the angular momentum about the growth direction makes it
impossible to flip the electron spin along $z$-axis and thus
impossible to complete an arbitrary quantum operation, unless the
light beam is incident with an angle~\cite{Flatte_spinControl} or the
symmetry is broken by a magnetic field with a non-zero in-plane component.
Since in the near-field optics, the
incident light is usually normal to the surface, we need a
static in-plane magnetic field applied (whose direction is
defined as $x$-axis). Under the strong magnetic field, the
electron spin states are split into two Zeeman levels
$|+x\rangle$ and $|-x\rangle$ with energy $\pm \omega_c/2$, respectively.
We use these two states as the basis $|0\rangle$ and $|1\rangle$ of a qubit.
The Zeeman splitting $\omega_c$ is in the order of 0.1$\sim$1~meV under a
magnetic field of a few Tesla. In GaAs fluctuation QDs, the hole states
are still near-degenerate, since the large heavy-light hole splitting
($\Delta_{\rm hl}$ is tens of meV) makes the hole spin splitting
$\sim\omega_c^3/\Delta_{\rm hl}^2$ negligible even under a field
as strong as a few Tesla, so the hole states can still be defined
by the magnetic quantum number $J_z$ as $|\pm 3/2\rangle$, which
will also be denoted by hollow arrows as $|\Downarrow\rangle$
and $|\Uparrow\rangle$, respectively. In this case, the optical transitions
can be separated by the selection rules even when the energy difference
due to the electron spin splitting is relatively small. Note that the near-degeneracy of the
two trion states is not a necessary condition in the schemes of single spin rotation to be discussed below.
Actually, in self-assembled dots where hole states can be split by an in-plane magnetic field
due to the large heavy-light hole mixing, the splitting between the trion states can be exploited
to separate desired transitions from unwanted ones by energy difference.

\begin{figure}[b]
\begin{center}
\includegraphics[width=6cm]{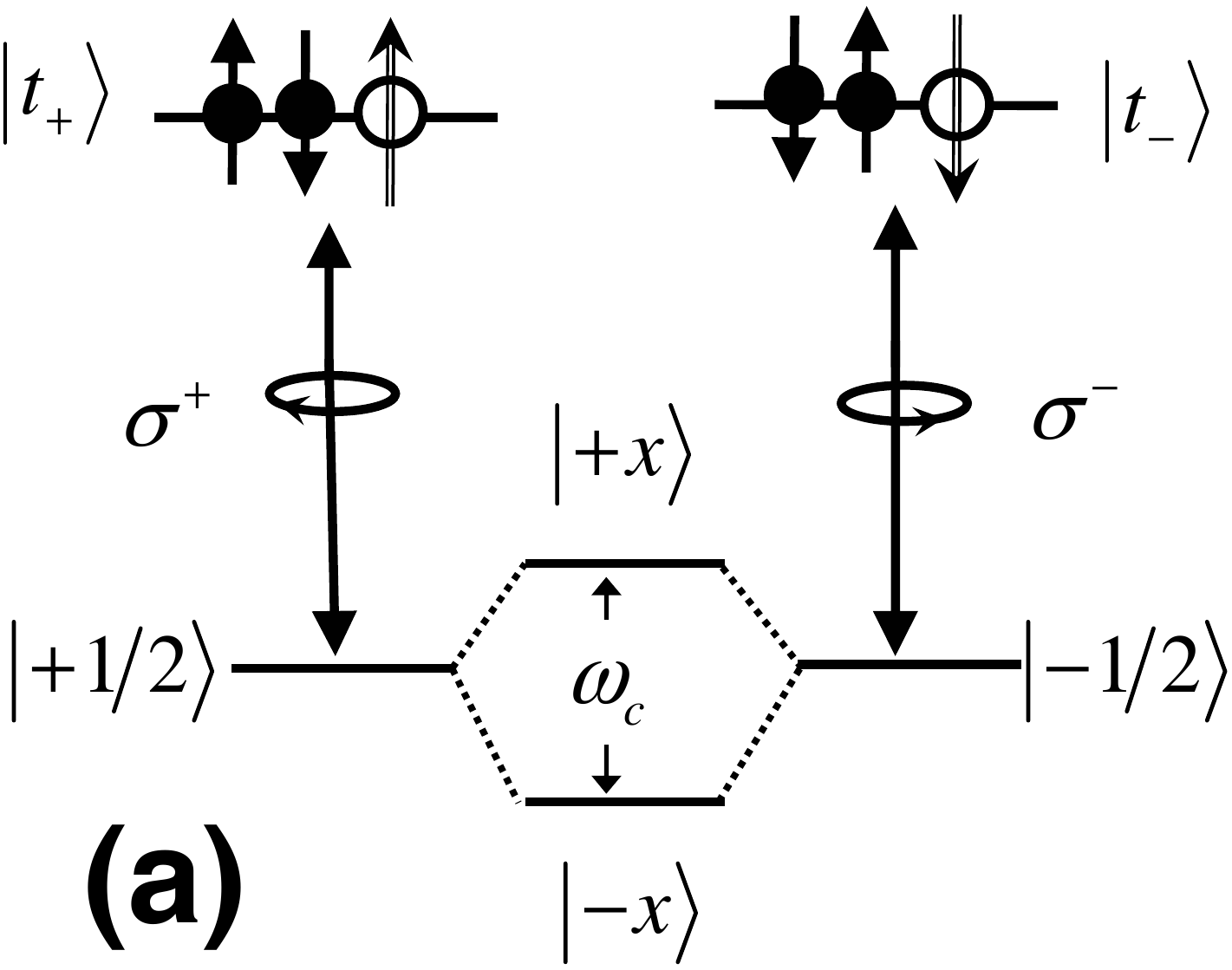} \hskip .5cm
\includegraphics[width=6cm]{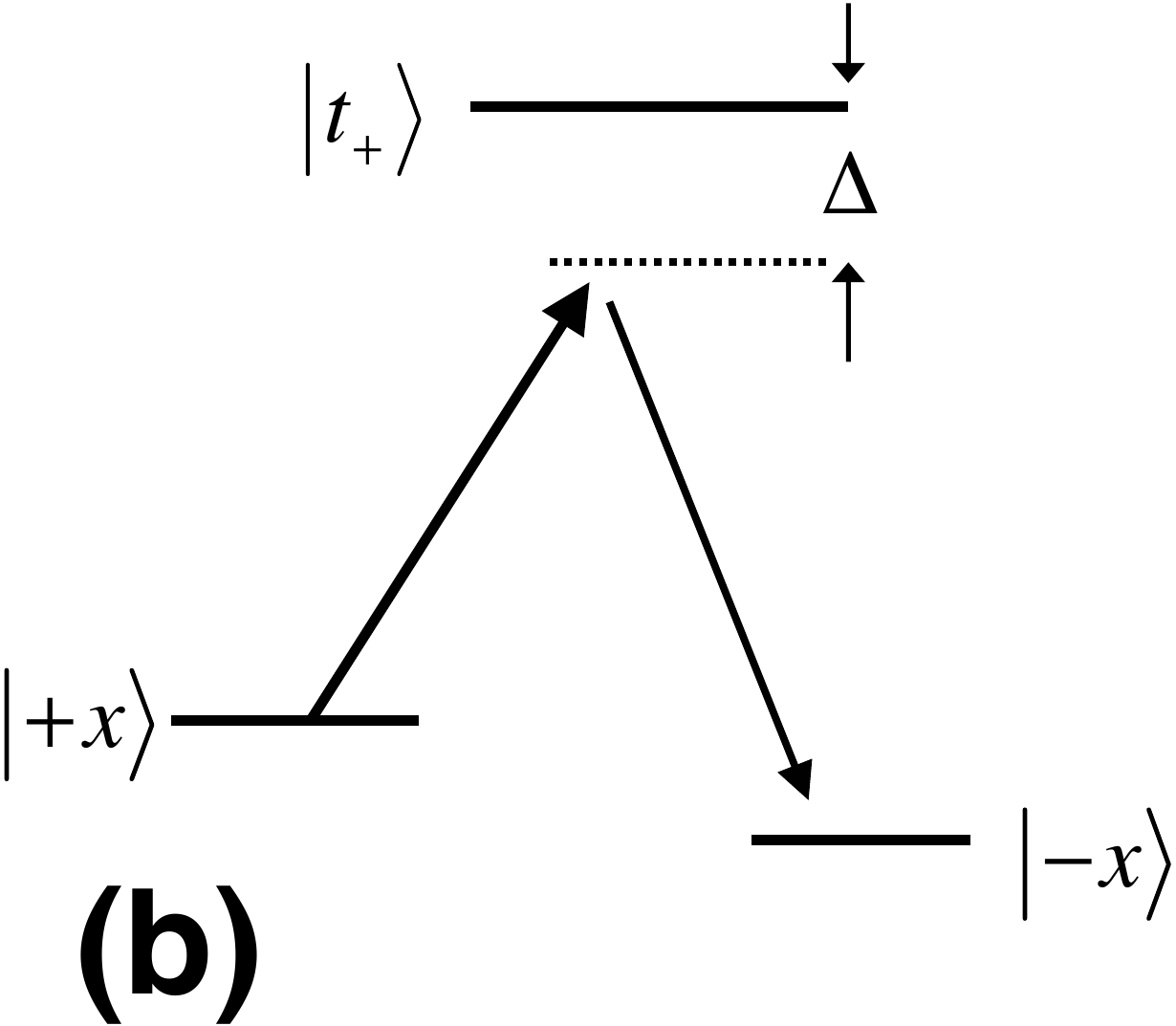}
\end{center}
\caption{(a) Optical selection rules for the electron-trion
transitions. (b) The Raman process in the $\Lambda$-type
three-level system formed by the electron spin states and the
trion states connected by a $\sigma^+$-polarized laser pulse.}
\label{fourlevel}
\end{figure}

To illustrate the essential physics underlying the control process,
we ignore the high-lying excited states and model the system
by four states consisting of the two split electron
spin states and two trion states $|t_{\pm}\rangle$
which are formed by two electrons in the singlet state
$|\uparrow\downarrow\rangle$ and one hole in the spin state$|\Uparrow\rangle$ or
$|\Downarrow\rangle$, respectively. According to the angular
momentum conservation about the $z$-axis, the selection rules of
the optical excitation, as depicted in Fig.~\ref{fourlevel}~(a),
is such that a light with circular polarization $\sigma^{\pm}$
will induce the transition from the electron states $|+1/2\rangle$
or $|-1/2\rangle$ to the trion states $|t_{\pm}\rangle$,
respectively.

So, under the excitation of an optical pulse with $\sigma^+$
polarization, the dynamics is reduced to a Raman rotation in
the $\Lambda$-type three-level system [as shown in Fig.~\ref{fourlevel}~(b)], which
is governed by the Hamiltonian
\begin{equation}
H=\omega_c s_x -\Omega^*(t)|+1/2\rangle\langle t_+|
-\Omega(t)|t_+\rangle\langle +1/2| -\Delta|t_+\rangle\langle t_+|,
\end{equation}
where $s_{x/y/z}$ is the electron spin operator along the $x$,
$y$, or $z$ direction, respectively, $\Omega(t)$ is Rabi
frequency of the laser pulse in the rotating frame, and $\Delta$
is the detuning of the laser relative to the trion state
$|t_+\rangle$. To eliminate the dynamic phase associated with
Zeeman splitting, the quantum operation
should be done in the frame rotating together with the free spins under the magnetic
field.  By the transformation $ S\equiv \exp\left(i\omega_c
ts_x \right),$ the Hamiltonian in the precessing frame is written in the matrix form
\begin{equation}
\tilde{H}=\left[\begin{array}{ccc}
0 & 0 & -{{\Omega^*}(t)e^{+i\omega_c t/2}/ \sqrt{2}} \\
0 & 0 & -{{\Omega^*}(t)e^{-i\omega_c t/2}/ \sqrt{2}} \\
-{{\Omega}(t)e^{-i\omega_c t/2}/\sqrt{2}} &
-{{\Omega}(t)e^{+i\omega_c t/2}/\sqrt{2}} & -\Delta
\end{array}\right],\end{equation}
in the basis of $e^{+i\omega_c t/2}|-x\rangle$, $e^{-i\omega_c
t}|+x\rangle$, and $|t_+\rangle$.

For large detuning ($|\Delta|$ much greater than the bandwidth of
the optical pulse and the Rabi frequency), the so-called
adiabatic approximation is justified and thus under a standard
canonical transformation, the off-diagonal terms in the
Hamiltonian between the electron spin states and the trion state
are eliminated up to the second order of the Rabi frequency. So
the transformed effective Hamiltonian is approximated as
\begin{equation}
\tilde{H}_{\rm eff}\approx\left[\begin{array}{ccc}
{|\Omega(t)|^2/(2\Delta)} & -{|\Omega(t)|^2e^{+i\omega_c t}/( 2\Delta)} & 0 \\
-{|\Omega(t)|^2e^{-i\omega_c t}/(2\Delta)} & {|\Omega(t)|^2/(2\Delta)} & 0 \\
0 &0 & -\Delta-{|\Omega(t)|^2/ \Delta}
\end{array}\right].
\end{equation}
This is equivalent to a magnetic field with strength
${|\Omega(t)|^2/( 2 g \mu_B \Delta)}$ precessing in the $x$-$y$
plane with the angular frequency $\omega_c$ (the time-dependent
optical Stark shift~\cite{Flatte_spinControl} of the electron energy
${|\Omega(t)|^2/(2\Delta)}$ contributes only a trivial global
phase-shift and can be ignored). In cases that the optical pulse is
much shorter than the spin precession period, the effective
magnetic field becomes an instantaneous pulse which can be
controlled in the femtosecond timescales (c.f. the GHz limit on
the control of external magnetic field). For $t\approx n\pi/(\omega_c)$ ($n$ is an integer, the magnetic field pulse is
effectively along the $z$-axis, and for time around
$(n+1/2)\pi/\omega_c$, the instantaneous rotation of the spin
is effectively along the $y$-axis. Thus we have two SU(2)
generators which can be combined to complete an arbitrary
rotation of the electron spin. To be specific, a spin rotation
with angle $\gamma$ along an axis defined by the Euler angles
$(\alpha,\beta)$, denoted as $R(\alpha,\beta,\gamma)$, can be
realized by at most 3 elementary rotations along
different axes in the $x$-$y$ plane, for
\begin{eqnarray}
R(\alpha,\beta,\gamma)
=R\left(\frac{\pi}{2},\frac{\pi}{2}+\beta,\frac{\pi}{2}-\alpha\right)
 R\left(\frac{\pi}{2},\beta,\gamma\right)
 R\left(\frac{\pi}{2},-\frac{\pi}{2}+\beta,\frac{\pi}{2}-\alpha\right)
,
\end{eqnarray}
where the three elementary rotations on the righthand side of the equation can be performed
in turn at $t=-\pi/(2\omega_c)+\beta/\omega_c$, $\beta/\omega_c$, and
$+\pi/\omega_c+\beta/\omega_c$, which can be
completed within half the precession period of the electron spin.
For an electron with Zeeman splitting of 1~meV, the duration of
an arbitrary operation of a single spin is less than 2~ps.

To complete a finite rotation of the spin at a time much shorter
than the precession period, the Rabi frequency should be large,
which, however, could demolish the condition for the adiabatic
approximation. For instance, if the Zeeman splitting is $0.1$~meV
and the detuning is $1$~meV, to complete a $\pi$ rotation within
a duration one tenth of the precession period, the Rabi frequency
is required to be about $3$~meV, greater than the detuning, which
makes the adiabatic approximation unjustified. In fact, to
validate the adiabatic condition for instantaneous operations, it
is required that $\omega_c\ll \Delta$, which means slow
operations for reasonably large detuning ($\Delta<10$~meV).
Alternatively, by shaping the control laser pulse, the rotation
can also be operated in the non-adiabatic regime or in a
non-instantaneous manner. No matter how intense and how fast the
optical pulse could be, the spin rotation in general cases,
however, is still limited by the precession period, as can be
seen from the equation for the spin polarization as
\begin{equation}
\partial_t \langle s_z\rangle=\omega_c\langle s_y\rangle-\partial_t\rho_{t},
\end{equation}
where $\rho_{t}$ is the population of the trion state. For a
complete operation, the residue population of the trion should be
zero, so the change of the spin momentum along the $z$-direction is
$\left|\delta\langle s_z\rangle\right|=\left|\omega_c\int_0^t\langle s_y\rangle dt\right| \le \omega_c t$,
which cannot be faster than the precession under the static external magnetic field.

In summary, under a moderate external magnetic field ($\lesssim
10$~T and $\omega_c\sim 1$~meV), an arbitrary spin rotation can
be accomplished well within 10~ps by up to three ultrashort
optical pulses (with simple shape and large detuning) or by one
pulse (with engineered shape). Remarkably, control of single electron
spins~\cite{Yamamoto_08NatureSpinControl,255wu:prl07,Berezovsky08}
and hole spins~\cite{Ramsay08PRLSpinControl}
in QDs in picosecond timescales have been recently realized in experiments.
Using the optical control, Yamamoto's group has demonstrated spin echo
for an ensemble of impurities in GaAs~\cite{Yamamoto_echo} and for a single QD spin~\cite{YamamotoQDecho}.
Greilich et al have realized optical rotation
of an ensemble of QD spins along arbitrary axes~\cite{Greilich2009NPhys}.

\subsection{Two-qubit gates by optical RKKY interaction}
\label{twobitgate}

To implement two-qubit quantum gates, optically induced RKKY
interaction (ORKKY) between electrons doped in QDs has been
proposed to couple two spins, laterally via continuum
excitons~\cite{ORKKY,Reinecke_indirect} or vertically via discrete
states~\cite{ORKKY_vertical}. The ORKKY interaction mediated by
continuum excitons~\cite{ORKKY,Reinecke_indirect}, due to the extension nature of the continuum
states, is less controllable in selectively coupling certain
spins. The ORKKY interaction between QDs vertically stacked~\cite{ORKKY_vertical}, on
the other hand, is not applicable to a scalable system with a planar
layout. For the sake of scalability, we consider to employ the
discrete excited states in laterally coupled QDs to induce the
ORKKY interaction, which, as described below, can be controlled to
selectively couple spins in designated adjacent QDs. The ORKKY
interaction is by nature a Raman process with the ground states
formed by the two-electron spin states and the intermediate states
by excitons charged with two excess electrons.

\begin{figure}[b]
\begin{center}
\includegraphics[width=5cm]{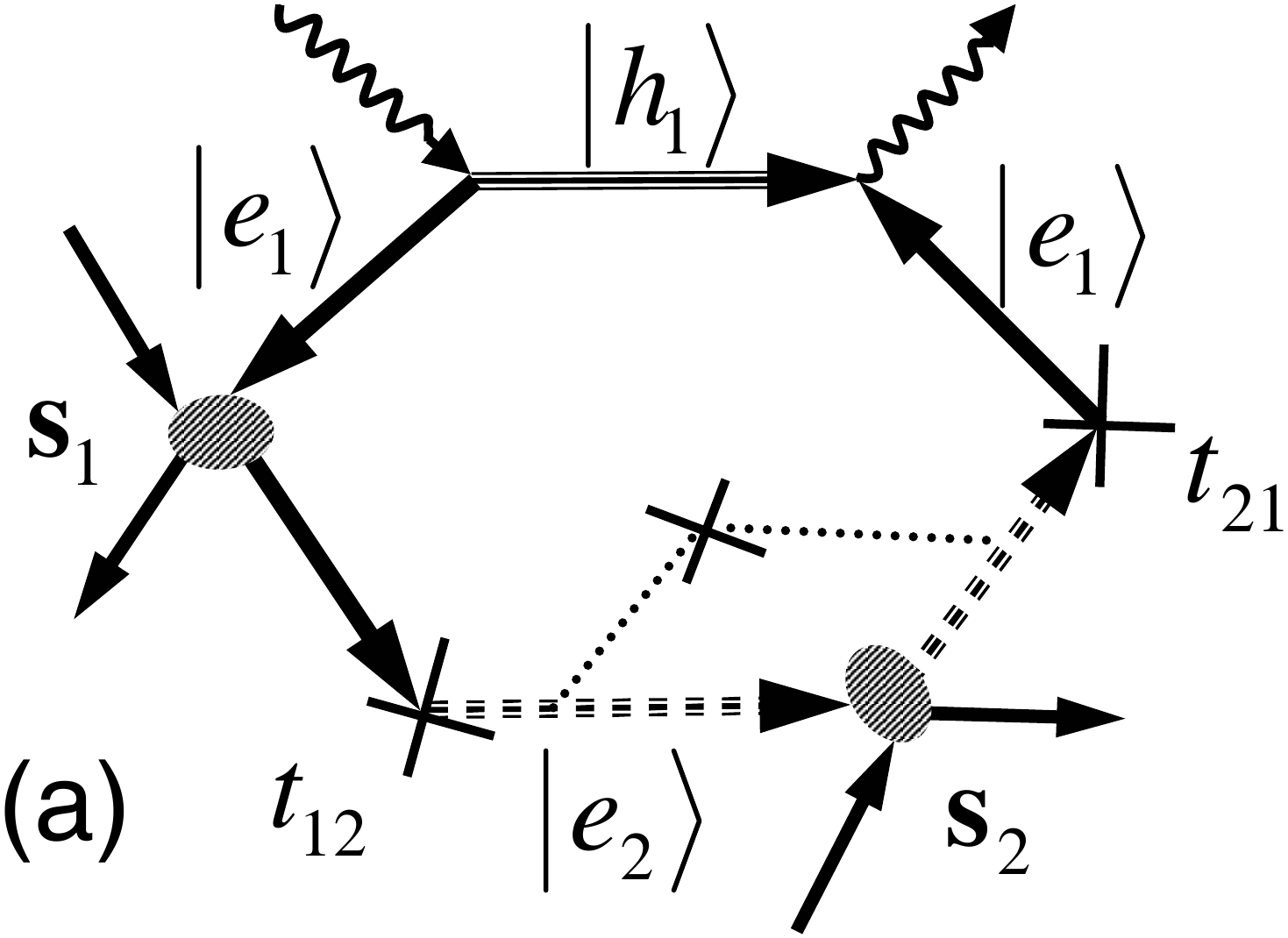} \hskip 0.5cm
\includegraphics[width=8cm]{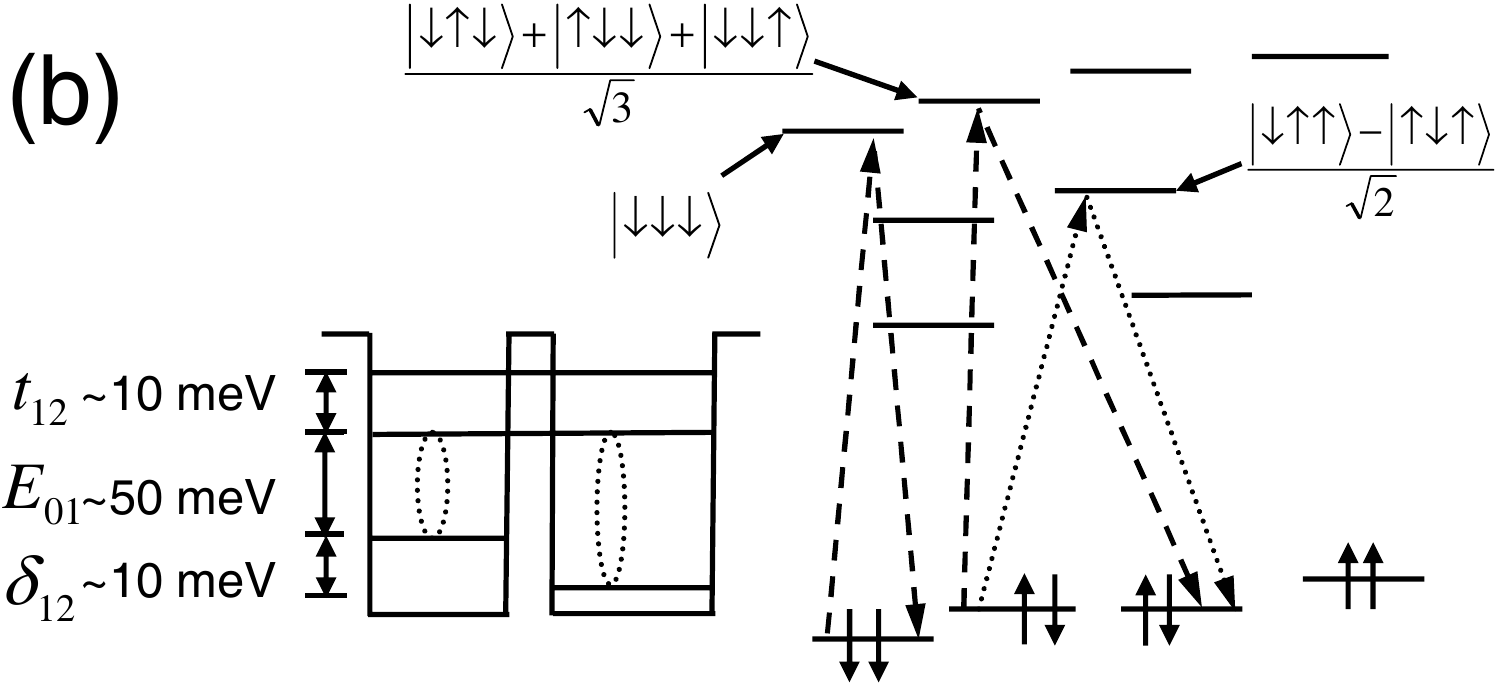}
\end{center}
\caption{(a) Feynman diagram for optical RKKY interaction. The
straight arrows are electron or holes lines, the wavy arrows are
photon lines, the tunneling is the cross vertices, the Coulomb
exchange is the scattering vertices, and the AC Stark field is the
dotted lines with cross. (b) Energy level schematics for two-qubit
gates. The two-spin states are split by the magnetic field,
and the exciton charged with two electrons (the hole is understood)
is split both by the Coulomb exchange energy and by the magnetic field. The
dotted arrows are optical excitation for a SWAP gate, which
together with the excitation represented by dashed arrows can also
accomplish the phase gate.} \label{FEYNMAN_ORRKY}
\end{figure}

The physical process of the optically induced RKKY interaction
via discrete intermediate states is depicted by the Feynmann
diagram in Fig.~\ref{FEYNMAN_ORRKY}~(a). The qubits under controls are the spins
of two electrons in the ground states of two neighboring QDs, denoted as ${\mathbf s}_1$ and
${\mathbf s}_2$. When the electrons are in the ground states, the inter-dot tunneling is negligible and
therefore the two spins have no direct exchange interaction. The basic elements of the ORKKY
process for mediated interaction between the two qubits can be described as:
\begin{enumerate}
\item an incident optical pulse
excites a direct electron-hole pair into the excited
electron and hole levels in, say, QD 1, denoted as
$|e_1\rangle\otimes|h_1\rangle$;
\item the optically excited electron will interact with the electron spin ${\mathbf s}_1$
with the strong exchange interaction ($\lambda_i\sim 5$~meV
in a typical InAs QD), while the electron-hole exchange
interaction is negligible in comparison;
\item with a strong quasi-cw optical field applied, the excited electron levels in
the two QDs can be tuned into resonance by the optical Stark
effect, and thus the electron in the excited state $|e_1\rangle$
can resonantly tunnel into the second QD with the tunneling rate
in the order of 10~meV for two QDs separated by 15~nm, while the
hole tunneling can be neglected due to the stronger confinement
and off-resonance condition;
\item
having tunnelled into the
excited level of the second QD, the optically excited electron
can exchange spins with the excess electron in the second QD
via the strong exchange interaction;
\item
after the exchange interaction with the
second qubit electron, the electron in the excited state can
tunnel back into the first QD;
\item
the electron back to the first
QD can be recombined with the hole by emitting a photon back into
the optical pulse, leaving the two excess electrons an effect of
indirect spin exchange. Since the laser frequencies can be
adjusted to selectively excite the exciton in one QD and to
selectively shift the level in another QD by the AC Stark effect,
a pair of adjacent QDs can be coupled on demand.
\end{enumerate}

Since the Coulomb and the exchange interactions could be
strong for electrons in the discrete states, as compared with the
optical Rabi frequency, it is better to treat the interactions
non-perturbatively by first exactly diagonalizing the states of
charged excitons, and use the eigenstates as the basis for
calculating the optically induced indirect exchange interaction.
In fact, when the intermediate states are discrete eigenstates of
the Coulomb interaction, the ORKKY process becomes equivalent to
the Raman process in the multi-level systems with the two-spin
states as  the ground states and the charged exciton states as
the intermediate states. Then the optical excitation can be
treated as a perturbation, similar to the case of the single-spin
rotation, and the adiabatic approximation may also be adopted when
the detuning is large.

As shown in Fig.~\ref{FEYNMAN_ORRKY}~(b), we assume the following
characteristic energies for two neighboring QDs of different
sizes, which can be realized by properly adjusting the QD sizes,
the separation, and the composite $x$ in In$_{1-x}$Ga$_x$As in
typical cases under consideration: The difference in transition
energy between the neighboring dots
is $\delta_{12}\sim$10~meV, the
distance between the ground and excited electron states
$E_{01}\sim 50$~meV, the tunneling strength between neighboring
excited states $t_{12}\sim 10$~meV, and the Coulomb exchange
energy between electrons $\lambda_{1/2}\sim 5$~meV. When the
electron-hole pair is excited into the excited single-particle
levels in the smaller QD, the difference between the binding
energy of the direct exciton (when both the electron and the hole
are in the smaller dot) and that of the indirect exciton (when
the electron has tunnelled into the large dot), which is $\sim
10$~meV, can already compensate most of the energy difference
($\delta_{12}$) between the excited single-particle levels in the
two dots. And if desired, a further optical pumping
can be used to fine-tune the energy levels in the dots so that near-resonance
tunneling can occur between the excited levels for electrons in
the two dots. The ground single-particle states for electrons and
the hole states can be virtually taken as localized due to the
relatively high barrier.

To illustrate the basic idea of using the discrete exciton states
in laterally coupled QDs for ORKKY interaction between the two
qubit spins, we assume that the hole is (virtually) created by the optical pulse only at the first
excited level in the smaller QD and the dynamics is essentially
determined by the interaction between the two qubit spins and the
optically excited electron. The role of the hole is then just
to impose the renormalization of the electron
energies. As mentioned above, both the tunneling and Coulomb
exchange are in the strong coupling regime, they should be
considered non-perturbatively. So we can first diagonalize the
Hamiltonian with the exchange energy and electron tunneling fully
included. As the tunneling is spin-independent, we consider only
the bond state of the QD molecule and treat the
anti-bond state as in far off-resonance (since it is about 10~meV
above). Thus the relevant spins are qubit 1 (${\mathbf s}_1$),
qubit 2 (${\mathbf s}_2$), and the spin of the bond molecular
state 3 (${\mathbf s}_3$). The Hamiltonian of these three spins is
\begin{eqnarray}
H& = & 2\lambda_{1}{\mathbf s}_1\cdot {\mathbf
s}_3+2\lambda_{2}{\mathbf s}_2\cdot {\mathbf
s}_3+2\omega_c(s_1^z+s_2^z+s_3^z) ,
\end{eqnarray}
where $\lambda_1$ and $\lambda_2$ are the exchange energy in each
dot. For simplicity, we assume $\lambda_1=\lambda_2=\lambda$. The
eigenstates  and the corresponding eigen energies can be worked
out as
\begin{eqnarray}
\begin{array}{ll}
\big|S,{1\over 2},-{1\over2}\big\rangle ={1\over
\sqrt{2}}\big(|\downarrow\uparrow\downarrow\rangle-|\uparrow\downarrow\downarrow\rangle\big)
, & \ \ \ \varepsilon=-\omega_c, \\
\big|S,{1\over 2},+{1\over2}\big\rangle ={1\over
\sqrt{2}}\big(|\uparrow\downarrow\uparrow\rangle-|\downarrow\uparrow\uparrow\rangle\big),
& \ \ \ \varepsilon=+\omega_c, \\
\big|P,{3\over 2},-{3\over2}\big\rangle
=|\downarrow\downarrow\downarrow\rangle , & \ \ \ \varepsilon=\lambda-3\omega_c, \\
\big|P,{3\over 2},-{1\over2}\big\rangle={1\over\sqrt{3}}\big(
 |\uparrow\downarrow\downarrow\rangle
+|\downarrow\uparrow\downarrow\rangle
+|\downarrow\downarrow\uparrow\rangle\big) , & \ \ \
\varepsilon=\lambda-\omega_c,
 \\
\big|P,{3\over 2},+{1\over2}\big\rangle
={1\over\sqrt{3}}\big(|\downarrow\uparrow\uparrow\rangle
+|\uparrow\downarrow\uparrow\rangle
+|\uparrow\uparrow\downarrow\rangle\big) , & \ \ \
\varepsilon=\lambda+\omega_c, \\
\big|P,{3\over 2},+{3\over2}\big\rangle
=|\uparrow\uparrow\uparrow\rangle
, & \ \ \ \varepsilon=\lambda+3\omega_c, \\
\big|P,{1\over
2},-{1\over2}\big\rangle={1\over\sqrt{6}}\big(|\uparrow\downarrow\downarrow\rangle
+|\downarrow\uparrow\downarrow\rangle
-2|\downarrow\downarrow\uparrow\rangle\big) ,& \ \ \
\varepsilon=-2\lambda-\omega_c,
 \\
\big|P,{1\over 2},+{1\over2}\big\rangle
={1\over\sqrt{6}}\big(|\downarrow\uparrow\uparrow\rangle
 +|\uparrow\downarrow\uparrow\rangle -2|\uparrow\uparrow\downarrow\rangle\big) ,
 & \ \ \ \varepsilon=-2\lambda+\omega_c,
\end{array}
\end{eqnarray}
as schematically shown in Fig.~\ref{FEYNMAN_ORRKY}~(b),
where the up/down arrows indicate in turn the spin states of qubit 1, qubit 2,
and the electron created by optical excitation, quantized along the external magnetic field direction.
Notice $\lambda\sim 5$~meV$\gg 2\omega_c \sim 1$~meV. The
two-qubit gates can just be realized by the Raman passages
between the two-spin ground states mediated by the charged
excitons formed by one hole plus three electrons in the spin
states shown above. To be specific, we discuss two frequently
used two-qubit gates as below.

\subsubsection{SWAP and $\sqrt{\text{SWAP}}$ gates}
\label{swapgates}

A SWAP gate interchanges the states of two qubits. Its matrix
form is shown in Fig.~\ref{gates}. In the Raman process, it just
flips the two states: $|\uparrow\downarrow\rangle$ and $|\downarrow\uparrow\rangle$. The intermediate state connecting
these two states by Raman process is
$|S,{1\over2},{1\over2}\rangle\equiv {1\over
\sqrt{2}}\big(|\uparrow\downarrow\uparrow\rangle-|\downarrow\uparrow\uparrow\rangle\big)$
(henceforth the hole state in the exciton has been omitted for the sake of simplicity),
which is optically coupled only to the two-spin singlet state ${1\over
\sqrt{2}}\big(|\uparrow\downarrow\rangle-|\downarrow\uparrow\rangle\big)$.
By performing a $2n\pi$ Rabi rotation between the intermediated
exciton state and the singlet ground state, a pure phase-shift
will be induced to the singlet state, whose value is determined
by the detuning. When such a phase-shift is controlled to be $\pi$
(which can be obtained by, e.g., setting the rotation angle to be
$2\pi$ and the detuning is zero), the two states
$|\uparrow\downarrow\rangle$ and $|\downarrow\uparrow\rangle$ are
just flipped and the SWAP gate is realized. In practice, the
optical pulse can be tuned off-resonance from the transition to
suppress the spontaneous decay. With proper polarization, the
optical pulse doesn't excite the nearby state $|S,{1\over
2},-{1\over2}\rangle$ and all other transitions are at
separated by an energy of $\lambda$. So this operation can be
accomplished in a period of time $\sim 10\times(2\pi
\lambda^{-1}) \sim 10$ ps. Considering the optical Stark pulse
used for tuning the resonant tunneling, a two-qubit SWAP gate in
such a scheme would require two optical pulses of duration of
about 10 ps.

By choosing a proper detuning, the phase-shift can also be
controlled to be $\pi/2$, and then the SWAP gate is performed
halfway, or the $\sqrt{\text{SWAP}}$ gate is realized, which has
the matrix form
\begin{equation}
U_{\text SS}=\left(\begin{array}{cccc}
1 & 0 & 0 & 0 \\
0 & \frac{1+i}{2} & \frac{1-i}{2} & 0 \\
0 & \frac{1-i}{2} & \frac{1+i}{2} & 0 \\
0 & 0 & 0 & 1 \end{array}\right).
\end{equation}
The $\sqrt{\text{SWAP}}$ gate can generate entanglement between
the two spins and can be used to realize the controlled phase
gate and the CNOT gate.

\subsubsection{Controlled phase gate}
\label{phasegate}

Another two-qubit gate which can be used in lieu of the CNOT gate
for universal quantum computation is the controlled phase gate which induces a phase-shift of
the target qubit depending on the state of the controlling qubit.
The general matrix form of the controlled phase gate is
\begin{equation}
\left(\begin{array}{cccc}
e^{i\phi_{\downarrow\downarrow}} & 0 & 0 & 0 \\
0 & e^{i\phi_{\downarrow\uparrow}} & 0 & 0 \\
0 & 0 & e^{i\phi_{\uparrow\downarrow}} & 0 \\
0 & 0 & 0 & e^{i\phi_{\uparrow\uparrow}}\end{array}\right) =
e^{i{\phi_{\downarrow\uparrow}+\phi_{\uparrow\downarrow} \over 2}}
\Big[R_x(\phi_{\uparrow\downarrow}-\phi_{\downarrow\downarrow})
\otimes R_x(\phi_{\downarrow\uparrow}-\phi_{\downarrow\downarrow})
\Big] \left(\begin{array}{cccc}
1 & 0 & 0 & 0 \\
0 & 1 & 0 & 0 \\
0 & 0 & 1 & 0 \\
0 & 0 & 0 & e^{i\phi}\end{array}\right),
\end{equation}
where the subscript of the phase-shift indicates the
corresponding qubit states, $R_x(\theta)$ is the single qubit
phase-shift of $\theta$, and $\phi\equiv
\phi_{\downarrow\downarrow}+\phi_{\uparrow\uparrow}
-\phi_{\downarrow\uparrow}-\phi_{\uparrow\downarrow}.$ When
$\phi=\pi$, the phase gate can be transformed into the CNOT gate
by applying certain single-spin gates, which are
usually assumed much easier than the two-qubit gates.

As shown in Fig.~\ref{FEYNMAN_ORRKY} (b), to obtain the
phase-shift, an optical pulse is applied to couple the state
$|\downarrow\downarrow\rangle$ to the exciton state $|P,{3\over
2}, -{3\over
2}\rangle\equiv|\downarrow\downarrow\downarrow\rangle$. The pulse
shifts the states by optical Stark effect and induces the phase
shift $\phi_{\downarrow\downarrow}$. This pulse will also couples
the electron state
$(|\downarrow\uparrow\rangle+|\uparrow\downarrow\rangle)/\sqrt{2}$
to the exciton state $|P,{3\over 2},- {1\over 2}\rangle\equiv
(|\downarrow\uparrow\downarrow\rangle+|\uparrow\downarrow\downarrow\rangle
 +|\downarrow\downarrow\uparrow\rangle)/\sqrt{3}$, inducing a phase
 shift to
 $(|\downarrow\uparrow\rangle+|\uparrow\downarrow\rangle)/\sqrt{2}$ and
 thus the rotation between $|\downarrow\uparrow\rangle$ and
 $|\uparrow\downarrow\rangle$. To obtain a pure phase gate,
 another pulse (with the same energy and polarization as in the SWAP
 gate) coupling the states
 $(|\downarrow\uparrow\rangle-|\uparrow\downarrow\rangle)/\sqrt{2}$
and  $|S,{1\over 2}, {1\over 2}\rangle\equiv
(|\downarrow\uparrow\uparrow\rangle-|\uparrow\downarrow\uparrow\rangle)/\sqrt{2}$
can be used to compensate the rotation. Finally, the conditional
phase-shift is $\phi\equiv \phi_{\downarrow\downarrow}
-\phi_{\downarrow\uparrow}-\phi_{\uparrow\downarrow}=-{1\over 3}\phi_{\downarrow\downarrow}.$ Similar to the case of the SWAP
gate, a phase-shift gate can be realized with three pulses
(including the optical Stark pulse) of duration of about 10 ps.

\subsection{Issues to be considered}
\label{issues}

In realistic cases, there are several issues which could degrade
the fidelity of the quantum gates.

The first one is the relaxation of the intermediate state by
spontaneous emission. In the stimulated
Raman adiabatic passage~\cite{Bergmann_1998}, the spin states can be flipped without
populating the exciton state, which, however, cannot
perform a general quantum gate since such a passage depends on the initial
state of the spin. To have a general quantum gate which transforms
a spin independent of the initial state, both the dark and bright
states should be employed. To suppress the spontaneous emission
and other scattering processes, it is preferable to have large
detuning so as to minimize the population of the intermediate
state~\cite{Calarco_OpticalQC,Chen_Raman,Piermarocchi05Control,Saikin_adiabatic}.

Another important effect affecting the fidelity is the
imperfection of the selection rules and the hole-mixing~\cite{Lovett_PRB05_SpinCouplebyXton}.
In realistic cases, the QDs would never have perfectly symmetric
shape and thus the lateral confinement could cause mixing between
states of different angular momentum (such as the heavy-light
hole mixing), which is worse for the excited states involved in the
two-qubit gates. Such effects, however, only induce systematic errors
or unwanted dynamics to the quantum gates designed for ideal condition.
In principle, the shapes, polarizations, and timings of the controlling
laser pulses can always be readjusted once the realistic system
parameters have been measured. It could also be possible to design pulses of certain
robustness against small deviations in the system
parameters~\cite{Liu_readwrite,Goswami_pulse}, and a scheme of using chirped pulse to implement quantum gates
robust against the mixing effect has been proposed~\cite{Lovett_PRB05_SpinCouplebyXton}.

For QDs with reflection symmetry with respect to the growth plane,
the imperfect selection rules may be tolerated by re-designing the polarization
of the control light.
In general cases, especially for small QDs which have irregular shapes,
the hole mixing may be used as a resource for quantum control.
In previous discussions in Sec.~\ref{Singlebitgate}, a static magnetic field applied along
a direction other than the growth direction has been required to
break the rotation symmetry so that an arbitrary rotation of a
single spin is possible. When the conservation of angular moment with respect to the growth
direction is not perfect and thus the ``forbidden'' transitions
would be made partially ``allowed'' due to the hole mixing,
an effective magnetic field along an arbitrary direction
for the spin could be induced by a properly polarized light beam
through the AC Stark effect and an arbitrary rotation of a
single spin could be realized even without a external magnetic field applied.
More discussions about the hole mixing and its effects on the optical control
are given in Appendix~\ref{mixing}.

The third problem with the scalability of the quantum computation is the
complexity in the energy level structure of multi-dot
systems. The analysis and characterization of the many levels
with a number of excess electrons require much effort
and furthermore~\cite{Bayer03PRLxfineQDs,Finley06PRLQDsspin,Petroff05PRLQDsphoton,
Gammon07PRBfineQDs,Gammon08NatPhysQDsMapping},
the optical pulses applied to a desired transition
would inevitably affect the other transitions in QDs nearby,
making the pulse design very demanding when the system
becomes large. In femtosecond chemistry, learning algorithms
haven been developed to design sophisticated pulses for
controlling the complex atomic and molecular
dynamics~\cite{Rabitz_1992,Warren_controlReview,rabitz00}. We expect
the quantum learning algorithm be a powerful tool
to deal with the design complexity in multi-dot systems.

With recent experimental progresses demonstrating the feasibility of optical
control of single spins, the optical control of two spins for implementing
two-qubit quantum gates is an immediate milestone for future experiments.
Indeed, the recent systematic investigation of the optical transitions in coupled
QD structures~\cite{Bayer01ScienceQDs,Bayer03PRLxfineQDs,
Petroff05PRLQDsphoton,Gammon07PRBfineQDs,Gammon08NatPhysQDsMapping} has laid a cornerstone for this target.

\section{Qubit Initialization}
\label{readandwrite}

A rapid and continuous supply of refreshed qubits is one criteria
for scalable quantum computation~\cite{DiVincenzo_Criteria_7}. Such a requirement is not only a prerequisite
for the initialization of a quantum computer, but also a key element
for quantum error correction where errors are
continuously generated during the operation in noisy environments and by imperfect control.
The initial preparation of a quantum computer could be done slowly, e.g.,
by simply cooling the system to very low temperature. For quantum error
correction, however, rapid reset of qubits is
crucial to recycle the spoiled qubits, otherwise a (infinitely) large number of
fresh qubits should be prepared and preserved before a quantum computation
commences so that the erroneous qubits could be replaced. The dynamical
recycling strategy is more economical than a static supply of
many qubits which deteriorate. It is desirable that the machine be as small as possible, with
operation cost as trade-off.

The essence of qubit initialization is preparing a pure quantum state
out of a mixed one. It amounts to cooling a qubit (ideally) to absolutely zero
temperature. Thus the key physical process is dumping entropy to
the environment. The aim is to build a quantum refrigerator in analogy
to a reverse Carnot cycle. In general, the cycle consists of the following
steps~\cite{Gammon_ini,Liu_readwrite,Friesen_ini,Imamoglu_QDSpinPrep,Emary07PRLspinIni,Xu07PRLspinIni,Gammon08PRLspinQDs}:
(1) pumping of the system to an excited state; and
(2) relaxation of the excited system with entropy dumped into the environment.
To initialize the spin qubit in an ultrafast timescale, a
quantum channel capable of dumping entropy rapidly is required.

In a QD, the only available thermal baths for dumping the entropy
of an electron spin are the nuclear spins, the host lattice (the
phonon bath), and the electromagnetic environment (the photon
bath). The coupling between the electron spin and the nuclear
spins is very weak (with the rate in the order of $10^{-6}$~sec$^{-1}$),
so only the phonon or photon bath could be used as entropy drain.
The direct coupling of an electron spin with either the lattice vibration or the electromagnetic
modes is known to be very weak. The solution is to transfer the spin state
into orbital states which couple to the phonon or photon bath strongly.
The entropy dissipation by rapid photon channel can be realized by optically
pumping the spin states to a trion state and coupling the trion
to a photon in a strongly coupled QD-microcavity-waveguide structure~\cite{Liu_readwrite}.
Or alternatively, excited trion states may provide an efficient entropy
channel realized by the rapid phonon emission in QDs.
Below we discuss these two possibilities.

\subsection{Initialization by entropy dumping to photon baths}

\subsubsection{Optical pumping}

\begin{figure}[b]
\begin{center}
\includegraphics[width=6cm]{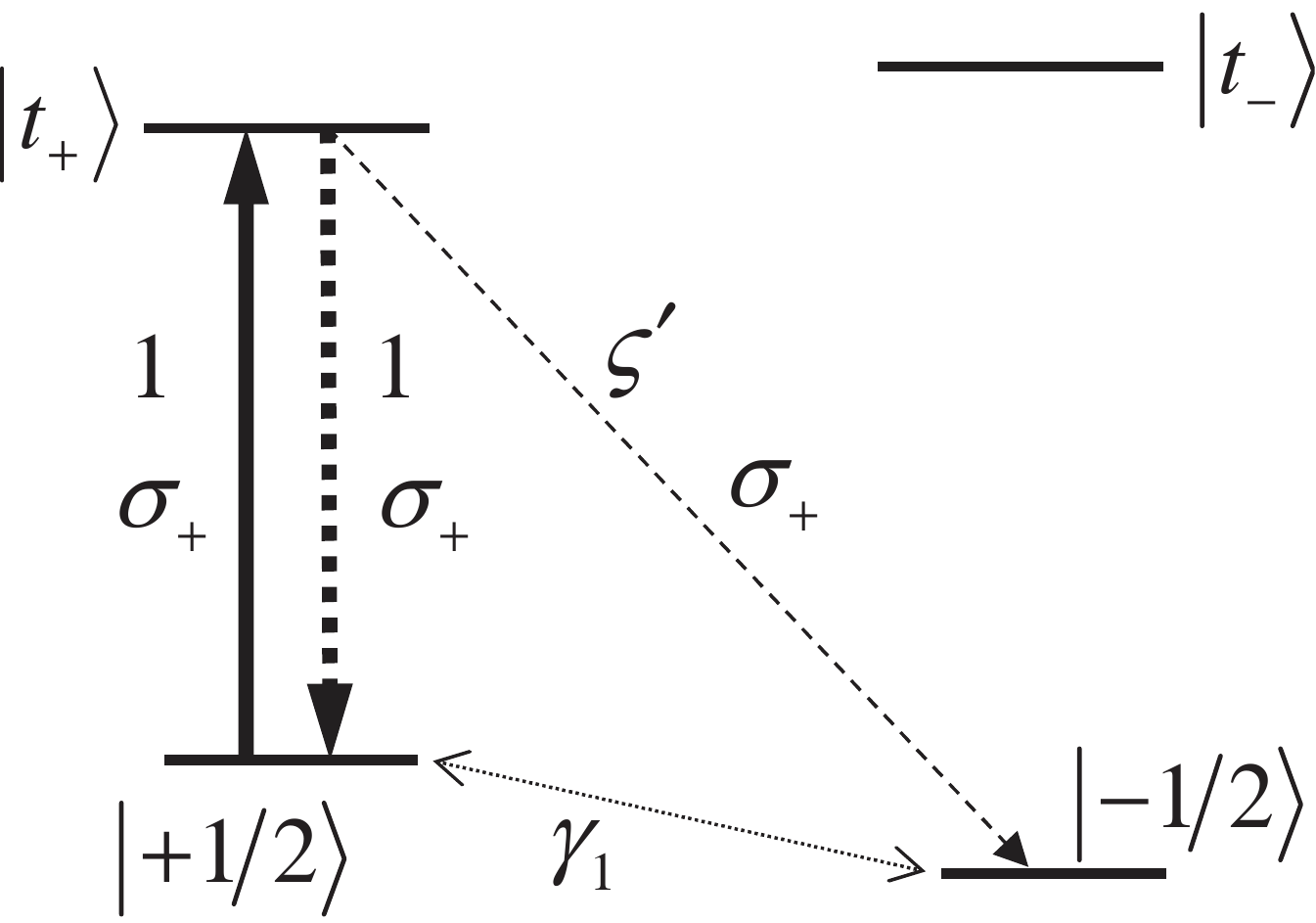}
\end{center}
\caption{Optical pumping of a spin in the Faraday configuration.
The solid line indicates the optical excitation, and the dotted lines
indicate the spontaneous emission. The polarization and the relative dipole matrix element
for each transition are indicated. The two spins states are connected by a spin-flip process with rate $\gamma_1$.}
\label{opticalpump}
\end{figure}

The idea of initializing a spin by optical pumping is illustrated in Fig.~\ref{opticalpump}:
An optical light brings one of the two spin states into a trion state,
and then the trion state relaxes to either spin state by spontaneous emission. After
sufficient cycling of the pumping process,
the electron spin will be in the state that is not coupled to the
trion state by the pump light, which has been demonstrated recently in
experiments for single electron spins in QDs~\cite{Imamoglu_QDSpinPrep,Xu07PRLspinIni}.
The scheme has also been applied to initialize single heavy-hole spins in QDs~\cite{Warburton_holeInitialization}.
With essentially the same physics, optical pumping
has also been used to initialize spins of NV centers in diamond~\cite{Jelezko2004_PRLNV,Hanson2006_PRLNV}.

To selectively excite only one electron spin state, one can choose the light polarization so that one
transition is ``forbidden'' due to the selection rules. The selection rules are based on some symmetry,
such as the (approximate) rotational symmetry about the growth direction.
To exploit the selection rules resulting from the rotational symmetry, one can use
a circularly polarized light. A $\sigma_+$ polarized light, e.g., couples the spin state $|+1/2\rangle$
to the trion state $|t+\rangle$ (see Fig.~\ref{opticalpump}). Then the trion state will relax to both spin states
by spontaneous emission. Now we have a dilemma: The selection rule prevents the spin state
$|-1/2\rangle$ to be reached. But remember the selection rules we have are approximate and the ``forbidden''
transitions are actually slightly allowed. According to the discussions in Appendix~\ref{mixing}, the trion state
$|t+\rangle$ has a small probability ($\sim {\zeta'}^2$) to relax to
the ``forbidden'' state $|-1/2\rangle$. Also, due to the approximate
selection rules, the state $|t-\rangle$ may be slightly excited,
and the spin state $|-1/2\rangle$ can be excited by the $\sigma_+$-polarized light to the two trion states,
which degrades the initialization efficiency. Furthermore, a magnetic field can be applied along the growth direction
(i.e., in the Faraday configuration), so that all the transitions except the desired ones as
depicted in Fig.~\ref{opticalpump} are off-resonance from the pump light.
Thus the pump process is characterized by several rates: The excitation
rate $G$, the spontaneous emission rates $\Gamma$ and $\tilde{\Gamma}\equiv \Gamma{\zeta'}^2$
(to the spin up and spin down states, respectively), and the longitudinal spin relaxation rate
$\gamma_1\equiv T_1^{-1}$.
The rate equations for the trion population $p_t$ and the electron populations in the
two spin states $p_{\pm}$ are established as
\begin{subequations}
\begin{eqnarray}
\dot{p}_t &=& -\left(\Gamma +\tilde{\Gamma}\right)p_t+G p_+, \\
\dot{p}_+ &=& -\gamma_1p_+ +\gamma_1 p_-  + \Gamma p_t- G p_+, \\
\dot{p}_- &=& -\gamma_1p_- +\gamma_1 p_+ +\tilde{\Gamma} p_t.
\end{eqnarray}
\end{subequations}
The quality of the initialization may be quantified by two
factors: the saturation time $T_s$ and the saturation spin
polarization $P_s$. In a typical GaAs or InAs QD at low temperature
and under a moderately strong magnetic field,
$\Gamma\sim 10^9$~s$^{-1}$, $\gamma_1\sim 10^3$~s$^{-1}$ (see Sec.~\ref{subsec_spinQubit_generaldecoherence}
for details), and $\zeta'\sim 1\%$ (for a unstrained dot~\cite{Loss_holespin}, the value
may be increased by strain~\cite{Chamarro_holespin}). Assuming the pumping rate $G\gg \Gamma$,
we obtain by the rate equations (note the trion population will eventually
becomes spin up population when the light is switched off)
\begin{subequations}
\begin{eqnarray}
T_s &\sim & 10/\tilde{\Gamma}\sim 0.1~{\rm ms}, \\
P_s &\cong & p_--p_+-p_t \approx  1-2\gamma_1/\tilde{\Gamma}\sim 98\%.
\end{eqnarray}
\end{subequations}
Such a high degree of electron spin polarization by optical pumping has been experimentally demonstrated~\cite{Imamoglu_QDSpinPrep}.
The initialization by optical pumping in the Faraday configuration, however, is rather slow and
the saturation polarization is limited by the spin flip rate relative to the ``forbidden''
spontaneous emission rate. Furthermore, the energy cost of the pump light is considerable.
To see the energy cost of the optical pumping in the Faraday configuration, we notice that
for just one useful photon emitted (which results in the target spin state),
the number of photon wasted (by spontaneous emission resulting in the original state)
is ${\zeta'}^{-2}\sim 400$ -- only $0.25\%$ energy of the pump light has been effective.
The limiting factor is the small dipole moment for the ``forbidden'' transition.

\begin{figure}[b]
\begin{center}
\includegraphics[width=6cm]{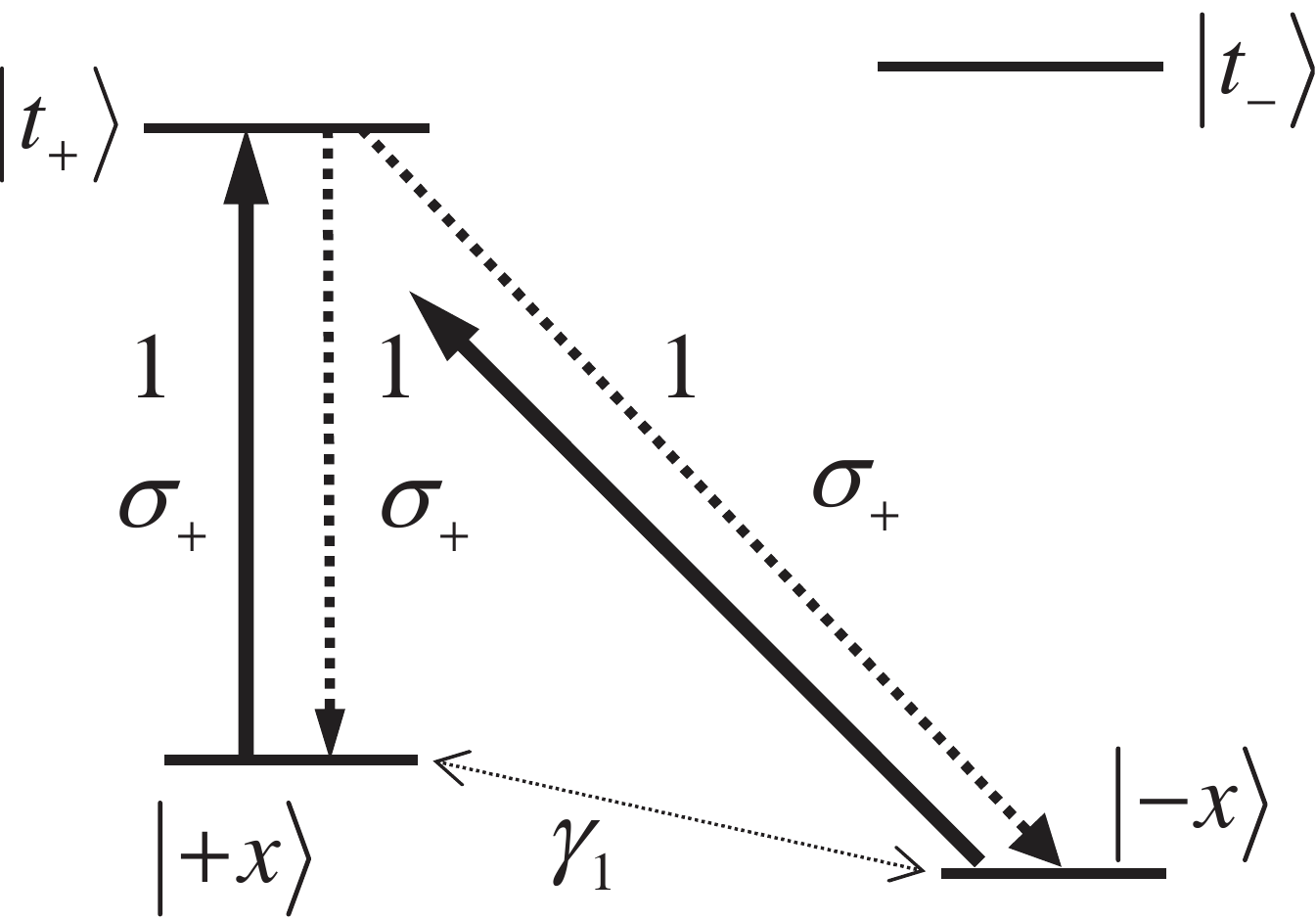}
\end{center}
\caption{The same as Fig.~\ref{opticalpump} except that the setup is in the Voigt configuration.}
\label{PumpVoigt}
\end{figure}

The solution is straightforward. One can work in the Voigt configuration in which a magnetic
field is applied in plane to quantized the electron spins to be $|\pm x\rangle$~\cite{Yao_phasegate,Emary07PRLspinIni,Xu07PRLspinIni}.
As shown in Fig.~\ref{PumpVoigt}, now both spin eigenstates
are connected to a trion state
by a $\sigma_+$-polarized excitation with the same dipole matrix element. The efficiency of optical
pumping in the Voigt configuration can be discussed similarly to that in the Faraday configuration.
But now there is another limiting factor of the saturation polarization: the backward optical excitation
process from the target spin state to the trion state which, though off-resonance,
is not negligible because of
the large dipole matrix element. The generation rate of the trion by the off-resonance excitation
is $\tilde{G}\equiv G\Gamma^2/\left(\Gamma^2+\omega_c^2\right)$ where $\omega_c$ is the
angular Lamor frequency of the electron
spin in the magnetic field. The rate equations for the trion population $P_t$ and the electron populations in the
two spin states $P_{\pm}$ are established as
\begin{subequations}
\begin{eqnarray}
\dot{p}_t &=& -2\Gamma p_t+G p_++\tilde{G}p_-, \\
\dot{p}_+ &=& -\gamma_1p_+ +\gamma_1 p_--Gp_+  + \Gamma p_t, \\
\dot{p}_- &=& -\gamma_1p_- +\gamma_1 p_+-\tilde{G}p_-+\Gamma p_t.
\end{eqnarray}
\end{subequations}
Under the condition that $G\gg\Gamma\gg \tilde{G}\gg \gamma_1$ (e.g., $G=10\Gamma$ and
$\omega_c=10^2\Gamma\sim 10^{11}$~s$^{-1}$), the efficiency of the initialization
by optical pumping in the Faraday configuration is characterized by
\begin{subequations}
\begin{eqnarray}
T_s &\sim & 10/{\Gamma}\sim 10~{\rm ns}, \\
P_s &\cong & p_--p_+ \approx  1-\tilde{G}/{\Gamma}\sim 99.9\%.
\end{eqnarray}
\end{subequations}
Owing to using the allowed transition in the Voigt configuration
instead of the forbidden transition in the Faraday configuration, the
spin initialization is faster by orders of magnitude.
Also, the saturating polarization is much closer to unity
since now the limiting factor is the off-resonance transition probability relative
to the resonant one, instead of the spin-flip rate relative to the
trion recombination rate due to the forbidden transition in the Faraday configuration.
The off-resonance transition could be
suppressed simply by enlarging the electron Zeeman splitting.
There are calculations~\cite{Emary07PRLspinIni} and experiments~\cite{Xu07PRLspinIni} demonstrating
efficient optical pumping of single electron spins in the Voigt configuration.

\subsubsection{Single-shot initialization with cavity enhancement}

In the previous discussion on spin initialization by optical pumping, we have seen
that it is crucial to have a rapid entropy dumping channel to have efficient
qubit cooling. The cooling duration of 10~ns is acceptable in many cases, but
it is still highly desirable to have even faster spin initialization so that
the qubit refreshing rate could catch up with the quantum gate and the error
generation rate. Suppose the error rate per qubit per operation is about the
quantum error correction threshold $10^{-3}$ and about 10 physical qubits are
used to code one logical qubit to incorporate the error correction within the
quantum logic. As we have discussed, the optical control for a simple
quantum gate should take about 10~ps. That means we should have a qubit reset
rate well above 1 qubit per nanosecond to avoid quantum computation being
held up due to the lack of refreshed qubits. To have such ultrafast spin initialization,
the entropy dumping channel should be specially engineered. One possible approach is
to enhance the coupling to the photon bath by increasing the local density of states
of photon modes. This enhancement is possible by putting a QD in the proximity
of a microcavity~\cite{Liu_readwrite}. As has been discussed earlier, a micro-photonic structure is
in any case needed to form a scalable large structure of QDs for distributed
quantum computation. Here we discuss how the in-situ cavity QED may be used to
selectively enhance the photon emission for entropy dumping and hence for ultrafast
spin initialization, following the procedure of Ref.~\cite{Liu_readwrite}.

\begin{figure}[b]
\begin{center}
\includegraphics[width=6cm ]{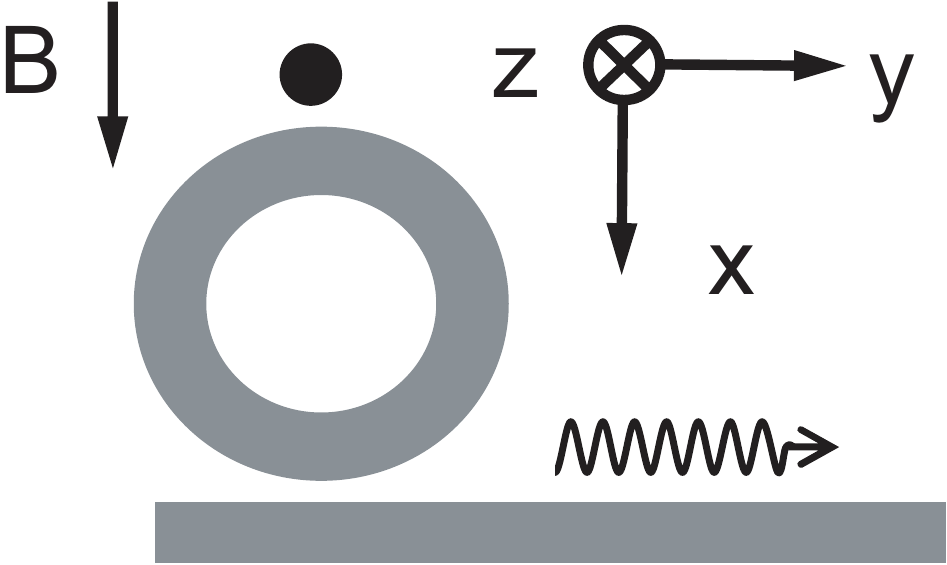}
\end{center}
\caption{Schematics of the dot-cavity-waveguide coupling structure.} \label{cooling_structure}
\end{figure}

The basic structure is depicted in Fig.~\ref{cooling_structure}.
To be specific, we consider a micro-ring coupled to a QD.
Such a structure may be constructed by etching a semiconductor
surface where QDs are located. Structures in photonic crystals
should be ideal alternatives (see Sec.~\ref{subsec_distributed_sructure}). The specific structure, however, is not
crucial to the physics to be discussed below.
The attached micro-cavity would strongly modify the electromagnetic
vacuum in the vicinity of the QD.
The coupling to the cavity mode (which is taken as a whispering gallery mode)
is realized due to the overlap between the QD and the evanescent
wave of the cavity mode. A  waveguide coupled to the
cavity serves as a quantum channel for a cavity photon to escape rapidly
to the environment. For the purpose of spin cooling, such a directional waveguide
is not necessary, and actually one could as well use a cavity with
large leakage (bad cavity situation). Since the emitted photon carries certain
information about the qubit, a guided channel
with cavity also enables such information to be retrieved,
either for quantum measurement (as will discussed later)
and for quantum error diagnosis. The incorporation of a waveguide along a cavity
then involves extra designing and fabricating cost.

\begin{figure}[b]
\begin{center}
\includegraphics[width=6cm]{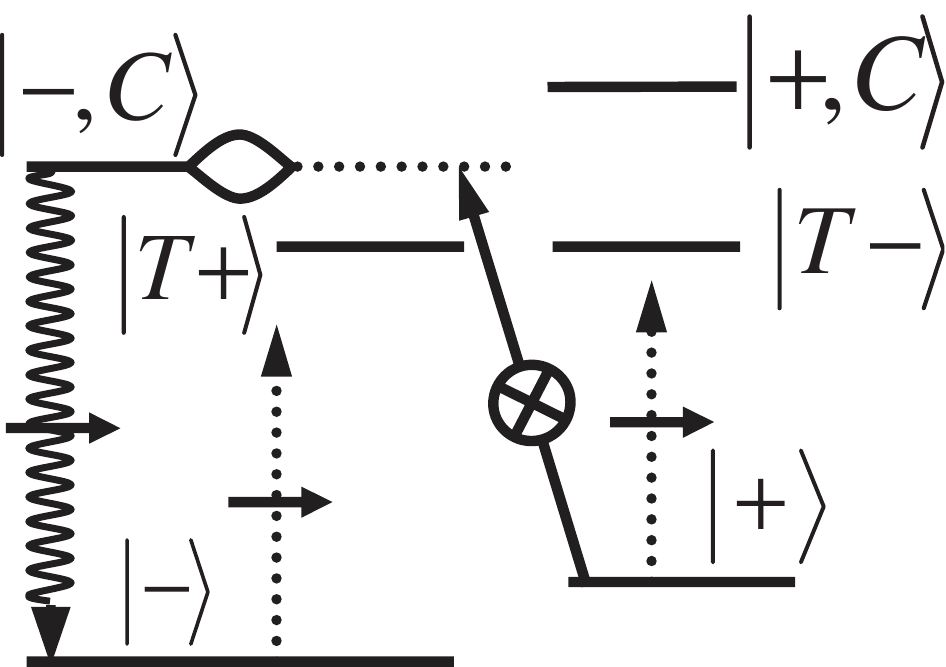}
\end{center}
\caption{Basic optical processes for initializing a spin by controlled cavity QED.
The dotted, solid, and wavy arrows represent the AC Stark pulse ($X$-polarized),
the tipping pulse ($Y$-polarized), and the spontaneous emission, respectively.}
\label{cooling_diagram}
\end{figure}

We consider the Voigt configuration. The spin eigenstates
under a static magnetic field in the $x$ direction are denoted
$|\pm\rangle$. The spin states may be flipped to the two
degenerate trion (exciton plus an electron) states $|T\mp\rangle$
or $|{T\pm}\rangle$ by an $X$- or $Y$-polarized tipping pulse,
respectively. The trion states are, by design, off-resonance
from the cavity modes. We assume such off-resonance condition for several considerations:
(1) It avoids the cavity-induced optical decoherence during quantum operations of the spin.
(2) A local node usually consists of a number
of QDs to a cavity, so it is unlikely to have all the QDs are in resonance with a cavity mode.
(3) The resonance coupling may be realized by transient control via optical Stark effect
which provides flexibility for selectively initialize a spin in a QD in the cluster.
The relevant cavity mode is denoted by $|C\rangle$. The evanescent wave of the
cavity mode is designed to be $X$-polarized in the vicinity of the
nanodot, so that when brought within resonance, the trion states
$|{T\pm}\rangle$ and the cavity states $|\mp,C\rangle$ are coupled
into two split trion-polariton states, respectively. This provides
a fast decay of the trion to a spin state by emitting a photon
into the quantum channel. The pump light is $Y$-polarized
so that a Raman pathway is formed from the spin state $|+\rangle$ to the
trion state $|T+\rangle$ (by the optical pumping),
to the cavity state $|-,C\rangle$ (by dot-cavity coupling), and to the spin state $|-\rangle$ (
by spontaneous photon emission into the waveguide).

\begin{figure}[b]
\begin{center}
\includegraphics[width=10cm]{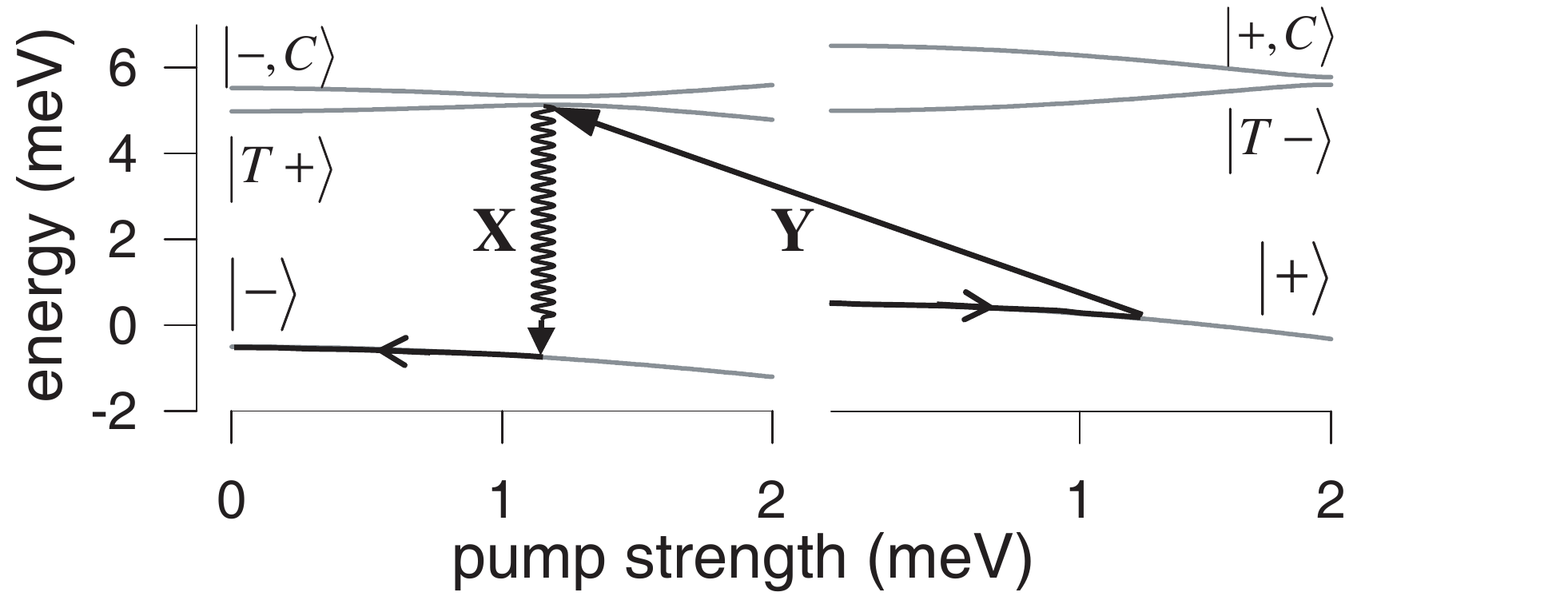}
\end{center}
\caption{(Extracted from Ref.~\cite{Liu_readwrite})
A quantum version of the reverse Carnot cycle for spin qubit initialization
in a QD. The grey curves are the energies of different states versus the Rabi
frequency of the AC Stark pulse, in the rotating frame.} \label{cooling_cycle}
\end{figure}

The optical cycle (similar to a reverse Carnot cycle) for cooling a spin
qubit is illustrated in Fig.~\ref{cooling_diagram}~\cite{Liu_readwrite}. Without loss of generality,
we assume that the electron before optical pumping is in an unpolarized state, i.e.,
${\hat{\rho}}(-\infty)=0.5|-\rangle\langle -|+0.5|+\rangle\langle +|$.
A cooling cycle consists of four basic steps:
\begin{enumerate}
\item An $X$-polarized AC Stark pulse
is adiabatically switched on, bringing the states $|{T+}\rangle$ and $|-,C\rangle$
into resonance by AC Stark effect.
\item
A $Y$-polarized pump pulse flips the spin
up state $|+\rangle$ to the polariton states formed by $|{T+}\rangle$ and $|-,C\rangle$.
\item
The polariton states relax to the spin down state $|-\rangle$ rapidly by emitting a
photon into the waveguide, dumping the spin entropy to the environment.
\item
The AC Stark pulse is adiabatically switched off. No photon-generation or spin-flip would take
place if the initial spin state is $|-\rangle$.
\end{enumerate}
Ideally, after one cooling cycle, the spin is fully polarized with the entropy mapped into the quantum channel,
and the final density matrix becomes
$|-\rangle\langle -|\otimes\left(0.5|0\rangle\langle 0|+0.5|1\rangle\langle 1|\right),$
where $|n\rangle$ is the $n$-photon waveguide state. This
is an idealization of the single-shot initialization of a spin qubit in a QD.
In reality, the single-shot initialization is subject to errors due to
the spontaneous emission of photons into free space, by which the
trion state may relax to either spin state depending on the polarization
of the emitted free-space photon, while the cavity photon couples only the
transition to the target spin state.

The whole system under the optical control is described by the Hamiltonian
\begin{eqnarray}
\hat{H}&\equiv& \Omega_C\hat{a}^{\dag}\hat{a}
\pm\frac{\omega_c}{2}|\pm\rangle\langle\pm|+\Omega_{T}|T\pm\rangle\langle
T\pm|
+ \left(g_{\rm cav}|T\pm\rangle\langle\mp|\hat{a}+{\rm H.c}\right)\nonumber \\
  &+&\left[\chi_t(t){\mathbf
e}_t+\chi_p(t){\mathbf e}_p\right]\cdot{\mathbf
e}_{X}\left(|T\pm\rangle\langle\mp|+r_C\hat{a}^{\dag}\right)+{\rm
H.c}
   \nonumber \\
  &+& \left[\chi_t(t){\mathbf
e}_t+\chi_p(t){\mathbf e}_p\right]\cdot{\mathbf
e}_{Y}|T\pm\rangle\langle\pm|+{\rm H.c},
\end{eqnarray}
where $\hat{a}$ is the cavity mode annihilation operator,
$\Omega_C$ is the cavity mode frequency, $\Omega_T$ is the bare trion state frequency,
$\omega_c$ is the electron Zeeman splitting,
$g_{\rm cav}$ is the coupling between the cavity and the QD, $\chi_p$ is the
Rabi frequency of AC Stark pulses with polarizations ${\mathbf e}_p$,
$\chi_t$ is the Rabi frequency of the pump pulse with polarization ${\mathbf e}_t$,
and $r_C$ is the strength of the direct coupling between the light
pulse and the cavity mode (relative to the coupling to the trion state).

\begin{figure}[b]
\begin{center}
\includegraphics[width=10cm]{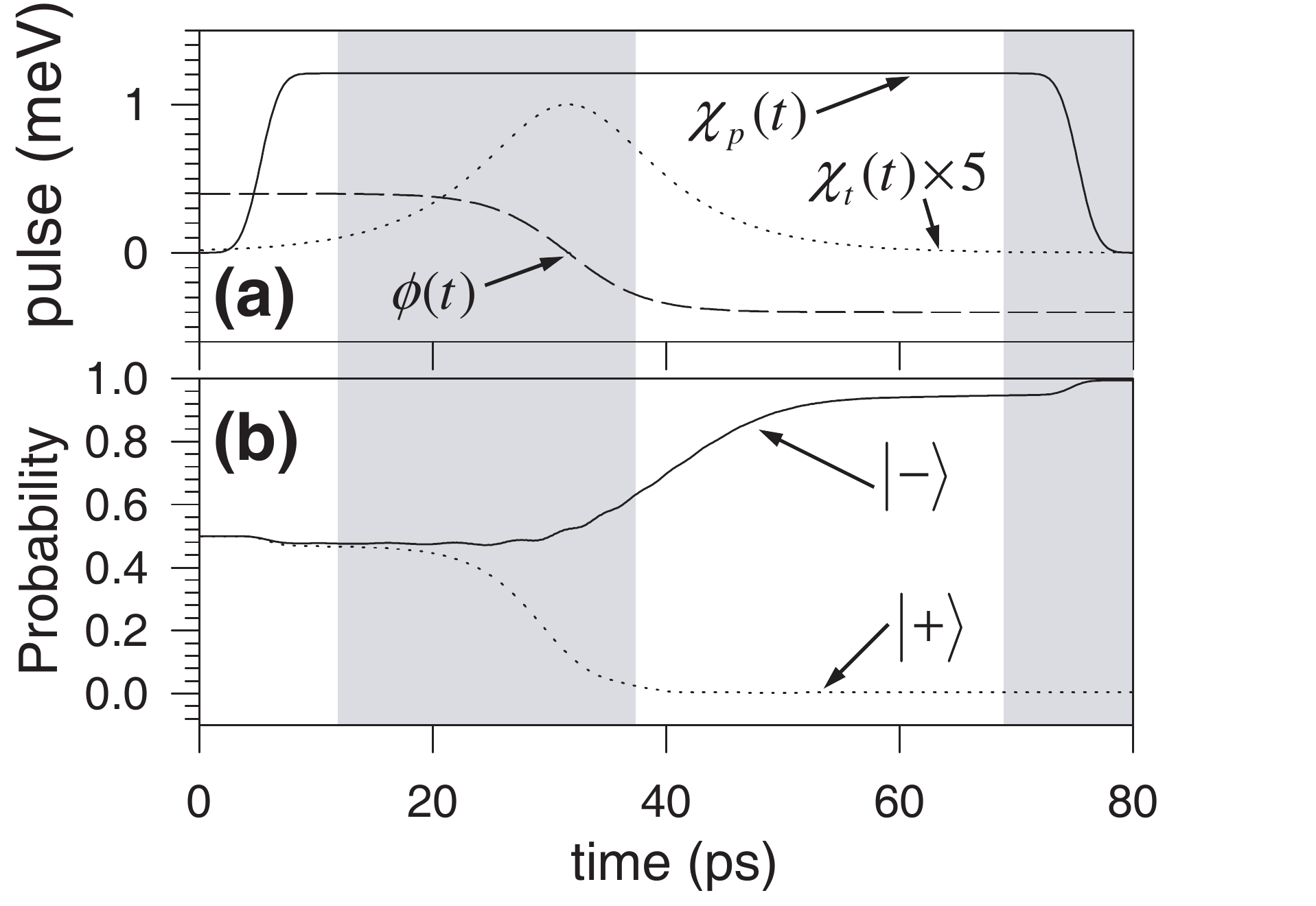}
\end{center}
\caption{(Extracted from Ref.~\cite{Liu_readwrite})
(a) The Rabi frequencies of the AC Stark pulse and the tipping pulse
(amplified by a factor 5), and the sweeping frequency of the
tipping pulse. (b) Probabilities of spin down and up. Different
steps of the cooling cycle, indicated by \textcircled{1}-\textcircled{4},
are distinguished by shadowed areas.} \label{cooling_curve}
\end{figure}

The dynamics in the cavity-dot system is rather complicated and some attention should be
paid to the designing of the controlling pulses. To bring the trion state into the resonance
with the cavity mode and to maintain the resonance, the AC Stark pulse is designed
to have an almost square profile. Also, the switch-on and off of the pulse should be made
smooth enough to avoid non-adiabatic excitation of the trion states from the target spin state.
We choose the pulse to be of the profile
\begin{equation}
\chi_{p}(t)=
{\chi_p}e^{-i\Omega_pt}
\left[\text{erf}\left({\sigma_p(t-t_1)}\right)-\text{erf}\left({\sigma_p(t-t_2)}\right)\right],
\end{equation}
[see Fig.~\ref{cooling_curve} (a)].
As the AC Stark pulse maintains the resonant cavity-dot coupling which
facilitates the photon escape to the quantum channel, the trion
state relaxes very fast (on the time-scale of $g_{\rm cav}^{-1}$
and $\gamma^{-1}$, $\sim10$~ps).
The flipping pulse should be a $\pi$-pulse for Rabi rotation
between $|+\rangle$ and $|{T+}\rangle$. Due to the dynamical nature
of the states (dressed by the AC Stark pulse)
and the rather small polariton splitting ($\sim 0.1$~meV), a
perfect $\pi$-rotation requires an extremely long pulse.
In principle, a full excitation of the spin up state to
the polariton states could be made much faster by pulse shaping.
One of such pulse shaping is to use geometrical control of the transition
which is robust against uncertainty of the polariton frequencies.
The geometrical control is realized by using a a chirped pulse
as $\chi_t(t)=\chi_t e^{-i\phi(t)-i\Omega_t t} {\rm sech}\left(\sigma_t (t-t_t)\right)$
with the phase sweeping rate $\dot{\phi}(t) =-\sigma_c
\tanh\left(\sigma_t (t-t_t)\right)$~\cite{Goswami_pulse}. The
frequency of the pulse now will sweep from $\sigma_c$ above
$\Omega_t$ to $\sigma_c$ below and the sweeping range
$\left[\Omega_t-\sigma_c,\Omega_t+\sigma_c\right]$ covers both of
the trion-polariton states. The initial spin state $|+\rangle$
will be brought adiabatically into a superposition of the two polariton
states, which relaxes rapidly to the target spin state
$|-\rangle$. Such a geometrical flip can also tolerate to some
degree laser fluctuations and uncertainty in the dipole moment,
transition energy, and selection rules.

The cooling process is simulated by numerically solving the
master equation of the dot-cavity system
\begin{eqnarray}
 \partial_t\hat{\rho}=-i\left[\hat{H},\hat{\rho}\right]-\frac{\gamma+\gamma'}{2}{\cal L}_{\hat{a}}\hat{\rho}-\frac{\Gamma}{2}{\cal}
\sum_{s,s'=\pm}{\cal L}_{|s\rangle\langle Ts'|}\hat{\rho},
\end{eqnarray}
where ${\cal L}_{\hat{o}}\hat{\rho}\equiv
2\hat{o}\hat{\rho}\hat{o}^{\dag}-\hat{o}^{\dag}\hat{o}\hat{\rho}-\hat{\rho}\hat{o}^{\dag}\hat{o}$
denotes a Lindblad super-operator, $\gamma$ is the cavity-waveguide escape rate, $\gamma'$ is
the cavity-free-space loss rate, $\Gamma$ is the trion decay rate
due to spontaneous emission into free-space. The multi-photon cavity states were
included in the numerical calculation, as they renormalize the AC
Stark shift (the real excitation of multi-photon states is
negligible due to the off-resonance condition). Inclusion of up to
3-photon states was found sufficient to obtain converged results.

We test the cooling efficiency with a set of realistic parameters~\cite{Liu_readwrite}.
The Zeeman splitting $\omega_c=1$~meV, $\gamma=0.2$~meV, $\gamma'=0.045$~$\mu$eV
(corresponding to an intrinsic $Q$-factor $\sim 3\times 10^7$),
the dot-cavity coupling $g_{\rm cav}=0.1$~meV, the cavity-trion
detuning $\Omega_C-\Omega_T-\omega_L/2=0.5$~meV, $\Gamma=1$
$\mu$eV, and $r_C=0.3$. For the parameters given above, the trion state $|{T+}\rangle$ and
the cavity state $|-,C\rangle$ are brought into resonance when the
AC Stark pulse strength $(2\chi_p)$ is maintained at 1.21~meV.
Maintaining the resonance for $t_2-t_1=70$ ps is found sufficient
for the total dissipation of the photon. The spectral width of the
AC Stark shift pulse ($\sigma_p=0.354$~meV) is set much smaller
than the detuning ($\Omega_T+\omega_L/2-\Omega_p=5.5$~meV),
so that the excitation due to non-adiabatic switch-on and
off is negligible. The flipping pump pulse has a frequency sweep range of
$\sigma_c=0.4$~meV, strength $\chi_t=0.2$~meV, and
duration $1/\sigma_t=6.58$~ps. The spin state $|+\rangle$ is flipped
to the polariton states with negligible error.

Figure~\ref{cooling_curve}~(b)  shows that a single cooling cycle
completed within 80 ps produces an almost 100\% polarized spin
from a maximally mixed state. The density matrix at the end of
the cycle is $\hat{\rho}=0.9945 |-\rangle \langle -|+0.0040|+\rangle\langle +|+\hat{\rho}_{\rm
err}$, where $\hat{\rho}_{\rm err}$ is the probability ($\approx$0.15\%) of the system remaining in the trion states
which results mainly from the non-adiabatic switching of the AC
Stark pulse. The extra error ($\approx 0.4\%$) comes mainly from the decay of the
trion with photon emission into free space.

\subsection{Initialization by entropy dump to phonon baths}
\label{initialization}

The phonon bath in a QD is
often taken as a source of
the qubit decoherence, but it can also be used as as a resource for ultrafast cooling.
The electronic energy levels in QDs are discrete, it has been argued that
the phonon emission would be much suppressed due to the lack of available
final states fulfilling the energy conservation~\cite{Bottleneck}.
This so-called phonon bottleneck effect, because of
its importance in QD lasers and detectors, has been extensively studied
both in experiments and in theories. Nonetheless,  experiments
have established various mechanisms for rapid relaxation of electrons from exited states to the ground states,
such as the Auger-process and the multi-phonon process~\cite{noBottleNeck}. The observed relaxation time
varies from tens of picoseconds to a few picoseconds~\cite{Rapidrelaxation1,Rapidrelaxation2}.
The spin relaxation, especially for holes due to the large spin-orbit coupling,
always accompanies with the energy relaxation, for the mixing in the excited states is
much stronger than in the ground states. Thus we can use the phonon emission as
fast entropy dumping channel for an electron spin in a QD.

\begin{figure}[b]
\begin{center}
\includegraphics[width=8cm]{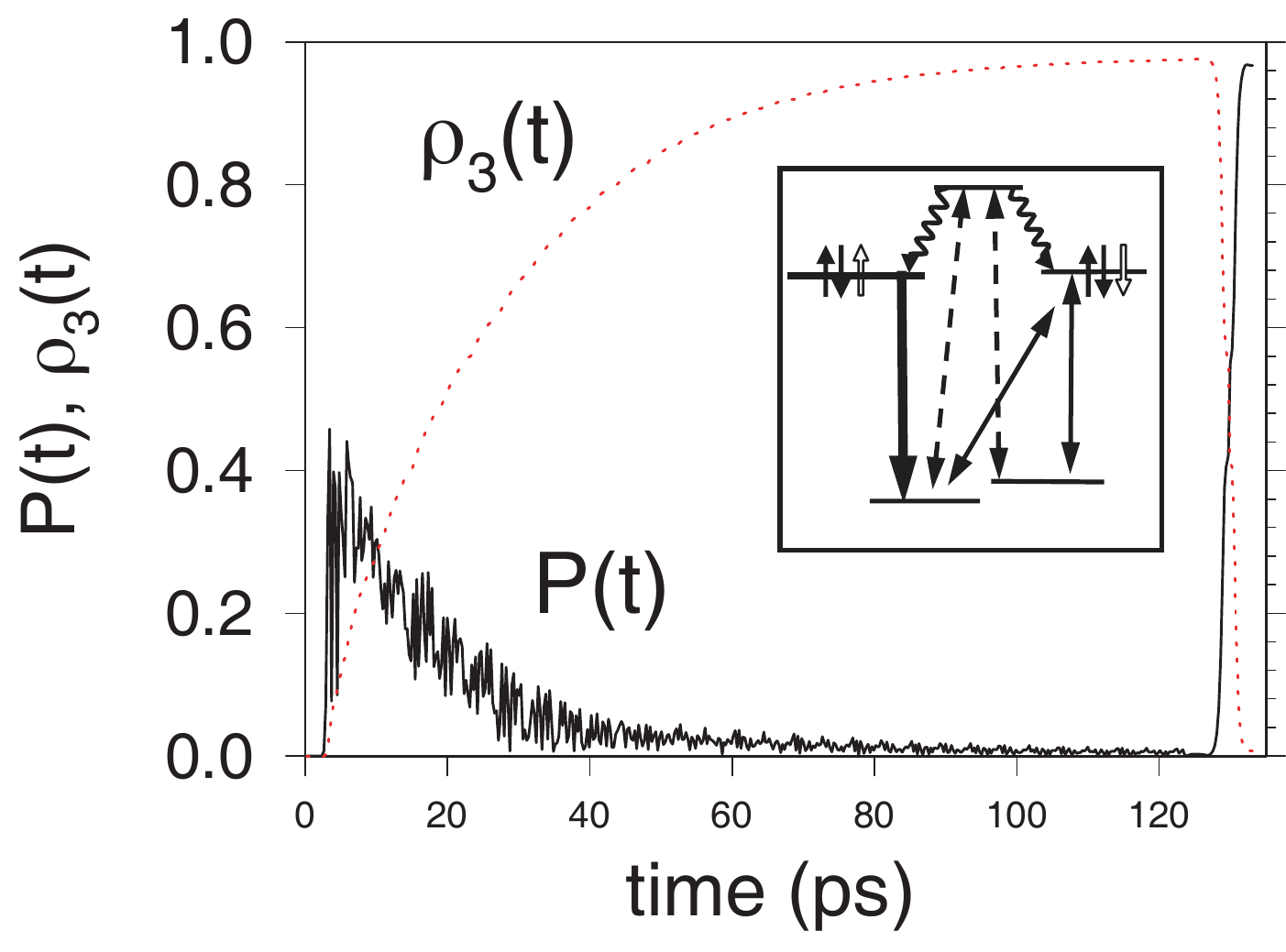}
\end{center}
\caption{{{}}  The numerical simulation of a typical
initialization cycle using phonon emission from excited trion states.
Inset: Schematics for the spin initialization process.}
\label{initializing}
\end{figure}

A cycle of the spin initialization is illustrated in the inset of
Fig.~\ref{initializing}:
\begin{enumerate}
\item
 A circularly polarized
laser pulse resonantly pumps the electron states to an excited
trion state.
\item
The excited trion state relaxes rapidly to the
ground trion states $|\uparrow\downarrow\Uparrow\rangle$ and
$|\uparrow\downarrow\Downarrow\rangle$. During the relaxation,
the hole spin is not conserved due to the strong heavy-light hole
mixing in the excited state, and without loss of generality we
assume the rate is the same for the relaxation to the two ground
trion states.
\item
To deplete the population of one of the ground
trion states, a circularly polarized laser pulse is applied to
resonantly couple, e.g., the trion state
$|\uparrow\downarrow\Downarrow\rangle$ to the electron spin state
$|\uparrow\rangle$.
\item
After a period of pumping and relaxation,
the population will be accumulated to the trion state
$|\uparrow\downarrow\Uparrow\rangle$, and, thus, an ultrafast
$\pi$-pulse can be applied to flip the trion state into a pure
electron spin state.
\end{enumerate}
The most time-consuming step of this laser
cooling process is the pump and relaxation processes which is
limited by the carrier relaxation rate and the pump strength.
Care should be taken to avoid the population being trapped in
some dark state which could be possible in the strongly driven
multi-level system (similar to the electromagnetically induced
transparency effect). Fig.~\ref{initializing} shows a typical
spin initialization cycle in which the spin is pumped from a
fully mixed state to a nearly fully polarized state in 120~ps
(the polarization at the end of the cycle is about 97\%), where
$\rho_3(t)$ is the population stored in the ground trion state
$|\uparrow\downarrow\Uparrow\rangle$, and $P(t)$ is the
polarization of the electron spin. The pump strength for the
excited and ground trion states has been set to be 2~meV and
0.5~meV, respectively, the carrier relaxation time is 2~ps, and
the spontaneous emission time is 1~ns. The $\pi$-pulse flipping
the trion state into the spin state is chirped for optimized
performance.

The phonon baths offer an alternative solution of ultrafast
spin initialization with comparable speed to a photon bath
enhanced by cavity QED. The real excitation of the excited electron
and hole states, however, could cause some complications. One such
issue is the trapping of electrons and holes in the dark states,
which cause the loss of a spin qubit. And in fluctuation QDs, it is not likely the excited state energy is high enough
for LO phonon emission.

\section{Quantum Non-demolition Measurement of Spin Qubits}
\label{readout}

The result of a quantum algorithm is obtained by quantum measurement of
specific qubits. To diagnose the errors generated in quantum error
correction, quantum measurement is also needed.

In current single-spin experiments in QDs~\cite{dotSteel,Nico_thesis,Imamoglu_QDSpinPrep,Atature:2007NaturePhysics,
Mikkelsen:2007,Yamamoto_08NatureSpinControl,Gammon08PRLspinQDs} and in diamond color centers~\cite{Jelezko:2002},
the measurement is usually achieved by cycling read of the spins,
which amounts to time-ensemble measurement. Remarkably, efficiency enhancement in measurement of NV center spins
in diamond has been made using ancillary nuclear spin qubits~\cite{L.Jiang10092009,Steiner10NVreadout},
but the single-shot measurement still remains illusive.

Eventually, single-shot measurement of quantum registers is demanded
for scalable quantum computation for large-scale problems. In Appendix~\ref{measurement_in_Shor},
we show that quantum non-demolition (QND) measurement, even with certain errors, may be converted to
a single-shot measurement and thus is scalable. To realize single-shot measurement,
the crucial issue is how to enhance the coupling between
the probe (photons, e.g.) and the spin qubits. Below we provide a possible solution of using cavity QED,
which may be implemented in-situ in a quantum network (discussed in Sec.\ref{sec-network}).
The discussion is based on QD cavity-QED systems, but may be readily extended to NV centers where cavity-QED have
been also demonstrated~\cite{Park:2006}.

\subsection{Scalability of quantum measurement in quantum computation}

In quantum computation (such as in Shor's algorithm~\cite{Shor94,NielsenChuang}), the quantum register could be in a
superposition state right before the measurement $|\Psi\rangle=\sum_x C_x|x\rangle,$ where
$|x\rangle\equiv |x_1,x_2,\ldots,x_N\rangle$ is a computational
basis state for the $N$-qubit register. The measurement should be
in the computational basis which returns any $|x\rangle$
contained in the superposition and the computation result is
derived from the measured $x$. If an algorithm requires the measurement of the
wavefunction $C_x$, i.e., the tomography of the quantum state, it would be
an analog computation instead of a digital one, and much worse, it would not be
scalable as the number of measurements would increase exponentially with the number of
qubits in the register to be measured~\cite{James_Tomography}.
Such measurements are ensemble measurements.

A point we would like to put forward here about the scalability of
quantum measurement in quantum computation is that an ensemble measurement
is not scalable in the sense that the size of the ensemble would increases
exponentially with the problem size (defined as the number of qubits
registering the computation result involving a quantum measurement)~\cite{Liu_readwrite}.
It has been well-known that an ensemble quantum computation is not scalable if the qubits cannot
be initialized to a pure state~\cite{Cory_ensemble}. The issue of scalability associated with the
ensemble measurement~\cite{Liu_readwrite}, however, has received less attention in spite of its
importance in quantum computation. This problem is briefly explained below, and more detailed
discussion is given in Appendix~\ref{measurement_in_Shor}.

First we notice that the
uncorrelated ensemble measurement cannot be used to read out the
quantum register in general algorithms (especially for those
terminating in superposition states such as Shor's
algorithm). In an uncorrelated measurement, the spins are measured
independently.  Thus, in general which basis states are in the superposition
cannot be deduced from the measured result. For example,
the superposition state $|000\rangle+|111\rangle$ will give the
same uncorrelated measurement result as the state
$|000\rangle+|011\rangle+|101\rangle+|110\rangle$.
In some algorithms such as Shor's algorithm for factorization, the
number of basis states in the superposition may increase exponentially
with the number of qubits, thus the number of possible
superposition states yielding the same uncorrelated measurement
result would increase exponentially with the number of qubits
measured.  So the quantum measurement has to be a correlated one
(e.g., the photon counting should be in coincidence) to be
scalable. If a coincidence measurement is a destructive one, the
procedure has to be run from the very beginning in each
repetition, and the superposition state can collapse into any
possible $|x\rangle$, which is in general different from one
cycle to another. To have a certain $x$ to be measured at least
twice for the sake of confidence, the number of repetitions to be
performed should be in the order of the number of basis states in
the superposition, which again could be an exponential function
of the problem size. In conclusion, the enhancement of signal-background
contrast by ensemble quantum computation is not a scalable solution.

In realistic cases, the signal of a single-shot measurement of a single quantum
object is usually too weak to be distinguished from noise.
Thus the signals are to be amplified either by simultaneously
measuring a large number of identical ``quantum computers'' running the same
quantum program or by repeating the quantum program under
identical conditions for a large number of times. This renders
the quantum computation to be an ensemble one and hence not scalable.

Using ensemble measurement as a solution to the detection efficiency problem
has been applied in various systems, including nuclear spins in liquid-phase NMR~\cite{Cory_ensemble}
or solid-state NMR~\cite{Ladd:2002PRL}. Here we consider quantum computing with optically controlled spins.
The spin-dependent absorption (or other optical methods such as spin-dependent
scattering and Faraday rotation) may be used to measure a single spin in a QD,
in which a probe pulse resonant with the spin-trion transition in a QD
measures the spin in the computational basis by detecting whether or not
a single photon of the probe pulse has been absorbed.
To measure a register with many qubits, the probe should be composed of
many pulses in which each pulse addresses a QD individually by spatial
and spectral resolution. In practice, the interaction between the probe pulse and a single spin
is very weak, and also photon collection and detection efficiency is less than one, so
the information obtained about a spin is on the average much less than one bit.
The probe measurement has to be repeated for many times to accumulate statistical
confidence in the measured result (or signal-to-noise ration).
Considering the fact that the probe process is destructive to the spin state
(since a trion state excited by the probe pulse may return to either spin state
regardless of the original spin state), the repetition has to be run from the very
beginning of the quantum computation which prepares the quantum computer to the same superposition
state. This makes the quantum computation an ensemble one, which in general is not scalable.

Nonetheless, the ensemble measurement could still be useful in
demonstrating quantum algorithms for small-size problems.
For instance, as will be shown later, to demonstrate
Shor's algorithm for factorizing 15 by optical control of electron spins
in QDs, the program would be completed in less than a few ns,
and the number of computational basis states in the superposition
is less than 10. A commercial Ti-Sapphire pulse laser with
repetition rate of about 100 MHz could be used to carry out an ensemble
of repeated running and measurement in a reasonably short time in
a pump-probe configuration, where the initialization and gate control
are viewed as a single complex pump pulse and the probe pulse is composed of
many frequency components (and detected in multi-channels.)

\subsection{Quantum non-demolition measurement via cavity quantum electrodynamics}

\begin{figure}[b]
\begin{center}
\includegraphics[width=6cm]{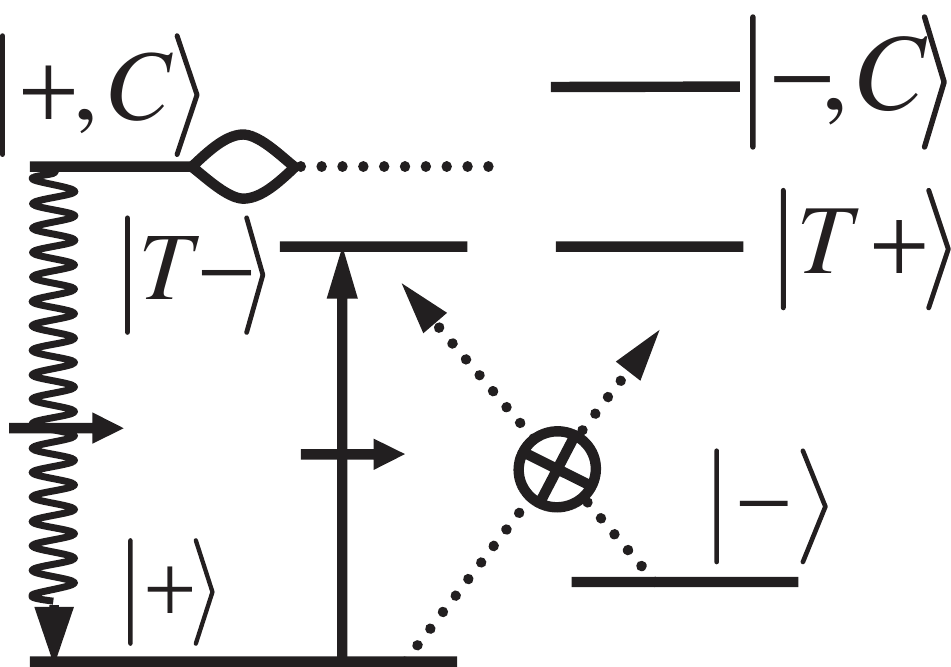}
\end{center}
\caption{Basic optical processes for QND measurement of a spin by controlled cavity QED.
The dotted, solid, and wavy arrows represent the AC Stark pulse ($Y$-polarized), the tipping pulse
($X$-polarized), and the spontaneous emission, respectively.} \label{QND_diagram}
\end{figure}

In essence, the realization of an efficient single-shot quantum measurement
involves two key elements: rapid quantum state entanglement of
the target qubit (here a spin in a QD) with a detectable information carrier (such as a photon), and
efficient and faithful collection of the information carrier (such as by a photon detector
with a high efficiency and a low dark count rate). Errors in the measurement,
due to the efficiency or dark count problems, e.g., can not be fully eliminated.
Thus cycling of single-shot measurement is required to accumulate statistical confidence.
As discussed in Appendix~\ref{measurement_in_Shor}, the cycling of single-shot measurement is scalable
when the measurement is a quantum non-demolition (QND) one, i.e., the qubit state after
the measurement is (ideally) an eigenstate in the measurement basis corresponding to
the measurement output.

The QND measurement being rapid is an essential element in the following sense:
The cycling of measurement has to be completed in a time much shorter
than the qubit is significantly disturbed by the environment.
So we should look for a rapid quantum information transfer between
the spin qubit and a medium to be detected. An ideal medium is photons.
The cavity-enhanced entropy dumping in the ultrafast initialization is
indeed a rapid information transfer (but there the information has been
viewed as noise). Thus the ultrafast spin initialization and
rapid QND may be integrated in the same micro-photonic structure~\cite{Liu_readwrite}.

\begin{figure}[b]
\begin{center}
\includegraphics[width=10cm]{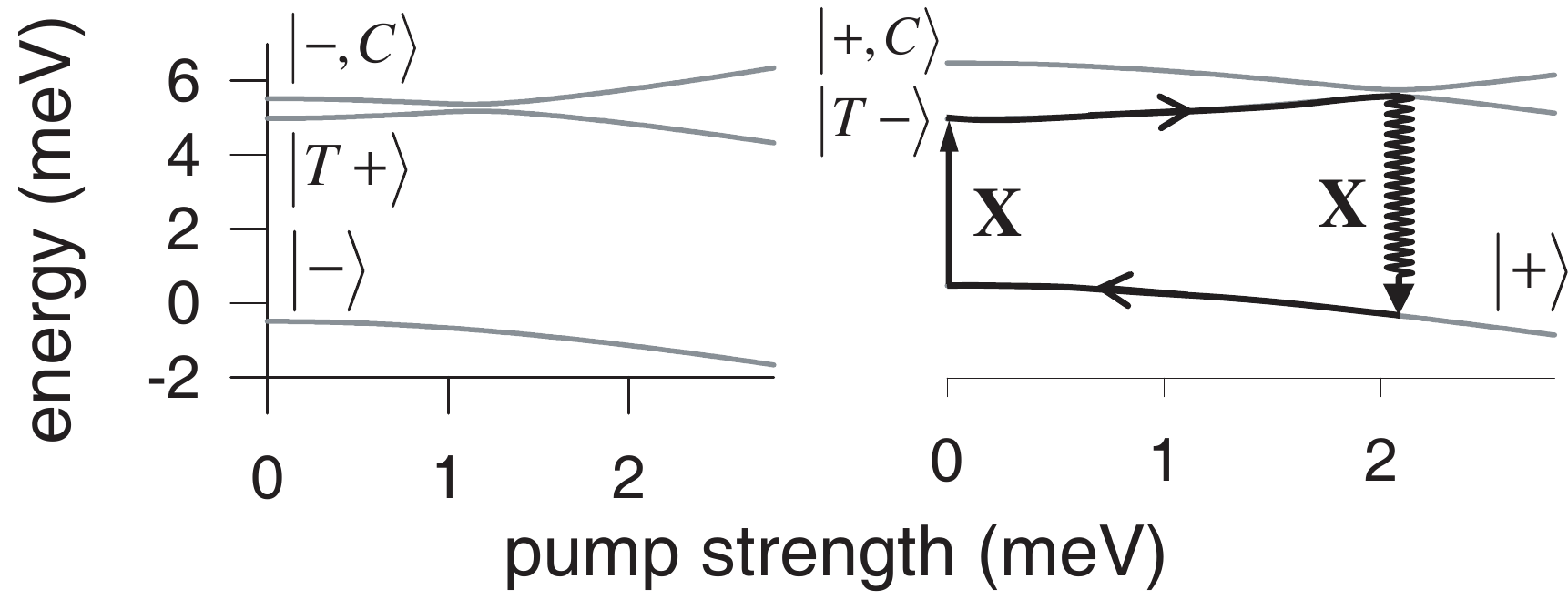}
\end{center}
\caption{(Extracted from Ref.~\cite{Liu_readwrite})
Detailed optical process for the measurement cycle.
The grey curves are the energies of different states versus the Rabi
frequency of the AC Stark pulse, in the rotating frame.} \label{QND_cycle}
\end{figure}

To ensure efficient detection of the transferred quantum information,
the photons escaped from the cavity should be directed
into a quantum channel, unlike in the initialization process where it
does not matter which direction the dumped entropy flows.
To realize such a directional information flow, a waveguide
could be fabricated in the proximity of the cavity, and the waveguide
may be coupled to a fiber which conducts the photon to a detector~\cite{microsphere1_vahala,microsphere2_vahala,
vahala_MCfiber,Painter_cavityFiber1,Painter_cavityFiber2,Painter_cavityFiber3,Vuckovic_cavityFiber}.

\begin{figure}[b]
\begin{center}
\includegraphics[width=10cm]{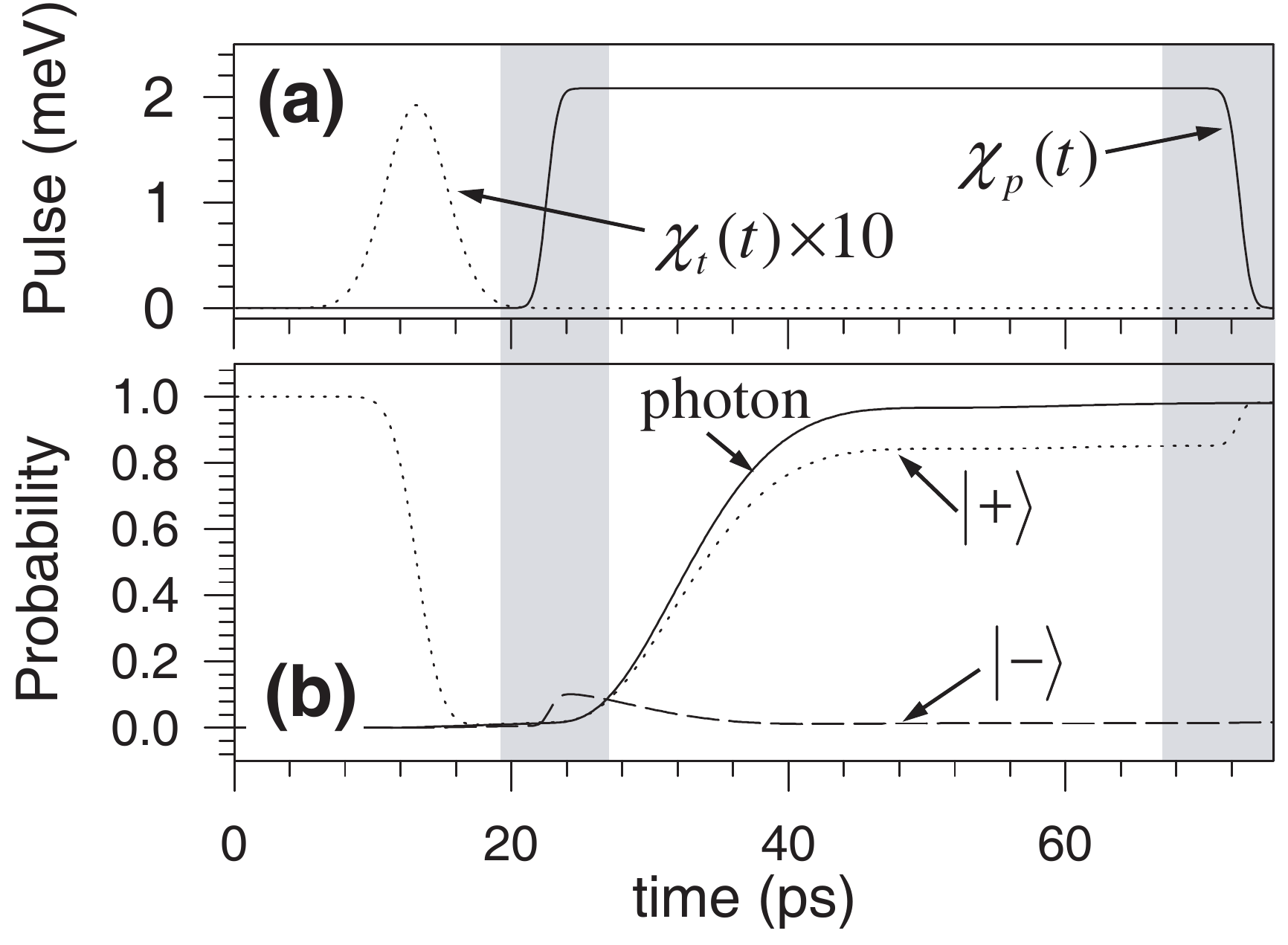}
\end{center}
\caption{(Extracted from Ref.~\cite{Liu_readwrite})
(a) The Rabi frequencies of the AC Stark pulse and the tipping pulse
(amplified by a factor 10). (b)  Probabilities of spin down
and up, and the number of waveguide photons, for a spin initially
polarized up. Different steps of the cooling cycle, indicated by \textcircled{1}-\textcircled{4},
are distinguished by shadowed areas.} \label{QND_curve}
\end{figure}

The optical control of the cavity-QED for a rapid measurement
is similar to that  for the ultrafast initialization~\cite{Liu_readwrite}.
But to enable measurement cycling, the measurement should be
non-destructive, i.e., the spin basis state should return back to
its initial state after a cycle of measurement, with close to unity
probability. Thus we need to switch the polarizations of the tipping and
the AC Stark pulses from $(Y,X)$ to $(X,Y)$,  respectively.
The energy diagram with optical transitions indicated is shown in
Fig.~\ref{QND_diagram}. A measurement cycle may be processed in four basic steps
(see Fig.~\ref{QND_cycle}):
\begin{enumerate}
\item
An $X$-polarized tipping pulse flips
the spin state $|+\rangle$ to the trion state $|{T-}\rangle$.
\item
A $Y$-polarized AC Stark pulse adiabatically switched on
drives the trion state into resonance with the cavity state
$|+,C\rangle$.
\item
The trion state resonantly tunnels into
the cavity state and relaxes rapidly back to the spin state
$|+\rangle$, leaving a photon emitted into the quantum channel.
\item
The AC Stark pulse is adiabatically switched off.
\end{enumerate}
Suppose that the spin state to be measured is
$\alpha|+\rangle+\beta|-\rangle$ and the channel is initially in the vacuum
state $|0\rangle$.
An ideal measurement process will transform the
system into the entangled state $\alpha|+\rangle|1\rangle+\beta|-\rangle|0\rangle$, so that the
detection of the photon projects the electron into a spin
basis state, providing a QND measurement of the spin.
Note that the pulse timing for measurement is different from that
for cooling [cf. Fig~\ref{cooling_curve} (a) and Fig.~\ref{QND_curve}
(a)]. In the measurement cycle, the flipping pulse need not be
chirped, since
here the Rabi flop occurs between stationary energy levels and the
transition between the spin state and the trion is well separated in frequency
from the cavity mode. Instead, a simple Gaussian $\pi$ pulse
$\chi_t(t)=\chi_te^{-\sigma_t^2(t-t_t)^2/2-i\Omega_t t}$ may be used.
The AC Stark pulse is chosen $Y$-polarized to avoid direct excitation of the cavity mode.

Numerically simulation is done to check the efficiency of the measurement cycle~\cite{Liu_readwrite}.
Since the initialization and the QND measurement are supposed to be operated
via the same photonic structure, the simulation is done with the same physical structure
as that in Fig.~\ref{cooling_curve}. The number of photons emitted into the waveguide
is calculated with $\partial_t n=\gamma\langle\hat{a}^{\dag}\hat{a}\rangle$.
The tipping and the AC Stark pulses are set such that $1/\sigma_t=2.19$~ps,
$\chi_t=0.192$~meV, $\Omega_t=\Omega_T-\omega_L/2$,
$\sigma_p=0.707$~meV, $2\chi_p=2.08$~meV,
$\Omega_T+\omega_L/2-\Omega_p=5.5$~meV, and the duration of the
pump pulse $t_2-t_1=50$~ps. After a single cycle of measurement,
an initial state $\hat{\rho}_0=|+\rangle \langle +|$ results in
the final state $ \hat{\rho}_1= 0.0161|-\rangle \langle
-|+0.9824|+\rangle\langle +|+\hat{\rho}_{\rm err} $ with the
number of photon emitted into the waveguide $n=0.9806$ [see
Fig.~\ref{QND_curve}~(b)], while an initial state
$\hat{\rho}_0=|-\rangle \langle -|$ results in the final state
$\hat{\rho}_1=0.9955 |-\rangle \langle -|+0.0040|+\rangle\langle
+|+\hat{\rho}_{\rm err}$ with $n=0.0015$ (not shown). The photon
emitted into the waveguide can be detected with high efficiency.
If the detector has a zero dark-count rate and an
efficiency of 50\%~\cite{Yamamoto_detector},
the POVM (positive operator-valued measures~\cite{NielsenChuang},
see Appendix~\ref{measurement_in_Shor} for more discussions)
for the measurement process can be defined as
\begin{subequations}
\begin{eqnarray}
\hat{P}_-&=& {0.9992}|-\rangle\langle-|+{0.5097}|+\rangle\langle+|, \\
\hat{P}_+&=&={0.0008}|-\rangle\langle-|+{0.4903} |+\rangle\langle+|,
\end{eqnarray}
\end{subequations}
for a non-click or click event, respectively.
Within 5 measurement cycles, e.g, the spin state is measured with accuracy
higher than 97\%, and the back-action noise to the spin is less
than 10\%, while the time duration is less than 0.4 ns, much
shorter than the spin decoherence time.

The QND measurement of single spins thus can be completed within 100~ps,
and the high-efficiency of collecting and detecting photons propagating
in waveguides enables a near unity
accuracy by repeating the single-shot measurement for only a few times. The measurement is still subject to
the problem of less than unity efficiency as well as back-action noise.
As discussed in Appendix~\ref{measurement_in_Shor}, quantum gates and error
tolerating coding can be combined to achieve a sufficiently faithful measurement
of a qubit without rewinding a quantum computing program.

\section{Networking Local Nodes \label{sec-network}}

In section \ref{subsec_distributed_sructure}, we have introduced the
dot-cavity-waveguide coupled structure for distributed quantum
information processing. Quantum nodes are formed by clusters of
singly charged QDs with electron spins as carriers of
stationery qubits. Single photons in optical waveguides or fibers can
transport quantum information between distant nodes. Interfacing
between single spins in QDs and single photons is made possible by the
strong photon confinement in solid state micro-cavities.

The separation between quantum nodes on a single chip which
can range from $\sim ~\mu$m to $\sim~$cm allows parallel optical control, and
intra-chip communication is realized by optical waveguides.
Inter-chip communication is possible by wiring chips together with
optical fibers~\cite{Painter_OE05,Painter_PRB04,Painter_APL04}. For intra-chip communication and short distance
inter-chip communication, the decoherence of the photon qubit is
negligible~\cite{PCWaveguideLoss,Sugimoto:04,FiberOptics_book05}. Thus, it is possible and also highly desirable to
perform inter-node operations in a deterministic way (to be
contrasted with most quantum cryptography and linear optics quantum
computation schemes based on projective measurement which renders
the logical controls probabilistic). The key component is a quantum
interface that allows the deterministic state transfer between spin
and photon qubit.

\begin{figure}[b]
\begin{center}
\includegraphics[width=10cm]{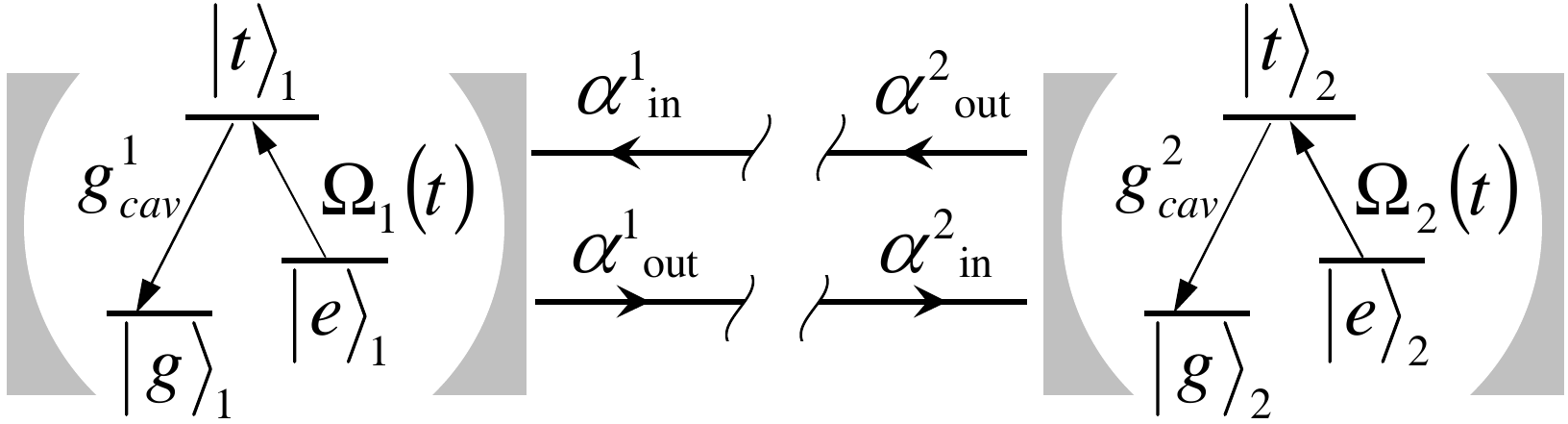}
\caption{Illustration of state transfer in a quantum network. The node is composed of a
cavity coupled to a three-level
$\Lambda$ system. The two ground states $|g\rangle$ and $|e\rangle$
of the three level systems form the Hilbert space for the stationary
qubits. State $|g \rangle$ is coupled to the intermediate $|t
\rangle$ by the cavity mode with strength $g_{\rm cav}$ and
$|e\rangle$ to $|t \rangle$ by a classical light with Rabi frequency
$\Omega(t)$. Direct excitation of cavities by the classical light is
assumed absent. The cavities themselves are coupled to the outside continuum
which forms a photonic channel. Two nodes are connected by
the photonic channel in the following way: the output of node 1 is
directed to node 2 as its input and vice versa.} \label{system}
\end{center}
\end{figure}

The prototype quantum interface for this purpose was proposed by
Cirac {\it et al.}~\cite{StateTransfer_Cirac_Kimble}. It is composed
of a cavity coupled to a three-level $\Lambda$ system, illustrated
in Fig.~\ref{system}. The two ground states, $|g\rangle$ and
$|e\rangle$, of the three level system form the stationary qubit.
State $|g \rangle$ is coupled to the intermediate state $|t \rangle$
by the cavity mode and $|e\rangle$ to $|t \rangle$ by the
external laser field. Direct excitation of cavity by the external laser is assumed
absent. Through the imperfect mirror, the cavity is coupled to the
electromagnetic continuum which forms a photonic channel. A
Raman path from $|e\rangle$ to $|g\rangle$ through the intermediate
state $|t\rangle$ is thus formed. If the three level system is
initially in state $|e\rangle$, an external laser pulse can bring it
to state $|t\rangle$ by a $\pi$ rotation which relaxes
to state $|g\rangle$ by spontaneous emission of a cavity photon. The
cavity photon then goes into the photonic channel forming a
single photon wavepacket. If the three level system is initially
in state $|g\rangle$, it will remain in this state provided the cavity is in its vacuum.
The quantum state carried by the three-level system is thus mapped into
the photon number subspace of the outgoing photon,
\begin{eqnarray}
\left(C_g|g\rangle+C_e|e\rangle\right) \otimes |{\rm vac}\rangle
\rightarrow |g\rangle\otimes\left[C_g|{\rm
vac}\rangle+C_e|\alpha_{\rm out}\rangle\right] .
\end{eqnarray}
where $|\alpha \rangle$ denotes a single photon wavepacket in the
photonic channel and $|{\rm vac}\rangle$ the channel vacuum. This
process forms the basis for the sending function (i.e. the mapping from a stationery
qubit to a flying qubit) of a quantum node. The
receiving function is the mapping from a flying qubit to a stationary
qubit and can be realized as the time reversal of a sending process,
\begin{eqnarray}
|g\rangle\otimes(C_g|{\rm vac}\rangle+C_e|\alpha_{\rm in}\rangle)
\rightarrow (C_g|g\rangle+C_e|e\rangle)\otimes |{\rm vac}\rangle.
\end{eqnarray}
With the output of the sending node directed as the input of the
receiving node (see Fig.~\ref{system}), transfer of qubits between
two distant nodes can be performed.

This quantum network by cavity QED was
originally proposed for quantum computation with atomic
systems~\cite{StateTransfer_Cirac_Kimble}. Critical experimental steps towards realizing such a
quantum interface in atom-cavity QED systems have been progressively
demonstrated~\cite{Rempe_atom_cavity,Kimple_atom_cavity,Keller_Ion_cavity},
including the initial demonstration of reversible state transfer
between photons and atoms~\cite{Kimble_reversiblemapping}. Via
similar cavity-assisted Raman processes, schemes for mapping between
motional states of single trapped atoms~\cite{Parkins_Kimble_transfer} or collective excitation of atomic
ensembles~\cite{Lukin_impedence_matching} and the quantum states of
single photons are also proposed.

For quantum computation with QD spins~\cite{Loss_QDspinQC},
deterministic quantum network control is indeed possible in the
distributed structure discussed previously in
section~\ref{subsec_distributed_sructure}. In subsection
\ref{subsec_solidstateinterface}, we will first show how to realize such
a prototype quantum interface between a QD electron spin and
a photon mode in a waveguide/fiber. In subsection \ref{subsec_interfacecontrol},
we will describe the exact solution to the
interface dynamics between a spin qubit and a photon qubit in the
prototype quantum node in the most general scenario. Control
schemes based on the exact solution form the basis of a variety of
inter-node operations in quantum networks as discussed in
Sec.~\ref{subsec_network}. The issue of unavoidable
inhomogeneity of solid-state quantum nodes is properly resolved with
the exact interface solution. In subsection
\ref{subsec_net_imperfect}, we study the effects of various sources
of errors on the interface operations. A summary and outlook are
given in subsection~\ref{subsec_QIPbyNemesis}.

\subsection{Dot-Cavity-Waveguide structure as spin-photon interface \label{subsec_solidstateinterface}}

\begin{figure}[b]
\begin{center}
\includegraphics[width=12cm]{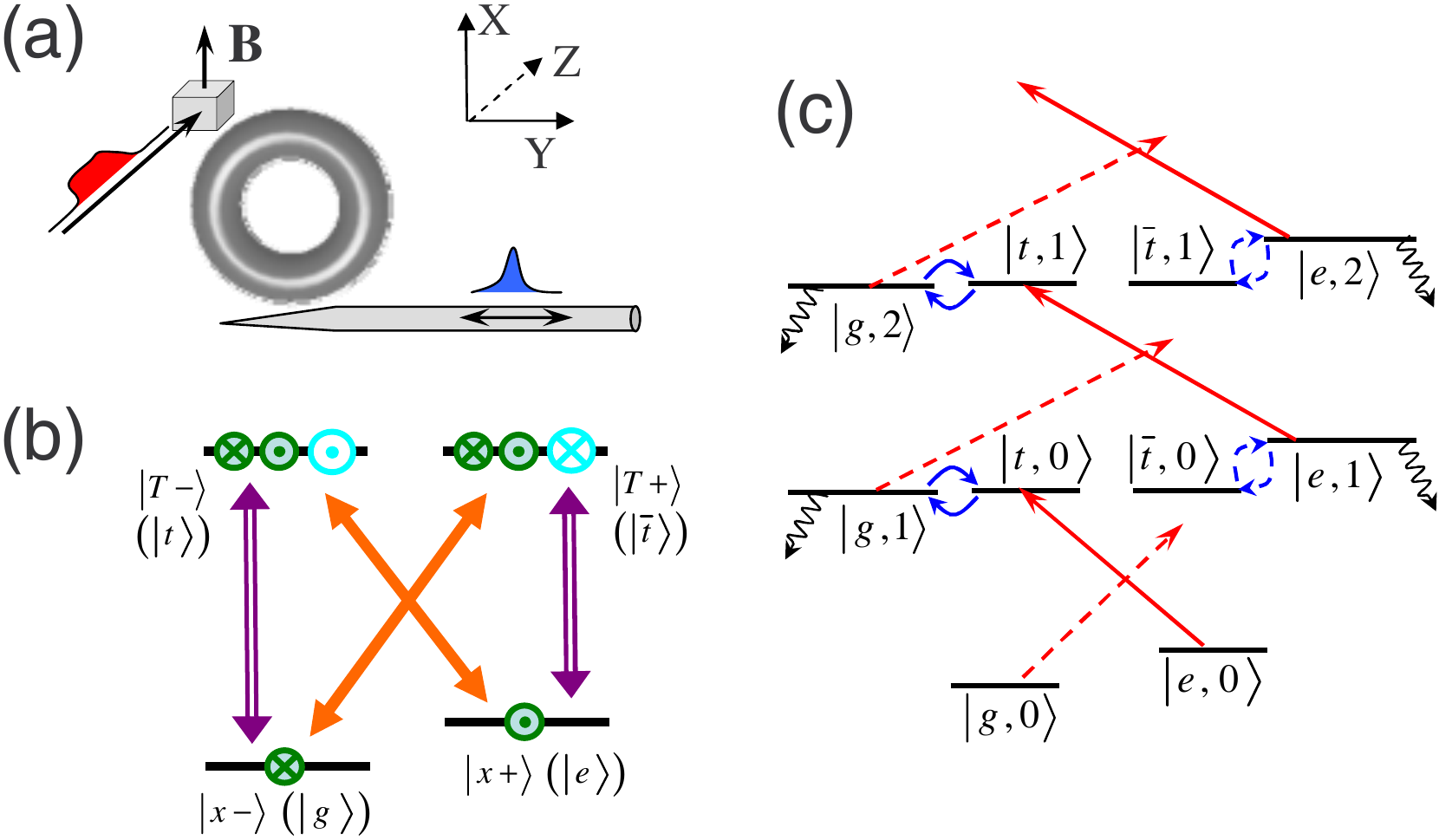}
\caption{(a) A high-$Q$ micro-ring coupling a
`tapered' waveguide and a singly charged QD. (b) Optical
selection rules in the basis where spins are quantize along the field
($x$) direction. The hollow two-headed arrow denotes $X$-polarized
light and the solid arrow denotes $Y$-polarized light. (c)
The level diagram and optical process. In $|s,n\rangle$, $s=g,e,t$
or $\bar{t}$ denotes an electronic state in the dot and $n$ denotes
the number of photons in the single cavity mode. Straight, curved,
and wavy arrows represent the laser
excitation, dot-cavity coupling, and cavity-fiber tunneling,
respectively.  The resonant and off-resonance
processes are represented by solid and dashed lines, respectively.}
\label{solidinterface}
\end{center}
\end{figure}

Here we show how to implement a dot-cavity-waveguide coupled structure
as a deterministic quantum interface for single spins in QDs and single photons in waveguides.

The coupled structure can be realized with any of the micro-cavity
systems discussed in section \ref{subsec_distributed_sructure}. In
Fig.~\ref{solidinterface}(a), a QD sitting in the
evanescence field of whispering gallery mode of a ring cavity is illustrated as an
example. There are two essential requirements for such a coupled
structure to be an efficient interface. First, the dot-cavity
coupling must have a large Purcell factor, so that the QD
optical transitions are dominantly coupled to the cavity field.
Second, the leakage of the cavity photon into free space should be
small as compared with the tunneling into the attached waveguide. If
both conditions are fulfilled, the evolution through the desired
quantum pathway, i.e. QD trion $\leftrightarrow$ cavity
photon $\leftrightarrow$ waveguide photon, occurs on a much faster timescale
than the leakage out of it~\cite{yao_network,Liu_readwrite}. Lowering of the
cavity $Q$-factor due to the coupling to the waveguide is part of
the process and has no deleterious effects on the quantum operation.
The cavity with the reduced $Q$-factor is {\em not} required to be
in the strong coupling regime~\cite{yao_network,Yao_networkJOB,Fattal_Qnet}. This will become
clear when we discuss the control of this interface in the following
subsections.

The qubit is represented by the two spin states $|x-\rangle$ and
$|x+\rangle$ and spin manipulations are mediated by the two trion
ground states $|T-\rangle$ and $|T+\rangle$ (see
section~\ref{subsec_spinQubit_levels}). In
the convention of the prototype quantum interface, the two spin states are also
denoted here as $|g\rangle$ and $|e\rangle$ respectively, and the
two trion states $|T+\rangle$ and $|T-\rangle$
as $|t\rangle$ and $|\bar{t}\rangle$, respectively.
While $|g\rangle$ and $|e\rangle$ have energies $\omega_g$ and
$\omega_e$ in a static magnetic field normal to the optical axis of
the dot (see Fig.~\ref{solidinterface}(a)), $|t\rangle$ and
$|\bar{t}\rangle$ typically have a much smaller energy splitting
($\omega_t \simeq \omega_{\bar{t}}$) in GaAs fluctuation dot because of its negligible
in-plane g-factor of the heavy holes~\cite{Gammon_PRB2002,Marie_holespinbeat}, but
a splitting comparable to the electron's in self-assembled InAs dot~\cite{Xu07PRLspinIni}.

By design, the selected cavity mode of frequency $\omega_c$ is
$X$-polarized in the vicinity of the QD and a $Y$-polarized
control laser of central frequency $\omega_L$ and complex Rabi
frequency $\Omega(t)$ directly couples to the QD
transitions~\cite{Shih_mollow_g1,Shih_mollow_g2}. Therefore, by the
selection rules shown in Fig.~\ref{solidinterface}, the cavity field
couples only to the straight transitions $|g\rangle \rightarrow
|t\rangle$ and $|e\rangle \rightarrow |\bar{t}\rangle$ , and the
controlling laser couples only to the cross transitions $|g\rangle
\rightarrow |\bar{t}\rangle$ and $|e\rangle \rightarrow |t\rangle$.
The laser light and cavity mode satisfy the two-photon resonance
condition: $\omega_L+\omega_{e}=\omega_{c}+\omega_{g}$. The Raman
detuning $\Delta \equiv \omega_t-\omega_L-\omega_{e}$ is also much
smaller than the electron Zeeman splitting $\omega_e-\omega_g$.
Thus, by the Zeeman splitting and the selection rules, the trion
state $|\bar{t}\rangle$ is off-resonance to the laser light and the
cavity mode [shown by dashed lines in the Fig.~\ref{solidinterface}(c)].

At a sending node, the Raman process consists in first the laser
field exciting the spin state $|e,0\rangle$ to the trion state
$|t,0\rangle$, then the trion state resonantly coupled to the cavity
state $|g,1\rangle$ which finally is rotated to the spin state
$|g,0\rangle$ forming a photon wave packet in the waveguide (here
$0$ and $1$ denote the number of photons in the single cavity mode).
The receiving mode is just the time-reversed process. Undesirable
dynamics involving the state $|\bar{t}\rangle$ is eliminated by
making the Zeeman splitting sufficiently larger than the cavity-dot
coupling and the Rabi frequency. The resultant optical process is
the cavity-assisted Raman process in a $\Lambda$-type three-level
system as required by the prototype quantum interface (see
Fig.~\ref{system}). Numerical calculations including the
non-resonance transitions and realistic decoherence have been
performed and high fidelity of desired operations at the quantum
interface is demonstrated (see subsection \ref{subsec_net_imperfect}
for details).

\subsection{Control of spin-photon interface\label{subsec_interfacecontrol}}

In the previous subsection, we have shown that the
dot-cavity-waveguide coupled structure could be an efficient solid
state realization of the prototype quantum interface. In order for
such quantum interface to be suitable for a quantum network, proper
control schemes are required.

The difficulty in realizing the network lies in the receiving end.
Instead of being trapped at the quantum node, the single photon
pulse can be reflected by the cavity unless the pulse shape of the
classic control laser matches the single photon wavepacket exactly,
known as quantum impedance
matching~\cite{Lukin_impedence_matching,Lukin_Adiabatic_interface}.
One way to deal with this requirement was provided
in~\cite{StateTransfer_Cirac_Kimble}. The central idea is that if a
laser pulse can be found for generation of an outgoing photon
wavepacket with time reversal symmetry, by using this laser control
at the sending node and its time reversal at the receiving
node, the time reversal symmetry will guarantee the photon
wavepacket to be completely trapped at the receiving node. A
solution for such a laser control pulse was given
in~\cite{StateTransfer_Cirac_Kimble}.

The time reversal symmetric control scheme requires the sending and receiving
quantum nodes to be identical in terms of optical transition frequencies and strength
of coupling between the components. Unfortunately, such requirement
is practically impossible to fulfill in solid state systems since
the fabricated QDs and micro-cavities naturally have shape variations
and size fluctuations, leading to inhomogeneity in optical
frequencies. The cavity field has a highly non-uniform profile and thus
cavity-dot coupling differ from dot to dot. For realization of
a quantum network with solid state structures, control schemes beyond
the time symmetric one is required. In the sending function, it is
straightforward to solve for the outgoing single photon wavepacket
$\alpha(t)$ if we know the form of the laser control pulse
$\Omega(t)$. The key problem is the inverse functional relation
$\Omega(t)=\mathcal{F}^{-1}[\alpha(t)]$, i.e., given an arbitrary
single photon wavepacket $\alpha(t)$, the exact laser pulse that can
generate this single photon wavepacket at a sending node, or
completely trap it at a receiving node. Knowing this relation allows
the sending and receiving operations to be separately addressed and
hence the construction of a quantum network with heterogeneous quantum nodes.

Solutions to $\Omega(t)=\mathcal{F}^{-1}[\alpha(t)]$ were first
given in the adiabatic
approximations~\cite{Lukin_impedence_matching,Lukin_Adiabatic_interface}.
Exact solutions in the non-adiabatic form were later found by the
authors~\cite{yao_network,Yao_networkJOB}, as briefly described
below.

The Hamiltonian including the interaction between the single-mode cavity
and the three-level system and the channel continuum is,
\begin{eqnarray}
H &=&\omega _{c}a^{\dagger }a+\omega _{t}\left| t\right\rangle
\left\langle t\right| +\omega _{e}\left| e\right\rangle \left\langle
e\right| +\int_{0}^{\infty }d\omega \omega b_{\omega }^{\dagger
}b_{\omega } +g_{\rm cav}\left( i\left| t\right\rangle
\left\langle g\right| a+{\rm H.c.}\right) \notag \\
&& + \frac{1}{2}\left[ i\Omega \left( t\right) e^{-i\omega
_{L}t}\left| t\right\rangle \left\langle e\right| +H.c.\right]
+\int_{0}^{\infty }d\omega \left( i\sqrt{\gamma /2\pi }b_{\omega
}^{\dagger }a+{\rm H.c.}\right),
\label{hamiltonian}
\end{eqnarray}
where $b_{\omega }$ is the annihilation operator for the mode of
frequency $\omega $ in the channel continuum and $a$ is the
annihilation operator for the cavity mode. The energy of state
$|g\rangle$ is set as zero. The $|g\rangle \rightarrow |t\rangle $
transition is coupled to the cavity mode with strength $g_{\rm cav}$.
The $|e\rangle \rightarrow |t\rangle $ transition is coupled by the
external control laser of time-dependent Rabi frequency $\Omega (t)$
and central frequency $\omega_{L}$. The coupling of the cavity mode
to the channel continuum is assumed constant:$\sqrt{\gamma /2\pi }$.
An ideal situation is assumed neglecting photon leakage into free
space through intermediate state $|t\rangle$ or the cavity sidewall.

We note that the system described by this Hamiltonian, under the
laser excitation and the cavity-dot and cavity-channel interaction,
has two invariant Hilbert subspaces, with the basis $\left\{ \left|
g,0\right\rangle \left| \mathrm{vac}\right\rangle \right\} $ and
$\left\{ \left| e,0\right\rangle \left| \mathrm{vac}\right\rangle
,\left| t,0\right\rangle \left| \mathrm{vac}\right\rangle ,\left|
g,1\right\rangle \left| \mathrm{vac} \right\rangle ,\left|
g,0\right\rangle \left| \omega \right\rangle \right\} $
(where in $|s,n\rangle$, $s=g,e$ or $t$ denotes the
state of the three-level system, $n$ denotes the number of photons in
the cavity mode, and $\left| \omega \right\rangle $ denotes the
one-photon Fock state of the channel mode of frequency $\omega$). So
the evolution of the system can be generally described by the state
$C_{g}\left| g,0\right\rangle \left| {\rm vac} \right\rangle
+C_{e}\left| \Psi ^{e}\left( t\right) \right\rangle $ in the
interaction picture, where
\begin{eqnarray}
\left| \Psi ^{e}\left( t\right) \right\rangle &=& \beta _{e}\left(
t\right) \left| e,0\right\rangle \left| \mathrm{vac}\right\rangle
+\beta _{t}\left( t\right) \left| t,0\right\rangle \left|
\mathrm{vac}\right\rangle \nonumber \\
&+& \beta _{c}\left( t\right) \left| g,1\right\rangle \left|
\mathrm{vac}\right\rangle +\int_{0}^{\infty }d\omega _{\omega}\alpha
_{\omega}\left( t\right) \left| g,0\right\rangle \left| \omega
\right\rangle .
\label{wavefunction}
\end{eqnarray}
The time evolution of the amplitudes in the interaction picture is
described by the following Schr\"{o}dinger equations,
\begin{subequations}
\begin{eqnarray}
\dot{\beta}_{e} &=&-\frac{\Omega^{\ast }}{2}e^{-i(\omega _{t}-\omega
_{L}-\omega _{e})t}\beta _{t},  \label{eom1_1} \\
\dot{\beta}_{t} &=&g_{\rm cav}e^{i(\omega _{t}-\omega _{c})t}\beta
_{c}+\frac{ \Omega }{2}e^{i(\omega _{t}-\omega _{L}-\omega
_{e})t}\beta _{e},
\label{eom1_2} \\
\dot{\beta}_{c} &=&-g_{\rm cav}e^{-i(\omega _{t}-\omega _{c})t}\beta
_{t}-\sqrt{ \gamma }\alpha _{in}\left( t\right) -\frac{\gamma
}{2}\beta _{c}
\label{eom1_3} \\
&=&-g_{\rm cav}e^{-i(\omega _{t}-\omega _{c})t}\beta
_{t}-\sqrt{\gamma }\alpha _{out}\left( t\right) +\frac{\gamma
}{2}\beta _{c}, \label{eom1_4}
\end{eqnarray}
\label{eom1}
\end{subequations}
where $\alpha _{\mathrm{in}}\left( t\right) \equiv \int d\omega
\alpha _{\omega }\left( t_{0}\right) e^{-i(\omega -\omega
_{c})t}/\sqrt{2\pi }$ with $t_{0}\rightarrow -\infty $ and $\alpha
_{\mathrm{out}}\left( t\right) \equiv \int d\omega \alpha _{\omega
}\left( t_{1}\right) e^{-i(\omega -\omega _{c})t}/\sqrt{2\pi }$ with
$t_{1}\rightarrow +\infty $ can be regarded as the incoming and
outgoing wavepacket of the photon in the quantum channel,
respectively. From Eqs.~(\ref {eom1_3}) and ~(\ref{eom1_4}), we note
that evolution of $\beta_c(t)$ is simply an instantaneous map of the
difference between the input and output field in the photonic
channel
\begin{equation}
\sqrt{\gamma }\beta_{c}\left( t\right)=\alpha _{\rm out}\left(
t\right) -\alpha _{\rm in}\left( t\right).  \label{bd}
\end{equation}
$\beta _{t}(t)$ is also readily expressed in terms of $\alpha
_{\rm in}(t)$ and $\alpha _{\rm out}(t)$ as,
\begin{eqnarray}
\beta _{t}&=&\frac{-\dot{\beta}_{c}-\sqrt{\gamma }\alpha _{in}\left(
t\right) - \frac{\gamma }{2}\beta _{c}}{g_{\rm cav}}e^{i(\omega
_{t}-\omega _{c})t}
\nonumber \\
&=& \frac{-(\dot{\alpha}_{\rm out}-\dot{\alpha}_{\rm
in})/\sqrt{\gamma}-(\alpha _{\rm in} (t) +\alpha _{\rm out}
(t))\sqrt{\gamma }/2 }{g_{\rm cav}}e^{i(\omega _{t}-\omega _{c})t}.
 \label{beta_t}
\end{eqnarray}
So as the amplitude of $\beta _{e}(t)$,
\begin{equation}
\frac{d}{dt}\left| \beta _{e}\right| ^{2}=-\frac{d}{dt}\left| \beta
_{t}\right| ^{2}+g_{\rm cav}\left[\beta _{c}^{\ast }\beta
_{t}e^{-i(\omega _{t}-\omega _{c})t}+\beta _{c}\beta _{t}^{\ast
}e^{i(\omega _{t}-\omega _{c})t}\right],  \label{beta_e_am}
\end{equation}
and the phase,
\begin{align}
\frac{d}{dt}\arg (\beta _{e})= & \frac{1}{2i}\left| \beta _{e}\right|
^{-2}\left( \dot{\beta}_{t}\beta _{t}^{\ast }-\beta_{t}\dot{\beta}_{t}^{\ast }\right)
\nonumber \\
&+\frac{ g_{\rm cav}}{2i}\left| \beta
_{e}\right| ^{-2}\left[\beta _{t}\beta _{c}^{\ast }e^{-i(\omega
_{t}-\omega _{c})t}-\beta _{c}\beta _{t}^{\ast }e^{i(\omega
_{t}-\omega _{c})t}\right].  \label{beta_e_ph}
\end{align}
Finally, from Eq.~(\ref{eom1_2}), the complex Rabi frequency of the
laser pulse $\Omega (t)$ can be expressed in terms of the amplitudes
that have been solved above,
\begin{equation}
\Omega \left( t\right) =2\frac{\dot{\beta}^{t}-g_{\rm cav}\beta
^{c}}{\beta ^{e}}.
\end{equation}
Thus the desired operation, with $\alpha_{\rm in}(t)$ and
$\alpha_{\rm out}(t)$ arbitrarily specified, can be generated on
demand as long as the normalization condition of the wavefunction is
not violated
\begin{equation}
\frac{d}{dt}\left(\left| \beta _{e}\right| ^{2}+\left| \beta _{t}\right|
^{2}+\left| \beta _{c}\right| ^{2}\right)=\left| \alpha _{\rm
in}(t)\right| ^{2}-\left| \alpha _{\rm out}(t)\right| ^{2}.
\label{normalization}
\end{equation}

The functions of this quantum interface can be classified into three
types:
\begin{enumerate}
\item
If there is no incoming photon,
the quantum interface generates an outgoing photon wavepacket of a specified shape;
\item
If there is an incoming photon wavepacket of a specified shape,
it is completely trapped by the quantum interface so that there is non
outgoing field;
\item
If there is an incoming photon wavepacket of a specified shape,
the quantum interface generates an outgoing photon wavepacket of another specified shape  -- a controlled scattering process.
\end{enumerate}
The first two types of control
form the basis for the quantum network operation. With
control of the type III, the quantum interface can act as a controllable
scatter or pulse shaper for single photon wavepackets. This control
can also be considered as the combination of consecutive controls of
type II and I. In the following subsection, we will discuss in more
details how to implement the first two types of controls for a
quantum network.

\subsection{Inter-node operations in a quantum network \label{subsec_network}}

The sending node of a quantum network is operated with control of
type I. The initial conditions are: $\alpha_{\mathrm{in}}(t)=0$,
$\beta _{c}(t_{0})=0$, $\beta _{e}(t_{0})=1$ and $\beta
_{t}(t_{0})=0$. The integral form of Eq.~(\ref{normalization}) becomes
\begin{equation}
\left|{\beta}_{e}\right| ^{2} = 1-\sin^2\theta\int_{t_0 }^{t} \left|
\tilde{\alpha}_{\rm out}\left( \tau \right) \right| ^{2}d\tau
-\left| {\beta}_{c}\right| ^{2} - \left| g_{\rm cav}\right|^{-2}
\left| \dot{{\beta}}_{c}+{\gamma } {\beta}_{c} /2 \right| ^{2},
\label{bd1}
\end{equation}
where $\tilde{\alpha}_{\mathrm{out}}$ is the normalized wavepacket
of the emitted photon, and $\sin ^{2}\theta $ is the average photon
number. For a photon number and a pulse shape arbitrarily specified,
the amplitude of the cavity mode is determined by Eq.~(\ref{bd}) as
$ \beta _{c}(t)= \sqrt{ \gamma }\tilde{\alpha}_{\mathrm{out}}(t)\sin
\theta $. If we pose the problem of finding the laser control pulse
to produce a specified shape of the outgoing photon wavepacket, the
fact that the right-hand side of Eq.~(\ref{bd1}) is positive
requires the specified output pulse be sufficiently smooth, i.e.,
the pulse generation process be slower than the cavity-channel
tunneling and the dot-cavity coupling rate (with time scales $\gamma
^{-1}$ and $g_{\mathrm{cav}}^{-1}$, respectively). At the remote
future time, $t_{1}\rightarrow +\infty $, the photon emission process
is completed, i.e., $\beta _{c}(t_{1})=\dot{\beta} _{c}(t_{1})=0$,
so $\beta _{e}(t_{1})=e^{i\phi }\cos \theta $ with the controllable
phase $\phi $ given by Eq.~(\ref{beta_e_ph}). The general form
of the photon generation process can be expressed as
\begin{eqnarray}
&& \left( C_{g}|g\rangle +C_{e} |e\rangle \right) \otimes
|\mathrm{vac}\rangle \nonumber \\
&\overset{\Omega(t)}{\longrightarrow }& C_{g}|g\rangle \otimes
|\mathrm{vac}\rangle + C_{e}\left[ e^{i\phi }\cos \theta |e \rangle
\otimes |\mathrm{vac} \rangle +\sin \theta |g\rangle \otimes
|\tilde{\alpha}_{ \mathrm{out}}\rangle \right].
\label{general}
\end{eqnarray}
A full Raman process corresponds to $\theta=\pi/2$ and
$\beta_e(t_1)=0$, where Eq.~(\ref{general}) is reduced to,
\begin{eqnarray}
\left(C_g|g\rangle+C_e|e\rangle\right) \otimes |{\rm
vac}\rangle\stackrel{\Omega(t)}{\longrightarrow}
|g\rangle\otimes\left(C_g|{\rm vac}\rangle+C_e|\tilde{\alpha}_{\rm
out}\rangle\right),
\end{eqnarray}
which results in the mapping of the stationary qubit onto the flying
qubit. If initially the three level system is entirely in state
$|e\rangle$, this mapping operation can function as a
deterministic generation of a single-photon wavepacket with any desired shape
$\tilde{\alpha}_{\rm out}(t)$. If the Raman cycle is controlled to
be partially completed ($\theta<\pi/2$), the state initially in
$|e\rangle\otimes |{\rm vac}\rangle$ is transformed into an
entangled state of the stationary spin and the flying photon
\begin{eqnarray}
|e\rangle \otimes |\mathrm{vac}\rangle
\overset{\Omega(t)}{\longrightarrow }
e^{i\phi}\cos\theta|e\rangle\otimes|{\rm
vac}\rangle+\sin\theta|g\rangle\otimes|\tilde{\alpha}_{\rm
out}\rangle.
\end{eqnarray}
The entanglement entropy $E=-\cos^2\theta \log_2
\cos^2\theta-\sin^2\theta \log_2 \sin^2\theta$ can be set at
any value between 0 and 1 depending on the rotating angle $\theta$.

The receiving node is operated with control of type II, typically as
a full Raman cycle, in the quantum network scheme. With the three
level system initially on state $|g\rangle$ and the incoming photon
$C_g|{\rm vac}\rangle+C_e|\alpha_{\rm in}(t)\rangle$, the mapping
transformation is,
\begin{eqnarray}
|g\rangle\otimes(C_g|{\rm vac}\rangle+C_e|\alpha_{\rm in}\rangle)
\stackrel{\Omega(t)}{\longrightarrow}
(C_g|g\rangle+C_e|e\rangle)\otimes |{\rm vac}\rangle \label{map2}.
\end{eqnarray}
As in the sending process, the incoming photon pulse $\alpha_{\rm
in}(t)$ can be arbitrarily specified, provided that it is smooth enough.

By combining the sending and receiving processes, the transfer of a
qubit from one node to another can be easily implemented, with the
outgoing photon from the sending node directed as the incoming
photon for the receiving node. As the photonic channel is linear,
two state-transfer operations with opposite directions can be
performed in parallel, and qubits at the two nodes will be swapped.
Swap operations can only be performed between nodes separated with
sufficiently large distance so that photon traveling time in the
channel is longer than the interface operation time.

If the operation at the sending node has been designed to produce an
entangled state of the stationary and the flying qubit, the mapping
process at the receiving node will just produce an entangled state
of the two nodes by the transformation,
\begin{eqnarray}
|e\rangle_{1}|g\rangle_{2} \otimes|{\rm vac}\rangle
&\stackrel{\Omega_1(t)} {\longrightarrow}&
e^{i\phi}\cos\theta|e\rangle_1|g\rangle_{2}\otimes|{\rm
vac}\rangle+\sin\theta|g\rangle_1|g\rangle_{2}\otimes|\tilde{\alpha}_{\rm
out}\rangle  \notag \\
&\stackrel{\Omega_2(t)}{\longrightarrow}&
\left[e^{i\phi}\cos\theta|e\rangle_{1}|g\rangle_{2}+\sin\theta|g\rangle_1|e\rangle_{2}
\right] \otimes|{\rm vac}\rangle. \label{entangle}
\end{eqnarray}
Non-local entanglement can thus be generated deterministically in
the quantum network.

With the exact solutions for the interface dynamics, the sending and
receiving functions can be separately addressed. This enables the
construction of a quantum network with heterogeneous quantum nodes,
which is essential for solid state realization. To illustrate this,
we give below an exemplary control strategy. The control laser pulse
at a sending node can have a general
shape $\Omega_1(t)e^{-i\omega_{L}^1 t}$ with the pulse area satisfying the specified
rotation angle $\theta$. The outgoing single photon wavepacket has a
definite shape $\alpha_{\rm out}^1(t)$ (in the rotating frame
defined by frequency $\omega_{c}^1$), determined by the laser
control $\Omega_1(t)$ and the parameters of the sending node $g_{\rm
cav}^1$, $\gamma_1$ and Raman detuning $\Delta_1 \equiv
\omega_{t}^1-\omega_{L}^1-\omega_{e}^1=\omega_{t}^1-\omega_{c}^1$.
Then we pose the problem of finding the optical control
$\Omega_2(t)e^{-i \omega_{L}^2 t}$ to trap the single photon
wavepacket $\alpha_{\rm in}^2(t) \equiv \alpha_{\rm out}^1(t)
e^{i(\omega_{c}^2-\omega_{c}^1)t}$ at the receiving node which may
have a different set of parameters $g_{\rm cav}^2$, $\gamma_2$,
$\omega_{t}^2$, $\omega_{e}^2$ and $\omega_{c}^2$. From
Eq.~(\ref{bd1}) and the discussion follows, it implies
that the tolerance of the node inhomogeneity is determined by the node
bandwidth, i.e. the dot-cavity coupling $g_{\rm cav}$ and
dot-waveguide tunneling $\gamma$. Thus, a large dot-cavity coupling
$g_{\rm cav}$ is essential. This tolerance was discussed more
explicitly by Fattal {\it et al} in Ref.~\cite{Fattal_Qnet}.

\subsection{Operations with imperfections \label{subsec_net_imperfect}}

In this subsection, we discuss the effects of various imperfections
that may occur in a realistic quantum network and the corresponding
mitigation.

\subsubsection{Intrinsic photon leakage into free space} The desired
quantum evolution is through the pathway of trion - cavity photon -
waveguide photon. The main causes of photon leakage out of this
quantum pathway is through the trion decay by spontaneous emission
and the cavity mode leakage other than the tunneling into the
waveguide. The waveguide and fiber loss is negligible on the
distance-scale of relevance for intra-chip communication and for
inter-chip communications if the chips are distributed in a spatial
range of $\lesssim10$~cm. As long as the photon leakage rate is much
smaller than the bandwidth of the desired quantum pathway
(determined by the dot-cavity coupling $g_{\rm cav}$ and the
cavity-waveguide tunneling $\gamma$), high fidelity operations can
be expected. Typical trion decay rates in self-assembled QDs
are $\Gamma \sim \mu$eV~\cite{Bayer_Linewidth_QD,Shih_mollow_g1}, and
the intrinsic loss rate of a high-$Q$ cavity (i.e. excluding
coupling to the dot and the waveguide) can be potentially achieved at
$\gamma_0\gtrsim 0.1$ $\mu$eV (corresponding to a $Q$-factor $\sim
10^7$)~\cite{Noda_PhotonicCrystal1,Noda_PhotonicCrystal2}.
The state-of-the-art dot-cavity coupling constant achieved is
$g_{\rm cav}=0.1$~meV~\cite{khitrova_rabi_splitting,forchel_rabi_splitting,Bloch_rabi_splitting},
while the cavity-waveguide tunneling rate is controlled in design by
the gap distance. Thus the bandwidth can be two orders larger than
the leakage rate.

\begin{figure}[b]
\begin{center}
\includegraphics[width=10cm]{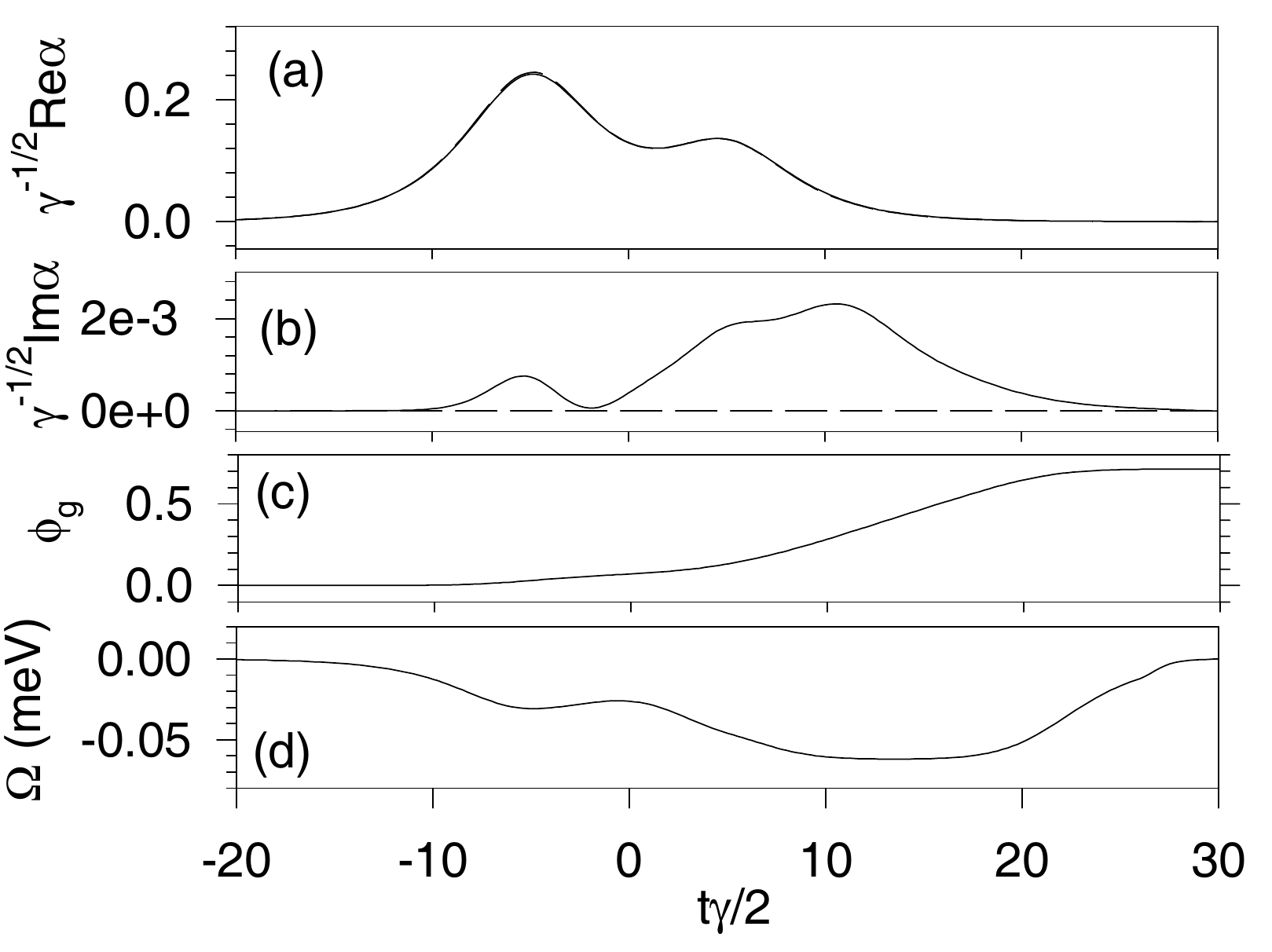}
\caption{(Extracted from~Ref.\cite{yao_network}) Generation of a single photon wavepacket
with an asymmetric double sech shape (see text). (a) Real part of
the dimensionless amplitude of the simulated single photon
wavepacket (solid line) as a function of the dimensionless time
$\gamma t/2$. The deviation from the target shape (dashed line) is not visible.
(b) Imaginary part of the simulated single photon wavepacket (solid
line) and that of the target one (dashed line). (c) Phase drift of the
component $|g,0\rangle$. (d) Rabi frequency of the control laser
pulse.} \label{gene}
\end{center}
\end{figure}

The simulation result of mapping a spin state to a flying photon
wavepacket with the shape targeted as an asymmetric superposition of
two sech-functions as $\alpha^{\rm
ideal}_{\text{out}}(t)=\text{sech}(\gamma t/6+5)+0.5
\text{sech}(\gamma t/6-5)$ is shown in Fig.~\ref{gene}. The trion
decay rate, based on experiment~\cite{Bayer_Linewidth_QD}, is set at
$\Gamma=3 \mu$eV, and the intrinsic cavity loss rate is assumed to
be $\gamma_0=0.1~\mu$eV~\cite{Noda_PhotonicCrystal2}. The
cavity-fiber tunneling rate is chosen to be $\gamma=0.2$ meV and the
dot-cavity coupling constant $g_{\rm
cav}=0.1$~meV~\cite{khitrova_rabi_splitting}. The fidelity of the
single photon generation $|\langle \alpha^{\rm ideal}_{\rm
out}|\alpha_{\rm out}\rangle|\approx 0.9912$. Because of the
non-adiabatic optical pumping and dot-cavity coupling, the whole
mapping process can be completed within 300 ps. The simulation of
the photon absorption process shows an overall fidelity greater than
$0.99$ as well. With the above chosen parameters, the cavity mode
broadening due to the coupling to waveguide is actually larger than
the dot-cavity coupling. The system is therefore {\it not} in the
strong coupling regime by the usual
definition~\cite{khitrova_rabi_splitting,forchel_rabi_splitting,Bloch_rabi_splitting}.
High fidelity is nonetheless guaranteed by the large Purcell factor.

\subsubsection{Unwanted coupling to energy levels beyond the 3-level $\Lambda$ system}

In the energy level structure of the dot-cavity
coupled system, non-resonance coupling to other energy levels could
lead to excitation out of the 3-level $\Lambda$ subspace and AC
Stark shift of the qubit states of interest (see
Fig.~\ref{solidinterface}(c)). These effects have been included in
the numerical simulation given in Fig.~\ref{gene}. As shown in
Fig.~\ref{gene}(c), AC stark shift induces a deterministic phase
drift between $|g\rangle$ and $|e\rangle$ . As the two excitation
pathways starting respectively from $|g\rangle$ and $|e\rangle$ are
independent of each other (see Fig.~\ref{solidinterface}(c)), this
phase drift is independent of the coefficients $C_{g}$ and $C_{e}$.
Therefore, it can be compensated by a single-qubit operation
irrespective of the quantum state being mapped. Leakage out of the
qubit subspace by the non-resonance excitation to multi-photon states
can be greatly suppressed if the Zeeman splitting is much larger
than the Rabi frequency and the cavity-dot coupling. For InAs
self-assembled QDs, Zeeman splitting $\sim $~meV can be
achieved in a moderately strong magnetic field ($\sim10$~T) due to
the large g-factor of InAs materials.

\begin{table}[b]
\caption{Effect of errors in coupling parameters on the fidelity of entanglement and of state transfer.} \label{table2}
\begin{tabular}{c|c|c|c|c}
 \hline  \hline  & no error & $10 \%$ $g$ error & $10 \%$ $\gamma $ error&
$10 \%$ $\Omega(t) $ error \\
 \hline Entanglement & 0.9912 & 0.9872 & 0.9894 & 0.9862 \\
 \hline Transfer & 0.9901 & 0.9870 & 0.9891 & 0.9879 \\
  \hline  \hline
\end{tabular}
\end{table}

\subsubsection{Unknown parameter offsets}

Solutions to the laser pulses for desired controls of the quantum interface are
based on the knowledge of the coupling strength $g_{\rm cav},\gamma$
and $\Omega(t)$. But in practice, there could be
various unknown errors on parameters due to imperfect
characterization of the system. The robustness of the control
schemes in presence of unknown system parameter errors is thus a
critical feature. In Table.~\ref{table2}, we list the effects of
unknown offsets from the assumed values of various parameters on the
fidelity of two typical quantum network operations: (i) entangling
two quantum nodes into state $e^{i\phi}|g\rangle _{1}|e\rangle _{2}+
|e\rangle_{1}|g\rangle _{2} $; (ii) transfer of the state $|g\rangle
+ |e\rangle$ between two nodes. In both cases, the target shape of the involved single
photon wavepacket is ${\rm sech}(\frac{\gamma t}{6})$ and
the design of the control laser pulses uses the assumed parameter values. The system
shows a surprising robustness: $10\%$ unknown errors on $g_{\rm
cav},\gamma$ or $|\Omega(t)|$ only reduce the fidelity by less than
$1 \%$. This intrinsic robustness against unknown parameter errors
paves the way for learning studies of the system parameters by trial
and error~\cite{Rabitz_1992,rabitz00,Rabitz_2001}, and
classical
feedback controls in the quantum
network~\cite{wiseman93,Mabuchi_realtime}.

\begin{figure}[b]
\begin{center}
\includegraphics[width=10cm]{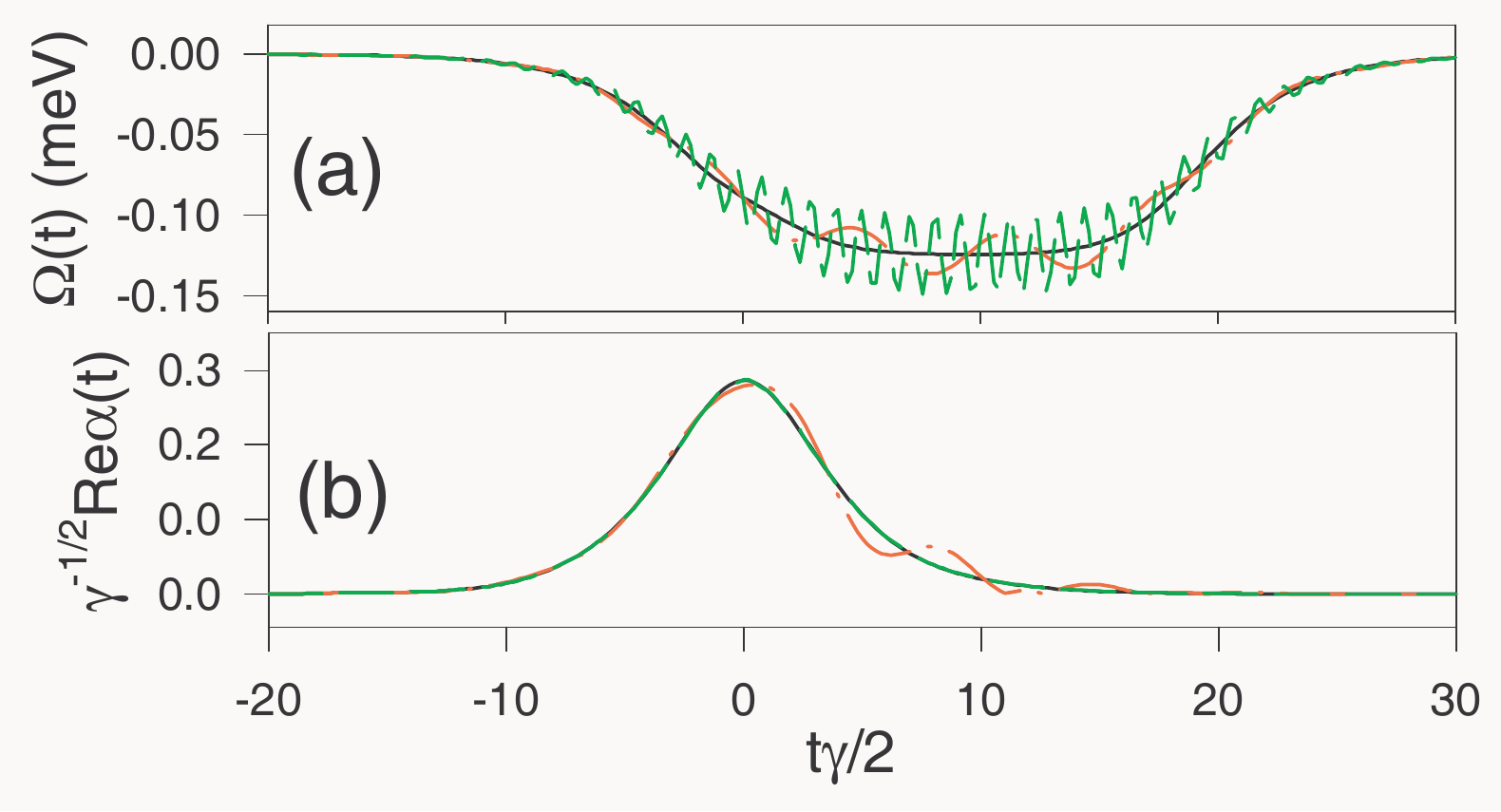}
\caption{(Extracted from Ref.~\cite{Yao_networkJOB})
Transfer of state $|g\rangle +|e\rangle$ between two nodes in
presence of amplitude fluctuations of the control lasers.
(a) The desired control laser pulse by design at the sending node
(solid curve) and the actually applied control pulse with slow fluctuations
(dashed-dotted curve) and fast fluctuations (light curve). The control laser pulse at the
receiving node has a similar error applied.
(b) The generated intermediate single photon wavepacket. The solid curve shows
the target shape of the single photon if the ideal control laser pulse is used
(solid curve in (a)). Deviation of the generated
single photon wavepacket is invisible in the case of fast laser
fluctuations. The fidelity of the transfer is $0.9912$ with slow
fluctuations and $0.9922$ with fast fluctuations. The parameters
used are: $\gamma=0.2$meV, $g_{\rm cav}=\gamma/2$, $\gamma_t=3
\mu$eV, and intrinsic cavity leakage $0.05~\mu$eV.} \label{shape_f}
\end{center}
\end{figure}

\subsubsection{Laser intensity fluctuations}

The $|\Omega(t)|$ error
studied in Table.~\ref{table2} is a global one on the amplitude,
e.g. induced by the QD being slightly out of focus from the
laser control field or an unknown offset from the assumed value of
the dot dipole moment. The actual control laser pulse may have shape
deviations from the desired ones, e.g. due to temporal fluctuations
in laser amplitude. The effects of this error source have been
studied in~\cite{Yao_networkJOB}, where the control scheme is found
to be immune against fast fluctuations (see Fig.~\ref{shape_f}).
This robustness is due to the finite bandwidth of the quantum
interface determined by the cavity-dot coupling strength
and the cavity-waveguide tunneling rate. Any temporal fluctuations in the control field with the
frequency higher than the interface bandwidth are effectively
averaged out. The time independent amplitude error listed
in Table.~\ref{table2} may also be considered as a special shape error
which is actually the worst scenario for the fidelity.

\subsubsection{Laser phase fluctuations}

The complex Rabi frequency $\Omega(t)$ can also have phase uncertainties due to laser phase
drift which is typically slower than the time scale of interface
operation. Assume that the laser control field at the moment of
sending operation has an unknown phase of $\varphi_{1}$ and, hence,
the Rabi frequency is $\Omega _{1}\left( t\right) e^{i\varphi
_{1}}$. From the form of the coupling term that involves the laser
field in the Hamiltonian Eq.~(\ref{hamiltonian}), the unknown phase
factor can be absorbed by redefining the state $\left|
\tilde{e}\right\rangle _{1} \equiv e^{-i\varphi _{1}}\left|
e\right\rangle _{1}$, so that,
\begin{equation}
\frac{1}{2}\left[ i\Omega _{1} ( t ) e^{i\varphi _{1}}e^{-i\omega
_{L}t} | t\rangle _{11} \langle e | +H.c.\right] \equiv \frac{1
}{2}\left[ i\Omega _{1} ( t ) e^{-i\omega _{L}t} | t
\rangle_{11}\langle \tilde{e}| +H.c.\right]
\end{equation}
and we can make the same transform at the receiving node where the
control laser has the unknown phase $\varphi_2$ at the moment the
single photon arrives. Starting from a general state $\left(
C_{g}|g\rangle _{1}+C_{e}|e\rangle _{1}\right) |g\rangle _{2}\otimes
|{\rm vac}\rangle \equiv C_{g}|g\rangle _{1}|g\rangle _{2}\otimes
|\mathrm{vac}\rangle +C_{e}e^{i\varphi _{1}}|\tilde{e}\rangle
_{1}|g\rangle _{2}\otimes |\mathrm{ vac}\rangle$, two-node
operations in the presence of laser phase uncertainties can be
generally expressed as,
\begin{eqnarray}
&& C_{g}|g\rangle _{1}|g\rangle _{2}\otimes |\mathrm{vac}\rangle
+C_{e}e^{i\varphi _{1}}|\tilde{e}\rangle _{1}|g\rangle _{2}\otimes
|\mathrm{vac}\rangle  \notag \\
&\overset{\Omega _{1}(t)}{\longrightarrow }& C_{g}|g\rangle
_{1}|g\rangle _{2}\otimes |\mathrm{vac}\rangle \notag \\
&& + ~C_{e}e^{i\varphi _{1}}\left[ e^{i\phi }\cos \theta
|\tilde{e}\rangle _{1}|g\rangle _{2}\otimes |\mathrm{vac} \rangle
+\sin \theta |g\rangle _{1}|g\rangle _{2}\otimes |\tilde{\alpha}_{
\mathrm{out}}\rangle \right]  \notag \\
&\overset{\Omega _{2}(t-\tau )}{\longrightarrow }& \left[
C_{g}|g\rangle _{1}|g\rangle _{2} + C_{e}e^{i\varphi _{1}}\left(
e^{i\phi }\cos \theta |\tilde{e}\rangle _{1}|g\rangle _{2}+\sin
\theta |g\rangle _{1}| \tilde{e}\rangle _{2}\right) \right]\otimes
|\mathrm{vac}\rangle.
\end{eqnarray}
The final state is equivalent to $\big[ C_{g}|g\rangle _{1}|g\rangle
_{2}+C_{e} ( e^{i\phi }\cos \theta |e\rangle _{1}|g\rangle
_{2}+e^{i\varphi _{1}}e^{-i\varphi _{2}}\sin \theta |g\rangle
_{1}|e\rangle _{2} ) \big] \otimes |\mathrm{vac}\rangle$. If the
control laser fields at the two nodes is phase locked
so that there is a certain relative phase between
$\Omega_1(t)$ and $\Omega_2(t-\tau)$, the two-node operation is well
protected from laser phase fluctuations.

\subsubsection{Deterministic phase and shape variations in photon propagation}

Unknown offsets from the assumed waveguide/fiber
dispersion relation or nonlinearity of the dispersion relation can
cause phase and shape variations of the single photon wavepacket
during the propagation. At low temperature where thermal
fluctuations are suppressed, such variations are deterministic and
can thus be incorporated in the design of the receiving node
control. Close-loop adaptive feedback
control~\cite{wiseman93,Mabuchi_realtime} or quantum learning
algorithms~\cite{Rabitz_1992,rabitz00,Rabitz_2001} can be
implemented to pre-characterize such variations.

\subsubsection{Loss and indeterministic fluctuation in photon propagation}

For inter-node operations between well separated nodes
(distance $\gtrsim$~m), photon losses and indeterministic
fluctuations during the propagation in the fiber could be
non-negligible. Error correction schemes dealing with such
propagation loss has been proposed using auxiliary stationary qubits
in the quantum node~\cite{vanEnk_ecc}. The idea of quantum repeaters
with nested purification schemes~\cite{Cirac_Qrepeater} might also
be incorporated into the quantum network design for protection
against the photon propagation loss and indeterministic
fluctuations.

\subsection{Summary: Coherent quantum manipulation by remote control
\label{subsec_QIPbyNemesis}}

In this section, we have discussed how to unite
clusters of QD spins into a network for distributed quantum information
processing. The advances in the fabrication of coupled structures of
QDs, semiconductor micro-cavities, and optical waveguides make
possible a high efficiency quantum interface between stationary spin
qubits and flying photon qubits, where the latter can be used for
communication between seperated clusters. As indicated in the
DiVincenzo criteria~\cite{DiVincenzo_Criteria_7}, such capability
greatly enhances the chance towards the construction of
fault-tolerant scalable quantum computers in these systems.

Counterintuitively, real excitations of the continuum modes in the
quantum channel act as a conduit for mediating coherent operations
rather than the cause of dissipation. A more abstract picture for the network structure here is the
3-level $\Lambda$ system coupled to a one-dimensional continuum
modes, formed by the cavity-waveguide coupled structure, considered
as a whole electromagnetic continuum. The important function of the cavity is in
providing a local resonance in spectrum with large spectral weight near the frequency of the QD
optical transition, which results in a large Purcell factor (i.e. the
ratio between the spectral weight of the cavity-waveguide modes and
the free-space modes). The unitary evolution dominates over the irreversible processes
within the coupled system of the QD and the
cavity-waveguide continuum. Temporal shaping of the laser control
pulse provides sufficient freedom to control such evolution. While
the relevant part of the spectrum of the cavity-waveguide continuum
is of a simple Lorentz shape with width $\gamma \ll \omega _{c}$,
this Markovian condition is not a mandatory requirement. Provided
that the 3-level $\Lambda$ system is coupled to a continuum where a
spectral weight peak (irrespective of shape) results in a high
Purcell factor at the relevant optical frequency, control schemes
for high efficiency network operation is
possible~\cite{Liu_network_nonMarkovian}. This opens up
possibilities for networking localized stationary qubit by continuum
modes without cavity QED and in the non-Markovian regime.

\section{Challenges}

We have discussed all the necessary elements for implementing scalable quantum computation
with electron spins in QDs under optical control. To reach the larger goals,
there remain many obstacles. Here we present an overview of what technologies may be at the top of the
required list to accomplish the goals. To have a concrete idea of the challenges for short- to mid-term
pursuit, we will also give an estimate of the resources
for a benchmark task: factorization of 15 with Shor's algorithm.

\subsection{Resource estimate for Shor's algorithm for factorizing 15\label{factorize}}

Undoubtedly, Shor's algorithm for factorizing integers
is the most important example demonstrating the superpower of
quantum computation~\cite{Shor94,NielsenChuang}.
It gives a solution with time consumption only polynomially
increasing with the problem size (the bit length of the number to
be factorized) and thus offers an exponential speedup over all
known classical counterparts. Historically,
Shor's algorithm has stimulated the exploding enthusiasm in quantum computation
by showing its computation power. An efficient factorization scheme can be used to break
the public-key encryptions such as the RSA protocol which is
widely used in internet communication. The factorization of the
first ``non-trivial'' number -- 15 by Shor's algorithm has been
realized with the liquid NMR spectroscopy~\cite{Chuang_factoring15},
which serves as a benchmark for quantum computation in other systems. Here we give an
estimate of the resources required to accomplish such a milestone in the
optically controlled spin-based QD system.

\begin{figure}[b]
\begin{center}
\includegraphics[width=7cm]{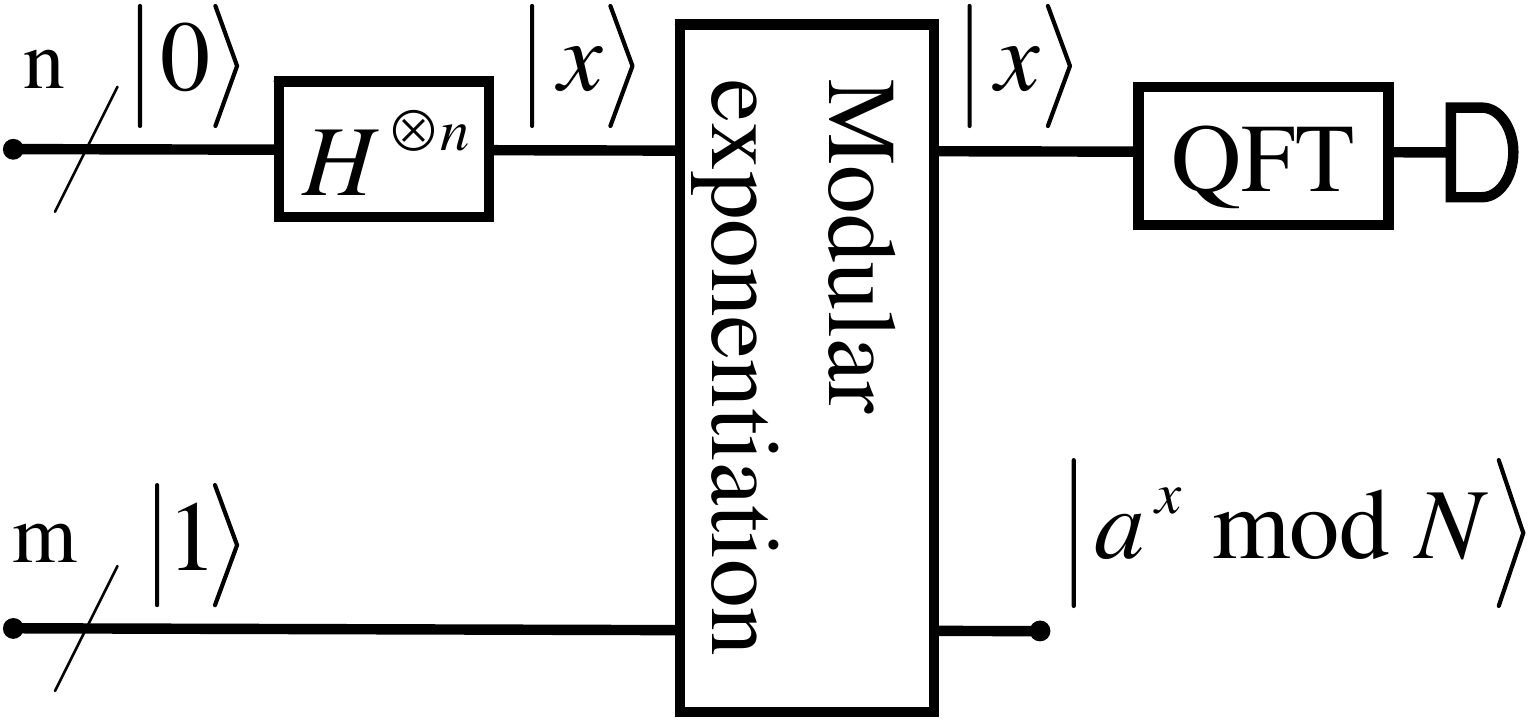}
\end{center}
\caption{Logic flow-chart of Shor's algorithm for order searching.} \label{orderfinding}
\end{figure}

The key step of Shor's algorithm for factorizing a number $N$ is to find
the smallest nonzero $r$ satisfying $a^{r}=1\ {\rm mod}(N)$
(i.e., the order of $a$ with respect to $N$), where the seed $a$ is co-prime to $N$ [i.e. the maximum
common factor of $a$ and $N$, denoted as $(a,N)$ is 1] and can be randomly selected. By the fact that
the order $r$ is the period of the modular exponentiation function $f(x)=a^x\ {\rm mod}(N)$, it
suffices to search for the period or to measure
the frequency of the function $f(x)$, which can be accomplished by the quantum
Fourier transformation (QFT). The process of finding the period
can be expressed as~\cite{Shor94,NielsenChuang}
\begin{eqnarray}
|0^{n}\rangle\otimes|0^m\rangle & \stackrel{{\text
QFT}^{-1}}{\longrightarrow} &
\sum_{x=0}^{2^n-1}|x\rangle\otimes|0^m\rangle
 \stackrel{f(x)}{\longrightarrow}  \sum_{x=0}^{2^n-1}|x\rangle\otimes|a^x\ {\text mod}(N)\rangle \nonumber \\
& \stackrel{{\text QFT}}{\longrightarrow} &
\sum_{y=0}^{2^n-1}\sum_{x=0}^{2^n-1}e^{i2\pi xy/2^n}|y\rangle
\otimes |a^x\ {\text mod}(N)\rangle,
\end{eqnarray}
where $|0^{n(m)}\rangle$ denotes a $n(m)$-qubit register set at
the zero state, $m=[\log_2 N]+1$ is the bit length of $N$, $x$ is
a binary number, and the normalization constants for the states
have been omitted. Both the QFT and the modular exponentiation
can be carried out with the number of elementary quantum gates
polynomially increasing with the problem size ($m$). A more detailed
review of the order-finding algorithm is given in Appendix~\ref{Shor_algorithm}.

\begin{figure}[b]
\begin{center}
\includegraphics[width=7.5cm]{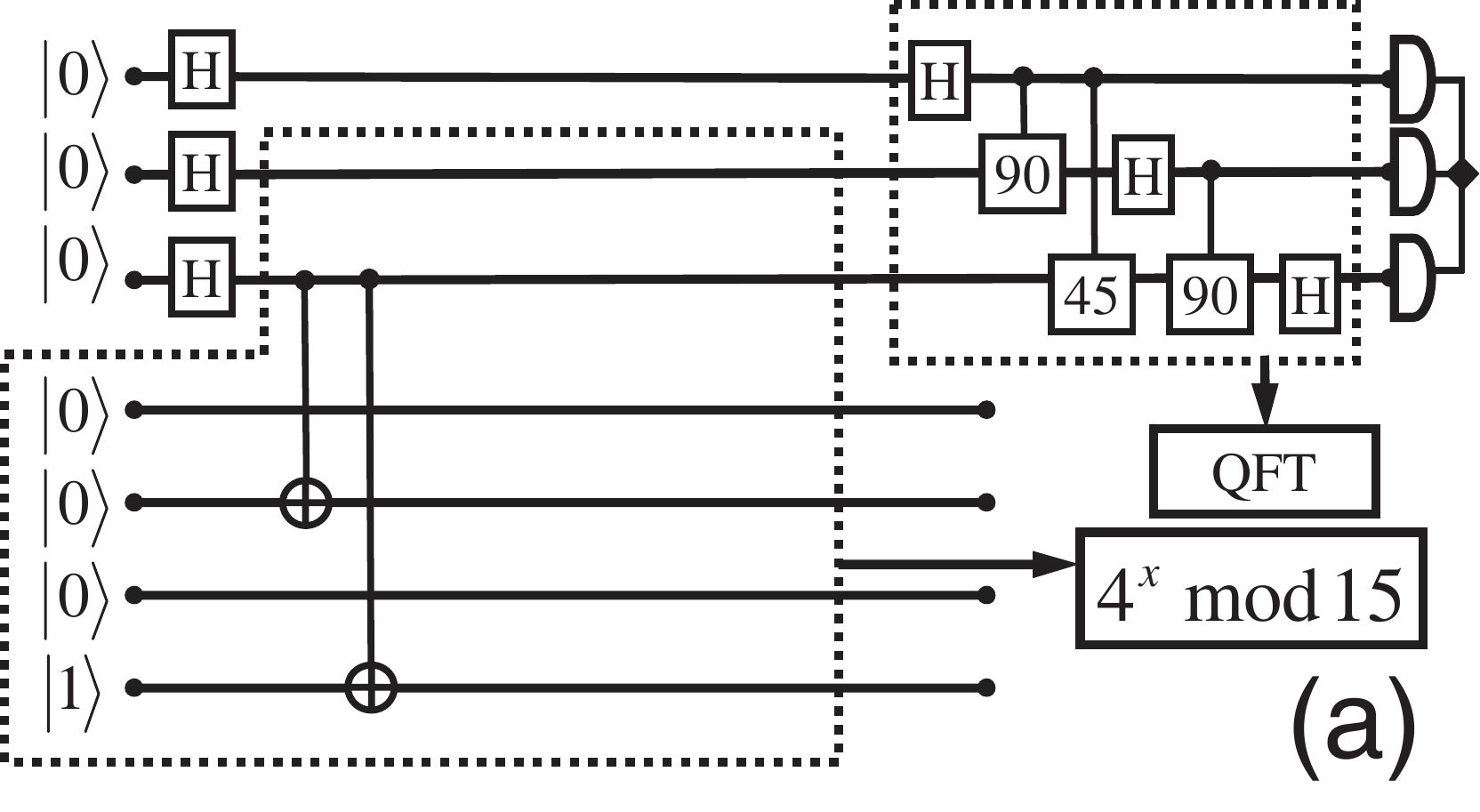}
\vskip .5cm
\includegraphics[width=7.5cm]{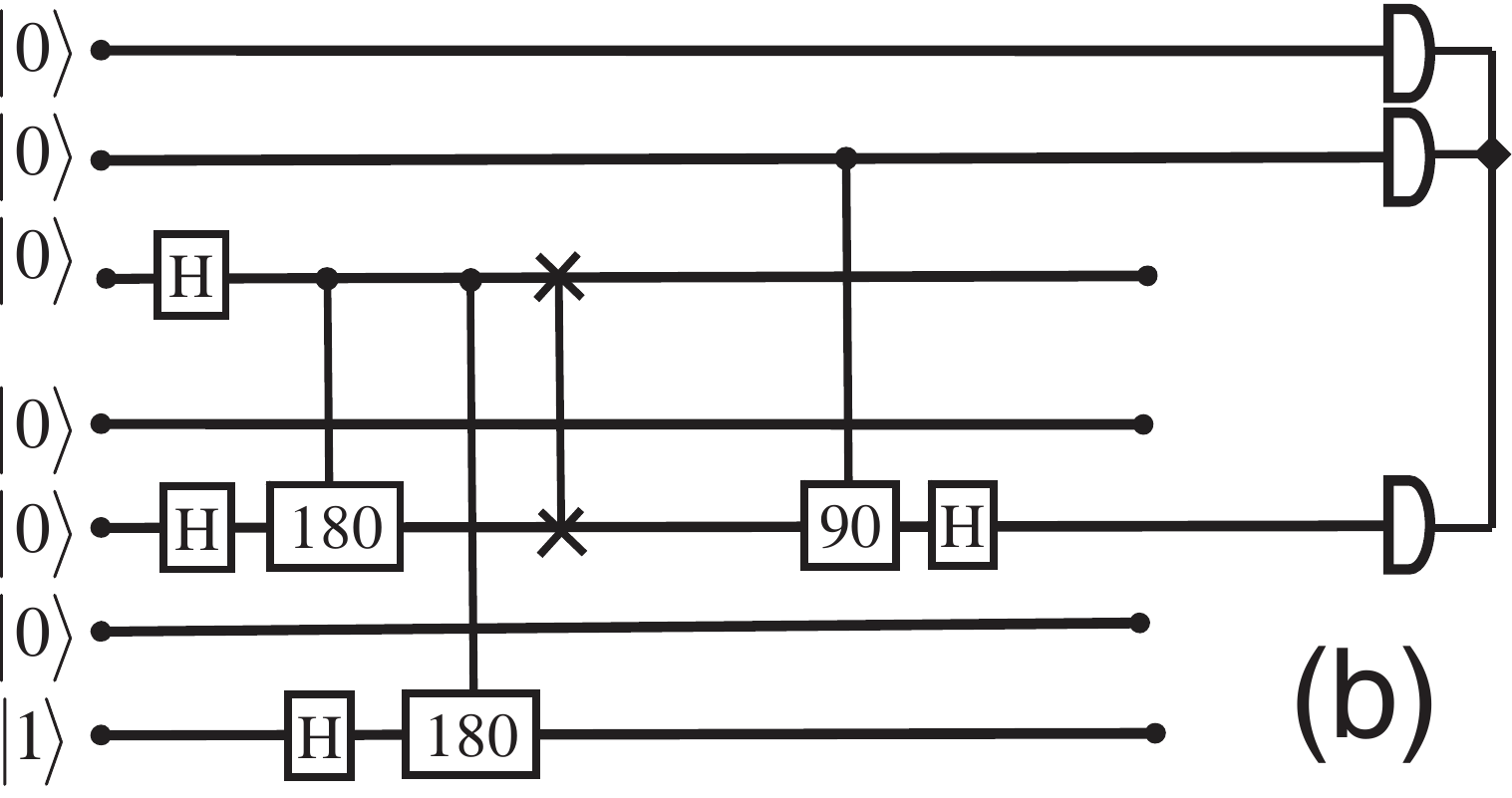}
\end{center}
\caption{(a) Quantum circuit for searching $r$ for $4^r=1$
mod(15), where the subprograms for
the modular exponentiation and the QFT are blocked together as indicated.
(b) The simplified circuit with only local two-qubit
gates.} \label{circuit1}
\end{figure}

As $f(x)$ is a periodic function with period $r$, the
population of the basis state of the first register at the end of
the program will be non-zero only for $y\approx 2^n c/r$ ($c=0, 1, \ldots, r-1$), for
the QFT will transform a period function
into sharp peaks at the multiples of the frequency $2^n/r$. Thus,
a quantum measurement of the first register in the computational
basis will result one of states in the superposition,
thus yielding the number $2^n c/r$ from which the order
$r$ can be derived and the number $N$ factorized. The order $r$ generally is
of the order of magnitude of $N\sim 2^m$, so the number
of basis states in the superposition is $\sim 2^{m}$, an
exponential function of the problem size, which explains why a
QND measurement is required to make
Shor's algorithm scalable for factorizing large numbers,
as discussed in Sec.~\ref{readout}.

The flow-chart of Shor's algorithm for order finding is shown
in Fig.~\ref{orderfinding}. The first QFT subroutine can be
simplified to a series of single-qubit Hadamard gates (see Appendix~\ref{Shor_algorithm}
for the definition of the elementary quantum gates)
since the initial state is set
to be $|0^n\rangle$. The second
register is initialized to be $|1\rangle$ to facilitate the
modular exponentiation. In principle, the length of the first
register $n$ should be large enough to reproduce the real number
$1/r$ with sufficient effective bits. To factorize 15, it turns out that
the orders of all co-primes of 15  are either 2 (for
$a$=4, 11, 14) or 4 (for $a$=2, 7, 8, 13), both of which are a
factor of $2^3$, so three-qubit register should be enough to
resolve $2^n/r$.
Including the second register, 7 qubits are sufficient
to demonstrate the algorithm for the first ``non-trivial'' target: 15.

\begin{figure}[b]
\begin{center}
\includegraphics[width=7cm]{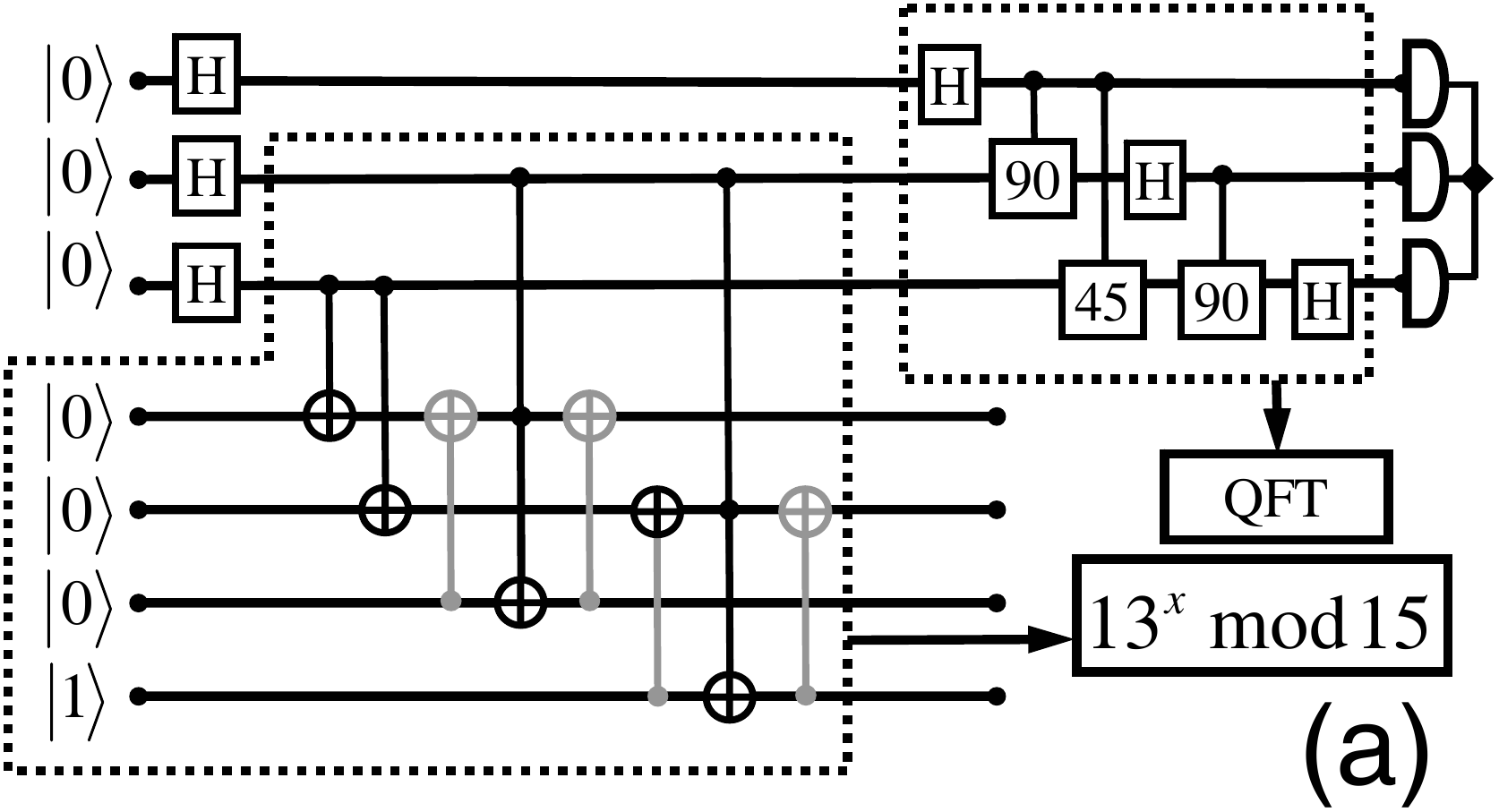}
\vskip 1cm
\includegraphics[width=13.5cm]{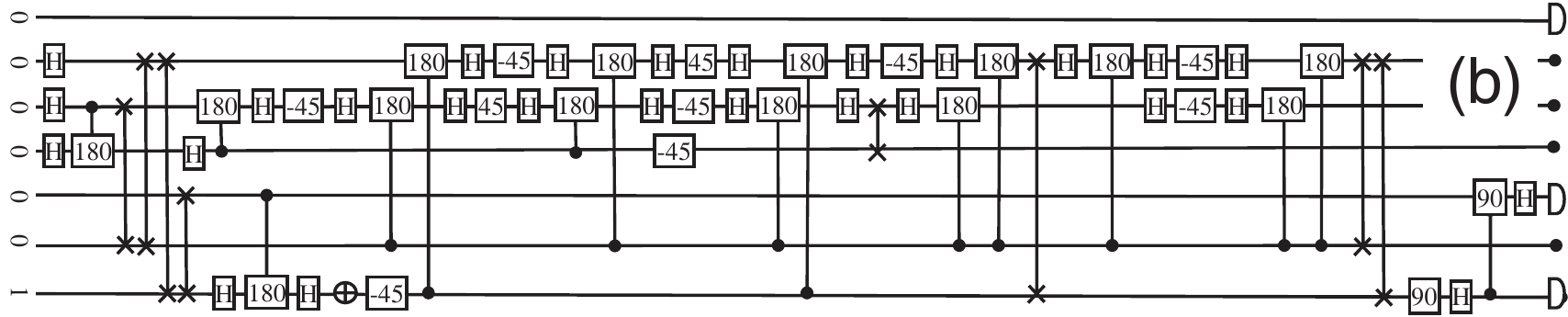}
\end{center}
\caption{(a) Quantum circuit for searching $r$ for $13^r=1$
mod(15). The faded components can be removed without changing the
computing result. (b) The simplified circuit with all gates
experimentally realizable.} \label{circuit2}
\end{figure}

Figure~\ref{circuit1} and \ref{circuit2} give the specific
quantum circuits constructed with the elementary quantum gates
(as defined in Appendix~\ref{elementgates}) to implement Shor's algorithm for
factorizing 15 with the seed number selected as $a=4$ (an easy
example) and $a=13$ (a difficult example). Compilation and
optimization have been performed to reduce the number of physical
operations:
\begin{enumerate}
\item To minimize the  number of SWAP gates for
coupling remote qubits in the linear configuration, the labelling
of the qubits in the register has been optimized, and at the end
of the program the qubits are not swapped back to their original
positions but re-labelled instead, as indicated in
Fig.~\ref{structure}~(a).
\item The quantum gates are omitted if
they act on the second register after the last gate controlled by
the first register, since such gates have no effect on the QFT of
the first register.
\item Whenever possible, the sequential
single-bit operations on the same qubit are combined into one
single-qubit gate (in experiment, all one-bit gates between two
controlled gates can be realized by a single-spin rotation).
\end{enumerate}
The numbers of gates and pulses and the running time needed in
different cases are estimated with the following considerations:
\begin{enumerate}
\item All sequential single-qubit gates on the same
qubit are combined as one single-spin rotation which can be
completed with three pulses within 10~ps.
\item
 Each SWAP gate can be completed by two pulses within 10~ps.
\item
Each controlled phase gate can be completed by three pulses in 10~ps.
\item
 Each qubit can be initialized by three pulses within 120~ps.
\item
To reduce the complexity of pulse designing, all operations within
the local module should be performed serially (to factorize
larger numbers, a large-scale quantum network of distributed
modules will be needed and in that case operations could be
performed in parallel in
separated modules).
\item
The time consumption for quantum measurement is not counted.
\end{enumerate}
The timescales of the single- and two-qubit gates are chosen in accordance with
the discussions in Sec.~\ref{quantumgates}. While there are so far no experimental demonstrations
of optically controlled two-qubit gates, the estimated timescales of the single-qubit gates
are similar to the experimentally realized
gates~\cite{Ramsay08PRLSpinControl,Yamamoto_08NatureSpinControl,Yamamoto_echo,Greilich2009NPhys}.
The estimated resource requirements are summarized in
Table~\ref{costsummary}. Alternatively, the controlled phase
gates can be replaced by the $\sqrt{\text{SWAP}}$ gates, and the
resource requirement is expected to be similar.

\begin{table}[t]
\caption{The numbers of gates and pulses
and the time required to factorize 15 by Shor's quantum algorithm with two typical examples of seed numbers
$a$ for $a^r=1\ \text{mod} (15)$.}
\begin{tabular}{|c||c|c|c|c|c|}
\hline
$a$ & \# 1-bit gates & \#  SWAP & \# phase gate &     \#  pulses      & time-cost \\
\hline
a=4 & 4 &  1 & 3 & 44 & 0.9 ns \\
\hline
a=13  & 20 &  8 & 15 & 142 &  1.3 ns  \\
\hline
\end{tabular}
\label{costsummary}
\end{table}

\subsection{Technologies most needed}

\subsubsection{Complex multi-pulse optics}

As listed in Table~\ref{costsummary}, to factorize
the small number 15, the number of optical pulses needed is in the order of 100. It is would be
extremely difficult to synchronize so many laser pulses and to stabilize the
relative phases, but all the pulses may be viewed as a single one with complex
frequency and phase design. The complex pulse may be formed from a single laser source
by pulse shaping techniques, such as acousto-optical modulation~\cite{Verluise_PulseShape,Warren_pulseShape}.
Design of such a complex pulse is also very challenging. Learning or genetic algorithms may be developed
to deal with the problem, with lessons learned from controlling complex
chemical processes by ultrafast optical pulses~\cite{Rabitz_1992,Warren_controlReview,rabitz00}.
The measurement involves many probe pulses interacting with many QDs.
The signals are to be analyzed through frequency
multiplexer as well as phase modulation with heterodyne detection~\cite{Nico_thesis}.
Thus multi-dimensional spectroscopies~\cite{Cundiff_2D,Cho_2Doptics,Mukamel_multiDoptics,Warren_2D}
are desirable for characterizing the systems and for implementing a small-scale benchmark demonstration.
Eventually, the measurement would have to be done with an efficient quantum channel like
a cavity-waveguide structure.

\subsubsection{System fabrication and characterization}

The design of the quantum computer and its operation
follow the sequence:
The physical system is constructed first, with certain uncontrollability.
Then it is characterized. The optical control will be designed according
to the system parameters. In such a procedure, we
do not require an ultimate control of the system fabrication (hardware) but
defer the difficult work to the control design (software)
stage. For example, we do not require, and actually do not desire that
all the QDs are almost identical. But of course, the system should be fabricated fulfilling certain
conditions. Basically, we need a system made of QD clusters, which should meet
the following requirements:
\begin{enumerate}
\item QDs are only locally coupled. For a system with coupling between remotely separated dots,
the design of the controlling pulses would require overhead increasing exponentially with the number
of qubits, since essentially the control design amounts to solving the Sch\"{o}dinger equation of the
whole system.
\item
The coupling between different QDs in a local node should be weak.
Otherwise, tunneling between different dots would make the local node rather a large QD molecule
 instead of a cluster of individual dots, making even a single-qubit gate
 as complex as a control of all the qubits in the local node.
\item
Even though we do not require all the QDs to be nearly
identical,  the size of different QDs in a cluster should be in a relatively small range.
  Otherwise, the optical control, especially the energy shift by the AC Stark effect, would
  be extremely difficult.
\item
The number of active electrons in each QD should be controlled to be one, either by
doping or by gate voltage control.
\end{enumerate}%
Once a QD system has been constructed, the characterization is not any less demanding.
The parameters to be determined include electron doping level, electron spin $g$-factors, frequencies of
the ground and excited state transitions, selection rules or relative dipole matrix elements for
different transitions, tunneling rates and exchange interaction of electrons in different levels and different dots, etc.
Identification of the exciton, bi-exciton, or trion transitions in two-dot systems have already been demonstrated
in two vertically coupled dots with the help of varying the gate voltage~\cite{Gammon06SciQDs}.
A pair of laterally coupled dots remains a challenge.
The characterization of a cluster of, for example, seven QDs, is
a tremendously difficult target for the current experiment capability.
In the long run, we would expect no full characterization
of the system be required. The solution may lie in, again, the learning algorithm, by which,
the controlling optical pulses are to be designed, self-adaptive to the fidelity of a certain set of quantum gates.

\subsubsection{Nano-photonics}

The requirement of nano-optics is two-folded: First, the control of a local node
requires near-field optical addressing.
Second, the ultrafast fast initialization and QND measurement of a qubit, and coupling between different nodes need
cavities and waveguides fabricated in-situ with the QDs.

To address individually each cluster of QDs, a micro-lens and micro-fiber may be used.
Each QD within a cluster is distinguished by its fingerprint transition frequencies.
The spatial resolution required is given by
the distances between clusters instead of the size of the clusters. Using high-index material for the micro-lens,
spatial resolution $\sim 0.1$~$\mu$m may be achievable.  In cases where the spatial resolution is not enough to single out
dot clusters, an alternative solution would be further pulse shaping (probably with a learning algorithm as well)
to eliminate the coupling between different clusters covered by one micro-lens.

To connect QDs with cavities and waveguides, two structures look promising.
One is obtained by etching the surface where the QDs are grown~\cite{vahala_review}.
Electron-beam lithography and chemical etching (sometimes plus some annealing) have already produced high-quality
microcavities and waveguides on semiconductor surfaces. Even strong coupling between a QD and
microcavity in such a system has been demonstrated~\cite{Bloch_rabi_splitting}. Photonic crystals are another
promising possibility. Point defects in photonic crystals can be made into nano-cavities with
Q-factor $\gtrsim 10^6$ and effective volume less than a half-wavelength
cubed~\cite{Noda_PhotonicCrystal1,Noda_PhotonicCrystal2}.
Strong coupling between a QD and a photonic crystal nano-cavity has also been
demonstrated~\cite{khitrova_rabi_splitting,Imamoglu_deterministic_strongcouple}.
Quantum electrodynamics of single QDs in nano-cavity in photonic crystals
can be engineered~\cite{Kress_PRB05,Finley_NJP2009,Finley_NJP2009b}.
Waveguides in photonic crystals can be made by line defects which may be
coupled to remote nodes by optical fibers~\cite{Vuckovic_cavityFiber}.
How to combine these photonic structures with the QD clusters, especially,
how to assemble them in proper layouts and positions, would demand a great
deal of progress in sample processing technologies.

We would like to point out that the photonic structure fabricated on the QD system
may also be used to individually address each local node.
Thus, we eventually may need no micro-lens or micro-fibers to attach each dot cluster.

A wilder conjecture is that the lasers be integrated in the photonic structure.
Micro- or nano-lasers made of photonic crystal cavities have actually come into being.
The remaining problem would be to make the laser emit into a waveguide, and to tailor it into wanted shapes and sequences.
How is one to control the laser on-chip then? Electrical gates may be used. So we come around a full circle
and find a point where different quantum computation strategies may be synthesized to achieve a common goal.

\section{Conclusion}
\label{summary}

We have discussed various aspects of a scalable scheme of quantum
computation based on optical control of electron spins in
semiconductor QDs. To implement such a scheme, a number
of outstanding challenges remain to be overcome.

We would like to add some remarks on two different
philosophies in implementing digital computing or more generally
automatic reasoning, which may give us some inspiration in the
journey to realize quantum computation. One is related to the von Neumann
structure of computers and the other is related to the Turing ACE structure,
both of which are based on Turing's insightful view of programs and data
as essentially the same for a universal computer. Turing's design is deeply
rooted on his finding of universal computers and thus has a hardware of minimal
instructions with complex functions to be implemented by software programming.
The von Neumann structure, while keeps as well the important role of programming,
tries to maximize the usage of hardware design to implement a large number of
mathematical functions which would otherwise be solved simply by programming.
We follow Turing's perception in describing the blueprint of a quantum computer based on
semiconductor QDs and optical control of spins in them, simply for one reasons:
In implementing quantum computation, the hardware part is far more formidable than the
design of optimal control. Thus we propose no need of perfectly controlled
arrays of almost identical QDs but a control scheme programmed
after a physical structure has been constructed and characterized.
The randomness in the system synthesis is not to be eliminated but to be utilized.
The requirement for a functional physical block is relatively simple: sufficient coupling for
a universal set of quantum gates to be programmed. Finally, we note
that Turing's philosophy is becoming more and more important nowdays in large-scale classical computers which tend
towards the RISC (reduced instruction set computer) architecture.

When will a quantum computer come into being of practical usage? We do not know.
But a hint may lies in the comparison between our present situation and the situation
something 60 years ago when engineers were working hard to maintain thousands of vacuum
tubes functioning together for a while before one or another went wrong. Scaling up of a
quantum computer may be not as rapid as classical computers have done, but just be aware
that adding one functioning qubit supposedly doubles the power of a quantum computer,
which is worth 18 hardworking months in the sense defined by Moore's law.

\section*{Acknowledgements}

This review is based on works done in long collaborations with many people, including
D. G. Steel, D. Gammon, P. Chen, C. Peirmarocchi, S. E. Economou, S. K. Saikin, C. Emary, W. Yang
and under financial supports from ARDA/ARO, ARO/NSA/LPS, QuIST/AFOSR, NSF and Hong Kong RGC.

\begin{appendix}

\section{Hole-mixing and selection rules in a quantum dot \label{mixing}}

\begin{figure}[b]
\begin{center}
\includegraphics[width=7cm]{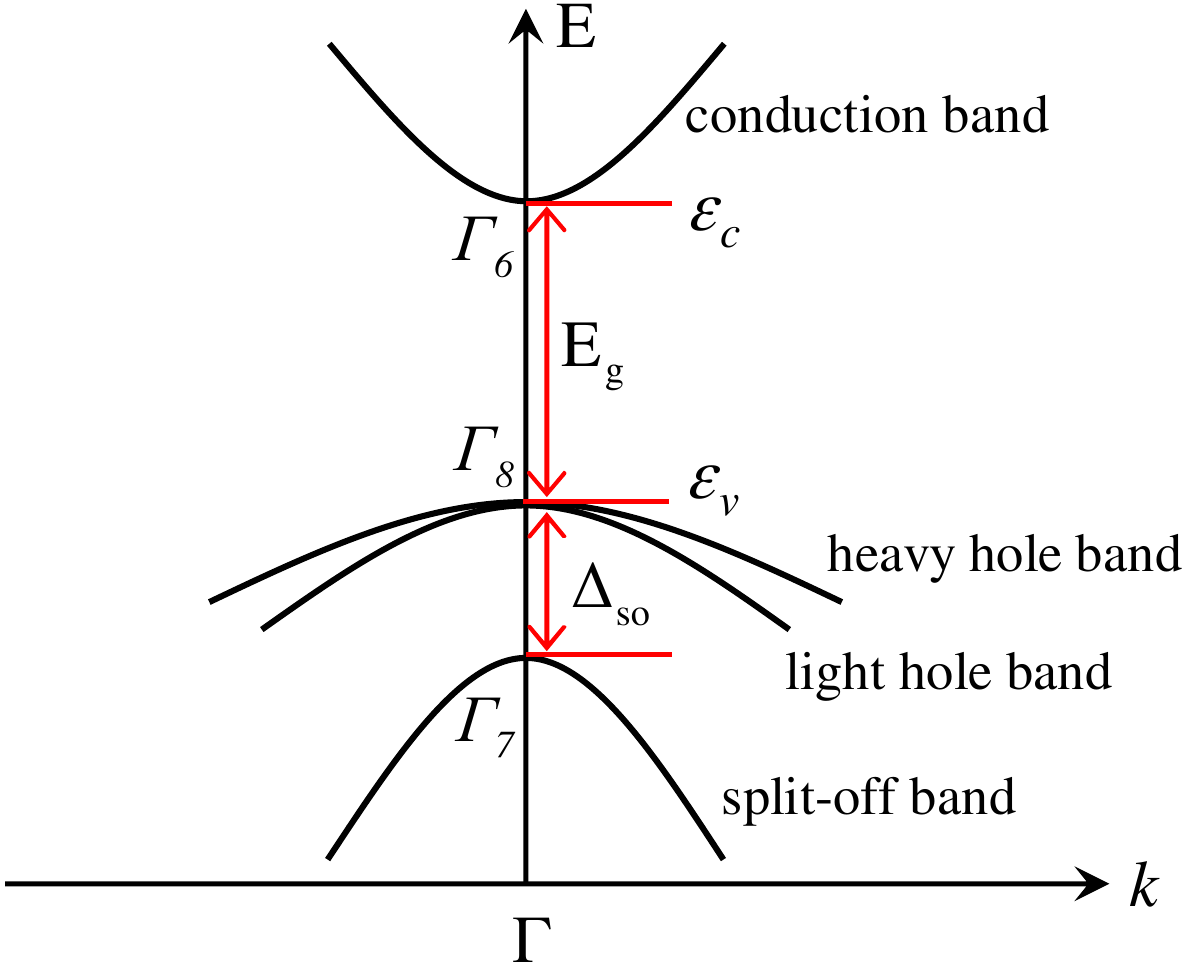}
\caption{Bulk band structures of direct bandgap III-V semiconductors.} \label{band_bulk}
\end{center}
\end{figure}

In the III-V bulk material, the top of valance band occurs at the
$\Gamma$ point of the Brillouin zone (see Fig.~\ref{band_bulk}). The spin-orbit coupling
of the $p$-orbit splits the $p$-type valance band to a quadruplet
with ${\it \Gamma}_8$ symmetry (or a total angular momentum $J=3/2$)
and a doublet with ${\it \Gamma}_7$ symmetry (or a total angular
momentum $J=1/2$). The $\Gamma$-7 band lies much lower in energy than the $\Gamma$-8 bands.
From now on, we focus on the bands derived from
the $J=3/2$ quadruplet, which are described by the Luttinger
Hamiltonian in the framework of effective-mass theory~\cite{Yu_Cardona}
\begin{eqnarray}
H_{L}=\frac{1}{2m_0}\left(
\begin{array}{cccc}
P_1 & Q   &  R   &  0  \\
Q^* & P_2 &  0   &  R  \\
R^* & 0   & P_2  & -Q \\
0   & R^* & -Q^* & P_1
\end{array}
\right),
\end{eqnarray}
expressed in a matrix form in the basis $\{|J_z=3/2 \rangle , |J_z=1/2
\rangle, |J_z=-1/2 \rangle, |J_z=-3/2 \rangle \}$, with
\begin{subequations}
\begin{eqnarray}
P_1 & = & \left(\gamma_1-2\gamma_2\right)k_z^2 +\left(\gamma_1+\gamma_2\right)\left(k_x^2+k_y^2\right), \\
P_2 & = & \left(\gamma_1+2\gamma_2\right)k_z^2 +\left(\gamma_1-\gamma_2\right)\left(k_x^2+k_y^2\right), \\
Q &= & -2\sqrt{3}\gamma_3 k_z\left(k_x-ik_y\right), \\
R &= & - \sqrt{3}\left[\gamma_2\left(k_x^2-k_y^2\right) -2i\gamma_3
k_xk_y\right],
\end{eqnarray}
\label{luttinger}
\end{subequations}
where $\gamma_{1,2,3}$ denote the Luttinger coefficients. The $|
J_z=\pm3/2 \rangle$ and $|J_z=\pm1/2 \rangle$ bands are usually
referred as the heavy hole and light hole bands respectively by
their different effective mass in the $z$ direction.

\begin{figure}[b]
\begin{center}
\includegraphics[width=12cm]{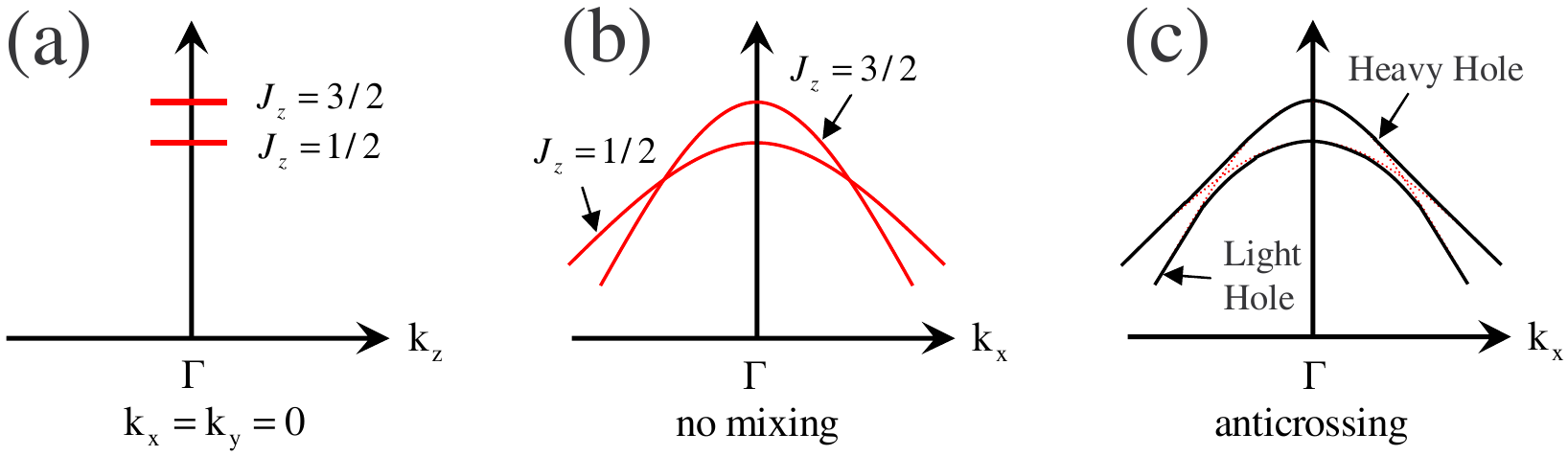}
\caption{Dispersion relation of the
$|J_z=3/2\rangle$ and $|J_z=1/2\rangle$ valence bands in a quantum
well~\cite{Yu_Cardona}. (a) Energy shift due to the confinement in
the growth direction. (b) In-plane dispersion of the two valance
bands when the ``off-diagonal" terms are neglected (without mixing).
(c) Level anti-crossing when ``off-diagonal" terms are included.}
\label{QW_hole}
\end{center}
\end{figure}

Hole mixing effects are caused by the QD confinement
potential $V_h(x,y,z)$. Here for simplicity, the growth direction
$z$ is assumed to be along the $[001]$ direction. For other growth
directions, the discussions below can be generalized
straightforwardly, and the essential conclusions would not be
changed. Strong quantum confinement in the growth ($z$) direction
lifts the four-fold degeneracy at $\Gamma$ point (see
Fig.~\ref{QW_hole}(a)). The light hole states actually have a larger
in-plane effective mass and therefore their dispersion tend to cross
into the heavy hole bands at finite in-plane wavevector $k_{\perp}$
(see Fig.~\ref{QW_hole}(b)). Spin-orbit coupling thus results in
anti-crossing and the mixture of the heavy-hole and light-hole bands
(see Fig.~\ref{QW_hole}(c))~\cite{Yu_Cardona}.

Let us consider first the zeroth-order approximation for the hole-mixing.
When the confinement size along the $z$-direction is much smaller than the
lateral size, we have $\langle k_z^2 \rangle \gg \langle k_x^2 \rangle, \langle k_y^2 \rangle$. If the off-diagonal couplings $Q$ and
$R$ are neglected, the Hamiltonian is diagonal in the basis of the angular momentum
quantized along the growth direction. The heavy and light holes, with kinetic energy $P_1$ and $P_2$,
are characterized by their angular momentum states as $|\pm 3/2\rangle$ and $|\pm 1/2\rangle$, respectively.
The heavy hole (HH) and the light hole (LH) are separated by an energy $\Delta_{HL}=2\gamma_2\langle k_z^2\rangle/m_0$~\cite{Yu_Cardona}.
The hole-mixing are induced by $Q$ and $R$ terms. The $Q$ terms couple $|\pm 3/2\rangle$ to $|\pm 1/2\rangle$,
and the $R$ terms couple $|\pm 3/2\rangle$ to $|\mp 1/2\rangle$.
As we usually have $\Delta_{HL}\gg \langle Q\rangle, \langle P\rangle$, the zeroth-order approximation
and the corresponding selection rules determined by the angular momentum conservation are often adequate
to understand the optical transitions.

For small QDs, the mixing may be important. Now let us consider the HH-LH mixing in
different situations.

First we consider a confinement potential with rotational symmetry about the growth direction.
If we assume $\gamma_2=\gamma_3$, the Hamiltonian has the rotational symmetry.
The HH and LH states coupled by the $Q$ and $R$ terms must have different orbital angular momentum.
So the light-hole components mixed into, e.g., the heavy hole ground state are not optical active.
In this case, the mixing has no effect on the optical transitions but a reduction of the dipole
matrix element. The problem comes from the fact that in reality we usually do not have
a cylindrical $V(x,y,z)$ or $\gamma_2\ne \gamma_3$ (since the lattice has no spherical symmetry).

\begin{figure}[b]
\begin{center}
\includegraphics[width=8cm]{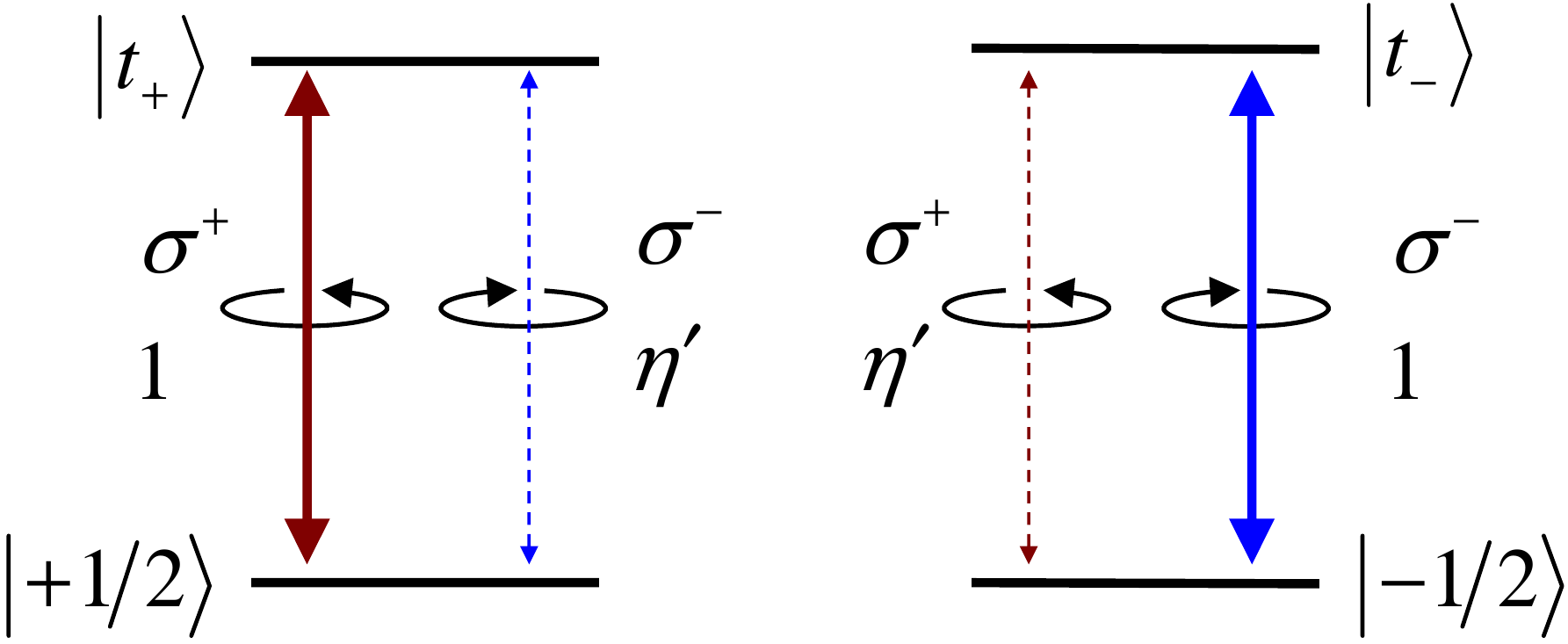}
\end{center}
\caption{Selection rules of optical transitions in a doped QD with inversion
symmetry along the growth direction. The solid lines denote the allowed transitions,
and the dashed lines denotes the ``forbidden'' transitions which have a relative
dipole matrix element $\eta'$.}
\label{mixing1}
\end{figure}

Let us still make the assumption that the confinement potential $V$ has inversion symmetry in the
$z$-direction, which is reasonably fulfilled in fluctuation QDs. Since the $Q$ terms are
linear in the $k_z$, they couple HH and LH states with different parity which are separated by large energies
due to the strong confinement in the $z$-direction.
In the even parity ground state, the hole mixing effect caused by the $Q$ terms is negligible, especially in optical transitions.
Thus, we need only to consider the coupling induced by the $R$ terms. Thus the optically active components
of the HH ground states are
\begin{eqnarray}
|H\pm\rangle & = & |\pm 3/2\rangle+\eta |\mp 1/2\rangle,
\end{eqnarray}
where $\eta\sim L_z^2/L_{x,y}^2$ (with $L_{z}$ and $L_{x,y}$ being the vertical and lateral
confinement sizes, respectively) and normalization is understood.
The dipole matrix elements between the electron spin states $|\pm 1/2\rangle$ and the
trion states $|t\pm\rangle$ which contains the HH in $|H\pm\rangle$ are (in arbitrary units)
\begin{subequations}
\begin{eqnarray}
\langle t\pm|{\mathbf d}\cdot {\boldsymbol \sigma}_{\pm}|\pm 1/2\rangle &=& 1, \\
\langle t\pm|{\mathbf d}\cdot {\boldsymbol \sigma}_{\mp}|\pm 1/2\rangle &=& \eta',\\
\langle t\mp|{\mathbf d}\cdot {\boldsymbol \sigma}_{\pm}|\pm 1/2\rangle &=& 0, \\
\langle t\mp|{\mathbf d}\cdot {\boldsymbol \sigma}_{\mp}|\pm 1/2\rangle &=& 0,
\end{eqnarray}
\end{subequations}
where ${\boldsymbol \sigma}_{\pm}$ is the circular polarization of a light normally propagating
and $\eta'\sim \eta$. The selection rules are shown in Fig.~\ref{mixing1}.
The two subspaces $\{|+1/2\rangle, |t+\rangle\}$ and $\{|-1/2\rangle, |t-\rangle\}$ are
disconnected under optical coupling, which means the electron spin along the growth direction is
still conserved unless an external magnetic field is applied along an in-plane direction.
With a magnetic field along the $x$-direction, we can choose a light with the polarization, e.g.,
\begin{eqnarray}
\tilde{\boldsymbol \sigma}_{+} &\equiv {\boldsymbol \sigma}_{+}-\eta' {\boldsymbol \sigma}_{-},
\end{eqnarray}
which couples the electron spin eigen states $|\pm x\rangle$ to the common trion state $|t+\rangle$ with
renormalized dipole matrix elements $\left(1\pm \eta'^2\right)$.
The theories for the case without hole mixing can be generalized simply by
replacing the dipole matrix elements there with the renormalized.

In systems without the inversion symmetry, which is often the case for InAs self-assembled QDs,
the mixing would make the two ground HH states to be
\begin{eqnarray}
|H+\rangle & = & |+3/2\rangle+\zeta |+1/2\rangle+\eta|-1/2\rangle+\xi|-3/2\rangle, \\
|H-\rangle & = & |-3/2\rangle+\zeta |-1/2\rangle+\eta|+1/2\rangle+\xi|+3/2\rangle,
\end{eqnarray}
where $\eta$ and $\zeta$ are small numbers ($\lesssim 10\%$ for typical self-assembled
QDs), and $\xi\sim \eta\zeta$. Now the dipole matrix elements are
\begin{subequations}
\begin{eqnarray}
\langle t\pm|{\mathbf d}\cdot {\boldsymbol \sigma}_{\pm}|\pm 1/2\rangle &=& 1, \\
\langle t\pm|{\mathbf d}\cdot {\boldsymbol \sigma}_{\mp}|\pm 1/2\rangle &=& \eta' ,\\
\langle t\mp|{\mathbf d}\cdot {\boldsymbol \sigma}_{\mp}|\pm 1/2\rangle &=& \zeta', \\
\langle t\mp|{\mathbf d}\cdot {\boldsymbol \sigma}_{\pm}|\pm 1/2\rangle &=& \xi',
\end{eqnarray}
\end{subequations}
with $\eta'\sim\eta$, $\zeta'\sim \zeta$, and $\xi'\sim\xi$, all being small numbers.
Interestingly, all the states now are optically connected. Thus it is possible to perform an
arbitrary spin rotation by optical control without applying an external magnetic field or
tilting the light beam from the normal direction.
(For a QD without a symmetry
axis, the normal direction is not special).
For example, if the light polarization is chosen to be
${\mathbf X}\equiv {\boldsymbol \sigma}_{+}+{\boldsymbol \sigma}_{-}$,
the dipole matrix elements between the spin states $|\pm x\rangle$ and the trion states
$|t\pm x\rangle\equiv |t+\rangle\pm |t-\rangle$  are
\begin{subequations}
\begin{eqnarray}
\langle t+x|{\mathbf d}\cdot {\mathbf X}|+x\rangle &=& 1+\eta'+\zeta'+\xi', \\
\langle t+x|{\mathbf d}\cdot {\mathbf X}|-x\rangle &=& 0 ,\\
\langle t-x|{\mathbf d}\cdot {\mathbf X}|+x\rangle &=& 0, \\
\langle t-x|{\mathbf d}\cdot {\mathbf X}|-x\rangle &=& 1+\eta'-\zeta'-\xi'.
\end{eqnarray}
\end{subequations}
Thus the AC Stark effect induces an effective magnetic field along the $x$-direction, which,
as compared the effective magnetic field along the growth direction induced by
a circularly polarized light, is reduced by a factor $\sim \left(\zeta'+\xi'\right)$.
To realize an arbitrary rotation without a static magnetic field, the optical field would
need to be much stronger than with a static magnetic field, unless the QD lateral size
is comparable to the vertical size (such as in the case nanocrystals or spherical QDs formed
by chemical deposition).

The heavy-light hole mixing coefficients $\zeta$ and $\eta$ can be extracted using the
polarization dependent absorption spectroscopy as described in Ref.~\cite{measureholemixing}.
For typical InAs self-assembled dots studied by Steel's group,  $\zeta$ is negligible and $\eta$
varies in the range $\sim 0.1-0.2$ for different dots~\cite{Xu07PRLspinIni,Steel_locking}.

\begin{figure}[b]
\begin{center}
\includegraphics[width=8cm]{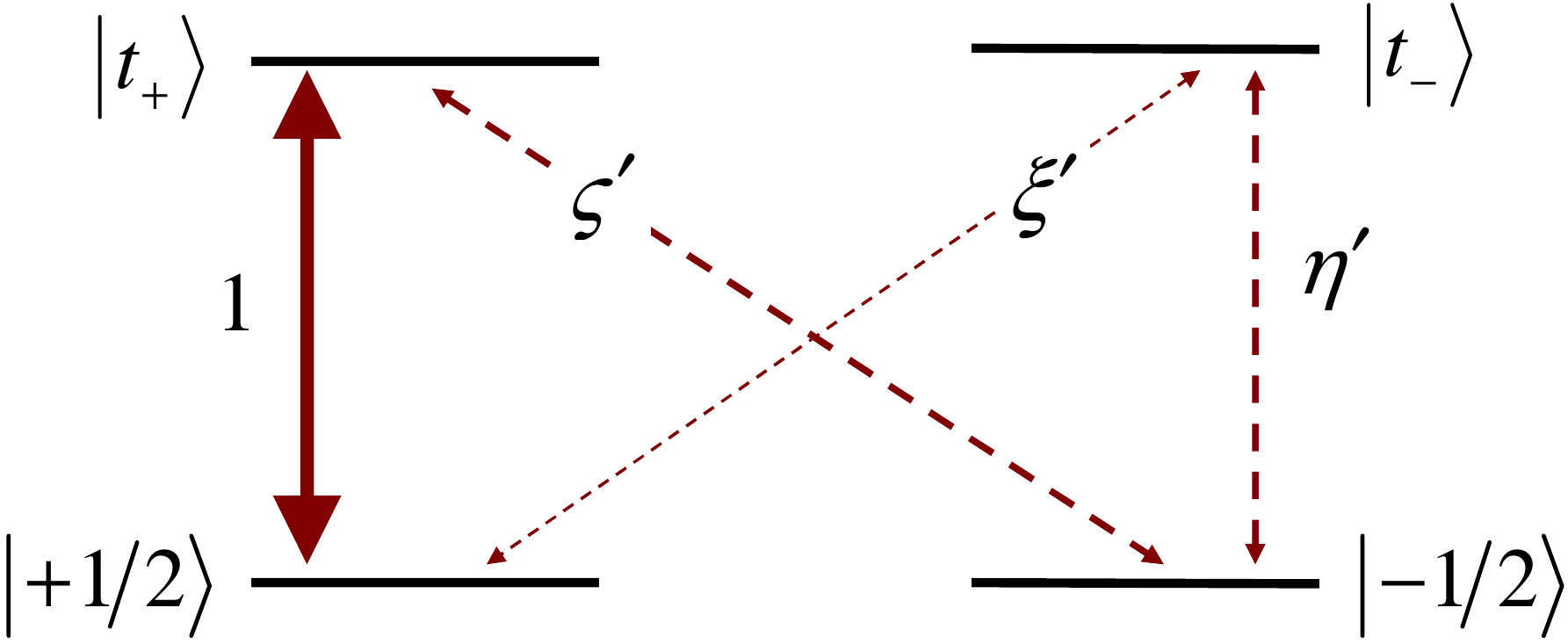}
\end{center}
\caption{The relative dipole matrix elements in a doped QD with irregular
shape but a large lateral to vertical size ratio. The light polarization is assumed to
$\sigma_+$. The solid lines denote the allowed transitions, and the dashed lines denotes
the ``forbidden'' transitions which have relative dipole matrix elements $\eta'$, $\zeta'$, and $\xi'$.}
\label{mixing2}
\end{figure}

\section{Theory of electron spin decoherence by interacting nuclear spins}
\label{decoherence_appendix}

In this appendix, we give the details on the theory of electron spin
decoherence by interacting nuclear spins in a strong magnetic field ($\geq$ 1 Tesla),
which was first formulated in Ref~\cite{Yao_Decoherence}.

The mesoscopic system consists of an electron with a spin vector
$\hat{{\mathbf S}}_{\rm e}$ and $N$ nuclear spins, $\hat{{\mathbf
J}}_{n}$, with Zeeman energies $\Omega_{\rm e}$ and $\omega_{n}$
under a magnetic field $B_{\rm ext}$, respectively, where $n$
denotes both positions and isotope types (e.g. $^{75}$As, $^{69}$Ga
and $^{71}$Ga in GaAs). The interaction can be separated as
``diagonal'' terms which involve only the spin vector components
along the field ($z$) direction and ``off-diagonal'' terms which
involve spin flips. Because the electron Zeeman energy is much
larger than the strength of the hyperfine interaction, the
off-diagonal term is eliminated by a standard canonical
transformation, with the second-order correction left as the
hyperfine-mediated nuclear interaction~\cite{Yao_Decoherence},
called the extrinsic interaction in Section~\ref{subsubsec-2}.
For the same reason, the off-diagonal part of the nuclear
interaction contributes only when the terms conserve the Zeeman
energies (so-called secular terms in the NMR terminology). Hence,
the non-secular terms are negligible. The total reduced Hamiltonian
is obtained for the limit of long longitudinal
electron spin relaxation time ($T_1 \rightarrow \infty$),
\begin{equation}
 \hat{H}_{\rm red}=\hat{H}_\text{e}+\hat{H}_\text{N}+\sum_{\pm}|\pm\rangle \hat{H}^{\pm} \langle\pm|,
 \label{eq-effH}
\end{equation}
with $\hat{H}_{\rm e}=\Omega_{\rm e} \hat{S}^z_{\rm e}$,
$\hat{H}_{\rm N}=\omega_{n}\hat{J}^z_{n}$, and the interaction
terms,
\begin{equation}
\hat{H}^{\pm}=\pm \hat{H}_A+\hat{H}_B+\hat{H}_D\pm \hat{H}_E,
\label{eq-Hpm}
\end{equation}
given by,
\begin{subequations} \begin{eqnarray}
\hat{H}_A &=&
  {\sum_{n\ne m}}'\frac{a_n a_m}{4  \Omega _{\rm e}}
  \hat{J}_n^{+}\hat{J}_m^{-}\equiv {\sum_{n \ne m}}' A_{n,m} \hat{J}_n^{+}\hat{J}_m^{-}, \ \ \ \   \label{HA} \\
\hat{H}_B &=&{\sum_{n \ne  m}}' B_{n,m} \hat{J}_n^{+}\hat{J}_m^{-},
\label{HB}  \\
\hat{H}_D &=&{\sum_{n < m}} D_{n,m} \hat{J}_n^{z}\hat{J}_m^{z},
\label{HD}  \\
\hat{H}_E &=&
 \sum_{n} \frac{a_n}{2}\, \hat{J}_n^{z}\equiv \sum_{n}E_n \hat{J}_n^{z},   \label{HE}
\end{eqnarray}  \label{Hamiltonian}
\end{subequations}
where $|\pm\rangle$ are the eigenstates of $\hat{S}^z_{\rm e}$, the
summation with a prime runs over only the homo-nuclear pairs, the
subscript $A$ denotes the extrinsic hyperfine mediated interaction,
$B$ the off-diagonal part of the intrinsic nuclear interaction, $D$
the diagonal part of the intrinsic interaction, and $E$ the diagonal
part of the contact electron-nuclear hyperfine interaction. The
hyperfine energy, determined by the electron wavefunction, has a
typical energy scale $E_n\sim a_n \sim 10^6$~s$^{-1}$ for a dot with
about $10^6$ nuclei~\cite{Paget}. The sum, $\mathcal{A} \equiv
\sum_n a_n$, is the {\it hyperfine constant} depending only on the
material. The intrinsic nuclear spin-spin interaction has the
near-neighbor coupling strength $B_{n,m}\sim {D}_{n,m} \sim
10^2$~s$^{-1}$. The extrinsic hyperfine-mediated interaction, which
is unrestricted in range within the QD and associated with
opposite signs for opposite electron spin states, has an energy
scale dependent on the field strength, $A_{n,m}\sim
1$--$10$~s$^{-1}$ for field $\sim 40$--$1$~T. Thus, the intrinsic
interaction dominates local pair-interactions, while non-local pairs
are driven by the extrinsic mechanism.

\subsubsection{Formal theory of decoherence in a nuclear spin bath}  \label{subsubsec-decoh}

After the initialization step of the electron spin qubit, the
electron-nuclear spin system is prepared in a product state
with the nuclear spins in a thermal state with temperature $T$,
described by the density matrix
\begin{eqnarray}
\hat{\rho}(0)=\hat{\rho}^{\rm e}(0)\otimes \hat{\rho}^{\rm N}.
\end{eqnarray}
The time evolution of the reduced density matrix of the electron
spin,
\begin{eqnarray}
\hat{\rho}^{\rm e}(t)={\rm Tr}_{\rm N} \hat{\rho}(t),
\end{eqnarray}
is
obtained by tracing over the nuclear spins, may be expressed in the
form,
\begin{eqnarray}
\rho^{\rm e}_{\mu,\nu}(t)=\sum_{\mu',\nu'}{\mathcal L}_{\mu,\nu;
\mu',\nu'}(t) \rho^{\rm e}_{\mu',\nu'}(0)
\end{eqnarray}
where $\rho^{\rm e}_{\mu,\nu}\equiv \langle \mu|\rho^{\rm
e}|\nu\rangle$, and $|\mu\rangle$, $|\nu\rangle\in\{|+\rangle,\
|-\rangle\}$. The superoperator or correlation function ${\mathcal
L}_{\mu,\nu; \mu',\nu'}$ can be expressed in terms of the evolution
operator and contains the information on the electron spin
relaxation and decoherence.

The Hamiltonians of Eq.~(\ref{eq-effH}) for the $T_1 \rightarrow
\infty$ limit conserves the electron $\hat{S}^e_z$ quantum number:
$[\hat{H},\hat{S}_z^e]=0$. Hence, the correlation function has
following properties,
\begin{subequations}
\begin{eqnarray}
{\mathcal L}_{\mu,\nu; \mu',\nu'}(t)&=& {\mathcal L}_{\mu,\nu}(t)\delta_{\mu,\mu'}\delta_{\nu,\nu'}, \\
{\mathcal L}_{\mu,\mu}(t)&=&1, \\
{\mathcal L}_{+,-}(t)&=& {\mathcal L}^*_{-,+}(t),
\end{eqnarray}
\end{subequations}
and the specific expression for the free-induction decay
(FID),
\begin{equation}
{\mathcal L}_{+,-}(t) =  e^{-i\Omega_e t}{\rm Tr}_{\rm N}
\left[\hat{\rho}^N e^{+i\hat{H}^- t}e^{-i\hat{H}^+ t}\right],
\end{equation}
which can be straightforwardly extended to dynamics under pulse
control.

The ensemble of nuclear spins, at temperature $T \gtrsim \omega_n
\gg A_{n,m}, B_{n,m}, D_{n,m}, E_n$, may be approximated by the
density matrix,
\begin{eqnarray}
\hat{\rho}^{\rm N} \approx e^{-\hat{H}_{\rm N}/T} = \sum_{\mathcal
J}P_{\mathcal J}|{\mathcal J}\rangle\langle {\mathcal J}|,
\label{thermal}
\end{eqnarray}
where $|{\mathcal J}\rangle\equiv \bigotimes_n |j_n\rangle$, $j_n$
being the quantum number of nuclear spin $n$ in the magnetic field
direction. $P_{\mathcal J}$ is the thermal distribution factor.
While {\it single-system dynamics}  (i.e., with the nuclear bath initially
in a pure state $|\mathcal{J}\rangle$) could be the ultimate
aim for quantum applications, we
note that all experiments to date are performed under the ensemble
scenario, either in a spatial ensemble of many
dots~\cite{Gurudev,Cheng_spinT1_phasemodulation,Greilich_lock} or
in a single dot with repeated measurements~\cite{T2star_Marcus,Gammon_t2star,Koppens_Rabi},
where the statistical average over the  initial configurations is needed.

The correlation function ${\mathcal L}_{+,-}(t)$ can then be
generally expressed as,
\begin{equation}
{\mathcal L}_{+,-}(t) = \sum_{\mathcal J} P_{\mathcal J}  e^{-i
\phi_{\mathcal J}(t)} \left|\langle \mathcal{J}^- (t) |
\mathcal{J}^+ (t) \rangle\right| \label{ensemble_original}
\end{equation}
In FID,
 $| \mathcal {J}^{\pm} (t) \rangle =
e^{-i \hat{H}^{\pm} t} | \mathcal{J} \rangle$ and $\phi_{\mathcal
J}(t)=(\Omega_e + \mathcal{E}_{\mathcal J}) t$ where ${\mathcal
E}_\mathcal{J}=\sum_n j_n a_n$ is the contribution to the electron
zeeman splitting from the Overhauser field in the nuclear
configuration $| \mathcal{J} \rangle$.

\subsubsection{Pair correlation approximation and pseudo-spin picture}
\label{subsubsec-pss}

The solution to the single-system evolution
$|\mathcal{J}^{\pm}(t)\rangle$ is key to both single spin
decoherence and ensemble decoherence behaviors under free induction
decay and pulse controls. Due to the slowness of the nuclear spin
interacting dynamics, this evolution is well described by the
pair-correlation approximation for the nuclear spin
bath~\cite{Yao_Decoherence,Saikin_Cluster}. Within a time $t$ much
smaller than the inverse nuclear interaction strength, the total
number of pair-flip excitations $N_{\rm flip}$ is much smaller than
the number of nuclei $N$. The probability of having pair-flips
correlated can be estimated to be $P_{\rm corr}\sim 1-e^{-q N_{\rm
flip}^2/N}$ ($q$ being the number of homo-nuclear nearest
neighbors), which is negligible in the relevant timescale of
electron spin
decoherence~\cite{Yao_Decoherence,Yao_DecoherenceControl,Saikin_Cluster,Liu_spinlong,Witzel:2006}.
Thus, the pair-flips as elementary excitations from the initial
state can be treated as independent of each other, with a relative
error $\epsilon \lesssim P_{\rm corr}$. Then the single-system
dynamics $|\mathcal{J}^{\pm} (t)\rangle$ can be described by the
excitation of pair-correlations as non-interacting quasi-particles
from the ``vacuum'' state $|\mathcal{J}\rangle$, driven by the
``low-energy'' effective Hamiltonian,
\begin{equation}
\hat{H}^\pm_{\mathcal J} = \sum_k \hat{\mathcal H}^{\pm}_k\equiv
\sum_k {\mathbf h}_k^{\pm} \cdot \hat{\boldsymbol \sigma}_k/2,
\label{pseudospin_Hamil}
\end{equation}
which has been written in such a way that the pair-correlations are
interpreted as $1/2$-pseudo-spins, represented by the Pauli matrix
$\hat{\boldsymbol \sigma}_k$, with $k$ labeling all possible
flip-pairs~\cite{Yao_DecoherenceControl}. The time evolution from
the initial state $| \mathcal{J} \rangle $ can be viewed as the
rotation of the pseudo-spins, initially all polarized along the $+z$
pseudo-axis: $\bigotimes_k |\uparrow_k\rangle $, under the effective
pseudo-magnetic field,
\begin{equation}
 {\mathbf h}^{\pm}_k \equiv (\pm
2A_k+2B_k,0, D_k \pm E_k), \label{pseudofield}
\end{equation}
where, for the electron spin state $|\pm\rangle$, $\pm A_k$ and
$B_k$ are the pair-flip transition amplitudes contributed by the
extrinsic nuclear interaction $\hat{H}_A$ and the intrinsic nuclear
interaction $\hat{H}_B$, respectively, and $D_k$ and $\pm E_k$ are
the energy cost of the pair-flip contributed by the diagonal nuclear
coupling $\hat{H}_D$ and the hyperfine interaction $\hat{H}_E$,
respectively. The decoherence then can be analytically derived as
\begin{eqnarray}
{\mathcal L}^s_{+,-}(t) =
\prod_k\left|\langle\psi^-_k(t)|\psi^+_k(t)\rangle\right|
\approx\prod_k e^{-\delta_k^2/2},
\end{eqnarray}
where $|\psi^{\pm}_k(t)\rangle$ are the two conjugated states of
pseudo-spin $k$ at time $t$ conditioned on the electron spin state
$|\pm\rangle$. In FID, $|\psi^{\pm}_k (t) \rangle \equiv
e^{-i\hat{\mathcal H}_k^{\pm} t}|\uparrow_k\rangle$; while with a
$\pi$ pulse to flip the electron at $t=\tau$, $|\psi^{\pm}_k (t
> \tau) \rangle \equiv e^{-i\hat{\mathcal H}_k^{\mp} (t-\tau)}
e^{-i\hat{\mathcal H}_k^{\pm} \tau}|\uparrow_k\rangle$.
$\delta_k^2\equiv 1-\left|\langle\psi^-_k|\psi^+_k\rangle\right|^2$
possesses a simple geometrical interpretation: the squared distance
between the two conjugate pseudo-spin states on the Bloch sphere,
which quantifies the entanglement between the electron spin and the
pseudo-spin.

A couple of justified simplifications can provide an understanding
of the effects of various mechanisms on the spin decoherence. First,
the energy cost by the diagonal nuclear coupling ($D_k$) can be
neglected as it is by three orders of magnitude smaller than that by
hyperfine interaction ($E_k$). Second, for near-neighbor pair-flips,
the intrinsic nuclear interaction is much stronger than the
hyperfine mediated one for the field strength under consideration.
Third, for non-local pair-flips, the intrinsic interaction is
negligible due to its finite-range characteristic. Thus we can
separate the flip-pairs into two subsets, $K_A$, which contains
$O(N^2)$ non-local flip-pairs driven by the effective
pseudo-magnetic field ${\mathbf h}^{\pm}_k\approx \left(\pm
2A_k,0,\pm E_k\right)$, and $\in K_B$, which
contains $O(N)$ near-neighbor flip-pairs driven by ${\mathbf
h}^{\pm}_k\approx \left(2B_k,0,\pm E_k\right)$. The conjugate
pseudo-spins will precess along opposite directions in the non-local
subset $K_A$, and symmetrically with respect to the $y$-$z$ plane in
the near-neighbor subset $K_B$. The decoherence can be readily
grouped by the two different mechanisms as
\begin{eqnarray}
{\mathcal L}^{\rm s}_{+,-}\cong \prod_{k \in
K_B}e^{-\frac{t^4}{2}E_k^2 B^2_k
 {\rm sinc}^4\frac{h_k t }{2}}
 \prod_{k\in K_A}e^{-2 t^2 A_k^2 {\rm sinc}^2 (h_k t)}, \label{twomechanism}
\end{eqnarray}
where $h_k = | {\mathbf h}^{\pm}_k|$ and ${\rm sinc}(x)\equiv
\sin(x)/x$.

In III-V QDs, because of the large number of nuclear spins, the lost of electron
spin coherence is much faster than the build-up of higher-order nuclear spin correlations.
The decoherence is therefore well described by the pair-correlation approximation as given
in Eq.~(\ref{twomechanism}). In other systems such as Si or diamond NV centers with a dilute
nuclear spin bath, corrections from higher-order nuclear spin correlations
will become important~\cite{Yang2008PRB,Yang2009PRB,Witzel:2006,Saikin_Cluster}.

Lattice distortion can result in local electric field gradients, inducing
the quadrupole interaction for nuclear spins with moment greater than 1/2.
Recent experimental works have indeed demonstrated signatures of quadrupolar
interactions for nuclear spins in InAs self-assembled dot~\cite{Amand_quadrupole,Imamoglu_quadrupole}.
The quadrupolar interaction can be well incorporated in the theory described in this appendix
as contributions to energy cost for nuclear pair-flips [i.e. the $D_k$ term in Eq.~(\ref{pseudofield})]
when reliable parameter is extracted from experiments. We expect that quadrupolar interaction
does not affect electron spin free induction decay and Hahn echo decay where the $D_k$
term is unimportant, but may affect Carr-Purcell echoes and spin echoes by complex
pulse sequences when the $D_k$ term plays a non-negligible
role~\cite{Yao_Decoherence,Yao_DecoherenceControl,Liu_spinlong,Yang2008PRB}.

\section{Quantum measurement in Shor's algorithm \label{measurement_in_Shor}}

We show by explicitly examining Shor's algorithm the crucial role of
quantum non-demolition measurement for quantum computation to be scalable.

\subsection{Order finding for Shor's algorithm \label{Shor_algorithm}}

For the reader's convenience, we give a brief review of the
algorithm for finding the order of a number, the core subroutine
for Shor's algorithm. For a comprehensive description of the algorithm,
the reader is referred to Ref.~\cite{NielsenChuang}.
The order $r$ of a number $x$ with respect
to a number $N$ is defined by the relation $x^r =1 \ {\rm mod} \
N$. For $2^{L-1}\le N<2^L$ and $1<x<N$, the task is to find $r$
with resources at most polynomial to $L$.

The observation is that $f(n)\equiv x^n \ {\rm mod}\ N$ is
periodic with period $r$, i.e., $f(n+r)=f(n)$, indicating a
quantum Fourier transformation (QFT) may be used to find the spectrum of
this function and thus to find $r$. The algorithm is outlined as follow.
\begin{enumerate}
\item Two registers with $t$ and $L$ qubits, respectively, are
zeroed initially, and thus the initial state is,
$$|00\cdots 0\rangle|00\cdots 0\rangle.$$
\item The QFT is applied to the first register to obtain the state,
$$\frac{1}{\sqrt{2^t}}\sum_{n=0}^{2^t-1}|n\rangle |00\cdots 0\rangle,$$
where $n=b_1b_2\cdots b_t$ is a binary number with $b_i=0$ or 1.
\item With the advantage of quantum parallelism,
one evaluation of the function $f(n)$ is added to the second register to reach the state
$$\frac{1}{\sqrt{2^t}}\sum_{n=0}^{2^t-1}|n\rangle |x^n\ {\rm mod}\ N\rangle.$$
By $f(n)=f(n+r)$, the state can be rewritten as,
$$\frac{1}{\sqrt{2^t}}\sum_{k=0}^{r-1}\sum_{l=0}^{\left[(2^t-k)/r\right]}|k+lr\rangle |x^k\rangle,$$
where $[n/m]$ denotes the greatest integer not greater than $n/m$.
Such an expression suggests a solution of the spectrum by QFT.
\item After an inverse QFT applied to the first register, the state becomes,
$$
\frac{1}{{2^t}}\sum_{j=0}^{2^t-1}\sum_{k=0}^{r-1}\sum_{l=0}^{\left[(2^t-k)/r\right]}e^{-i2\pi\frac{(k+lr)j}{2^t}}|k+lr\rangle |x^k\rangle.
$$
\end{enumerate}%
If $2^t$ happens to be an integer multiple of $r$, the terminating state is just,
$$
\frac{1}{r}\sum_{s=0}^{r-1}\sum_{k=0}^{r-1}e^{-i2\pi k s/r}|2^t s/r\rangle |x^k\rangle,
$$
only for $j=2^t(s/r)$ will the amplitude be nonzero. Generally,
$r$ would not divide $2^t$, but if $2^t$ is much larger than $r$,
the spectrum after the QFT will be distribution
composed of peaks around $2^t(s/r)$ for $(0\le s<r)$. The larger is the first register, the
sharper will the peaks be. The probability of the state of the
first register being away from $2^t s/r$ by
a distance $2^p$ is calculated to be
less than $1/\left(2^{p+1}-4\right)$, so with
probability greater than $1-1/\left(2^{p+1}-4\right)$, the
fraction $s/r$ can be determined up to the first $t-p$ bits by a
measurement of the first $t-p$ bits of the first register. If
$t-p$ is chosen to be $2L+1\equiv N$, $r$ can be determined from the
first $N$ bits of the binary fraction number $s/r$, i.e., $[2^{N}s/r]$,
by continued fraction.

\subsection{Issues with the measurement}

\begin{figure}[b]
\begin{center}
\includegraphics[width=12cm]{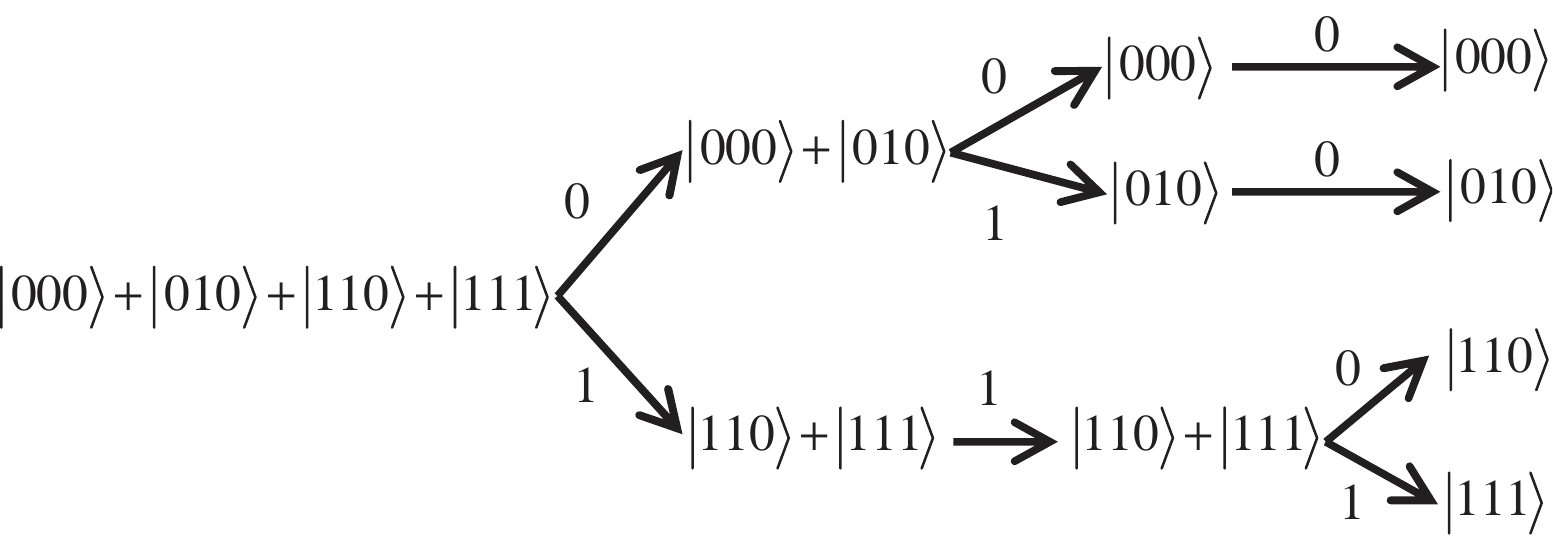}
\end{center}
\caption{The collapse (quantum jump) of a multi-qubit
state
under a measurement of the qubits
in sequence. The number associated with
each arrow indicates the output of the measurement.}
\label{cascade}
\end{figure}

The key feature of Shor's algorithm is that, though the terminating state is a superposition of
many computational basis states $\sum_x |x\rangle$ (where $|x\rangle=|b_1b_2\cdots b_t\rangle$
with $b_i=0$ or 1), it is not necessary to know all the amplitudes to solve the problem.
Actually, an ideal measurement on the computational basis will
project the superposition state into an arbitrary basis state which has
a nonzero amplitude, and with high probability,
the fraction $s/r$ can be determined up to $2L+1$ bits. If the measurement is performed
in a single shot, the register may be read out bit by bit, and the superposition state will
collapse in a cascade manner, so the resources required by the whole readout step is
$O(L)$ and thus the measurement is scalable. The cascading readout can be illustrated by the
example of reading the state $|000\rangle+|010\rangle+|110\rangle+|111\rangle$,
in which the state collapse may follow the steps shown in Fig.~\ref{cascade}.
Only $N$ single-bit measurements are required to have a $N$-bit superposition state collapsed into
a basis state and only $N$ bits of classical memory are needed to record the measurement result.
So a single-shot measurement on a single quantum register needs less resources than polynomial
to the problem size in Shor's algorithm.

\subsection{Ensemble measurement}

Now we will show, that, in Shor's algorithm, an ensemble measurement requires resources
increasing exponentially with the size of the problem.
The terminating state of Shor's algorithm can be written as,
$$\frac{1}{r}\sum_{k=0}^{r-1}\sum_{s=0}^{r-1} \left|\widetilde{s/r}\right\rangle
 \otimes \left|R(s,k)\right\rangle \otimes \left|x^k\right\rangle, $$
where $\widetilde{s/r}$ denotes the first $N$ binary bits of $s/r$, and $R(s,k)$ denotes
the state of the last $p$ bits of the first register which are not accurate in
describing $s/r$. Only the first $N$ bits of the first register need to be measured.
The detected probability is uniformly distributed among the $r$ states
$\left|\widetilde{s/r}\right\rangle$, which are spaced almost equally by the distance
$2^t/r$.

\subsubsection{Correlated measurement}

Suppose the first $N$ qubits of the first register are measured with coincidence counting.
Each basis state $|b_1b_2\cdots b_{N}\rangle$ could lead to a click in a corresponding channel.
To accumulate confidence in an ensemble measurement, a channel should get at least two clicks.
In each single-shot trial of the measurement ensemble, the state would collapse into different
states. Each output has to be recorded and stored before one of them is confirmed.
So the number of recording channels and the size of classical memories used for data
storage scale as $2^N$, exponentially increasing with the problem size in Shor's algorithm.
This requires exponentially increasing physical resources such as spectral resolution or
spectrometer bandwidth in spectroscopy.

If the number of available channels and classical memory registers are limited by $C$.
We may randomly or uniformly choose $C$ $x$'s from the $2^N$ possible numbers. What is the chance
that we can receive a signal at any one of these $C$ channels? As the probability of having
a signal in an arbitrary channel is $\sim 2^{-N}$, the probability of having signal in
any of the $C$ channels is $P_C=2^{-N}C$.
So, for any finite probability $P_C$, the
number of channels needs to be $C\sim 2^N P_C$, increasing exponentially with the problem size.

\subsubsection{Uncorrelated measurement}

In an uncorrelated measurement of an $N$-qubit register, the ensemble is
first divided into $4^g(N/g)$ portions ($g$ is a fixed small integer),
and every $4^g$ portions can be used for $4^g$ independent $g$-qubit
measurements to obtain the density matrix of these $g$ qubits. Once
the reduced density matrices for all the $g$-qubit subsystems have
been obtained. One could apply some classical algorithm to re-construct
the density matrix of the $N$ qubits. The re-construction, of course,
could not be certain, but the uncertainty nonetheless does not exclude
the possibility that one might search for a correct result from all possible states
which give the reduced matrix elements. The number of measurements, the number of
recording channels, and the size of storage, all these resources scale only linearly
with the problem size.

However, there is a fundamental problem underlying the uncorrelated measurement strategy:
to yield an $N$-qubit output, essentially $N$ bits of information have been generated.
To generate the same amount of information, the $g$-qubit
reduced density matrices have to be measured with $N-\ln(N/g)$ bit accuracy.
By the rule of thumb in experimental physics, measuring any physical quantity
with $N$-qubit precision would require resources scaling as $2^{N}$.
Only in exceptional cases, may the register state be derived from the knowledge of the
reduced density matrices. For instance, if all one-qubit density matrices
are pure states, the register state is obviously the outer product of all of them.
But in general, it is difficult
to determine the register state from the one-qubit density
matrices. For example, if all the $N$ qubits are maximally entangled with other qubits,
such as in the states,
$$|0000\rangle+|0011\rangle+|0110\rangle+|1001\rangle+|1100\rangle+|1111\rangle,$$
and,
$$|0000\rangle+|0101\rangle+|1010\rangle+|1111\rangle,$$
the uncorrelated measurement would turn out to be $N$ maximally mixed density matrices in both cases,
from which little information can be obtained about the register state. To determine the
order $r$, one has to search from all possibilities, of which the number is $\sim 2^N$.

Below we will show that,
for the first $N$ qubits in Shor's algorithm, there are $\sim 2^N$ possible
terminating states which would produce almost the same one-qubit reduced density matrix.
\begin{lemma}
For a state $\sum_{l=0}^{\left[(2^N-1)/k\right]}|lk\rangle$
($k$ is an odd number greater than 1), the reduced density matrix of any qubit has
no off-diagonal term in the computational basis.
\end{lemma}
{\bf Proof} \ \
If there is an off-diagonal term $\langle 0|\rho|1\rangle$ for the $j$th
qubit, there have to be at least two states $|x\rangle$ and $|x'\rangle$ in the superposition,
which are different only at the $j$th bit. So we have that $\left|x-x'\right|=2^{n-j}$ is divided
by $k$, which is impossible since $k$ is an odd number $\square$
\begin{corollary}
For the state $\sum_{l=0}^{\left[(2^N-1)/(2^pk)\right]}|l2^pk\rangle$
($k$ is an odd number greater than 1), the reduced density matrix of any qubit has
no off-diagonal term in the computational basis, and the reduced density matrices of the
last $p$ qubits are all $|0\rangle\langle0|$.
\end{corollary}
\begin{corollary}
 For the state $\sum_{l=0}^{\left[(2^N-1)/2^p\right]}|l 2^p\rangle$,
 the reduced density matrices of the first $n-p$ qubit  are all $\left(|0\rangle+|1\rangle\right)
{1\over 2}\left(\langle0|+\langle1|\right),$ and the reduced density matrices of the
last $p$ qubits are all $|0\rangle\langle0|$.
\end{corollary}
\begin{lemma}
For the state $\sum_{l=0}^{\left[(2^N-1)/k\right]}|lk\rangle$,
if the odd number $k<2^{\alpha N/2}$ for a specific number $\alpha \in(0,1)$,
the states $|x\rangle$ with $x_j=0$ or $1$ have the same probability up to $cN$-bit precision
to occur in the superposition, where $c$ is a constant less than 1.
\end{lemma}
{\bf Proof} \ \  The integer numbers $x=\{ x_1x_2\cdots x_N \}$ with $x_j=0$ or $1$
form alternatively $2^j$ segments with length $2^{N-j}$. For an arbitrarily chosen
number $\beta\in(\alpha,1)$, the segment length $2^{N-j}$ is either less or
greater than $2^{\beta N/2}$. If $2^{N-j}>2^{\beta N/2}$, as the number of the
multiples of $k$ in a segment is greater than $2^{(\beta-\alpha)N/2}$. The numbers
of multiple $k$ in two neighboring segments differ by at most 1. So the occurring
probability of $x_j=0$ is different from that of $x_j=1$ by at most
$1/2^{(\beta-\alpha)N/2}$. If $2^{N-j}\le 2^{\beta N/2}$, we observe that the first $x$
of every $k$ segments is $k 2^{N-j}$, a multiple of $k$. As $k$ is an odd number,
each $k$ segments starting with $x_j=0$ will be followed by $k$ segments starting with
$x_j$=1, and vice versa, until the end of all segments. So, in every $2k$ segments, the
numbers of $lk$'s with $x_j=0$ or $1$ are exactly the same. As the difference of the
occurring numbers in 2 neighboring segments is at most 1, the difference in $k$ segments
is at most $(k+1)/2$. So, the difference of the occurring probability in all the segments
is at most $(k+1)2^{N-j}/2^{N+1}<1/2^{N(1-\alpha/2-\beta/2)}.$
Let $c=\min (1-\alpha/2-\beta/2, \beta/2-\alpha/2)$, we have the probability of
occurring of $x_j=0$ is the same as that of occurring of $x_j=1$, accurate up to
$cN$ bits $\square$

From the Lemma 1 and 2, we have directly the following theorem.
\begin{theorem}
For all states $\sum_{l=0}^{\left[(2^N-1)/k\right]}|lk\rangle$ [$k$ is an odd
number, and $k<2^{\alpha n/2}$ for a specific number $\alpha \in(0,1)$],
all one-qubit reduced density matrices obtained by uncorrelated measurements are the
same up to $cN$ significant bits.
\end{theorem}
\begin{corollary}
For all states $\sum_{l=0}^{\left[(2^N-1)/(2^pk)\right]}|l 2^pk\rangle$
[$k$ is an odd number, and $k<2^{\alpha N/2}$ for a specific number $\alpha \in(0,1)$],
all one-qubit reduced density matrices obtained by uncorrelated measurements are the
same with $cN$ significant bits.
\end{corollary}
The theorem and corollary above are consistent with a theorem recently proved by Popescu
et al~\cite{Popescu06NatPhys}:
Almost all $N$-qubit states would give almost the same $g$-qubit reduced density matrix,
as long as $N$ is large and $N\gg g$.

The terminating states of the register to be measured in Shor's algorithm have
the form of the superposition states in the theorem above, with at most one-bit deviation.
So, unless at least the density of matrix of one qubit is measured with $O(N)$ effective bits,
there are $\sim 2^N$ possible superpositions corresponding to the same set of one-qubit reduced
density matrices. On the one hand, searching the correct one from all those possibilities
needs resources $\sim 2^N$ in all known classical or quantum algorithms. On the other hand,
determining the density matrix of a qubit with $O(N)$-bit accuracy also requires
resources $\sim 2^N$.
So uncorrelated single-qubit ensemble measurement is provably unscalable for Shor's algorithm.
Though there is no proof in general cases, it would be rather
surprising that some ensemble measurement scheme is scalable for Shor's algorithm.

\subsection{Single-object measurement with error}

In general, scalable quantum computation needs to be performed on a single quantum object
(rather than an ensemble) with single-shot measurement. In reality, detectors used in
the readout procedure have unavoidable inefficiency or errors. Thus, the Kraus operators~\cite{NielsenChuang} for
a POVM (positive operator-valued measure) of a certain qubit can be written as
\begin{subequations}
\begin{eqnarray}
A_0 &=& \sqrt{1-d}|0\rangle\langle 0|+\sqrt{1-e} |1\rangle\langle 1|, \\
A_1 &=& \sqrt{d}|0\rangle\langle 0|+ \sqrt{e}|1\rangle\langle 1|,
\end{eqnarray}
\end{subequations}
with detector efficiency $e$ and dark count rate $d$. For a state $|\psi\rangle$, the
probability and the resultant state for the output 0 and 1 are, respectively,
\begin{subequations}
\begin{eqnarray}
P_0=\langle \psi |A^{\dag}_0A_0|\psi\rangle & {\rm and}& A_0|\psi\rangle, \\
P_1=\langle \psi |A^{\dag}_1A_1|\psi\rangle & {\rm and}& A_1|\psi\rangle,
\end{eqnarray}
\end{subequations}
Suppose that the detecting error
rate at each qubit is greater than a finite number $\varepsilon$, the probability
of reading out a $N$-qubit register is less than $(1-\varepsilon)^N$, exponentially
small with increasing the size of the register. A single-shot measurement then is
insufficient for scalable quantum computation. Thus, if there exist detector errors as always,
the measurement has to be repeated for enough times
to obtain sufficient confidence in a readout result. Furthermore, if the measurement is
destructive, the quantum computation has to be rewound from the very beginning of the
algorithm, making the measurement equivalent to an ensemble measurement.

In an uncorrelated measurement, different qubits of the register are measured and recorded
independently, and the error rate at each bit is finite, so the density matrices of each
qubit of the register can be measured by repeating the quantum computation for
a number of times proportional to
the register size. But the problem here is the same as discussed for the
uncorrelated measurement in the previous section.

In a correlated measurement, the probability of correctly reading out
the
projected
state is exponentially small [$\sim (1-\varepsilon)^N$],
and yet the probability of the terminating state collapsing into the
same basis state is also exponentially small ($<1/r$). So before a
readout result is repeated once for accumulating sufficient confidence,
the quantum computation has to be repeated
a number of times which increases exponentially
with the problem size.

\subsection{Quantum non-demolition measurement}

In a QND measurement, the state will remain unchanged after the
projection into the measurement basis (which is also the computational basis).
So the readout can be repeated many times in every qubit to accumulate
confidence of the readout result, without rewinding the whole algorithm from the
beginning.

Now we calculate the resources required in reading out the state of
a $N$-qubit register. If the error rate in reading out each qubit by an
$M$-shot QND measurement is $\varepsilon_M$,
the probability of successfully reading out the register is
$(1-\varepsilon_M)^N$. To have a finite success probability, we require
$s_N\equiv(1-\varepsilon_M)^N>s$ where $s$ is a finite number smaller than 1. When
$\varepsilon_M$ is small, $s_N\approx e^{-N\varepsilon_M}$, so we require the
error rate of a $M$-shot QND measurement $\varepsilon_M<-(1/N)\ln s$.

To obtain the error rate of a $M$-shot measurement, we define its POVM's.
The Kraus operators for the POVM of a $M$-shot measurement giving $m$ photon counts
can be derived as,
\begin{eqnarray}  A_{M,m} & =& \left(\begin{array}{c} M \\ m \end{array}\right)^{1/2}(1-d)^{(M-m)/2}d^{m/2}|0\rangle \langle 0|
\nonumber \\ &  + &
\left(\begin{array}{c} M \\ m \end{array}\right)^{1/2}(1-e)^{(M-m)/2} e^{m/2}|1\rangle \langle 1|
\nonumber \\
&\equiv & \sqrt{d_{M,m}}|0\rangle\langle 0|+\sqrt{e_{M,m}} |1\rangle\langle 1|.
\end{eqnarray}
When $d_{M,m}<e_{M,m}$, it is more probable that the qubit is in the state $|1\rangle$, and vice versa.
As
\begin{equation}
d_{M,m}/e_{M,m}=\left(\frac{1-d}{1-e}\right)^M \left[\frac{d(1-e)}{e(1-d)}\right]^m.
\end{equation}
monotonically decreases with $m$, we can define a $m_0$ so that all $A_{M,m<m_0}$ are
indicators of $|0\rangle$ and all $A_{M,m>m_0}$ are
indicators of $|1\rangle$, with the $m_0$ given by $d_{M,m_0}/e_{M,m_0}=1$ or,
\begin{equation}
m_0=\frac{M \ln\frac{1-d}{1-e}}{\ln\frac{1-d}{1-e} +\ln\frac{e}{d}}\equiv\alpha M.
\label{alpha}
\end{equation}
As $d_{M,m}$ and $e_{M,m}$ as functions of $m$ have peaks at $dM$ and $eM$, respectively,
we have,
\begin{equation}
d<\alpha <e.
\end{equation}
Now the POVM can be calculated from,
\begin{eqnarray}
&& P^{(M)}_0 \equiv \sum_{m<m_0} A_{M,m}^{\dag}A_{M,m} \equiv (1-d_M)|0\rangle \langle 0| + (1-e_M)|1\rangle \langle 1|,
\nonumber \\
&& P^{(M)}_1 \equiv \sum_{m\ge m_0} A_{M,m}^{\dag}A_{M,m} \equiv d_M|0\rangle \langle 0| + e_M|1\rangle \langle 1|, \nonumber
\end{eqnarray}
where,
\begin{eqnarray}
d_M & = & \sum_{m\ge m_0} d_{M,m} = \sum_{m\ge m_0}\left(\begin{array}{c} M \\ m \end{array}\right)(1-d)^{M-m}d^{m}  \nonumber \\
&= & \frac{M!}{(M-m_0)!(m_0-1)!}\int^1_{1-d} t^{M-m_0}(1-t)^{m_0-1} dt \nonumber \\
& < & \frac{M!(1-d)^{M-m_0}d^{m_0}}{(M-m_0)!(m_0-1)!} \nonumber \\
& \sim & \sqrt{\frac{\alpha M}{2\pi(1-\alpha)}}
\left[\frac{(1-d)^{1-\alpha}d^{\alpha}}
 {(1-\alpha)^{1-\alpha}\alpha^{\alpha}}\right]^M,
\end{eqnarray}
and similarly,
\begin{eqnarray}
1-e_M < \sqrt{\frac{\alpha M (1-e)^2}{2\pi(1-\alpha)e^2}}
\left[\frac{(1-e)^{1-\alpha}e^{\alpha}}
 {(1-\alpha)^{1-\alpha}\alpha^{\alpha}}\right]^M.
\end{eqnarray}

The error rate is defined as $\varepsilon_M=\max({d_M,1-e_M})$, so the number of required
QND measurements per qubit $M\sim \ln N$, and the total
number of measurement is proportional to $N\ln N$. So a
QND is scalable. In experiment, $e$ and $d$ cannot be determined exactly, but fortunately
$\alpha$ need not be determined exactly  and whenever it is between $d$ and $e$,
the results for $d_M$ and $e_M$ are unchanged in the equations above.

Note that with the aid of entanglement gates and a supply of fresh qubits,
a destructive measurement can be converted into a QND one. The idea is based on the
transformation of a qubit and $M$ zeroed auxiliary qubits into an entangled state by $M$
entanglement gates:
$$\left(\alpha|0\rangle+\beta|1\rangle\right)\bigotimes_{m=0}^{M-1}|0\rangle \Longrightarrow
 \alpha|0\rangle\bigotimes_{m=0}^{M-1}|0\rangle+\beta|1\rangle\bigotimes_{m=0}^{M-1}|1\rangle.
 $$
The $M$ auxiliary qubits are to be read out. As all these qubits are entangled, once one
qubit is collapsed into a basis state $|0\rangle$ or $|1\rangle$, all the qubits will be
collapsed into the same state. So even a destructive measurement with detecting error
can be used to read out the qubit.

\begin{figure} [b]
\begin{center}
\includegraphics[width=10cm]{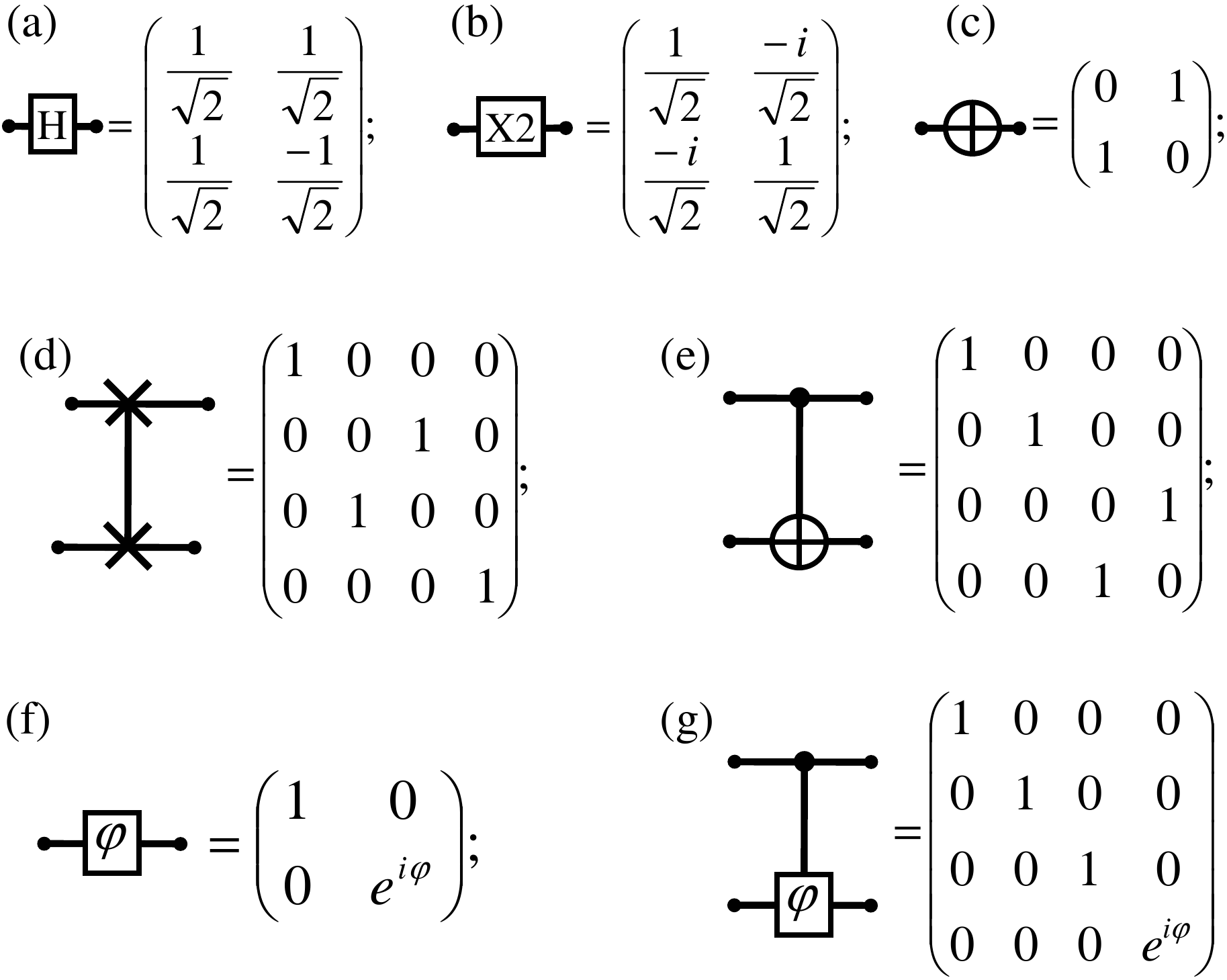}
\end{center}
\caption{Some elementary quantum gates and their
matrix representation, including (a) Hadamard
($H$), (b) rotation of $90$ degrees about the $x$-axis ($X2$),
(c) NOT ($N$), (d) SWAP ($W$), (e) Controlled NOT ($C$), (f) single-bit phase-shift
($S_{\phi}$), and (g) Controlled phase-shift ($P_{\phi}$).} \label{gates}
\end{figure}

How about a QND measurement with back-action noise, i.e.,
reaction to measurement that disturbs the state after a measurement cycle? With the idea above for converting
destructive measurement into a QND one, we can employ the concept of
{\em fault-tolerant measurement} to deal with this problem.  In the so-called
fault-tolerant measurement,  a qubit is first encoded into a stabilizer
code, after a single measurement, any
back-action noise will be diagnosed and corrected
using the error syndrome since this noise acts only
on a single qubit (by assumption). After the
error correction, another measurement would be performed, and so on.
The fault-tolerant measurement thus allows an imperfect QND measurement to
read out the result of a quantum algorithm with polynomial resources.

\section{Elementary quantum gates}
\label{elementgates}

\begin{figure}[t]
\begin{center}
\includegraphics[width=8 cm]{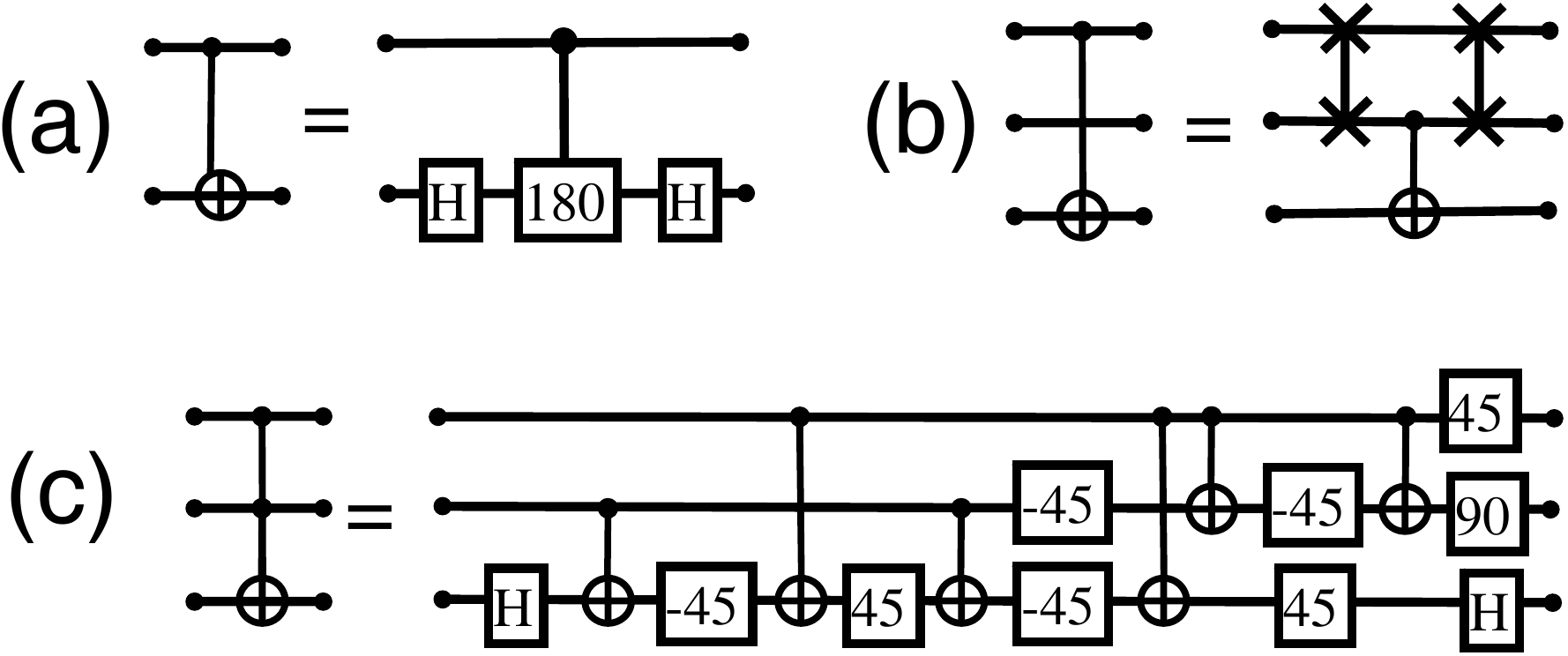}
\end{center}
\caption{Realization of several control gates, including (a) CNOT,
(b) remote CNOT, and (c) Toffoli ($C_2$).} \label{gates2}
\end{figure}

Here we define a series of elementary quantum gates used in the
quantum circuits presented in Sec.~\ref{factorize}. The quantum gates of interest are
defined in Figs.~\ref{gates} and \ref{gates2}.
The Hadamard gate can be
realized by the spin rotation operations up to a trivial global
phase, as the transformation operator for the Hadamard gate can
be expressed in terms of spin rotations as $H=e^{i\pi/2} e^{i\pi
s_z} e^{i\pi s_y/2}.$ The CNOT gate can be realized by the
controlled $\pi$-phase gate together with two Hadamard gates. The
CNOT gate on two remote bits can be realized by a local CNOT gate
plus some SWAP gates. The doubly controlled NOT (Toffoli) gate
can be realized by six CNOT gates plus some single-qubit
gates. The controlled $\pi$-phase gate and  the
CNOT gate can be realized by two $\sqrt{\text{SWAP}}$ gates plus
some single-spin rotations. As any controlled
phase gate can be realized by two CNOT gates plus some
single-qubit gates, the circuits in Figs.~\ref{circuit1}
and \ref{circuit2} can be alternatively
realized with $\sqrt{\text{SWAP}}$ gates as well.

\end{appendix}


\bibliography{ReferencesRev1}

\end{document}